\documentclass[12pt]{article}
 \usepackage{xcolor}
\usepackage{amssymb}
\usepackage{amsmath}
\usepackage{amstext}
\usepackage{graphicx,epsfig}
\usepackage{epsfig}
\usepackage{verbatim} 
\usepackage{slashed}
\usepackage{fancyhdr}
\usepackage{fancybox}
\usepackage{bbold}
\usepackage{enumitem}
\usepackage{subfigure}
\usepackage{bbm}
\usepackage{parskip}
\usepackage{cite}
\usepackage{slashed}

\newcommand{\Comment}[1]{{}}
\definecolor{MyDarkBlue}{rgb}{0.15,0.15,0.45}
\usepackage[linktocpage=true]{hyperref}
\hypersetup{
colorlinks=true,
citecolor=MyDarkBlue,
linkcolor=MyDarkBlue,
urlcolor=MyDarkBlue,
}

\usepackage{genyoungtabtikz}
\Yboxdim{13pt} 
\Ylinethick{.8pt}

\newcommand{\be}{\begin{equation}}
\newcommand{\ee}{\end{equation}}
\newcommand{\bea}{\begin{eqnarray}}
\newcommand{\eea}{\end{eqnarray}}
\newcommand{\beas}{\begin{eqnarray*}}
\newcommand{\eeas}{\end{eqnarray*}}
\newcommand{\nn}{\nonumber}
\newcommand{\la}{ {\langle}} 
\newcommand{\ra}{ {\rangle}}

\newcommand{\rd}{{\rm d}}
\newcommand{\half}{{\frac{1}{2}}}

\newcommand*\dualarrow{\ \raisebox{-1.5pt}{\epsfig{file=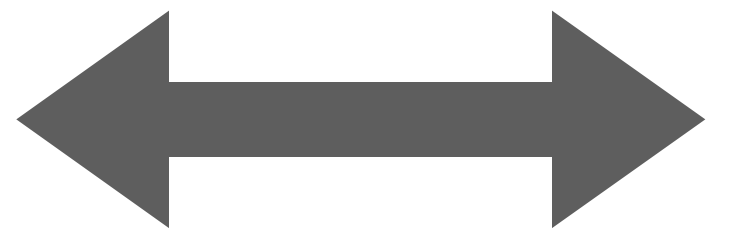,height=0.12in,width=0.25in}}\ }

\newcommand*\dualarrowcurved{\ \raisebox{-0.0pt}{\epsfig{file=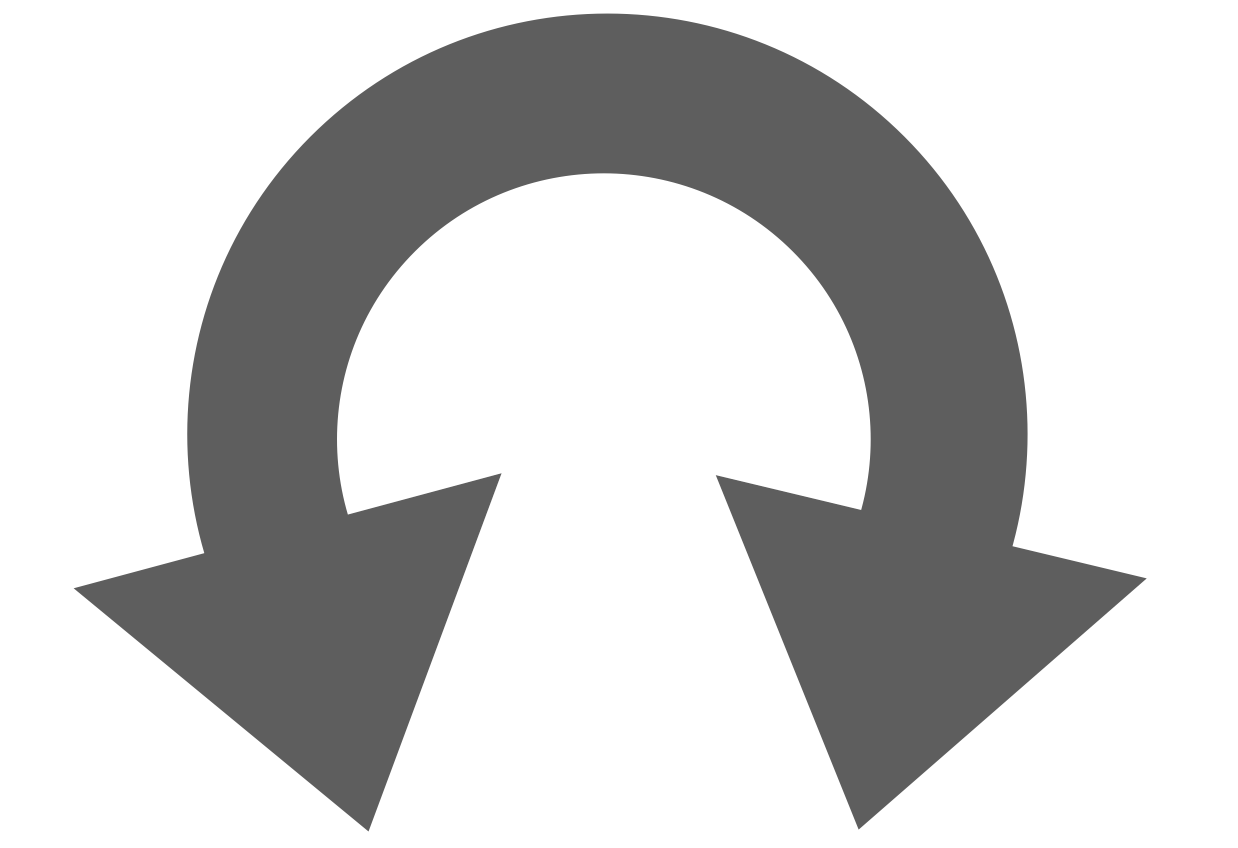,height=0.18in,width=0.26in}}\ }

\newlength{\fdagwidth}
\newlength{\diagupwidth}
\newlength{\stepback}
\newcommand{\fdag}[2][\diagup]{\text{$#2$\settowidth{\fdagwidth}{$#2$}\settowidth{\diagupwidth}{$#1$}\setlength{\stepback}{0.5\fdagwidth}\hspace{-\stepback}\hspace{-0.5\diagupwidth}$#1$\hspace{\stepback}\hspace{-0.5\diagupwidth}}}

\numberwithin{equation}{section}

\begin{document}


\begin{center}
{\Large \bf{De Sitter Representations}}
\end{center} 
 \vspace{1truecm}
\thispagestyle{empty} \centerline{
{\large {Kurt Hinterbichler}}\footnote{E-mail: \Comment{\href{mailto:kurt.hinterbichler@case.edu}}{\tt kurt.hinterbichler@case.edu}} 
}

\vspace{1cm}

\centerline{{\it CERCA, Department of Physics,}}
\centerline{{\it Case Western Reserve University, 10900 Euclid Ave, Cleveland, OH 44106}}

\begin{abstract} 

We review the representations of $\frak{so}(1,D)$, the algebra of isometries of $D$ dimensional de Sitter space.  We cover the representations in all $D$, including mixed symmetry representations and fermionic representations, and connect them to the various types of fields that can propagate on de Sitter space.   The presentation is from a physics point of view, favoring concrete constructions over abstract considerations.  

\end{abstract}

\newpage

\thispagestyle{empty}
\tableofcontents
\newpage
\setcounter{footnote}{0}

\parskip=5pt
\normalsize

\section{Introduction}

Our universe is entering a stage of accelerated expansion that appears to be driven by a positive cosmological constant \cite{SupernovaSearchTeam:1998fmf}.  If it is indeed a cosmological constant, and if there are no long distance surprises in the way gravity operates, then the far future of our universe will be described by de Sitter (dS) space.  At the other end of the timeline, the leading paradigm for the very early universe is inflation \cite{Guth:1980zm}, and if this paradigm is correct, then the geometry of the very early universe was also very nearly that of a dS space.  This means that the whole history of the universe as we know it may be a transitory interpolation between two dS phases.  Should this be the case, an understanding of our universe necessarily requires an understanding of the physics of dS space.

As was realized long ago by Wigner \cite{Wigner:1939cj}, physics on a spacetime can be organized according to irreducible representations of the isometry group of that spacetime.  On $D$ dimensional flat spacetime, this leads, along with positivity of the energy, to the familiar classification of particles: particles are either massive or massless, in the massive case they have spin (which is a representation of the massive little algebra $\frak{so}(D-1)$), and in the massless case they either have helicity (which is a representation of the compact part of the massless little algebra, $\frak{so}(D-2)$), or they live in more exotic infinite-spin representations \cite{Wigner:1939cj,Bargmann:1948ck} (these are often called continuous spins; they have not been seen in nature but they are being actively explored as a possibility, see \cite{Bekaert:2017khg} for a review).  See \cite{Brink:2002zx,Bekaert:2006py} for reviews of the representation theory of the Poincar\'e group in all dimensions.

$D$ dimensional de Sitter space, dS$_D$, is the maximally symmetric $D$ dimensional spacetime of positive curvature.  The isometry algebra of dS$_D$ is $\frak{so}(1,D)$, which has dimension $D(D+1)/2$, the same dimension as that of the Poincar\'e group of $D$ dimensional flat space.  De Sitter space is therefore just as symmetric as flat space, and so representation theory is just as powerful in constraining physics on dS space as it is on flat space.   

This review covers the representations of $\frak{so}(1,D)$.  As we will see, this representation theory is rich, and dS space allows for exotic types of particles, such as partially massless (PM) and shift symmetric particles, that have no flat space counterparts.  This review can be considered as a companion to the review \cite{kurtrachelreviewtoappear}, where the PM and shift symmetric fields are reviewed from the field theoretic point of view.   Here, we describe the dS representations in their own right, and describe how they match up to these various fields.

The study of physics on dS has been of recent interest primarily in the context of the cosmological correlators program, which seeks to systematize the computation of inflationary observables and better understand their structure.  This program subdivides more-or-less into works that attempt to understand and improve upon perturbative computations using dS Feynman diagrams, and ``bootstrap" approaches that seek to non-perturbatively extract results by maximally exploiting the constraints imposed by fundamental requirements such as unitarity, locality, and dS symmetry.
The study of dS correlators  gained in popularity following \cite{Arkani-Hamed:2015bza}; a selection of some of the works is \cite{Meerburg:2016zdz,Arkani-Hamed:2017fdk,Kehagias:2017cym,Chen:2017ryl,Arkani-Hamed:2018kmz,Goon:2018fyu,Benincasa:2018ssx,Baumann:2019oyu,Benincasa:2019vqr,Sleight:2019hfp,Sleight:2019mgd,Wang:2019gbi,Sleight:2020obc,Baumann:2020dch,Goodhew:2020hob,DiPietro:2021sjt,Melville:2021lst,Hogervorst:2021uvp,Jazayeri:2021fvk,Baumann:2021fxj,Goodhew:2021oqg,Sleight:2021plv,Bonifacio:2022vwa,Heckelbacher:2022hbq,AguiSalcedo:2023nds,Stefanyszyn:2023qov,Arkani-Hamed:2023kig,Loparco:2023rug,Penedones:2023uqc,Albayrak:2023hie,Green:2023ids,Benincasa:2024leu,Werth:2024aui}, and some overviews are \cite{Baumann:2022jpr,Benincasa:2022gtd}.

Besides cosmological correlators, the formal study of other types of physics and observables on dS space has benefited from an understanding of dS representation theory.
For example, Euclidean partition functions of fields propagating on dS can be written in terms of characters of the dS representations \cite{Anninos:2020hfj,Law:2020cpj,Law:2021hwc,Sun:2021rrs,Law:2023ohq,Ball:2024xhf,Law:2025ktz}.
Some earlier works that employed a group theoretic approach towards de Sitter field theory are \cite{Bros:1994dn,Bros:1995js,Bros:1998ik,Joung:2006gj,Joung:2007je}, and some more recent work is \cite{Sun:2020sgn,Penedones:2023uqc,Loparco:2023rug,Loparco:2025azm}.

The representation theory of the dS algebra is of interest for reasons other than physics on dS.  For example, particles on flat space, which are described using unitary irreducible representations of the Poincar\'e group, can be described in a basis adapted to the Lorentz subgroup of the Poincar\'e group, which is a de Sitter group (see \cite{Iacobacci:2024laa} and references therein), and this description is natural for celestial amplitudes (see \cite{Strominger:2017zoo,Raclariu:2021zjz,Pasterski:2021raf,McLoughlin:2022ljp} for reviews).   The de Sitter algebra is also the conformal algebra for Euclidean maximally symmetric spaces, so it becomes relevant for the study of Euclidean conformal field theory (CFT).  The modern application of this representation theory to CFT follows \cite{Mack:2009mi} and has led to many lines of investigation such as \cite{Costa:2012cb,Penedones:2015aga,Caron-Huot:2017vep}.  The dS algebra is also the isometry algebra of Euclidean anti-de Sitter space (AdS), and a connection can be made between observables in dS and Euclidean AdS \cite{Sleight:2020obc,DiPietro:2021sjt,Sleight:2021plv}.  In $D=2$, the algebra of isometries of dS$_2$ is the same as that of Lorentzian AdS$_2$ and governs the low energy behavior of the Sachdev–Ye–Kitaev (SYK) model \cite{Sachdev:1992fk}, a proposed dual of quantum gravity in AdS$_2$ \cite{Maldacena:2016hyu} (see \cite{Jha:2025rrz} for a recent review).   The algebra is also useful in studying compact hyperbolic manifolds that are quotients of Euclidean AdS \cite{Bonifacio:2023ban}.

The history of the study of dS representation theory dates to the early days of quantum field theory.  The study of unitary representations of the Lorentz algebra $\frak{so}(1,3)$, which is the $D=3$ dS algebra, goes back to Dirac \cite{Dirac:1945cm}, followed by Harish-Chandra \cite{9fea073f-7ff0-3fb5-984e-cb6ac72836f7}, Bargmann \cite{Bargmann:1946me} (which also covers the $D=2$ case) and Gel’fand and Naimark \cite{zbMATH03056355}.  The generalization to $D=4$ was studied by Thomas \cite{26fe42d6-bd6e-37c0-b45d-8bb27c0f126e}, Newton \cite{0a992498-b0aa-38af-b240-75ee70eb334a} and Dixmier \cite{Dixmier1961}.  The generalization to arbitrary dimensions was studied by Hirai \cite{10.3792/pja/1195523460,10.3792/pja/1195523378} and Takahashi \cite{BSMF_1963__91__289_0}.  For other early work, see also \cite{cmp/1103840604,10.1063/1.1666496,f4e767c0-b144-3046-b19f-3576a516cc62,61dbdb7d-9ac1-36a8-83ad-1eb6b8331554,Gavrilik1976,Higuchi:1986wu,10.1063/1.1665471}.  The theory of dS representations is now encompassed by the more general theory of representations and harmonic analysis for non-compact simple Lie algebras, due primarily to Harish-Chandra.   Books on the subject include \cite{gelfand2018representations,Dobrev:1977qv,Vilenkin1993}.

There have been several works in recent years that contain excellent reviews of dS representation theory, see e.g. \cite{Boers:2013pba,Basile:2016aen,Sun:2021thf,Sengor:2022lyv,Sengor:2022kji,Enayati:2022hed,Letsios:2023voc,RiosFukelman:2023mgq,Schaub:2024rnl}.  Our review aims to be complementary to these in several respects.   Our primary goal is completeness in covering all the possible representations: we have aimed to give a detailed account of all the representations in all dimensions, including both bosonic and fermionic representations,  mixed symmetry representations that occur in higher dimensions,  special features and equivalences that occur in lower dimensions, and single-field non-unitary representations that are usually discarded on physical grounds.   Another goal is presentation: we aim at an audience of physicists rather than mathematicians, and strive to make the presentation both pedagogical and useful as a self-contained reference.  We try to achieve this by favoring concrete and intuitive constructions over abstract considerations and proofs.  There are several downsides to this approach.  We will necessarily make many statements without proof, and the reader will have to consult the other reviews or the original specialist mathematical literature in order to find proofs for many of the statements.  We also limit our presentation to essentially listing the representations and understanding their structure in the simplest way, leaving out many other very interesting aspects of the representation theory that are better covered in other places, topics such as the decomposition of product representations, branching rules, the theory of Harish-Chandra characters \cite{HarishChandra1956}, and the more general theory of harmonic analysis.

\subsection{Conventions\label{conventionssec}}   

We will be working on dS space of spacetime dimension $D$, denoted as dS$_D$.  The dS Hubble scale is $H$, the dS radius is $1/H$.  We use the mostly plus metric and the curvature conventions of Carroll's general relativity textbook \cite{Carroll:2004st}.  The spatial slices of dS will have dimension $d\equiv D-1$.

Tensors are symmetrized and anti-symmetrized with unit weight, {e.g.} $t_{[\mu\nu]}=\half \left(t_{\mu\nu}-t_{\nu\mu}\right)$, $t_{(\mu\nu)}=\half \left(t_{\mu\nu}+t_{\nu\mu}\right)$, and we use $(\cdots)_T$ to represent the symmetric and fully traceless part of the enclosed indices. 

\textbf{Young tableaux:}  We make extensive use of Young tableaux to indicate the index symmetries of a tensor.   A Young tableau is made from stacking boxes in left-justified rows, where the number of boxes in a row is less than or equal to the number of boxes in the row above it.  Each box represents an index of the tensor and the shape of the diagram gives the index symmetries of the tensor: indices in a given column are fully anti-symmetric under interchange, and fully anti-symmetrizing all the indices in a given column along with any single index from any other column to the right of the given column yields zero. (This is the anti-symmetric convention for Young symmetries; there is also an alternative symmetric convention which we do not use.)

We denote a Young tableau either by drawing the boxes, or by giving a list of the numbers of boxes in each row.  A tableau with $p$ rows is thus denoted $[s_1,s_2,\ldots,s_p]$, where $s_i$ gives the number of boxes in the $i$-th row, $s_1\geq s_2\geq\ldots\geq s_p>0$.  We will sometimes use the shorthand of giving an exponent to denote multiple rows with the same length, e.g. $[5,3^2,1^3]\equiv [5,3,3,1,1,1]$.
For us, tableaux will always represent tensors which are completely traceless in all indices.  For each tableau, there is a Young projector, denoted by ${\cal Y}^T_{[s_1,s_2,\ldots,s_p]}$, which projects a generic tensor onto a tensor with the symmetries of the indicated traceless tableau.   The action of the projector is fully determined by first symmetrizing the indices of the tensor within each row, then anti-symmetrizing the indices within each column, and then subtracting all traces, with an overall normalization determinined by demanding the projector square to one.  For more details on Young tableaux, see \cite{Fulling:1992vm,Bekaert:2006py} and the books \cite{Tung:1985na,fulton1997young,hamermesh1989group}.   

In the case of fermionic representations we will be working with spinor-tensors, which have a single Dirac spinor index in addition to some number of tensor indices.  In odd $D$, where there is a unique Dirac spinor type, we denote the spinor-tensor by $[s_1,s_2,\ldots,s_p]_{1\over 2}$, where the tableau $[s_1,s_2,\ldots,s_p]$ gives the symmetries among the tensor indices, and where the tensor is always taken to be fully traceless and gamma-traceless.  In even $D$, there are two inequivalent spinor types, distinguished by having eigenvalue $\pm 1$ under the chiral gamma matrix, and we denote the spinor-tensor of each by $[s_1,s_2,\ldots,s_p]_{\pm {1\over 2}}$.

\section{De Sitter Space and its Algebra of Isometries\label{dSsection}}

We start by reviewing the geometry of dS space, the global coordinate system we will use to describe the representations (reps), the isometries, and their algebra.
Some other reviews covering the geometry of dS space are \cite{Spradlin:2001pw,Kim:2002uz,Moschella:2006pkh,Anninos:2012qw,Pascu:2012yu}.

\subsection{De Sitter space}

$D$ dimensional de Sitter space, dS$_D$, $D\geq 2$, is the maximally symmetric\footnote{There is a slight caveat in $D=2$: the construction given here gives a spacetime with the topology of a spatial circle cross time, which we will take to be dS$_2$, with the isometry group $SO(1,2)$.  But we can unwrap the spatial circle to give the universal cover of  dS$_2$, whose symmetry group is larger (though the symmetry algebra is the same).}  Lorentzian spacetime of spacetime dimension $D$.  It can be described as a surface embedded into a $D+1$ dimensional Minkowski space of signature $(1,D)$.  Let $X^{A}$, $A\in 0,1,2,\ldots,D$ be Cartesian coordinates of this embedding space, in which its metric is
\be \eta_{AB}=\left(\begin{array}{cc}-1 & 0 \\0 &  \delta_{IJ} \end{array}\right)\, ,\label{ambientemetriceede}\ee
with $I,J\in 1,\ldots,D$ and $\delta_{IJ}$ the Kronecker delta.  dS$_D$ of radius $1/H$ is the surface defined by the equation
\be \eta_{AB}X^AX^B={1\over H^2}\, .\label{dshyperbdefee}\ee
It is a connected manifold with the topology of a cylinder, ${\mathbb S}^d\times {\mathbb R}$, where $\mathbb{S}^d$ is the $d$-sphere, $d\equiv D-1$.

Let $x^\mu$ be intrinsic coordinates on the dS$_D$ surface, so that the embedding is $X^A(x)$.  The induced metric is
 \be g_{\mu\nu}={\partial X^A\over \partial x^\mu}{\partial X^B\over \partial x^\nu}\eta_{AB}\,,\label{inducedmetricfforme}\ee
 and is the dS$_D$ metric in the coordinates $x^\mu$.
 As a maximally symmetric space, the Riemann tensor, Ricci tensor and Ricci scalar of dS$_D$ are given by 
\be R_{\mu\nu\rho\sigma}=H^2\left(g_{\mu\rho}g_{\nu\sigma}-g_{\mu\sigma}g_{\nu\rho}\right),\ \ \ R_{\mu\nu}=(D-1)H^2g_{\mu\nu}\, ,\ \ \ R={D(D-1)H^2}\,, \ee
and dS$_D$ is a solution to the Einstein equations $R_{\mu\nu}-{1\over 2}Rg_{\mu\nu}+\Lambda g_{\mu\nu}=0$ with a cosmological constant $\Lambda$ given by
\be  \Lambda={(D-1)(D-2)\over 2}H^2\,.\ee

\subsection{Global coordinates\label{globalcoordinatessec}}

We will use global coordinates on dS$_D$, defined through the embedding functions
\bea 
&&X^{0}={1\over H} \sinh\left( H  t\right)\,,  \nn\\ 
&&X^{I}={1\over H}\cosh\left( H t \right) \, \hat X^I(\theta) \, , \ \ I\in 1,\ldots, D \,.\label{globalcoordsinemee}
\eea
Here $t \in (-\infty,\infty)$ is a time coordinate, and there is a set of $d=D-1$ intrinsic angular coordinates $\theta^i$, $i\in 1,\ldots,d$, on the unit $d$-sphere ${\mathbb S}^d$.  $\hat X^I(\theta)$ describes an embedding of this unit $d$-sphere into $D$ dimensional Euclidean space, 
\be \delta_{IJ}\hat X^I\hat X^J=1\,.\ee

The dS$_D$  metric in these coordinates is obtained by using \eqref{globalcoordsinemee} in \eqref{inducedmetricfforme}, 
\be ds^2=-dt^2+{1\over H^2}\cosh^2\left(H  t \right) \,  d\Omega_{d}^2\, ,\ee
where $d\Omega_{d}^2=g_{ij}d\theta^id\theta^j$, 
\be g_{ij}={\partial \hat X^I\over \partial \theta^i}{\partial \hat X^J \over \partial \theta^j}\delta_{IJ}\,,\label{spheremetrice}\ee
is the metric on ${\mathbb S}^d$ obtained by pulling back the flat metric $\delta_{IJ}$ of the sphere's embedding space to the unit sphere.  

\begin{figure}
\begin{center}
\epsfig{file=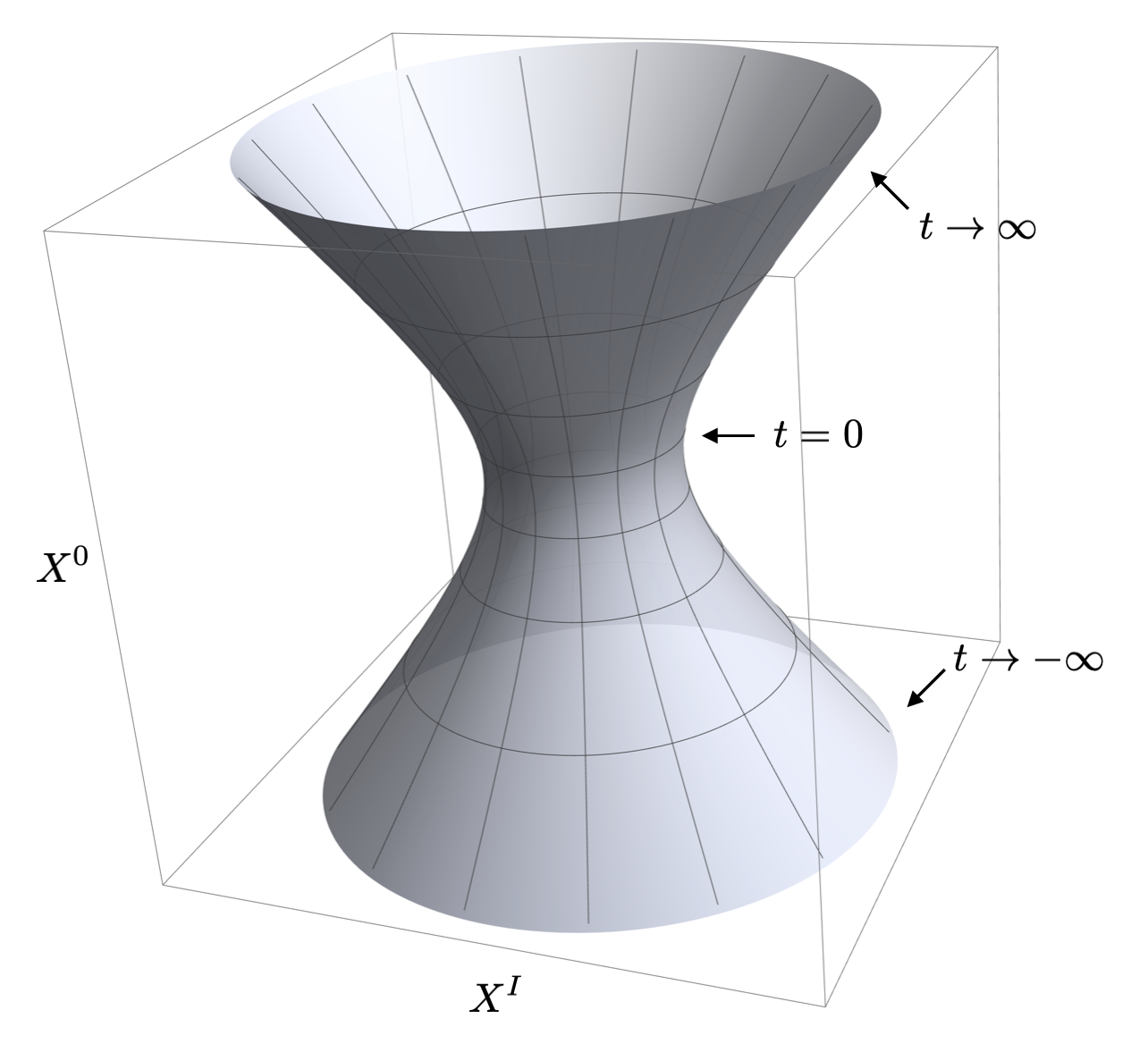,width=4in}
\caption{\small Global coordinates on de Sitter space, as seen in embedding space.}
\label{dsglobalcoords}
\end{center}
\end{figure}

As the name suggests, these global coordinates cover all of dS$_D$.  They describe a foliation of dS$_D$ into space-like $d$-spheres, given by the surfaces of constant $t$.  The spheres reach a minimum radius of $1/H$ at $t=0$ and grow exponentially to infinite size as $t\rightarrow \infty$ and as $t\rightarrow-\infty$, see figure \ref{dsglobalcoords}.

\subsection{Isometries and their $\frak{so}(1,D)$ algebra\label{isomalgsec}}

From the definition of dS$_D$ as the surface \eqref{dshyperbdefee} in an ambient $D+1$ dimensional Minkowski space, it is manifest that the isometry group is the Lie group $O(1,D)$ of Lorentz transformations in this ambient space, i.e. linear transformations of the $X^A$ that preserve the ambient metric \eqref{ambientemetriceede}.  This is the subgroup of isometries of the embedding space that preserve the dS$_D$ surface.  The group $O(1,D)$ has four connected components.  Starting from the component connected to the identity, the other components are reached by making, respectively, a spatial reflection, a time reversal, or both.  In global coordinates, a spatial reflection is made by applying an identical reflection to each spatial ${\mathbb S}^d$ (a reflection of ${\mathbb S}^d$ is a reflection through any plane in the sphere's embedding space that passes through the origin), and a time reversal is made by taking $t\rightarrow -t$.

We will be interested in reps of the Lie algebra rather than the group.  These are reps that exponentiate to reps of the covering group of the connected part of $O(1,D)$.  This will allow us to include the fermionic reps (and more exotic anyon-like reps in $D=2$), but it will preclude a detailed account of the possible actions of the discrete operations of spatial reflection and time reversal.

The Lie algebra of $O(1,D)$ will be denoted $\frak{so}(1,D)$.  It is realized by the algebra of Killing vectors of dS$_D$ with the Lie bracket as the commutator.
The Killing vectors of the embedding space which correspond to the Lorentz transformations are
\be {\cal M}^{AB}\equiv X^B\partial^A-X^A\partial^B.\label{killingenbeddinge}\ee
Under the Lie bracket they satisfy the $\frak{so}(1,D)$ algebra,
\be  \left\{ {\cal M}^{AB},{\cal M}^{CD}\right\} = \eta^{AC}{\cal M}^{BD}-\eta^{BC}{\cal M}^{AD}+\eta^{BD}{\cal M}^{AC}-\eta^{AD}{\cal M}^{BC}\,.\label{sod12algberaadse}\ee

It will be useful to split these into the following two sets,
\be  {\cal K}^I\equiv {\cal M}^{I,0}\,,\ \ \ {\cal M}^{IJ}\, , \ \ \ {I,J}\in 1,2,\ldots D\,.\label{mhifdhdgefee}\ee
The commutators \eqref{sod12algberaadse} upon making this split become
\bea
&&\left\{ {\cal K}^I ,{\cal K}^J\right\} = -{\cal M}^{IJ} \, , \nn\\
&&\left\{ {\cal M}^{IJ},{\cal K}^K\right\} =\delta^{IK}{\cal K}^J-\delta^{JK}{\cal K}^I  \, , \nn\\
&&  \left\{ {\cal M}^{IJ},{\cal M}^{KL}\right\} = \delta^{IK}{\cal M}^{JL}-\delta^{JK}{\cal M}^{IL}+\delta^{JL}{\cal M}^{IK}-\delta^{IL}{\cal M}^{JK}\, . \label{sod12algberaadse2e}\
\eea
This writing of the algebra manifests the fact that the ${\cal M}^{IJ}$ span a $\frak{so}(D)$ subalgebra of $\frak{so}(1,D)$, and that the ${\cal K}^I$ transform as a vector under this $\frak{so}(D)$ subalgebra.  The fact that a vector is an irreducible representation of $\frak{so}(D)$ implies that  $\frak{so}(D)$ is a maximal subalgebra of $\frak{so}(1,D)$.

Given a vector field ${\cal V}^A\partial_A$ in the ambient space, we can, after using the intrinsic and ambient metrics to appropriately move indices, pull it back to the dS$_D$ surface using the embedding $X^A(x)$ to obtain the dS$_D$ vector field $V^\mu\partial_\mu$ with $V^\mu=g^{\mu\nu} {\partial X^A\over \partial x^\nu} {\cal V}^B \eta_{AB}$. 
Using this to project the Killing vectors to the dS$_D$ surface using the global coordinates \eqref{globalcoordsinemee} gives the intrinsic Killing vectors in global coordinates,\footnote{In terms of intrinsic angular coordinates $\theta^i$ on the sphere $\mathbb{S}^d$, we should interpret ${\partial\over \partial \hat X_I}\rightarrow g^{ij} {\partial \hat X^I\over \partial\theta^j}\partial_i$, where $g_{ij}(\theta)$ is the metric on the sphere.  This means e.g. ${\partial\over \partial \hat X_I}\hat X^J=\delta^{IJ}-\hat X^I\hat X^J$.}
\bea
&& {\cal K}^I=  {1\over H}\hat X^I{\partial\over\partial t}+\tanh\left(H t \right) {\partial\over \partial \hat X_I}\, ,   \label{killingvectorsgede1} \\
&& {\cal M}^{IJ}=\hat X^J{\partial\over \partial \hat X_I}-\hat X^I{\partial\over \partial \hat X_J}\, . \label{killingvectorsgede2}
\eea
From here we see that the $\frak{so}(D)$ subalgebra spanned by the ${\cal M}^{IJ}$ is the subalgebra that preserves the spatial $d$-spheres of the global coordinates, and generates the $\frak{so}(D)$ isometries of them.  The ${\cal K}^I$ do not preserve the spheres, and we will call them boost transformations.

\section{General Structure of the Representations}

The general structure of the representations of the de Sitter algebra $\frak{so}(1,D)$ can be described as follows.  The vector space ${\cal F}$ on which a representation acts will be the infinite dimensional complex vector space of complex valued\footnote{One can also ask for representations on real vector spaces, but the complex ones are of primary interest in the quantum mechanical context, so we will not consider the interesting question of when the representations can be real.} fields of a specific type living on the spatial $d$-sphere, ${\mathbb S}^d$, $d=D-1$.  This sphere can be thought of as the future boundary of dS$_D$, the spatial sphere at $t\rightarrow\infty$ in global coordinates.   
The fields on the sphere can be thought of as the boundary values of fields on the entire dS$_D$ that satisfy a Klein-Gordon equation with a particular mass.  As we will review in the sections to follow, the isometries of dS$_D$ become the conformal symmetries of the boundary sphere, and under the action of the isometries the boundary fields transform as conformal primary fields.  The conformal transformation rules of the fields at the boundary then give the action of $\frak{so}(1,D)$ on ${\cal F}$.  In this action, the $\frak{so}(D)$ subalgebra acts by isometries of the sphere, and the boosts ${\cal K}^I$ act as the conformal transformations that are not isometries.  The action of ${\cal K}^I$ involves a complex parameter $\Delta$, the conformal weight of the field, which is related to the mass in the Klein-Gordon equation of the bulk field.  A representation will be characterized by a choice of a field which has irreducible index symmetries (i.e. it is traceless symmetric, fully anti-symmetric, or a more general traceless mixed symmetry tensor, and for the fermionic reps, it will be a spinor-tensor with appropriate gamma-tracelessness constraints to make it irreducible) along with a value of $\Delta$ to be used in the boost transformations.

A convenient basis of ${\cal F}$ will be obtained by breaking up the space into irreducible representations of the subalgebra $\frak{so}(D)$ of rotations of the sphere, i.e. the spherical harmonics of the field.  These $\frak{so}(D)$ reps will be ordinary finite dimensional unitary reps of the rotation group (these are reviewed in Appendix \ref{sorepsappendix}).  The action of the boost generators ${\cal K}^I$ will then move us among the $\frak{so}(D)$ reps, grouping them together into the larger $\frak{so}(1,D)$ rep.  
At a generic value of $\Delta$, we will be able to reach any given $\frak{so}(D)$ rep in ${\cal F}$ from any other by acting with the ${\cal K}^I$, so the space ${\cal F}$ will give an irreducible $\frak{so}(1,D)$ rep. 
But there will be some discrete values of $\Delta$ where this breaks down and we get reducible reps with invariant subspaces.  These subspaces must be factored out before arriving at an irreducible rep.  These types of reps will be those corresponding to gauge fields, i.e. massless, partially massless and shift symmetric fields, and those corresponding to finite dimensional reps.
In addition, for a given tensor field, some values of $\Delta$ will be equivalent to other values of $\Delta$, and in any given dimension $D$, there may be equivalences that relate some types of tensor fields to other types.

For each ${\cal F}$, some values of $\Delta$ will allow for the introduction of an invariant positive definite inner product on the vector space ${\cal F}$ (thus making it into a Hilbert space) under which the action of the algebra elements can be made Hermitian.  These will be the unitary reps.   Other values of $\Delta$ will give non-unitary reps.   

Almost all the dS reps will be infinite dimensional, and so they will contain an infinite number of the finite dimensional $\frak{so}(D)$ reps.  In particular, all the non-trivial unitary reps will be infinite dimensional (since $\frak{so}(1,D)$ is a non-compact Lie algebra, this follows from the general theorem \cite{weyl1925theorie1,weyl1926theorie2,weyl1926theorie3} that all non-compact Lie algebras have no non-trivial finite dimensional unitary reps).
There will be a few finite dimensional reps of $\frak{so}(1,D)$ containing a finite number of $\frak{so}(D)$ reps; these are non-unitary and come from the Wick rotations of the familiar finite dimensional unitary $\frak{so}(D+1)$ reps.

Ranging through all the types of tensor fields on the sphere, taking all of the above subtleties into account, we will arrive at a classification of all the inequivalent irreducible reps of $\frak{so}(1,D)$, unitary and (modulo a ``single field'' caveat to be discussed in section \ref{conclusionssec}) non-unitary.

Through a stereographic projection and a Weyl transformation, the space of conformal fields on the sphere can equally well be described as a space of conformal fields on the flat Euclidean plane.  The fact that the functions must be well defined on the sphere will lead to specific fall-off conditions that the functions on the plane must satisfy.   The action of the dS generators on functions on the plane is that of a $d$-dimensional Euclidean CFT.  This formulation in terms of the plane is naturally arrived at when using planar inflationary coordinates on dS, and this is the setting typically used in the study of cosmological correlators. Throughout the bulk of this review we stick to the sphere formulation because it is most natural for expressing the structure of the reps in terms of their $\frak{so}(D)$ content, however in Appendix \ref{flatslicing} we give the details of the flat formulation and how to translate the results into it.

In the remainder of this review, we will describe the structure of the reps case by case, starting with the simplest case of scalar fields and then working up progressively through the more complicated cases of spinning fields and fermionic fields.  In each case, the story will be essentially the same, as outlined above: starting with a space of fields on the sphere, we break it up into its $\frak{so}(D)$ spherical harmonics and then ask how these are linked together via the boosts into a $\frak{so}(1,D)$ rep.  We then identify the values of $\Delta$ for which the reps become reducible and describe how they reduce in each case.  Finally, we identify the equivalences between various values of $\Delta$ and describe the cases for which a unitary inner product can be defined.   For illustrative purposes, we have elected to repeat these steps for each type of field, at the price of making the review longer, more repetitive, and more redundant than it might have been.  For example, sections \ref{vectorsection} and \ref{spin2sec}, covering vector and spin 2 tensor reps, are redundant with and encompassed by section \ref{spinssec} on the spin $s$ tensor reps.  Readers may elect to skip the sections with field types that are not of interest. 

For each kind of field, there is a generic set of reps that exist for any $D$ once $D$ is large enough, but there are various subtleties in these reps, and equivalences between reps coming from different types of fields, that can occur for smaller values of $D$.
In each case, we will therefore start out by describing the various reps for generic $D$, ignoring these lower dimensional subtleties and equivalences (meaning everything we say will be true for $D$ large enough), and at the end of each section we  then go back and detail the various subtleties that occur for the lower values of $D$.  In section \ref{lowDsection}, we go through the various lower dimensional cases explicitly in order to spell out all the subtleties and equivalences that occur between types of reps.  In section \ref{unitarylistsection} we review the standard abstract summary of the unitary reps and describe how these match up to the field reps as we describe them.  Taking proper account of the lower dimensional subtleties and equivalences is essential for understanding how the list of abstract unitary reps matches up with the list of possible fields on dS.  Finally, there are many novel phenomena and simplifications that occur in the lowest dimension $D=2$, including rep types that have no higher $D$ counterparts, so we relegate this case to section \ref{D2section}, and restrict the discussion in all the remaining sections to $D\geq 3$.

\section{Bosonic Fields and Transformations}

We start with the bosonic reps of $\frak{so}(1,D)$.  These are the reps that will be realized on spaces of tensor fields on the sphere ${\mathbb S}^d$, transforming under the action of conformal transformations.  In this section, we review how this action comes about from the boundary action of isometries acting on on-shell tensor fields on dS$_D$.

\subsection{Fields on dS$_D$ and their boundary values on ${\mathbb S}^d$}

Consider a rank $r$ complex tensor field $\Phi_{\mu_1\ldots \mu_r}$ on dS$_D$, with indices that have permutation symmetries corresponding to some general mixed symmetry tableau $[s_1,\ldots,s_p]$, $r=\sum_{i=1}^p s_i$.
This tensor field will be on-shell, meaning it satisfies a Klein-Gordon equation with some mass\footnote{We use $\tilde m$ to indicate the bare mass entering this Klein-Gordon equation.  In the case of symmetric tensors and $p$-forms, this differs from the usual notion of mass $m$ that vanishes when the fields are at their traditional ``massless'' values, i.e. where the gauge symmetry is maximal, and which we will use in the sections below.  For the symmetric rank $s$ fields with $s\geq 1$, the relation between these definitions of mass is
\be \tilde m^2=m^2 +\left[s+D-2-(s-1)(s+D-4)\right]H^2\, .\label{spinsoffsete}\ee
For the anti-symmetric $p$-form fields with $p\geq 1$, the relation is
\be \tilde m^2=m^2 +p(D-p)H^2\, .\label{pformoffsete}\ee
For the scalar field we have $\tilde m^2=m^2$.
For more general mixed symmetry tensor types, there is no natural notion of ``massless'' and we use $\tilde m^2$.
}
$\tilde m$,
\be \left(\nabla^2-\tilde m^2\right)\Phi_{\mu_1\ldots \mu_r}=0\, ,\label{KGequationfre}\ee
is fully traceless, and is also fully transverse, i.e. it vanishes when contracting $\nabla^\mu$ with any index.  This on-shell field can be thought of as describing some generic kind of free bosonic particle on dS$_D$.

In global coordinates, the transversality and tracelessness constraints will generally constrain all the components of the tensor with any index in the $t$ direction.  Ignoring these for now, the components of the Klein-Gordon equation in the sphere directions become
\bea 
\left(\nabla^2-\tilde m^2\right) \Phi_{i_1\ldots i_r}&= &H^2\bigg[  -{1\over H^2}\partial_t^2 +{1\over H}({2r-d}) \tanh\left(H t\right) \partial_t \nn\\
&&\ \ \ \ \ \ \ +{1\over  \cosh^2\left(H t\right)}\nabla^2_\Omega +\left({r}+ {r(d-r)}  \tanh^2\left(H t\right)-{\tilde m^2\over H^2}\right)\bigg] \Phi_{i_1\ldots i_r} \,,\nn\\ \label{sphericalKGequee}
\eea
where $\nabla^2_\Omega$ is the Laplacian on ${\mathbb S}^d$, and $i_1,i_2,\ldots$ are indices on ${\mathbb S}^d$.

At late times $t\rightarrow \infty$, the equation \eqref{sphericalKGequee} can be solved as a power series in $e^{-Ht}$.   We look for a leading order solution $\Phi_{i_1\ldots i_r} (t,\hat X) \underset{t\rightarrow \infty}{\rightarrow} e^{-\left( \Delta-r\right) Ht}\phi_{i_1\ldots i_r} (\hat X)$, where $\Delta$ is some complex constant and $\phi_{i_1\ldots i_r} (\hat X)$ is the boundary value on the sphere.  Plugging this into \eqref{sphericalKGequee} and expanding everything in powers of $e^{-Ht}$, we find at leading order the equation
\be {\tilde m^2\over H^2}=-\Delta(\Delta-d)+r \,,\label{deltamregexe} \ee
which has the two solutions
\be  \Delta_\pm ={d\over 2}\pm \sqrt{{d^2\over 4}+r-{\tilde m^2\over H^2}}\,.\label{dscfttmassrelatione}  \ee
These solutions are related by
\be \Delta_-+\Delta_+=d\, .\ee
They give the leading parts of the asymptotic values of the two independent solutions of the second order Klein-Gordon equation,
\be \Phi_{i_1\ldots i_r}(t,\hat X)  \underset{t\rightarrow \infty}{\rightarrow} e^{-\left( \Delta_\pm-r\right) H t}\phi_{i_1\ldots i_r} (\hat X)\, .\label{ssassymprosole}\ee 
Given these two boundary data,\footnote{In cosmology we are instead usually interested in the situation where we fix some initial condition at early times, corresponding to choosing some initial vacuum state in the quantum theory.  The solution will then generically have both boundary solutions in the far future, and we will not be free to fix the ratio between them.} the rest of the solution is determined recursively as a power series\footnote{The recursion relation may fail for some discrete values of $\Delta$.  This indicates the presence of logarithmic terms in the power series expansion, and once these are added the recursion can proceed.} in $e^{-Ht}$.  Coming back to the components of the field with indices in the $t$ direction, they will also appear in \eqref{sphericalKGequee}, as well as in the equations coming from \eqref{KGequationfre} with components in the $t$ directions.  These $t$ components of the field will behave like $\sim e^{-(\Delta_\pm-r)Ht}e^{-r_t Ht}$, where $r_t$ is the number of indices in the $t$ direction, so these will be subleading.  They do not affect the leading relation \eqref{deltamregexe}, and they are fixed by the boundary data on the purely spatial components.

Note that there is a natural threshold where ${\tilde m^2\over H^2}= {d^2\over 4}+r$.
When the mass exceeds this threshold, $\Delta_\pm$ become imaginary, and the late-time solutions are oscillatory in time.  Below this threshold, $\Delta_\pm$ are real and the late-time solutions are exponentially growing or decaying in time.

\subsection{Action of the isometries on the boundary values\label{isometriessubsection}}

Consider now the action of the dS$_D$ Killing vectors \eqref{killingenbeddinge} 
on the tensor field $\Phi_{\mu_1\ldots \mu_r}$.  The change in a tensor field under an active infinitesimal transformation generated by a Killing vector is given by minus the Lie derivative ${\cal L}$ with respect to the Killing vector, so the transformations corresponding to \eqref{killingenbeddinge} are
\be \delta_{{\cal M}^{AB}}\Phi_{\mu_1\ldots \mu_r}=-{\cal L}_{{\cal M}^{AB}}\Phi_{\mu_1\ldots \mu_r}\,.\label{soondafctionee}\ee
The Lie derivative satisfies $\left[{\cal L}_{{\cal M}^{AB}}, {\cal L}_{{\cal M}^{CD}} \right]= {\cal L}_ {\left\{ {\cal M}^{AB},  {\cal M}^{CD}\right\}}$,  so the transformations \eqref{soondafctionee} satisfy the same $\frak{so}(1,D)$ algebra \eqref{sod12algberaadse},
\be  \left[\delta_{{\cal M}^{AB}},\delta_{{\cal M}^{CD}}\right] = \delta_{\left\{ {\cal M}^{AB},  {\cal M}^{CD}\right\}}\,. \ee

The ${\frak so}(1,D)$ transformations \eqref{soondafctionee} commute with the Laplacian and preserve the transversality condition and the index symmetry and tracelessness conditions of $\Phi_{\mu_1\ldots \mu_r}$.  They thus preserve the space of solutions of the Klein-Gordon equation \eqref{KGequationfre} and give an action of the dS algebra on this space of solutions.  Additionally, it does not mix $\Delta_+$ solutions with $\Delta_-$ solutions, and so each of these gives a separate space of solutions with an action of the dS$_D$ algebra.  It is these spaces of solutions which will be used to form the irreducible reps.

For the ${\frak so}(D)$ rotations ${\cal M}^{IJ}$ in \eqref{killingvectorsgede2}, the action is $\delta_{{\cal M}^{IJ}}\Phi_{\mu_1\ldots \mu_r}=-{\cal L}_{{\cal M}^{IJ}}\Phi_{\mu_1\ldots \mu_r}$, and the rotations do not act on $t$, so restricting to the spatial components, expanding at late times and using the asymptotic solution \eqref{ssassymprosole}, we find the action
\be  \delta_{{\cal M}^{IJ}}\phi_{i_1\ldots i_r}=-{\cal L}_{{\cal M}^{IJ}}\phi_{i_1\ldots i_r}\,, \label{latetimesheactiont}\ee
on the boundary fields.  We see that under the rotations, the late time field transforms as an ordinary tensor field on the sphere.  Now doing the same with the boosts ${\cal K}^I$ in \eqref{killingvectorsgede1}, at late times we get $\partial_t\rightarrow -H(\Delta-r)$, with $\Delta$ either of $\Delta_\pm$, and we find the action
\be    \delta_{{\cal K}^{I}}\phi_{i_1\ldots i_r}=\left[ (\Delta-r)\hat X^I -{\cal L}_{\partial \over \partial \hat X_I}\right] \phi_{i_1\ldots i_r}\,.\label{latetimesheactiont2}\ee
Here ${\partial \over \partial \hat X_I}$ is the vector field on the sphere whose intrinsic components are $g^{ij}{\partial \hat X^J\over \partial{\theta^j}}\delta_{IJ}$.

These actions \eqref{latetimesheactiont}, \eqref{latetimesheactiont2} satisfy the same commutators as in \eqref{sod12algberaadse2e}.  They are in fact precisely the conformal transformations on the unit sphere ${\mathbb S}^d$, where $\phi_{i_1\ldots i_r}$ transforms as a conformal primary field of weight $\Delta$.  We can see this as follows.  

Under a Weyl transformation on a manifold, the metric $g_{ij}$ and a tensor field $\phi_{i_1\ldots i_r}$ of Weyl weight $\Delta_W$  transform as $g_{ij}\rightarrow e^{2\sigma}g_{ij},\  \phi_{i_1\ldots i_r} \rightarrow e^{-\Delta_W\sigma} \phi_{i_1\ldots i_r}$, for arbitrary scalar gauge parameter $\sigma$ depending on the coordinates of the manifold.  The infinitesimal version of this Weyl transformation, $\delta_W$, reads
\be \delta_W g_{ij}=2\sigma g_{ij}\, ,\ \ \ \delta_W \phi_{i_1\ldots i_r} =-\Delta_W \sigma \phi_{i_1\ldots i_r}\, .\label{weyltransfere}\ee
Under infinitesimal diffeomorphisms, $\delta_D$, generated by a vector field $\xi^i$, the metric and tensor field transform as
\be \delta_D g_{ij}=-{\cal L}_\xi g_{ij}=-\left(\nabla_i\xi_j+\nabla_j\xi_i\right)\, ,\ \ \ \delta_D \phi_{i_1\ldots i_r} =-{\cal L}_\xi \phi_{i_1\ldots i_r} \, .\ee
The conformal transformations with respect to the metric $g_{ij}$ are those combined Weyl and diffeomorphism transformations that leave the metric invariant, $\left(\delta_W+\delta_D\right)g_{ij}=0$.  This imposes $\nabla_i\xi_j+\nabla_j\xi_i=2\sigma g_{ij}$, which upon taking a trace implies that 
\be \sigma={1\over d}\nabla\cdot \xi\, ,\label{sigmaeqe}\ee
and thus that
\be \nabla_i\xi_j+\nabla_j\xi_i-{2\over d}\nabla\cdot \xi g_{ij}=0\,.\label{CKEejee}\ee
This is the conformal Killing equation with respect to the metric $g_{ij}$, and any $\xi^i$ satisfying it is a conformal Killing vector of this metric.  Under the combined transformation $\delta=\delta_W+\delta_D$ with $\xi^i$ a conformal Killing vector, the tensor field $\phi_{i_1\ldots i_r}$ then transforms as
\be \delta \phi_{i_1\ldots i_r}=-\left( {\cal L}_\xi +{\Delta_W\over d} \nabla\cdot \xi \right)\phi_{i_1\ldots i_r}\,.\label{confgenesret}\ee
The Killing vectors are those conformal Killing vectors also satisfying $\nabla\cdot \xi =0$.  For these, the second term on the right hand side of \eqref{confgenesret} vanishes, and the tensor transforms only with the Lie derivative.

In our case, the manifold is ${\mathbb S}^d$ and the metric is the round sphere metric \eqref{spheremetrice}.  Its Killing vectors are the rotations \eqref{killingvectorsgede2}, whose components are ${\cal M}^{IJ}_i =2\hat X^{[I}\partial_i \hat X^{J]}$.  These satisfy $\nabla^i {\cal M}^{IJ}_i =0$, so \eqref{confgenesret} reproduces \eqref{latetimesheactiont}.
The additional conformal Killing vectors on ${\mathbb S}^d$ which are not Killing vectors are ${\cal K}^{I}_i=\partial_i\hat X^I$.  These satisfy $\nabla^i{\cal K}^{I}_i=-d\,\hat X^I$ 
and \eqref{confgenesret} then reproduces \eqref{latetimesheactiont2} for 
\be \Delta_W=\Delta-r \, .\label{weylnormdrele}\ee
The dS algebra thus acts on fields at infinity precisely as the conformal algebra on the sphere,\footnote{Note that the relation \eqref{weylnormdrele} also produces the usual flat space conformal transformations from \eqref{confgenesret} on ${\mathbb R}^d$ when the metric is flat, and the Euclidean flat space conformal algebra is also $\frak{so}(1,D)$.  This is manifest in the flat inflationary coordinates of dS, see appendix \ref{flatslicing}.} and both algebras are $\frak{so}(1,D)$.
 
 \subsection{Quadratic Casimir operator\label{casimirsec}}
 
 The action of the ${\frak so}(1,D)$ algebra has an associated quadratic Casimir operator, ${\cal C}_2$, which is often useful.  It is given by
\be {\cal C}_2 = -\half {\delta}_{{\cal M}^{AB}} {\delta}_{{\cal M}_{AB}}= -\half {\delta}_{{\cal M}^{IJ}} {\delta}_{{\cal M}_{IJ}} + {\delta}_{{\cal K}^{I}} {\delta}_{{\cal K}_{I}}  \,,\ee
and it commutes with the action of all the algebra elements, 
\be \left[ {\cal C}_2 , {\delta}_{{\cal M}^{AB}} \right]=0\,.\ee
Acting on the field $\Phi_{\mu_1\ldots \mu_r}$ on dS$_D$, it becomes minus the Laplacian up to the addition of a constant,
\be {\cal C}_2  =  -{1\over H^2}\nabla^2 + \sum_{i=1}^p s_i\left(s_i+D-2i\right) \, ,\ee
(where we recall that $[s_1,\ldots,s_p]$ is the Young tableau in which the indices of the field live), and it takes the following values on the on-shell and asymptotic fields, respectively:
 \bea 
{\cal C}_2  &=&-{\tilde m^2\over H^2}+\sum_{i=1}^p s_i\left(s_i+D-2i\right)=\Delta(\Delta-d)+\sum_{i=1}^p s_i\left(s_i+d-2i\right) \,.\label{c2gnsmdee}
 \eea

\section{Bosonic Representations\label{bosonsection}}

In this section we go through the bosonic reps, those carried by spaces of bosonic tensor fields on the boundary sphere transforming under $\frak{so}(1,D)$ as conformal fields of weight $\Delta\in {\mathbb C}$ as in \eqref{latetimesheactiont}, \eqref{latetimesheactiont2}, which we reproduce here for convenience:
\be \delta_{{\cal M}^{IJ}}\phi_{i_1\ldots i_r}=-{\cal L}_{{\cal M}^{IJ}}\phi_{i_1\ldots i_r}\,,\ \ \ \delta_{{\cal K}^{I}}\phi_{i_1\ldots i_r}=\left[ (\Delta-r)\hat X^I -{\cal L}_{\partial \over \partial \hat X_I}\right] \phi_{i_1\ldots i_r}\,.\label{latetimesheactiont3}\ee

 We start with the scalar reps in section \ref{scalarsec}, then build up through the vectors in \ref{vectorsection}, spin 2 tensors in \ref{spin2sec}, spin $s$ tensors in \ref{spin2sec}, $p$-forms in \ref{pformsec}, and culminating in the most general case of mixed symmetry reps in \ref{mixsymsec}.

\subsection{Scalar representations\label{scalarsec}}

We start with the simplest case, the scalar field reps.  In this case, using \eqref{ssassymprosole}, \eqref{deltamregexe}, \eqref{dscfttmassrelatione} with $r=0$, the Klein-Gordon equation $\left(\nabla^2-m^2\right)\Phi=0$ has the late time behavior $ \Phi \sim e^{-\Delta_\pm {H t}}$
where 
\be {m^2\over H^2}=-\Delta(\Delta-d)\, ,\ \ \ \Delta_\pm={d\over 2}\pm \sqrt{{d^2\over 4}-{ m^2\over H^2}}\,,  \label{dscfttmassrelatione32} \ee
and the $\frak{so}(1,D)$ action \eqref{latetimesheactiont3} becomes
\bea
&& \delta_{{\cal M}^{IJ}}\phi =\left( \hat X^I{\partial\over \partial \hat X_J}-\hat X^J{\partial\over \partial \hat X_I}\right)\phi \, , \ \  \delta_{{\cal K}^{I}}\phi = \left( \Delta\, \hat X^I-{\partial\over \partial \hat X_I}\right)\phi \, . \label{latetimesheactionm1} 
\eea

The vector space carrying the scalar $\frak{so}(1,D)$ reps will be the space of square integrable complex scalar functions on the $d$-sphere ${\mathbb S}^d$, with the action of the generators as given in \eqref{latetimesheactionm1}.  Call this space ${\cal F}^{[0]}_\Delta$,
\be {\cal F}^{[0]}_\Delta:\ \ {\rm complex\ scalar\ functions\ on\ } {\mathbb S}^d \,.\label{scalarspaceofsdjfe}\ee
The superscript $[0]$ stands for the empty Young tableau and refers to the fact that these are scalar functions on the sphere; this label will become important later on once we generalize to more complicated tensor and spinor reps.  The value of $\Delta$ enters only through the action of the boosts ${\cal K}^I$ in \eqref{latetimesheactionm1}.  As we will see, for generic $\Delta$ this space describes an irreducible rep, whereas for specific values of $\Delta$ it will develop invariant subspaces which must be factored out before the rep becomes irreducible.

As noted above, the ${\cal M}^{IJ}$ act as the rotations on the sphere ${\mathbb S}^d$, forming an $\frak{so}(D)$ maximal subalgebra of $\frak{so}(1,D)$, and it will be useful to break up the space ${\cal F}^{[0]}_\Delta$ into subspaces transforming as irreducible $\frak{so}(D)$ reps, i.e. spherical harmonics.  The scalar spherical harmonics can be taken as the following symmetric traceless tensors in the sphere's embedding space, 
\be Y_l^{I_1I_2\ldots I_l}(\hat X)\equiv \hat X^{(I_1}\hat X^{I_2}\cdots \hat X^{I_l)_T} \, ,\ \ \ l=0,1,2,\ldots \, ,  \label{scalarphserarmoe} \ee
where $Y_0(\hat X)\equiv 1$.  They are eigenfunctions of the Laplacian on ${\mathbb S}^d$, with the following eigenvalues,
\be -\nabla_\Omega^2  Y_l^{I_1\ldots I_l}=  l (l+d-1) Y_l^{I_1\ldots I_l}\,.\ee
They are linearly independent and complete, forming a basis of the space of functions \eqref{scalarspaceofsdjfe}.

The $l$-th set of spherical harmonics transforms in the irreducible symmetric traceless rank $l$ tensor rep of the rotation algebra $\frak{so}(D)$, represented by a Young tableau $[l]$ with one row of $l$ boxes.  The $\frak{so}(D)$ content of the space of functions ${\cal F}^{[0]}_\Delta$ thus consists of one copy of each symmetric tensor rep.  We can illustrate this content in terms of Young tableaux pictures as follows:
\be \raisebox{-10pt}{\epsfig{file=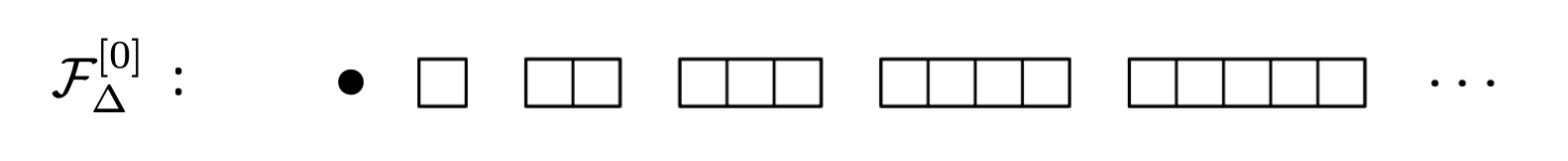,width=5.0in}}  \label{scalarsocontent}\, \ee
The dot at the beginning stands for the $l=0$ spherical harmonic $Y_0(\hat X)= 1$, which is the trivial rep of $\frak{so}(D)$.

Now consider the action \eqref{latetimesheactionm1} of the boost generators ${\cal K}^I$ on the spherical harmonic basis.  The spherical harmonics transform into each other as follows\footnote{Some useful intermediate results are 
\be \hat X^IY_l^{I_1\ldots I_l}=Y_{l+1}^{I I_1\ldots I_l} +\frac{l}{d+2 l-1} \delta^{I(I_1}Y_{l-1}^{I_2\ldots I_l)}-\frac{(l-1) l}{(d+2 l-3) (d+2 l-1)} \delta^{(I_1I_2}Y_{l-1}^{I_3\ldots I_l)I}\, ,    \ee
\be {\partial\over \partial \hat X_I} Y_l^{I_1\ldots I_l}=-lY_{l+1}^{I I_1\ldots I_l} +\frac{l (d+l-1)}{d+2 l-1} \delta^{I(I_1}Y_{l-1}^{I_2\ldots I_l)}-\frac{(l-1) l (d+l-1)}{(d+2 l-3) (d+2 l-1)} \delta^{(I_1I_2}Y_{l-1}^{I_3\ldots I_l)I}\, ,  \label{intanrese2de}  \ee
where in \eqref{intanrese2de} we have used that ${\partial\over \partial \hat X_I} \hat X^J=\delta^{IJ}-\hat X^I\hat X^J$.
},
\be \delta_{{\cal K}^I}Y_l^{I_1\ldots I_l} =\left(\Delta+l\right) Y_{l+1}^{I I_1\ldots I_l}  -\frac{l (d-\Delta +l-1)}{d+2 l-1} \delta^{I(I_1}Y_{l-1}^{I_2\ldots I_l)}+\frac{(l-1) l (d-\Delta +l-1)}{(d+2 l-3) (d+2 l-1)} \delta^{(I_1I_2}Y_{l-1}^{I_3\ldots I_l)I}\, . \label{klactione}\ee
(For $l=0,1$ these should be taken as
\be \delta_{{\cal K}^I}Y_0 =\Delta Y_{1}^{I},\ \ \  \delta_{{\cal K}^I}Y_1^{I_1} =\left(\Delta+1\right) Y_{2}^{I I_1}  -\frac{ d-\Delta }{d+1} \delta^{II_1}\,,
\ee
so vanishing denominators in \eqref{klactione} when $d=3$ are not an issue.)  

Looking at \eqref{klactione}, we can split the ${{\cal K}^I}$ transformation into two pieces ${{\cal K}^{I\pm}}$ that raise and lower the value of $l$ and do not depend on $\Delta$,
\be \delta_{{\cal K}^I}=\left(\Delta+l\right) \delta_{{\cal K}^{I+}}+ (d-\Delta +l-1) \delta_{{\cal K}^{I-}} \, ,\label{kpmmdda}
\ee
where 
\bea && \delta_{{\cal K}^{I+}}Y_l^{I_1\ldots I_l}=Y_{l+1}^{I I_1\ldots I_l}\,,\ \ \ \nn\\
&& \delta_{{\cal K}^{I-}}Y_l^{I_1\ldots I_l}= -\frac{l }{d+2 l-1} \delta^{I(I_1}Y_{l-1}^{I_2\ldots I_l)}+\frac{(l-1) l }{(d+2 l-3) (d+2 l-1)} \delta^{(I_1I_2}Y_{l-1}^{I_3\ldots I_l)I}\, . \label{kpmmdda2}\nn\\
\eea
We can think of the $\delta_{{\cal K}^{I\pm}}$ as ladder operators \cite{Letsios:2024snc} that move us up and down the $\frak{so}(D)$ reps present in ${\cal F}_{\Delta}^{[0]}$.  For generic $\Delta$, the coefficients in \eqref{kpmmdda} never vanish and so we can reach any $l$ state starting from any other, and thus the rep is irreducible.  In these cases we can illustrate the action of ${\cal K}^I$ as follows,
\be \raisebox{-10pt}{\epsfig{file=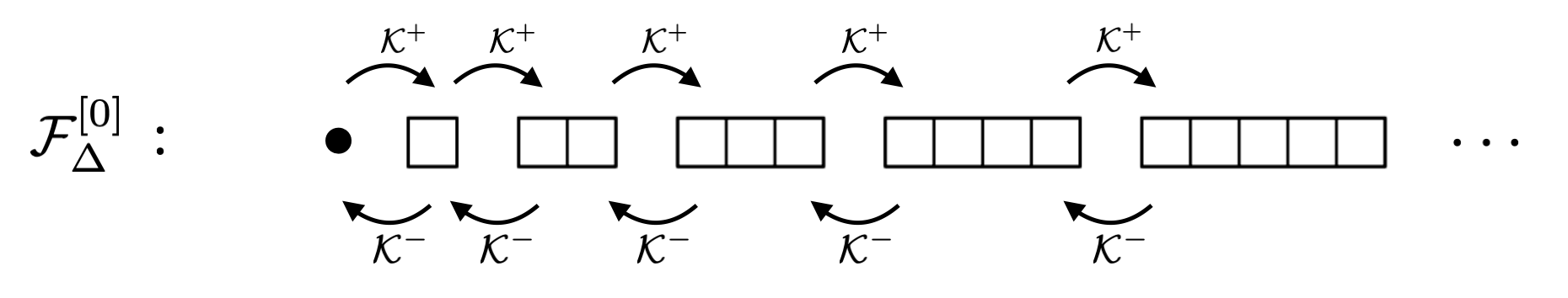,width=4.5in}}  \label{scalarsocontent2}\, \ee
These irreducible reps describe generic massive scalar fields on dS$_D$, with the mass determined in terms of $\Delta$ through \eqref{dscfttmassrelatione32}.

\textbf{Reducible cases:} 
There are, however, specific values of $\Delta$ where the coefficients in \eqref{kpmmdda} vanish:  
\begin{itemize}

\item Shift symmetric points:
\be \Delta=d+k\,, \ \  k=0,1,2,\ldots\ \ .\ee
If $\Delta$ is one of these values, the coefficient in front of $\delta_{{\cal K}^-}$ in  \eqref{kpmmdda} vanishes for $l=k+1$, and so starting in a state with $l\geq k+1$, we can never reach any of the states with $l<k+1$.  This means the states with $l\geq k+1$ form an irreducible infinite dimensional sub-representation.   We call this sub-rep 
\be {\cal D}_{d+k}^{[0]}\,. \label{shifscldge}\ee

We can illustrate this as follows in the example with $k=2$,
\be \raisebox{-10pt}{\epsfig{file=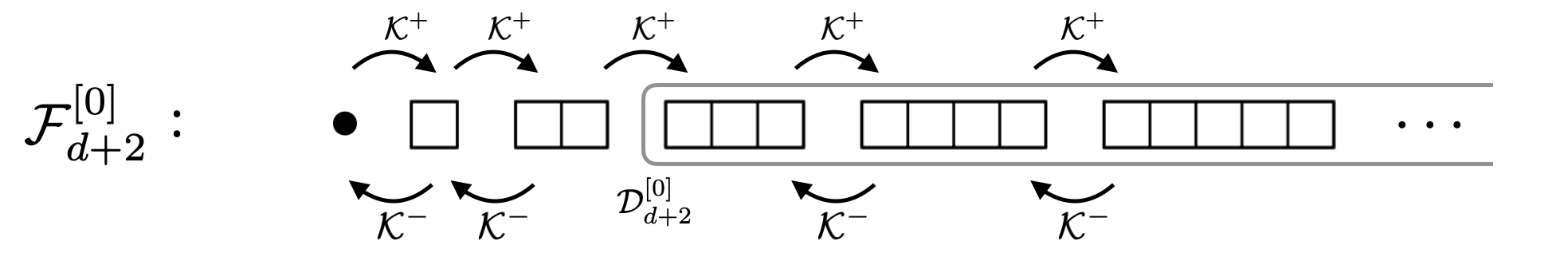,width=4.5in}}  \label{scalarsocontent4}\, \ee
Here we see the missing ${\cal K}^-$ link, and a grey box encloses the irreducible sub-rep ${\cal D}_{d+2}^{[0]}$.   

${\cal D}_{d+k}^{[0]}$ describes the physical modes of the level $k$ (using the classification of \cite{Bonifacio:2018zex}) shift symmetric scalar on dS$_{d+1}$ when the shift symmetry is gauged.

\item Finite points:
\be \Delta=-k\, ,\ \  k=0,1,2,\ldots \ \ .\ee
If $\Delta$ is one of these values, the coefficient in front of $\delta_{{\cal K}^+}$ in  \eqref{kpmmdda} vanishes for $l=k$, and so starting in a state with $l\leq k$, we can never reach any of the states with $l>k$.  This means the states with $l\leq k$ form an irreducible sub-rep, which we see is finite dimensional.  We call this sub-rep 
\be {\cal S}_{-k}^{[0]}\,.\ee

We can illustrate this as follows in the example with $\Delta=-2$,
\be \raisebox{-10pt}{\epsfig{file=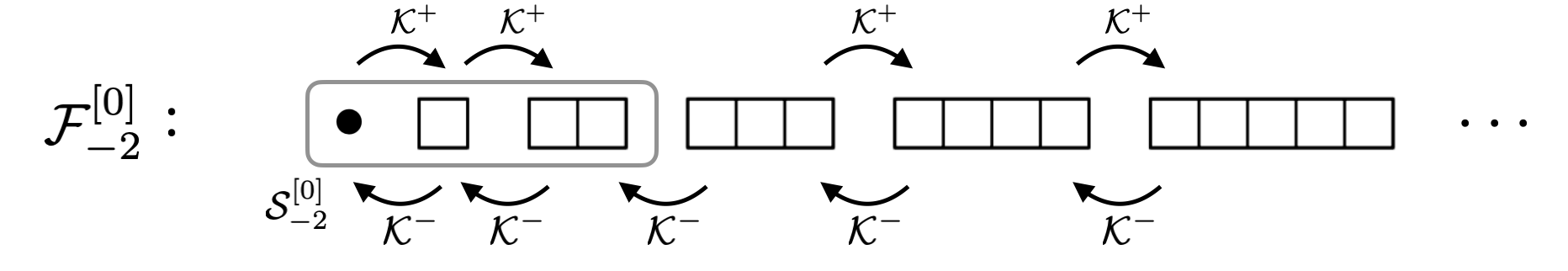,width=4.5in}}  \label{scalarsocontent3}\, \ee
Here we see the missing ${\cal K}^+$ link, and a grey box encloses the irreducible sub-rep ${\cal S}_{-2}^{[0]}$.    

${\cal S}_{-k}^{[0]}$ is nothing but the finite dimensional rank $k$ symmetric traceless tensor rep of $\frak{so}(1,D)$.   Restricting this to the $\frak{so}(D)$ content we are working with presently using the branching rules in appendix \ref{branchingappendix}, we see that it indeed splits up into the symmetric traceless tensors of rank $k,k-1,\ldots,0$.  It describes the parameters of the shift symmetries of the level $k$ shift symmetric scalar, i.e. the unphysical modes that are removed when the shift symmetry is gauged.

\end{itemize}

We have called these special sets of $\Delta$ values the shift symmetric and finite and points, respectively.  Note that in both of these cases, the reps ${\cal F}_{\Delta}^{[0]}$ are reducible (because they contain invariant subspaces) but not decomposable (because they do not split into a direct sum of invariant subspaces).

\textbf{Equivalences:} 
The reps at different values of $\Delta\in {\mathbb C}$ are not all inequivalent.  For generic $\Delta$, the rep ${\cal F}_{\Delta}^{[0]}$ is equivalent to its {\it shadow} rep ${\cal F}_{\bar\Delta}^{[0]}$ \cite{Ferrara:1972uq,Dobrev:1977qv,Simmons-Duffin:2012juh},
\be {\cal F}_{\Delta}^{[0]}\simeq {\cal F}_{\bar \Delta}^{[0]}\,,\label{scalarshadowsee}\ee
where
\be \bar\Delta\equiv d-\Delta\, .\ee
The shadow value $\bar \Delta$ of a given $\Delta$ is simply the other solution in \eqref{dscfttmassrelatione} to the mass relation \eqref{deltamregexe}, and so both $\Delta$ and $\bar \Delta$ correspond to the same mass value of the field.

The equivalence is seen via a linear intertwining operator $S_\Delta^{[0]}$, the {\it shadow transform}, that maps between the two spaces,
\be S_\Delta^{[0]} :\ {\cal F}_{\Delta}^{[0]}\rightarrow {\cal F}_{\bar\Delta}^{[0]}\,.\label{nsscendinee}\ee
This map commutes with the $\frak{so}(D)$ rotations, and it satisfies
\be \delta_{{\cal K}^I_{\bar\Delta}}S_\Delta^{[0]}=S_\Delta^{[0]} \delta_{{\cal K}^I_{\Delta}}\,.\label{Kcommsees} \ee
Since $S_\Delta^{[0]}$ commutes with the $\frak{so}(D)$ rotations, it can be diagonalized on the spherical harmonic basis \eqref{scalarphserarmoe}, so that we have 
\be S_\Delta^{[0]}Y_l^{I_1\ldots I_l}=s_{\Delta,l} Y_l^{I_1\ldots I_l}\, ,\label{sphericljhsksagdce}\ee
for some constants $s_{\Delta,l}$.
Acting with each side of \eqref{Kcommsees} on $Y_l$, using \eqref{sphericljhsksagdce}, writing out the ${\cal K}$'s in terms of the ${\cal K}^\pm$ as in \eqref{kpmmdda}, \eqref{kpmmdda2}, and equating the resulting $Y_{l+1}$ parts, we obtain a recursion relation (we get the same equation from the $Y_{l-1}$ parts),
\be ( \Delta+l)s_{\Delta,l+1}=( \bar\Delta+l)s_{\Delta,l}\, .\label{kpmrecursioneree}\ee
This is solved by
\be s_{\Delta,l}={\Gamma(\bar\Delta+l)\over \Gamma( \Delta+l)}{\Gamma(\Delta)\over \Gamma(\bar\Delta)}\, ,\label{Smatrixelementeesee}\ee
where we have chosen to normalize $S_\Delta^{[0]}$ so that $s_{\Delta,0}=1$.  For generic $\Delta$, the $s_{\Delta,l}$ never vanish and so the operator $S_\Delta^{[0]}$ is invertible, with the matrix elements of the inverse given simply by $1/s_{\Delta,l}$, so this map gives the equivalence \eqref{scalarshadowsee}. 

However, this equivalence breaks down at the special values of $\Delta$ given above where we have the reducible reps.
In the cases $\Delta=d+k$, $k=0,1,2,\ldots$, so that $\bar\Delta=-k$, where we have the infinite dimensional invariant subspace ${\cal D}^{[0]}_{d+k}$, the right hand side of \eqref{kpmrecursioneree} vanishes when $l=k$.  The recursion thus truncates, and we have 
\be s_{\Delta,l}=0\ {\rm for}\ l>k\, .\ee
The map $S_{d+k}^{[0]}$ is therefore not invertible in this case, and has an infinite dimensional kernel spanned by the states with $l>k$.  This kernel is precisely the invariant subspace ${\cal D}^{[0]}_{d+k}$, and the image is the invariant subspace ${\cal S}^{[0]}_{-k}$.  We thus have the isomorphisms
\be {\cal S}^{[0]}_{-k} \simeq {\cal F}^{[0]}_{d+k}/{\cal D}^{[0]}_{d+k}\,, \ \ k=0,1,2,\ldots \,.\label{scaiso2msfee}\ee

In the cases $\Delta=-k$, $k=0,1,2,\ldots$, so that $\bar\Delta=d+k$, where we have the finite dimensional invariant subspace ${\cal S}^{[0]}_{-k}$, the left hand side of \eqref{kpmrecursioneree} vanishes when $l=k$.  We thus cannot continue the recursion relation past this point starting from a finite $s_{-k,0}$.  We can instead start the recursion by taking $s_{-k,l=k+1}=1$.  The values of $s_{-k,l}$ for all smaller $l$ will then be fixed to be zero,
\be  s_{-k,l}=0\ {\rm for} \  l\leq k\, ,\ee
whereas $s_{-k,l}$ for $l>k$ will be non-vanishing. 
The map $S_{-k}^{[0]}$ given in this way is therefore not invertible, and has a finite dimensional kernel spanned by the states with $l\geq k$.  Note that this kernel is precisely the invariant subspace ${\cal S}^{[0]}_{-k}$, and the image is the invariant subspace ${\cal D}^{[0]}_{d+k}$.  We thus have the isomorphisms
\be {\cal D}^{[0]}_{d+k} \simeq {\cal F}^{[0]}_{-k}/{\cal S}^{[0]}_{-k}\,, \ \ k=0,1,2,\ldots \ .\label{firstscsisoe}\ee

We illustrate these maps and isomorphisms as follows in the example with $k=2$:
\be \raisebox{-20pt}{\epsfig{file=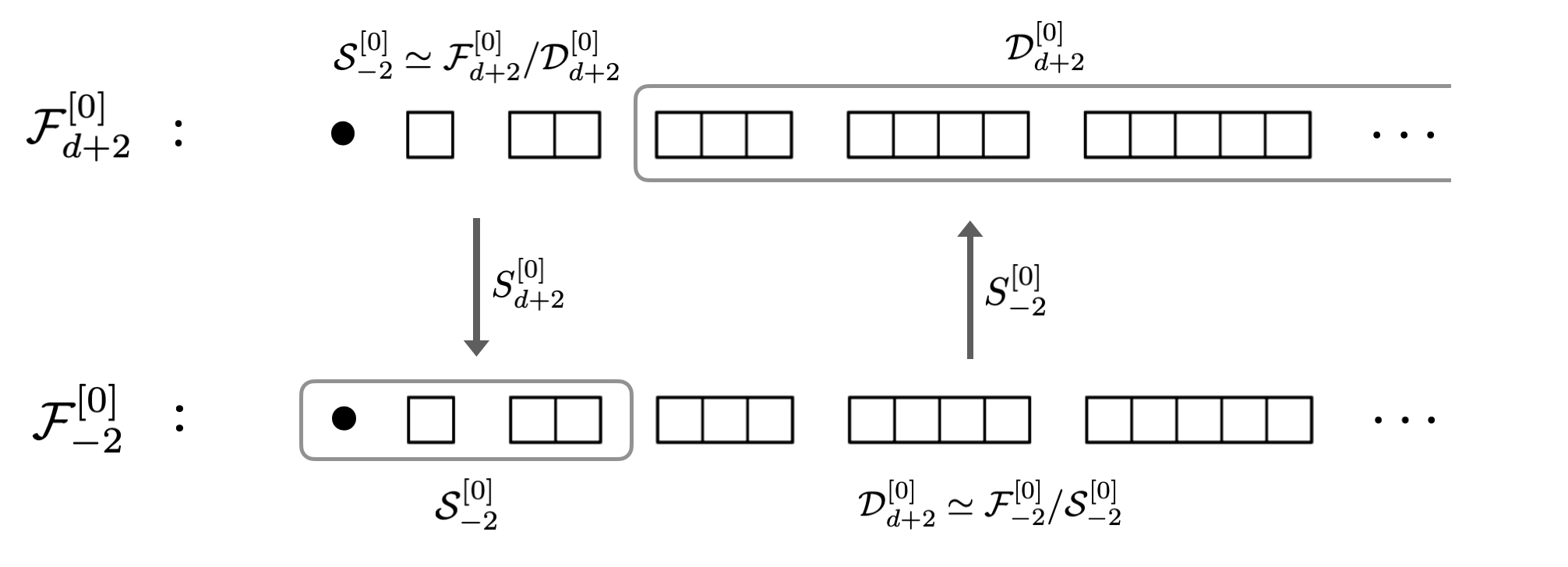,width=4.8in}}  \label{scalarsocontent5}\, \ee
The invariant subspaces are in the grey boxes, and they are both the kernel and image of the shadow maps.  

Apart from \eqref{scalarshadowsee}, and the above described subtleties in how it behaves at the reducible values of $\Delta$, there are no other equivalences among the reps.

\textbf{Unitarity:} We now turn to the issue of inner products and unitarity of these reps.  For generic $\Delta$, there is no $\frak{so}(1,D)$ invariant inner product on the space ${\cal F}_\Delta^{[0]}$.
There is, however, always an invariant bilinear form pairing the two spaces ${\cal F}_\Delta^{[0]}$ and ${\cal F}_{\bar \Delta}^{[0]}$ \cite{Vilenkin1993}.
Given $\phi_1\in {\cal F}_{\bar\Delta}^{[0]}$ and $\phi_2\in {\cal F}_\Delta^{[0]}$, the bilinear form is
\be (\phi_1,\phi_2)\equiv \int d^d\Omega\, \phi_1(\hat X)\phi_2(\hat X)\,.\label{bilinearformee}\ee
This integral is diffeomorphism invariant, and Weyl invariant under \eqref{weyltransfere} where $\phi_1$ transforms with Weyl weight $\bar\Delta$ and $\phi_2$ transforms with Weyl weight $\Delta$.   It is thus also invariant under the conformal transformations \eqref{latetimesheactionm1}, i.e. under the action of any of these we have $(\delta\phi_1,\phi_2)+(\phi_1,\delta\phi_2)=0$.  

The pairing \eqref{bilinearformee} does not immediately give an invariant inner product, since it takes arguments from spaces with two different actions of the algebra and is linear in both of its arguments, whereas an invariant inner product should take both arguments from spaces with the same action of the algebra, and it should be anti-linear in one of its arguments and linear in the other.

If, however, we have $\Delta^\ast=\bar\Delta$, then  $\phi^\ast \in {\cal F}_{\bar \Delta}^{[0]}$ if  $\phi \in {\cal F}_{ \Delta}^{[0]}$, so we can use \eqref{bilinearformee} to form an invariant inner product on ${\cal F}_\Delta^{[0]}$ as follows:
\be \la \phi_1 | \phi_2 \ra\equiv (\phi_1^\ast,\phi_2)\, ,\ \ \ \Delta^\ast=\bar\Delta\, , \ \ \ \phi_1,\phi_2\in {\cal F}_{ \Delta}^{[0]}\, .\label{innerprodprise} \ee
The condition $\Delta^\ast=\bar\Delta$ is equivalent to $\Delta={d\over 2}+i\nu$ with $\nu\in {\mathbb R}$.  In these cases, the inner product \eqref{innerprodprise} is invariant and manifestly positive definite, so these reps are all unitary (in our conventions, the algebra operators are anti-Hermitian).  These are known as the {\it principal series} reps,
\be {\rm scalar\ principal\ series:}\ \ \Delta={d\over 2}+i\nu\, ,\ \ \ \nu\in {\mathbb R}\,.\ee
Note that the rep with $\Delta={d\over 2}+i\nu$ is equivalent to the one with $\Delta={d\over 2}-i\nu$ by the shadow transform equivalence \eqref{scalarshadowsee}.  Other than this equivalence, all the principal series reps are distinct.

In the case where $\Delta$ is real, it is unaffected by conjugation and we can use the operator $S^{[0]}_{\Delta}$ to move a state from ${\cal F}_{ \Delta}^{[0]}$ to ${\cal F}_{\bar \Delta}^{[0]}$ and form an inner product on ${\cal F}_{ \Delta}^{[0]}$ as follows:
\be \la \phi_1 | \phi_2 \ra\equiv (S^{[0]}_{\Delta}\phi_1^\ast,\phi_2)\, ,\ \ \ \Delta^\ast=\Delta\, , \ \ \ \phi_1,\phi_2\in {\cal F}_{ \Delta}^{[0]}\,.\label{innerprodpriscsdee} \ee 
The positivity of this inner product is now equivalent to whether the matrix elements $s_{\Delta,l}$, given by \eqref{Smatrixelementeesee}, are positive for all $l\geq 0$.  Ignoring the reducible cases $\Delta,\bar\Delta=0,-1,-2,\ldots$ for now, the strongest constraint comes from $c_{\Delta,l=1}$, which is only positive in the range $0<\Delta<d$.
These are called the {\it complementary series} reps, and they are unitary,
\be {\rm scalar\ complementary\ series:}\  0<\Delta<d\, .\ee  
The point $\Delta=d/2$, where $\Delta=\bar \Delta=\Delta^\ast$, intersects the principal series: at this point, the shadow map \eqref{nsscendinee} becomes the identity, as can be seen from \eqref{Smatrixelementeesee}.  For now, and in the sections that follow, we will remain noncommittal about whether to include this point in the principal series or the complementary series.  We will make a choice in section \eqref{unitarylistsection} when we list all the distinct unitary representations.

The complementary series reps with $\Delta$ and $d-\Delta$ are equivalent due to \eqref{scalarshadowsee}.  Other than this equivalence, they are all distinct.

Now come back to the reducible cases, starting with the finite points $\Delta=-k$, $k=0,1,2,\ldots$.  For these we can use $S^{[0]}_{-k}$ in the inner product, and this has the finite dimensional kernel consisting of the  states with $l\leq k$, the sub-rep ${\cal S}^{[0]}_{-k}$.   All of these states will therefore have a norm of zero in the inner product, and are thus null states.  Apart from these null states, all the norms are positive, so if these null states are factored out, we will be left with a unitary rep.  The rep obtained after this factoring is ${\cal D}^{[0]}_{d+k}$, realized through the isomorphism \eqref{firstscsisoe}, so these are unitary reps.  These are the shift symmetric reps: they correspond to the physical modes of the shift symmetric fields, and so these fields are unitary (as argued from the field theoretic point of view in \cite{Bonifacio:2018zex}).  Later on we will see how they correspond to what are more commonly called {\rm exceptional series} reps of higher spins (and in $D=2$, we will see in section \ref{D2section} that they behave differently and there they are called discrete series reps).

For the shift symmetric points $\Delta=d+k$, $k=0,1,2,\ldots$, we use $S^{[0]}_{d+k}$ in the inner product, and this has the infinite dimensional kernel consisting of the states with $l> k$, which are the sub-rep ${\cal D}^{[0]}_{d+k}$.   All of these states will therefore have a norm of zero in the inner product, and are thus null states.  Apart from these null states, the nonzero norms do not all have the same sign, so when the null states are factored out, we will be left with a finite dimensional non-unitary rep.  The only exception is $k=0$, which has only one state and is therefore the trivial rep, which is the only finite dimensional unitary rep.   These finite dimensional reps are ${\cal S}^{[0]}_{-k}$, realized through the isomorphism \eqref{scaiso2msfee}.
As noted earlier, this rep is the symmetric traceless rank $k$ tensor rep of $\frak{so}(1,D)$.  The shift symmetries of a level $k$ scalar consist of precisely these states (the form of the shift symmetry is a rank $k$ tensor in the $(D+1)$-dimensional embedding space of the dS$_D$ space, and thus forms the rank $k$ symmetric traceless tensor rep of the dS algebra $\frak{so}(1,D)$ \cite{Bonifacio:2018zex}).

Apart from the unitary cases discussed here, there is no way to construct an invariant inner product on ${\cal F}_\Delta^{[0]}$, so all the other reps in the complex $\Delta$ plane are non-unitary.   (In the case of dS$_2$, i.e. $d=1$, there are additional subtleties and novelties that we discuss in section \ref{D2section}.)

Note that the spherical harmonics \eqref{scalarphserarmoe} satisfy
\be \int d^d\Omega \, Y_l(\hat X)_{ I_1\ldots I_l} Y_{l'}(\hat X)^{J_1\ldots J_{l'}}= {l!\over 2^{l-1}}{\pi^{d+1\over 2}\over \Gamma\left({d+1\over 2}+l\right)} \delta^{(J_1}_{(I_1}\cdots \delta^{J_l)_T}_{I_l)_T}\delta_{ll'}\,, \ee
and this implies that in all the unitary reps above, they form an orthogonal basis with respect to the inner product.  An analogous comment applies in all the more general cases discussed in later sections.

\textbf{Summary:} 
The scalar reps are summarized here in the complex $\Delta$ plane:
\be \raisebox{-60pt}{\epsfig{file=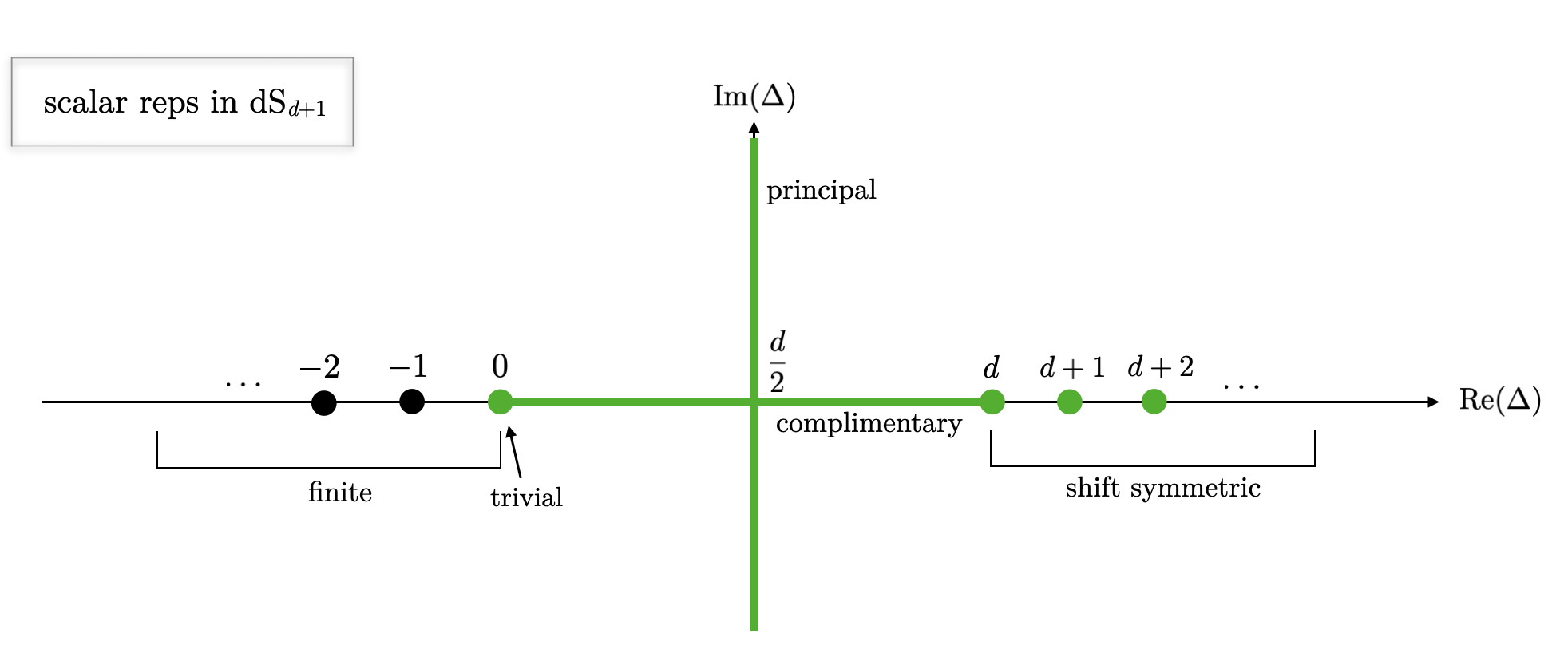,width=6.4in}}  \label{dsreps1}\, \ee
Each point in the plane is a rep.  Points in green are unitary reps.  The circles are the special points where the reps become reducible and thus shortened.   All of the reps, with the exception of finite reps, i.e. the black dots and the trivial rep at $\Delta=0$, are infinite dimensional.  The equivalence \eqref{scalarshadowsee} among the non-shortened reps is given by reflecting through the point $\Delta=d/2$ (due to the breakdown of the intertwiner map \eqref{nsscendinee}, the shortened reps represented by the circles are inequivalent despite the reflection).  Other than this equivalence, all the reps are distinct. 

The value of the quadratic Casimir operator ${\cal C}_2$ of section \ref{casimirsec} on the scalar reps is given by
\be {\cal C}_2= \Delta(\Delta-d) = -{m^2\over H^2}\,. \ee

We have the following correspondence between scalar fields and the non-trivial scalar unitary reps in dS$_{d+1}$:
\bea \begin{cases}
 {\rm principal:\ }\quad  \Delta={d\over 2}+i\nu \, , \ \ \nu\in {\mathbb R}\,, & {\rm heavy \ scalars:\ } {m^2\over H^2 }\geq {(D-1)^2\over 4}\,, \\
 {\rm complementary:\ } \quad 0<\Delta<d \, ,\ \   & {\rm light \ scalars:\ } 0< {m^2 \over H^2 }\leq  {(D-1)^2\over 4}\,, \\
  {\rm shift\ symmetric:\ } \quad \Delta=d+k \, , \ \ k=0,1,2,\ldots \,, &  {\rm shift\ symmetric \ scalars:\ } {m^2\over H^2 }=-k(k+D-1)\,.  
\end{cases} \nn
\eea
Note that the shift symmetric scalars have tachyonic masses with $m^2<0$, but they are nevertheless unitary reps.  This can be understood from the fact that the shift symmetries are to be gauged, i.e. modded out, in order to obtain an irreducible rep; the tachyonic instabilities are associated with the gauge modes, and so they are removed from the physical rep.

\subsection{Vector representations\label{vectorsection}}

We now turn to the next most complicated case, the vector, or spin 1, reps.  For vector fields, we typically use the mass $m^2$ that vanishes when the field becomes gauge invariant, which from \eqref{spinsoffsete} is offset from the mass $\tilde m^2$ in the Klein-Gordon equation by $\tilde m^2=m^2 +\left(D-1\right)H^2$.   From \eqref{deltamregexe}, \eqref{dscfttmassrelatione}, the relations between $\Delta$ and $m^2$ in the late time behavior $\sim e^{-\left( \Delta_\pm-1\right) {H t}}$ of \eqref{ssassymprosole} read
\be {m^2\over H^2}=-\left(\Delta-1\right)\left(\Delta-d+1\right)\,,\ \   \Delta_\pm={d\over 2}\pm\sqrt{{\left(d-2\right)^2\over 4}-{m^2\over H^2}}\, . \label{vectmassrelationere}\ee

The vector space carrying the vector reps will be the space of square integrable complex vector\footnote{We use the terminology ``vector'' despite the lowered index; there is of course no essential difference because of the metric on the sphere.} fields $\phi_i(\hat X)$ on the sphere ${\mathbb S}^{d}$, with the action of the generators of $\frak{so}(1,D)$ as given in \eqref{latetimesheactiont3} with $r=1$. 
Call this space ${\cal F}^{[1]}_\Delta$,
\be {\cal F}^{[1]}_\Delta:\ \ {\rm complex\ vector\ fields\ on\ } {\mathbb S}^d \,.\label{spacevecjee}\ee
The superscript $[1]$ denotes the single box Young tableau corresponding to the single index on the field $\phi_i$.  As with the scalars, for generic $\Delta$ this space will describe an irreducible rep, whereas for specific discrete values of $\Delta$ it will develop invariant subspaces which must be factored out before the rep becomes irreducible.

It will be useful to first break up the space ${\cal F}^{[1]}_\Delta$ into subspaces transforming as irreducible $\frak{so}(D)$ reps.  To start, we use the Hodge decomposition theorem to break up a general vector field into its transverse and longitudinal parts (there is no harmonic part on the sphere when $d\geq 2$, and the $d=1$ case is saved for section \ref{D2section}),
\be \phi_i=\chi_i+\nabla_i \chi\, ,\ \ \ \nabla^i\chi_i=0\,.\ee
The scalar $\chi$ in the longitudinal part can be expanded in the scalar spherical harmonics \eqref{scalarphserarmoe}.  However, the $l=0$ mode is not present, since it would be annihilated by $\nabla_i$, so we only have the spherical harmonics with $l\geq 1$.  The transverse vector part $\chi_i$ can be expanded in a basis of transverse vector spherical harmonics on ${\mathbb S}^d$.   These take the form \cite{rubin1984eigenvalues,rubin1985symmetric,Higuchi:1986wu}
\be Y^{I_1\ldots I_l,J}_{l,i}(\hat X)= \partial_i \hat X^{[J}\, \hat X^{I_1]}\hat X^{I_2}\cdots\hat X^{I_l}-{\rm traces}\, ,\ \ \ l=1,2,3,\ldots\ . \label{vecotrspherharmonicese}\ee
These have the $\frak{so}(D)$ index symmetries of traceless $[l,1]$ tableaux.  They form a basis of the space of transverse vectors on ${\mathbb S}^d$.  The natural Laplacian on the space of vectors is the Hodge Laplacian given by 
\be \Delta_H=-\nabla^2_\Omega+d-1\, ,  \ee 
and the transverse vector harmonics are eigenfunctions of it, with the following eigenvalues \cite{rubin1984eigenvalues,rubin1985symmetric,Higuchi:1986wu},
\be \Delta_H Y^{I_1\ldots I_l,J}_{l,i}= (l + d-2) (l+1)  Y^{I_1\ldots I_l,J}_{l,i}\, .\ee

Including these transverse vector harmonics and the $l\geq 1$ scalar harmonics, we can illustrate the $\frak{so}(D)$ content of the space \eqref{spacevecjee} as follows:
\be \raisebox{-10pt}{\epsfig{file=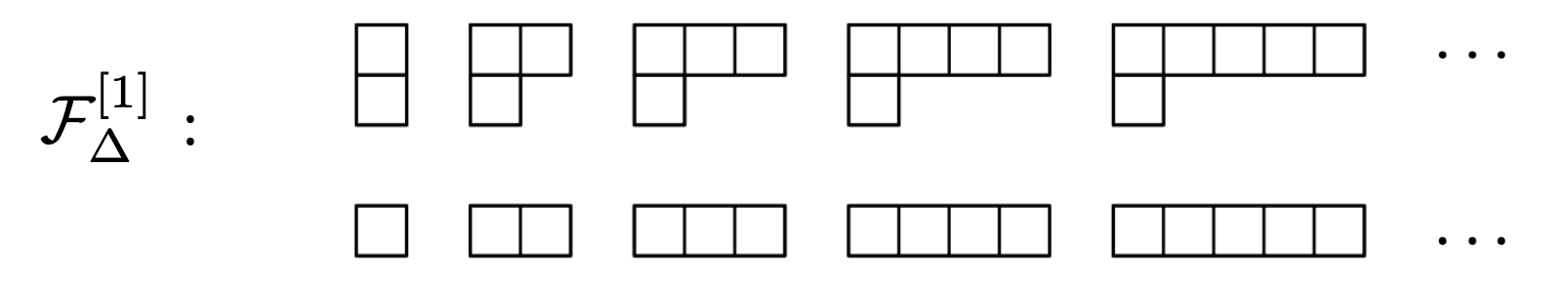,width=4.2in}}  \label{vectorsocontent}\, \ee
The top row in \eqref{vectorsocontent} consists of the reps of the transverse vector spherical harmonics \eqref{vecotrspherharmonicese} making up the transverse part $\chi_i$, and the bottom row consists of the reps of the scalar spherical harmonics \eqref{scalarphserarmoe}, minus the zero mode, making up the longitudinal part $\chi$.

Under the action of ${\cal K}^I$, a given $\frak{so}(D)$ rep will transform into a combination of others with either one box removed or one box added.  We can thus define operators, analogous to the ${\cal K}^{I \pm}$ ladder operators in the scalar case, that move us among the $\frak{so}(D)$ reps.  The actions form a lattice of nearest neighbor interactions, as illustrated here:
\be \raisebox{-10pt}{\epsfig{file=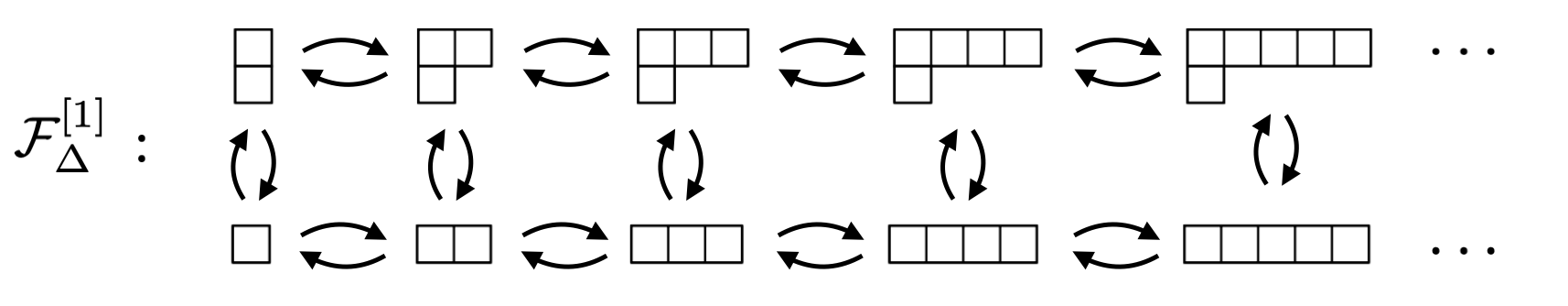,width=4.2in}}  \label{vectorsocontent2}\, \ee
Each arrow in this picture stands for an operator defined as follows: starting with the spherical harmonic at the base of the arrow, act with $\delta_{{\cal K}^I}$ using \eqref{latetimesheactiont3}, then project using the Young projector (described in section \ref{conventionssec}) corresponding to the Young diagram at the end of the arrow.  If the diagram at the end of the arrow has one fewer box than the diagram at the base, then we are contracting the index of ${\cal K}^I$ with the index of the spherical harmonic corresponding to the box that is removed, before projecting.  If the diagram at the end of the arrow has one more box than the diagram at the base, then we are keeping the index of ${\cal K}^I$ distinct from the indices of the spherical harmonic before projecting, with the index of ${\cal K}^I$ corresponding to the index of the added box.  Any other way of acting with ${\cal K}^I$ is either trivial or equivalent to one of the arrows shown.

\textbf{Reducible cases:} For generic $\Delta$, none of the actions represented by any of the arrows will give zero, and so we can move from any given $\frak{so}(D)$ rep to any other, and the full $\frak{so}(1,D)$ rep is irreducible.  But there are specific discrete values of $\Delta$ where some of the arrow operators shown above give zero, and the corresponding pathways are blocked.  In these cases, the reps ${\cal F}_{\Delta}^{[1]}$ are reducible but not decomposable.  They occur as follows:
\begin{itemize}

\item  Shift symmetric points: 
\be \Delta=d+k+1\, ,\ \  k=0,1,2,\ldots \ \ .\ee  
The arrows leading to the left from the $(k+2)$-th column in \eqref{vectorsocontent2} to the $(k+1)$-th column vanish, and so the columns from the $(k+2)$-th onward form an infinite dimensional sub-rep which we call 
\be {\cal D}_{d+k+1}^{[1]}\, .\ee

This is illustrated here for the case $k=2$, with the sub-rep enclosed in the grey boundary:
\be \raisebox{-10pt}{\epsfig{file=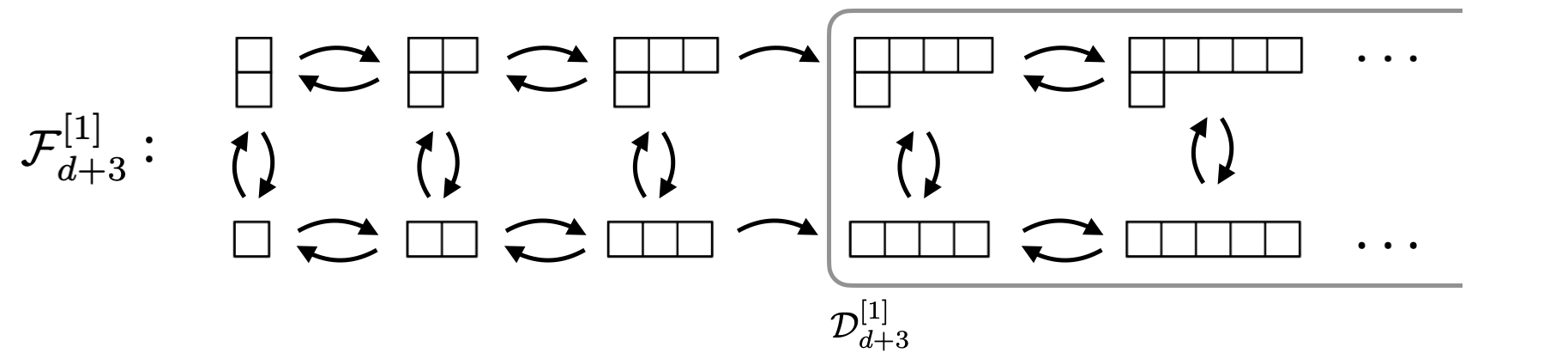,width=4.4in}}  \label{vectorsocontent4}\, \ee

We call these the shift symmetric reps; they correspond to the physical modes of the level $k$ shift symmetric vector field \cite{Bonifacio:2018zex,Bonifacio:2019hrj}.

\item Finite points:
\be \Delta=-k-1\, ,\ \  k=0,1,2,\ldots\ \ .\ee 
The arrows leading to the right from the $(k+1)$-th column in \eqref{vectorsocontent2} to the $(k+2)$-th column vanish, and so the first $k+1$ columns form a finite dimensional sub-rep which we call 
\be {\cal S}_{-k-1}^{[1]}\, .\ee

This is illustrated here for the case $k=2$, with the sub-rep enclosed in the grey boundary:
\be \raisebox{-10pt}{\epsfig{file=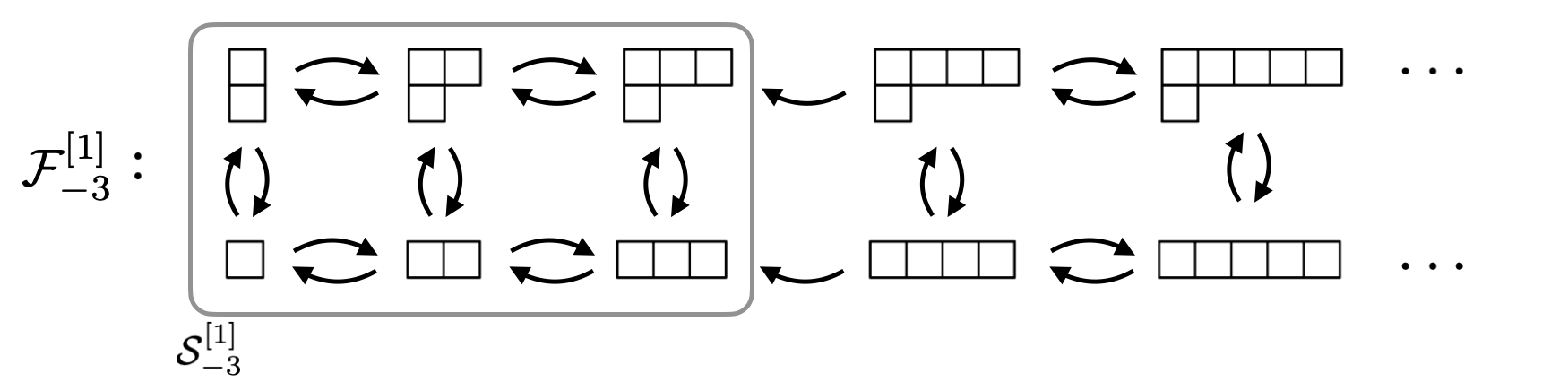,width=4.4in}}  \label{vectorsocontent3}\, \ee

The rep ${\cal S}_{-k-1}^{[1]}$ is nothing other than the $[k+1,1]$ tensor rep of $\frak{so}(1,D)$.  Note that the tableau for this tensor appears in the upper right corner of the enclosed grey box indicating ${\cal S}_{-k-1}^{[1]}$.  It and the other $\frak{so}(D)$ reps present within the box are obtained by branching $[k+1,1]$ from $\frak{so}(1,D)$ to $\frak{so}(D)$ using the branching rules in appendix \ref{branchingappendix}.  

We call these the finite reps. They are the shift symmetries of the level $k$ shift symmetric vector; indeed the shift symmetry is parametrized by a $[k+1,1]$ tensor in the dS's embedding space \cite{Bonifacio:2018zex}.

\item Massless point:
\be \Delta=d-1\,.\ee 
For this value, the arrows leading from the top row to the bottom row in \eqref{vectorsocontent2} vanish, so that the top row forms an irreducible sub-rep that we call 
\be {\cal V}_{d-1}^{[1]}\, .\ee

This is illustrated here, with the sub-rep enclosed in the grey boundary:
\be \raisebox{-10pt}{\epsfig{file=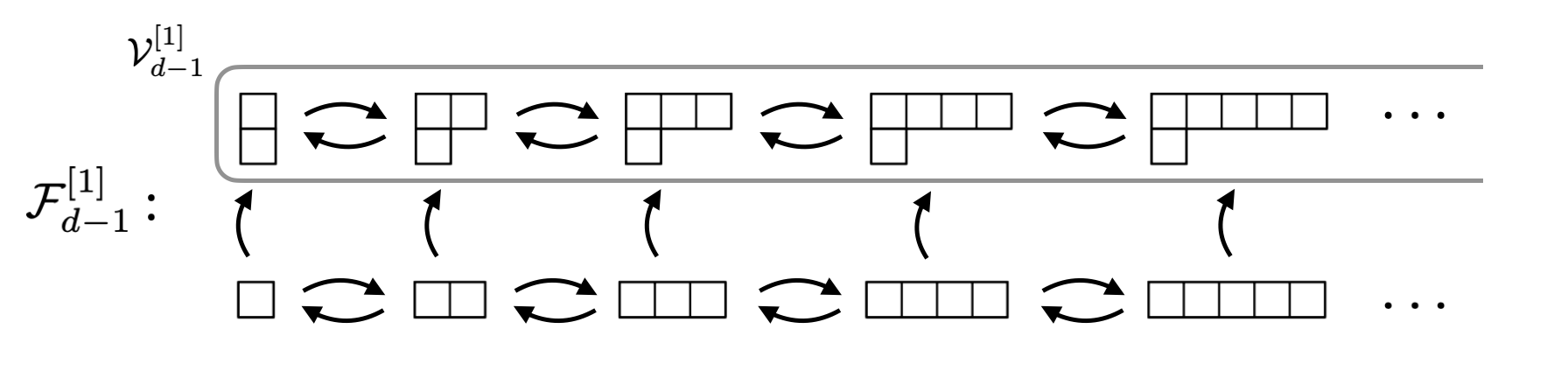,width=4.4in}}  \label{vectorsocontent5}\, \ee

We call this the massless rep.  From \eqref{vectmassrelationere}, it corresponds to the case $m=0$ where the vector field on dS gets a gauge symmetry, and it gives the physical modes of the massless vector field. 

\item Gauge point:
\be \Delta=1\,.\ee 
For this value, the arrows leading from the bottom row to the top row in \eqref{vectorsocontent2} vanish, so that the bottom row forms an irreducible sub-rep that we call 
\be {\cal U}_{1}^{[1]}\, .\ee

This is illustrated here, with the sub-rep enclosed in the grey boundary:
\be \raisebox{-10pt}{\epsfig{file=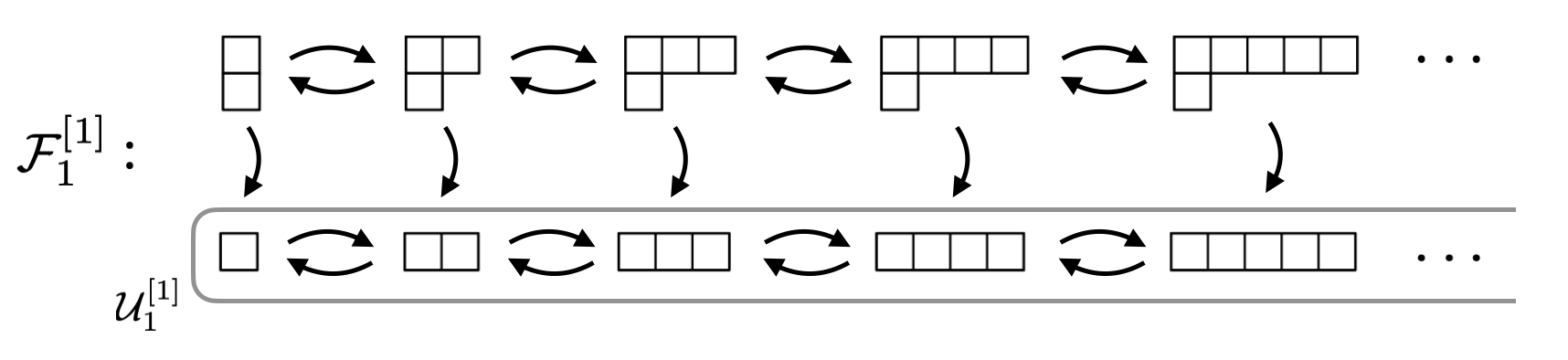,width=4.4in}}  \label{vectorsocontent6}\, \ee

We call this the gauge rep.  It corresponds to the pure gauge modes of the $m=0$ vector field. 
\end{itemize}

\textbf{Equivalences:} As with the scalar reps, there is an intertwining operator, the shadow transform, that connects the $\Delta$ and $\bar\Delta\equiv d-\Delta$ vector reps, 
\be S_\Delta^{[1]} :\ {\cal F}_{\Delta}^{[1]}\rightarrow {\cal F}_{\bar\Delta}^{[1]}\,.\label{shadcovee}\ee
It commutes with the $\frak{so}(D)$ rotations and satisfies $\delta_{{\cal K}^I_{\bar\Delta}}S_\Delta^{[1]}=S_\Delta^{[1]} \delta_{{\cal K}^I_{\Delta}}$.
For generic $\Delta$, it is invertible and the $\Delta$ and $\bar\Delta$ reps are equivalent to each other,
\be {\cal F}_{\Delta}^{[1]}\simeq {\cal F}_{\bar\Delta}^{[1]}\,.\label{vecequnfee}\ee
 But for the special values of $\Delta$ given above where ${\cal F}_{\Delta}^{[1]}$ develops a sub-rep, $S_\Delta^{[1]}$ develops a kernel which is always precisely this sub-rep: the kernel of $S_{d+k+1}^{[1]}$ is ${\cal D}_{d+k+1}^{[1]}$, the kernel of $S_{-k-1}^{[1]}$ is ${\cal S}_{-k-1}^{[1]}$, the kernel of $S_{d-1}^{[1]}$ is ${\cal V}_{d-1}^{[1]}$, and the kernel of $S_{1}^{[1]}$  is ${\cal U}_{1}^{[1]}$.  Similarly, the image of these maps is the sub-rep of the target space:  the image of $S_{d+k+1}^{[1]}$ is ${\cal S}_{-k-1}^{[1]}$, the image of $S_{-k-1}^{[1]}$ is ${\cal D}_{d+k+1}^{[1]}$, the image of $S_{d-1}^{[1]}$ is ${\cal U}_{1}^{[1]}$, and the image of $S_{1}^{[1]}$  is ${\cal V}_{d-1}^{[1]}$.

Between the shift symmetric and finite points, these maps induce the isomorphisms
\be {\cal S }^{[1]}_{-k-1}\simeq{\cal F }^{[1]}_{d+k+1}/{\cal D }^{[1]}_{d+k+1}\, ,\ \ \ {\cal D }^{[1]}_{d+k+1}\simeq{\cal F }^{[1]}_{-k-1}/{\cal S }^{[1]}_{-k-1}\, . \ee
This is illustrated here for the case $k=1$:
\be \raisebox{-0pt}{\epsfig{file=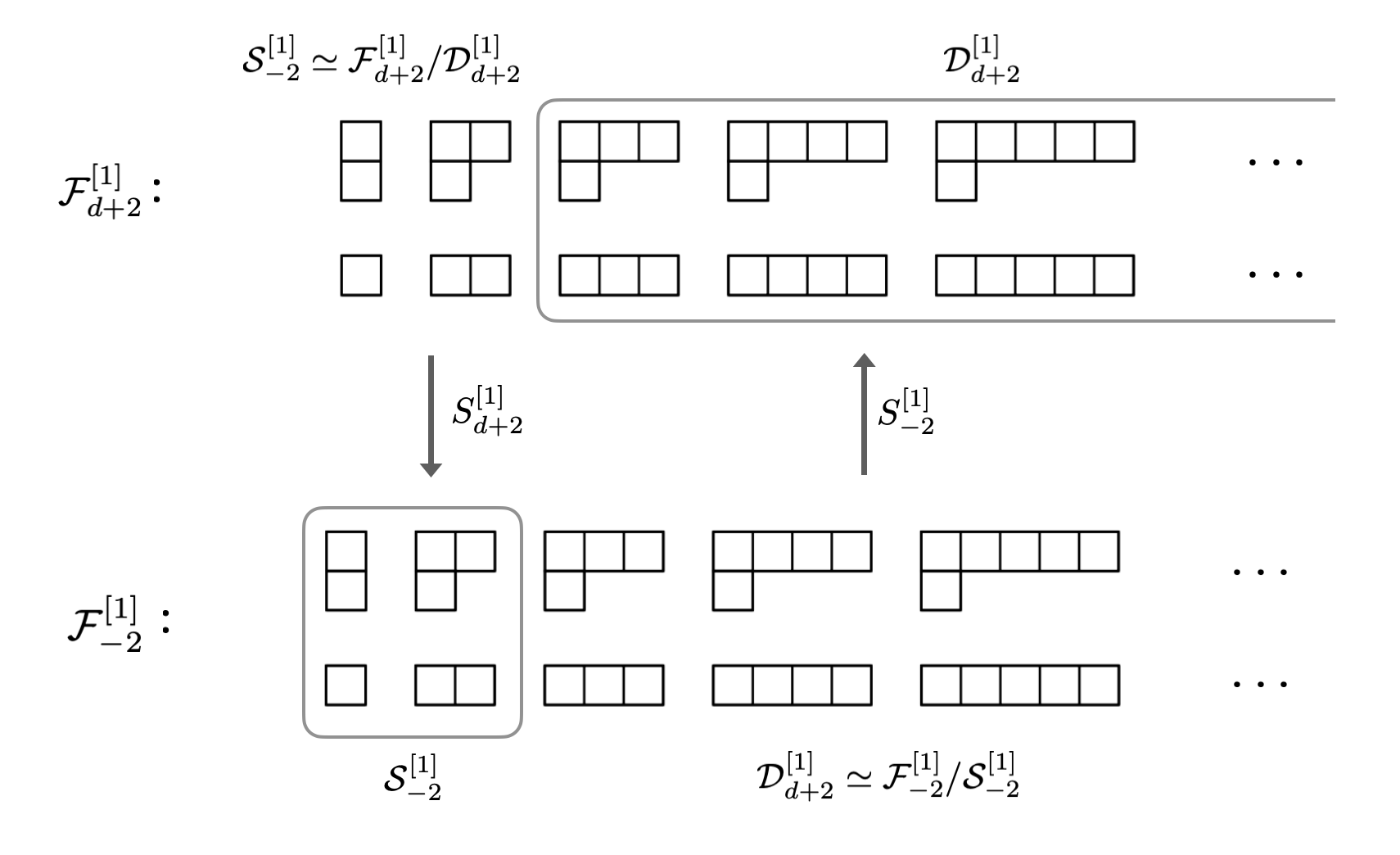,width=4.4in}}  \label{vectorsocontent7}\, \ee
Between the massless and gauge points, these maps induce the isomorphisms
\be {\cal U}^{[1]}_{1}\simeq {\cal F}^{[1]}_{d-1}/{\cal V}^{[1]}_{d-1}\, ,\ \ \ {\cal V}^{[1]}_{d-1}\simeq {\cal F}^{[1]}_{1}/{\cal U}^{[1]}_{1}\,. \ee
This is illustrated here:
\be \raisebox{-10pt}{\epsfig{file=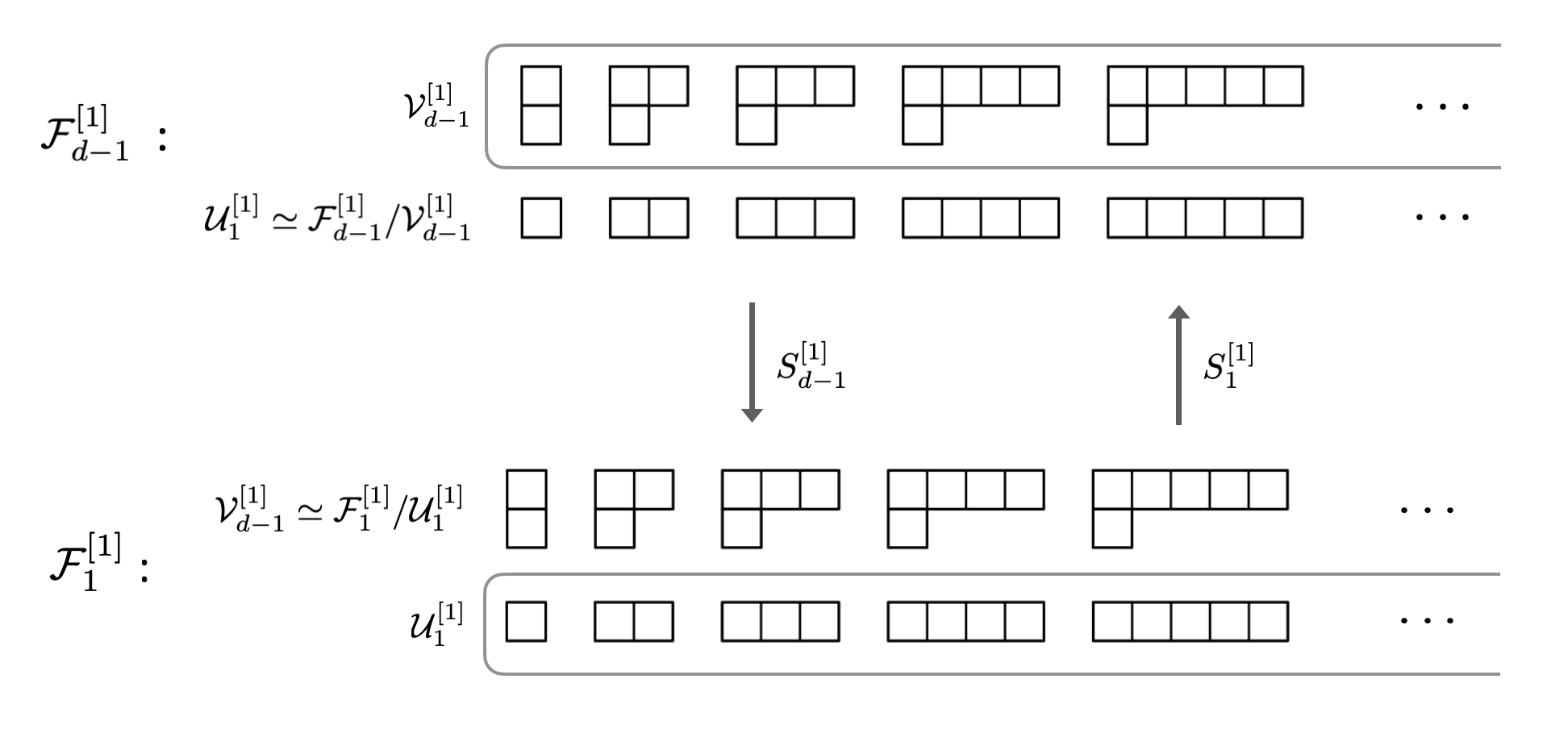,width=4.4in}}  \label{vectorsocontent8}\, \ee
In the pictures \eqref{vectorsocontent7},\eqref{vectorsocontent8}, a grey box encloses the sub-rep, which is simultaneously the image of the incoming map and the kernel of the outgoing map.

There are two additional maps that are of interest.  These are maps between vector reps and scalar reps.  The first is the gradient, which is an equivariant map between the following scalar and vector reps,
\be {\rm grad}: \  {\cal F}^{[0]}_{0} \rightarrow {\cal F}^{[1]}_{1} \, ,\ \ \phi\rightarrow \nabla_i\phi\,.
\ee
If $\phi$ transforms under ${\cal K}^I$ with weight $\Delta=0$, then $\nabla_i\phi$ transforms correctly with weight $\Delta=1$, so the gradient map is equivariant and adds one unit of $\Delta$.
The kernel of the gradient is the space of constant functions, i.e. the $l=0$ spherical harmonic, which is the sub-rep ${\cal S}^{[0]}_{0}$.  The image is precisely the lower row of \eqref{vectorsocontent}, which is the sub-rep ${\cal U}^{[1]}_{1}$

The other map of interest is the divergence, which is an equivariant map between the following vector and scalar reps,
\be {\rm div}: \  {\cal F}^{[1]}_{d-1} \rightarrow {\cal F}^{[0]}_{d} \, ,\ \ \phi_i \rightarrow \nabla^i\phi_i\,.
\ee
Acting between these spaces, the divergence preserves the transformation properties and raises the value of $\Delta$ by one.  Its kernel is the space of transverse vectors, which is precisely the top row of \eqref{vectorsocontent}, and is the sub-rep ${\cal V}^{[1]}_{d-1}$.  Its image is the space spanned by the $l>0$ spherical harmonics, which is the sub-rep ${\cal D}^{[0]}_{d}$.

The four spaces ${\cal F}^{[1]}_{d-1}$, ${\cal F}^{[1]}_{1}$, ${\cal F}^{[0]}_{d}$, ${\cal F}^{[0]}_{0}$ can be joined together into the following commutative diagram:
\be \raisebox{-20pt}{\epsfig{file=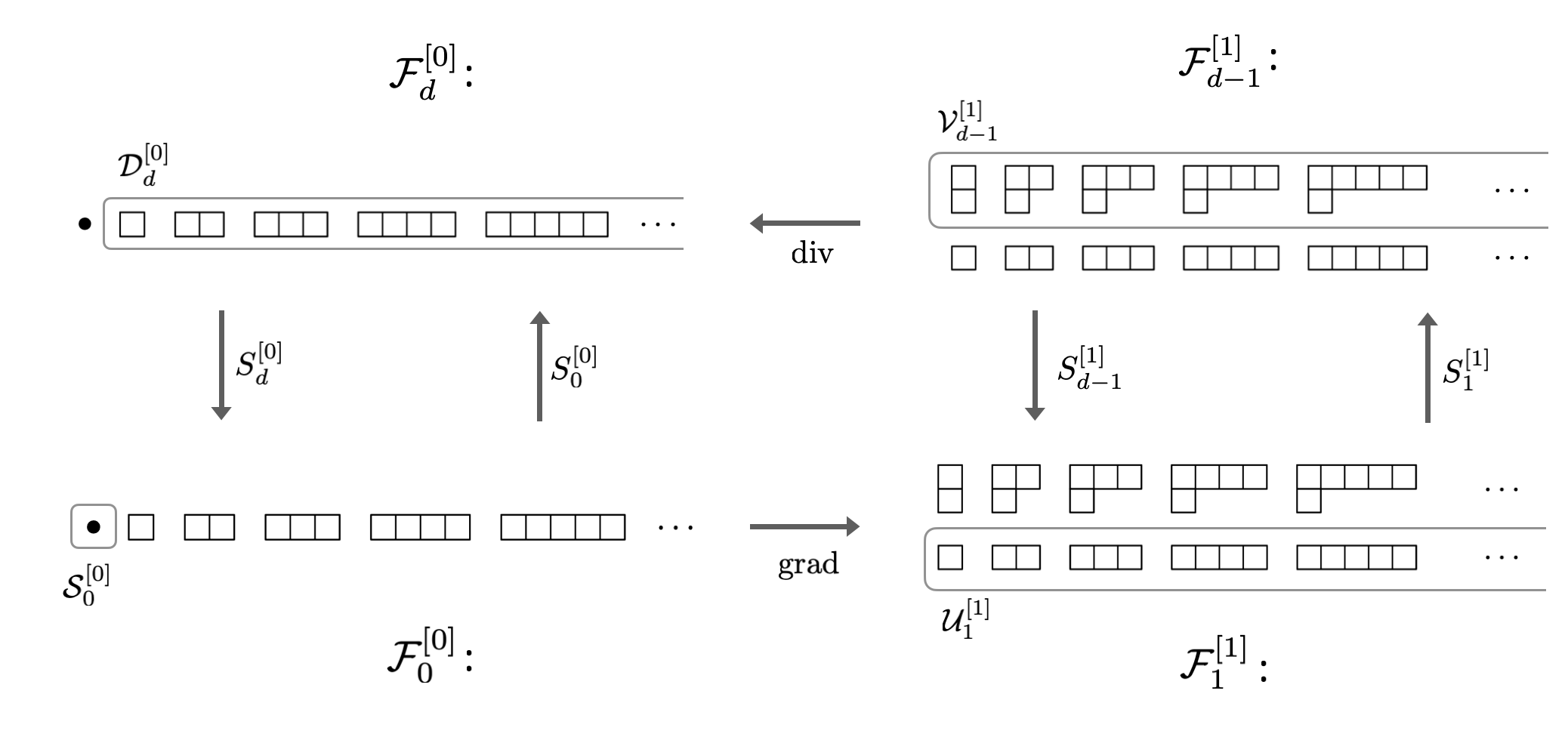,width=6.1in}}  \label{vectorsocontent9}\, \ee
The grey squares enclose the various sub-reps; these are simultaneously the kernel of every outgoing map, and the image of every ingoing map, and going through any two arrows consecutively gives zero.  From this picture, we get the isomorphism
\be { \cal D}^{[0]}_d\simeq { \cal U}^{[1]}_1\, .\label{vecshirdgahee}\ee
This expresses the statement that the gauge modes of a massless photon are precisely the physical modes of a $k=0$ shift symmetric (i.e. massless) scalar.

\textbf{Unitarity:}  There is a diffeomorphism and Weyl invariant\footnote{In the Weyl transformations we must remember to use the Weyl weight $\Delta_W=\Delta-1$, as indicated by \eqref{weylnormdrele}.  The difference between $\Delta_W$ and $\Delta$ is absorbed by the transformation of the inverse metric that is used to contract the two vectors in \eqref{bilinearformvee}.}, and thus conformally invariant, bilinear form pairing the two spaces ${\cal F}_\Delta^{[1]}$ and ${\cal F}_{\bar \Delta}^{[1]}$,
\be (\phi_1,\phi_2)\equiv \int d^d\Omega\, \phi_1^i(\hat X)\phi_{2\, i}(\hat X)\,,\ \ \ \  \phi_{1\, i}\in {\cal F}_{\bar\Delta}^{[1]},\ \ \phi_{2\, i}\in {\cal F}_{\Delta}^{[1]}\, . \label{bilinearformvee}\ee

In the case where $\Delta^\ast=\bar\Delta$ we can use this to make a manifestly positive definite, invariant inner product on ${\cal F}_\Delta^{[1]}$ via:
\be \la \phi_1 | \phi_2 \ra\equiv (\phi_1^{\ast},\phi_2)\, ,\ \ \ \Delta^\ast=\bar\Delta\, ,\ \ \ \phi_{1\, i},\phi_{2\, i}\in {\cal F}_\Delta^{[1]} \,. \label{innerprodvprise} \ee
This gives the vector principal series reps, which are all unitary:
\be {\rm vector\ principal\ series:}\  \Delta={d\over 2}+i\nu\,  ,\ \ \ \nu\in {\mathbb R}\,.\ee

In the case where $\Delta$ is real, we can use $S^{[1]}_{\Delta}$ to move a state from ${\cal F}_{ \Delta}^{[1]}$ to ${\cal F}_{\bar \Delta}^{[1]}$ and form an inner product on ${\cal F}_{ \Delta}^{[1]}$ as follows,
\be \la \phi_1 | \phi_2 \ra\equiv (S^{[1]}_{\Delta}\phi_1^\ast,\phi_2)\, ,\ \ \ \Delta\in {\mathbb R}\, ,\ \ \   \ \ \phi_{1\, i},\phi_{2\, i}\in {\cal F}_\Delta^{[1]}\,. \label{innerprodpriscsvdee} \ee 
The positivity of this inner product is now equivalent to whether the matrix elements of $S^{[1]}_{\Delta}$ are positive for all the different $\frak{so}(D)$ reps.  Away from the discrete special cases described above, it turns out that they are positive only in the range
\be {\rm vector\ complementary\ series:}\  1<\Delta<d-1\,. \label{veccomprangee}\ee
The point $\Delta={d/ 2}$ intersects the principal series, and,  as in the scalar case, we will remain agnostic as to which series to assign this point to until we get to section \ref{unitarylistsection}.
The vector principal and complementary series reps with $\Delta$ and $d-\Delta$ are equivalent due to \eqref{vecequnfee}.  Other than this equivalence, they are all distinct.

Now consider the cases of $\Delta$ where the reps become reducible.  For the finite points $\Delta=-k-1$, $k=0,1,2,\ldots$, we use $S^{[1]}_{-k-1}$ in the inner product \eqref{innerprodpriscsvdee}, and this has the finite dimensional kernel consisting of the sub-rep ${\cal S}^{[1]}_{-k-1}$.   All of these states will have zero norm in the inner product.  Apart from these null states, other states also have negative norm, so even once these null states are factored out, we will be left with non-unitary reps.  The rep obtained after this factoring is nothing but ${\cal D}^{[1]}_{d+k+1}$, realized through the quotient ${\cal D}^{[1]}_{d+k+1}\simeq {\cal F}^{[1]}_{-k-1}/{\cal S}^{[1]}_{-k-1}$,  so these are non-unitary reps. They correspond to the shift symmetric vectors (which were seen to be non-unitary using field theoretic methods in \cite{Bonifacio:2018zex}).

For the shift symmetric points $\Delta=d+k+1$, $k=0,1,2,\ldots$, we use $S^{[1]}_{d+k+1}$ in the inner product \eqref{innerprodpriscsvdee}, and this has the infinite dimensional kernel consisting of the states of the sub-rep ${\cal D}^{[1]}_{d+k+1}$.   All of these states will therefore be null in the inner product.  Apart from these null states, some of the norms are always still negative, so if the null states are factored out, we will be left with a finite dimensional non-unitary rep.  This is the rep ${\cal S}^{[1]}_{-k-1}$, realized through the quotient ${\cal S}^{[1]}_{-k-1}\simeq  {\cal F}^{[1]}_{d+k+1}/{\cal D}^{[1]}_{d+k+1}$.  

For the gauge point $\Delta=1$, we use $S^{[1]}_{1}$ in the inner product \eqref{innerprodpriscsvdee}, and this has as kernel the sub-rep ${\cal U}^{[1]}_{1}$.   All of these states will have zero norm in the inner product.  Apart from these null states, the other states all have the same sign norm, which can be taken to be positive by adjusting an overall sign if necessary.  So once the null states are factored out, we will be left with a unitary rep, which is the rep ${\cal V}^{[1]}_{d-1}$, realized through the quotient ${\cal V}^{[1]}_{d-1}\simeq {\cal F}^{[1]}_{1}/{\cal U}^{[1]}_{1}$.  These correspond to the physical states of the photon.

For the massless point $\Delta=d-1$, we use $S^{[1]}_{d-1}$ in the inner product \eqref{innerprodpriscsvdee}, and this has as kernel the sub-rep ${\cal V}^{[1]}_{d-1}$.   All of these states will have zero norm in the inner product.  Apart from these null states, the other states all have the same sign norm which can be taken to be positive by adjusting an overall sign if necessary, so once the null states are factored out, we will be left with a unitary rep, which is nothing but ${\cal U}^{[1]}_{1}\simeq {\cal D}^{[0]}_{d} $, realized through the quotient ${\cal U}^{[1]}_{1}\simeq {\cal F}^{[1]}_{d-1}/{\cal V}^{[1]}_{d-1}$.  These correspond to the longitudinal gauge modes of the photon, which are equivalent to the shift symmetric $k=0$ scalar rep.

We will see later in section \ref{unitarylistsection} that the unitary reps occurring among the reducible points are classified among the so-called exceptional series reps.  For all but the unitary cases discussed here, there is no other way to construct an invariant positive definite inner product on ${\cal F}_\Delta^{[1]}$, so all the other reps in the complex $\Delta$ plane are non-unitary.  

\textbf{Summary:} 
The vector reps are summarized here in the complex $\Delta$ plane:
\be \raisebox{-40pt}{\epsfig{file=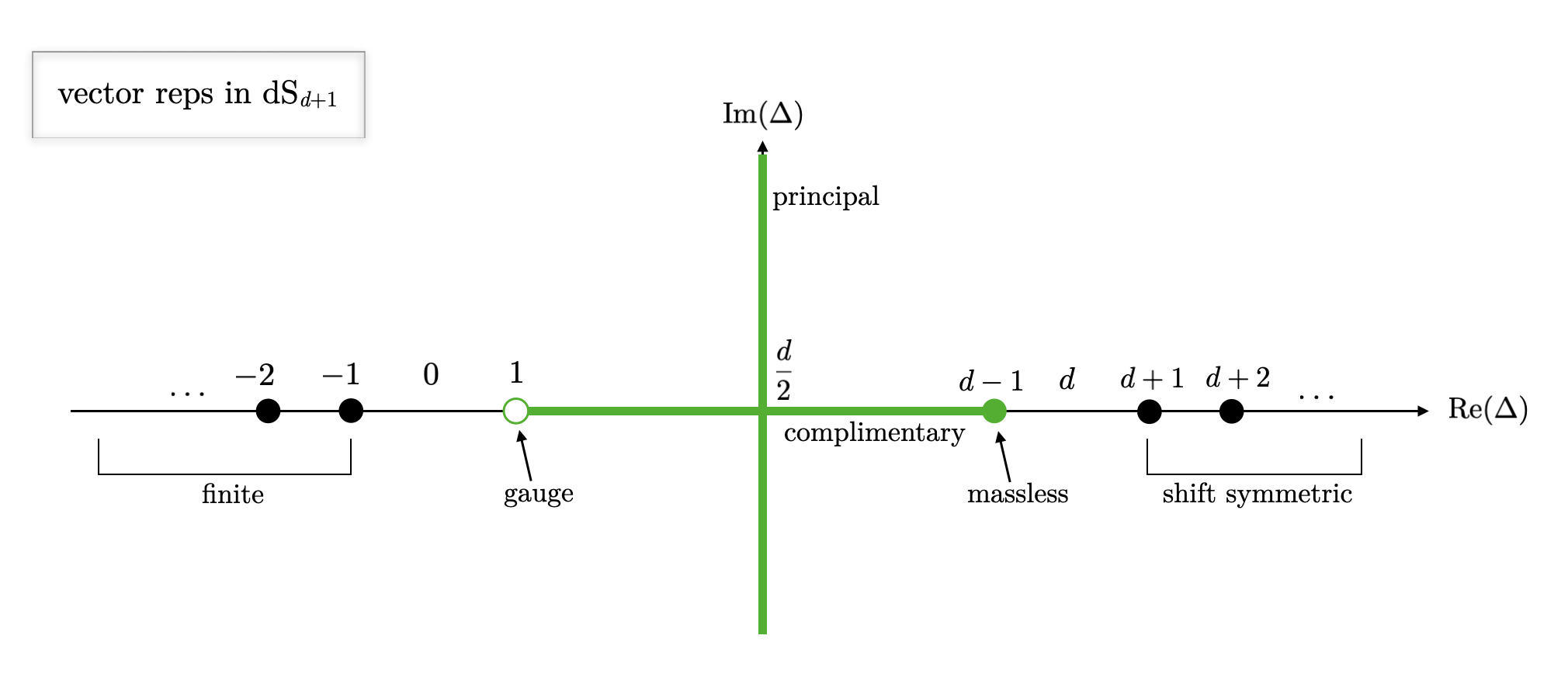,width=6.4in}}  \label{dsreps2}\, \ee
 Points in green are unitary reps.  The dots are the special points where the rep becomes reducible. The dot at $\Delta=1$ is hollow to indicate that it is equivalent to the $\Delta=d$ scalar rep through the equivalence \eqref{vecshirdgahee}, and is thus already accounted for among the scalar reps shown in \eqref{dsreps1}.
 The equivalence \eqref{vecequnfee} of the non-special reps is given by reflecting through the point $\Delta=d/2$, and due to the breakdown of the intertwiner map \eqref{shadcovee}, the special reps with dots are inequivalent despite the reflection.  Other than this equivalence, all the reps are distinct .

The value of the quadratic Casimir operator ${\cal C}_2$ of section \ref{casimirsec} on the vector reps is given by
\be {\cal C}_2= (\Delta-1)(\Delta-d+1)= -{m^2\over H^2}\, .\ee

We have the following correspondence between vector fields on dS$_{D}$ and the vector unitary reps:
\bea \begin{cases}
 {\rm principal:\ }\quad  \Delta={d\over 2}+i\nu \, , \ \ \nu\in {\mathbb R}\, , & {\rm heavy \ vectors:\ } {m^2\over H^2 }\geq {(D-3)^2\over 4}\, , \\
 {\rm complementary:\ } \quad 1<\Delta<d-1 \, , & {\rm light \ vectors:\ } 0< {m^2\over H^2 }\leq  {(D-3)^2\over 4}\, , \\
  {\rm massless:\ } \quad \Delta=d-1 \, , &  {\rm massless \ vector:\ } m^2=0\,.  
\end{cases} \nn
\eea
The shift symmetric vectors have tachyonic masses with $m^2<0$, but, unlike the scalars, this renders them non-unitary.   This is because a tachyonic mass implies ghost-like instabilities in the longitudinal mode that are not removed upon gauging the shift symmetry, as can be seen from e.g. the St\"uckelberg formulation \cite{Hinterbichler:2011tt}.

The vector reps have some additional subtleties that occur in the lower dimensions $D=3,4$, which we turn to next ($D=2$ is reserved for section \ref{D2section}).

\subsubsection*{$D=3$:}

 Consider, in $D=3$, the $\frak{so}(3)$ content of the vector rep for generic $\Delta$ depicted in \eqref{vectorsocontent}.  We can use the three dimensional epsilon tensor $\epsilon_{IJK}$ to dualize all the $\frak{so}(3)$ reps in the top row into single row reps, by contracting two of the indices of $\epsilon_{IJK}$ with the anti-symmetric index pair as described in appendix \ref{sorepsappendix}.  Once this is done, the top row looks identical to the bottom row.  Then, after taking a suitable linear combination of the two rows, it can be seen that they split apart into two separate irreducible reps, which we call ${\cal F}^{[1]_\pm}_\Delta$, 
\be {\cal F}^{[1]}_\Delta={\cal F}^{[1]_+}_\Delta\oplus {\cal F}^{[1]_-}_\Delta,\ \ D=3\,,\label{d3vecsplite}\ee
so ${\cal F}^{[1]}_\Delta$ in $D=3$ is in fact reducible and decomposable into two irreducible pieces.  
This is illustrated here:
\be \raisebox{-0pt}{\epsfig{file=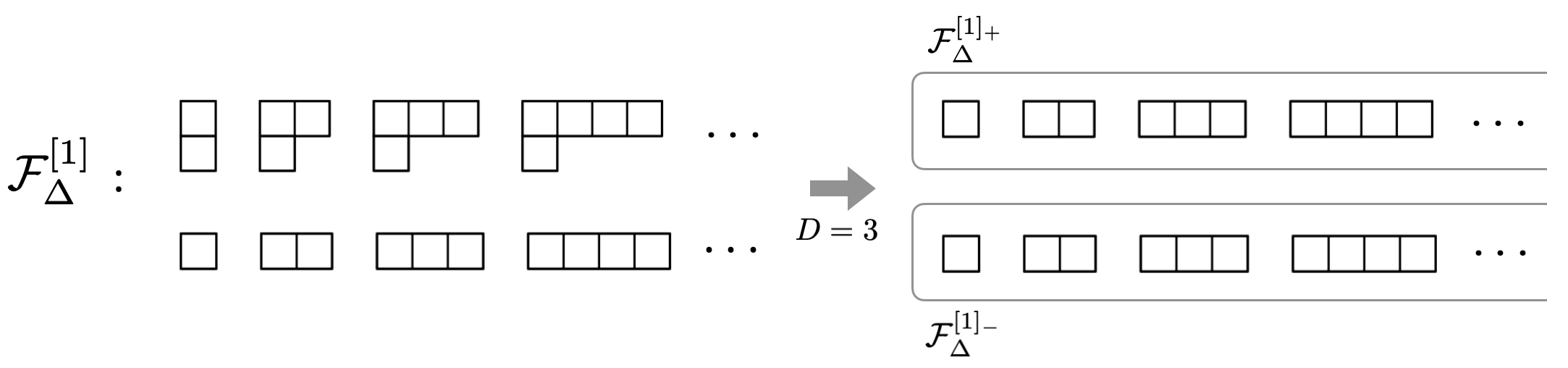,width=5.5in}}  \label{vectorsocontent10}\, \ee
The arrow indicates dualizing the top row as well as taking appropriate linear combinations of the two rows so that they separate under the action of ${\cal K}^I$.  Both ${\cal F}^{[1]_\pm}_\Delta$ contain the same $\frak{so}(3)$ content $[1],[2],[3],\ldots$, but they form distinct $\frak{so}(1,3)$ reps.

In $D=3$, the representation space of ${\cal F}^{[1]}_\Delta$ is the space of vectors on the 2-sphere.  The split \eqref{d3vecsplite} is nothing but the fact that this space can be broken up into (imaginary) self-dual and anti-self-dual vectors with respect to the $d=2$ epsilon tensor $\epsilon_{ij}$ on the sphere.  This reflects the fact that massive spin 1 fields on dS$_3$ come in two independent chiralities (just as on flat space, where the massive little group is $U(1)$ and massive fields have independent positive and negative chiralities with respect to it).  These reps are realized field theoretically by Chern-Simons terms that split the masses between the two chiralities \cite{Deser:1981wh}.  

The same splitting also occurs for the shortened reps ${\cal D}_{k+3}^{[1]}$ corresponding to the shift symmetric fields,
\be{\cal D}_{k+3}^{[1]}={\cal D}_{k+3}^{[1]_+}\oplus {\cal D}_{k+3}^{[1]_-}\, ,\ \ D=3\,,\label{d3vecsplite2}\ee
illustrated here for $k=1$,
\be \raisebox{-0pt}{\epsfig{file=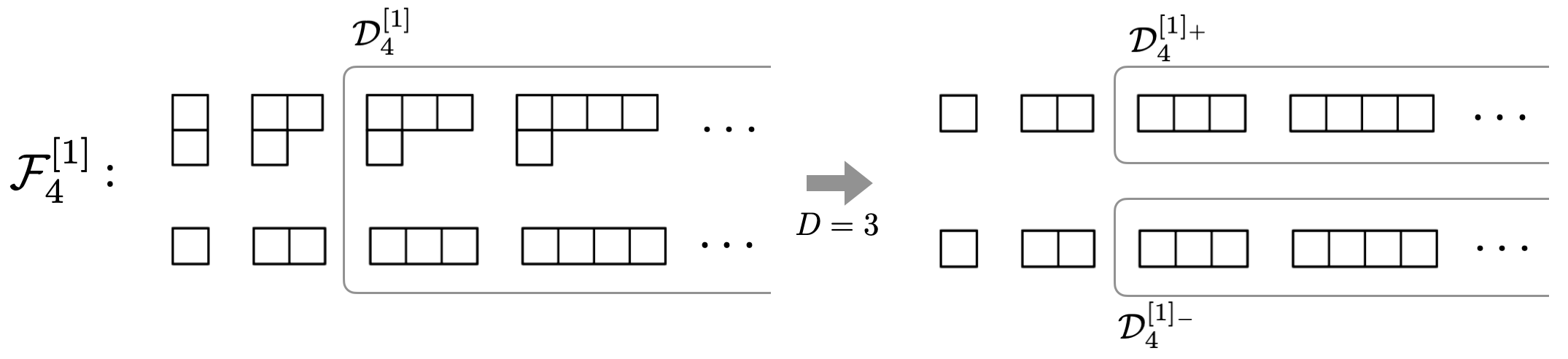,width=5.5in}}  \label{vectorsocontent11}\, \ee
It also occurs for the finite dimensional reps ${\cal S}_{-k-1}^{[1]}$ corresponding to the shift symmetries,
\be {\cal S}_{-k-1}^{[1]}={\cal S}_{-k-1}^{[1]_+} \oplus  {\cal S}_{-k-1}^{[1]_-} \, ,\ \ D=3\,,\label{d3vecsplite22}\ee
illustrated here for $k=1$,
\be \raisebox{-0pt}{\epsfig{file=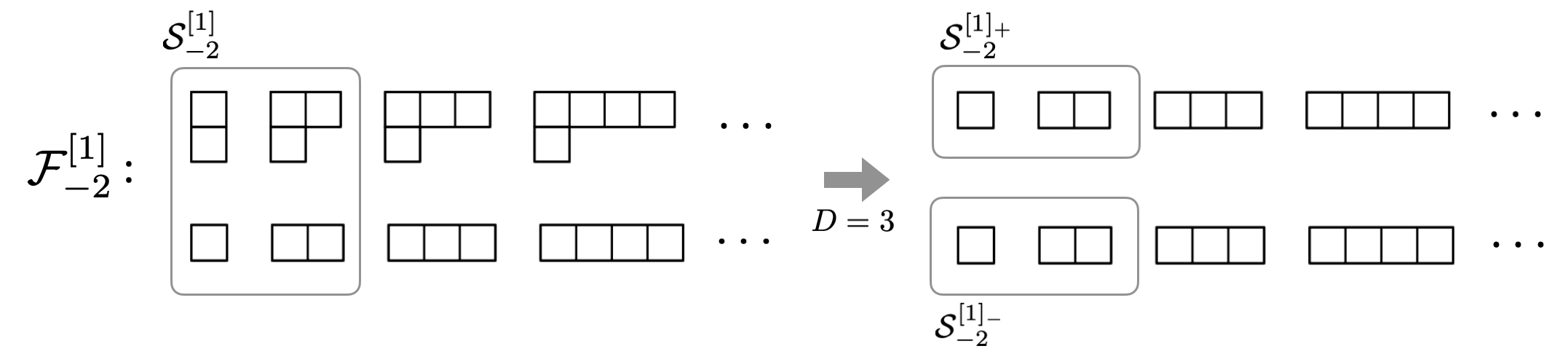,width=5.5in}}  \label{vectorsocontent12}\, \ee
These are the finite dimensional chiral reps $[k+1,1]_\pm$ of $\frak{so}(1,3)$, which when branched to $\frak{so}(3)$ using \eqref{evenDbranchinge} each give the reps $[1],[2],\ldots,[k+1]$.

The shadow transform that relates $\Delta$ and $\bar \Delta$ flips the chirality (this is the action of the so-called ``restricted Weyl group'' \cite{8cb4b67c-34de-3625-831e-f71092751ec0,Knapp1980,Dobrev:2012mea,Kravchuk:2018htv}), and so we have
\be {\cal F}^{[1]_+}_\Delta\simeq {\cal F}^{[1]_-}_{\bar \Delta}\, , \ \  D=3\,. \label{F1d3ejdxe} \ee
Note that this means that at $\Delta=d/2=1$, the two chiral reps are equivalent, ${\cal F}^{[1]+}_1\simeq {\cal F}^{[1]-}_1$.  At the shift symmetric and finite points, this flip in chirality tells us that we have
\be  {\cal D}^{[1]_\pm }_{k+3} \simeq {\cal F}^{[1]_\mp}_{-k-1}/{\cal S}^{[1]_\mp}_{-k-1}\,, \ \ {\cal S}^{[1]_\pm }_{-k-1} \simeq {\cal F}^{[1]_\mp}_{k+3}/{\cal D}^{[1]_\mp}_{k+3} \,,\ \ D=3\,.\ee 

In $D=3$, the range \eqref{veccomprangee} degenerates and there is no complementary series for the vector.  The point $\Delta=d-1$, which would have been where the complementary series ends at the massless rep, is also the point $\Delta=d/2$ which is where the complementary series intersects with the principal series.  

Turning now to the massless rep, if we look at the top row sub-rep ${\cal V}^{[1]}_1$ of ${\cal F}^{[1]}_1$ as illustrated in \eqref{vectorsocontent5}, we can dualize this top row and obtain a rep which is identical to that of the $k=0$ shift symmetric scalar, ${\cal D}^{[0]}_{2}$, and we have the isomorphism ${\cal V}^{[1]}_1\simeq {\cal D}^{[0]}_{2}$.
This reflects the fact that a massless vector in $D=3$ is dual to a massless ($k=0$) scalar, with the scalar's shift symmetry gauged.  From \eqref{vecshirdgahee}, the rep ${\cal U}^{[1]}_1$ representing the gauge modes is also isomorphic to ${\cal D}^{[0]}_{2}$, so in $D=3$ we see that the physical and gauge modes of a massless vector both carry the same rep.   Note that the point in \eqref{vectorsocontent5} is the same as the point in \eqref{vectorsocontent6} when $D=3$, and so in this case the massless vector modes and gauge modes actually split off into a direct sum of (equivalent) reps, so the rep is reducible and decomposable, rather than the factorizations and invariant subspaces (i.e. reducible but not decomposable), that occur in general $D$.  In summary, we have 
\be {\cal F}_1^{[1]_+}\simeq {\cal F}_1^{[1]_-} \simeq {\cal V}_1^{[1]}\simeq {\cal U}_1^{[1]} \simeq {\cal D}_2^{[0]}\,,\ \ \ D=3\,.\ee  
This degeneration and splitting of the massless and gauge rep in $D=3$ is illustrated here
\be \raisebox{-0pt}{\epsfig{file=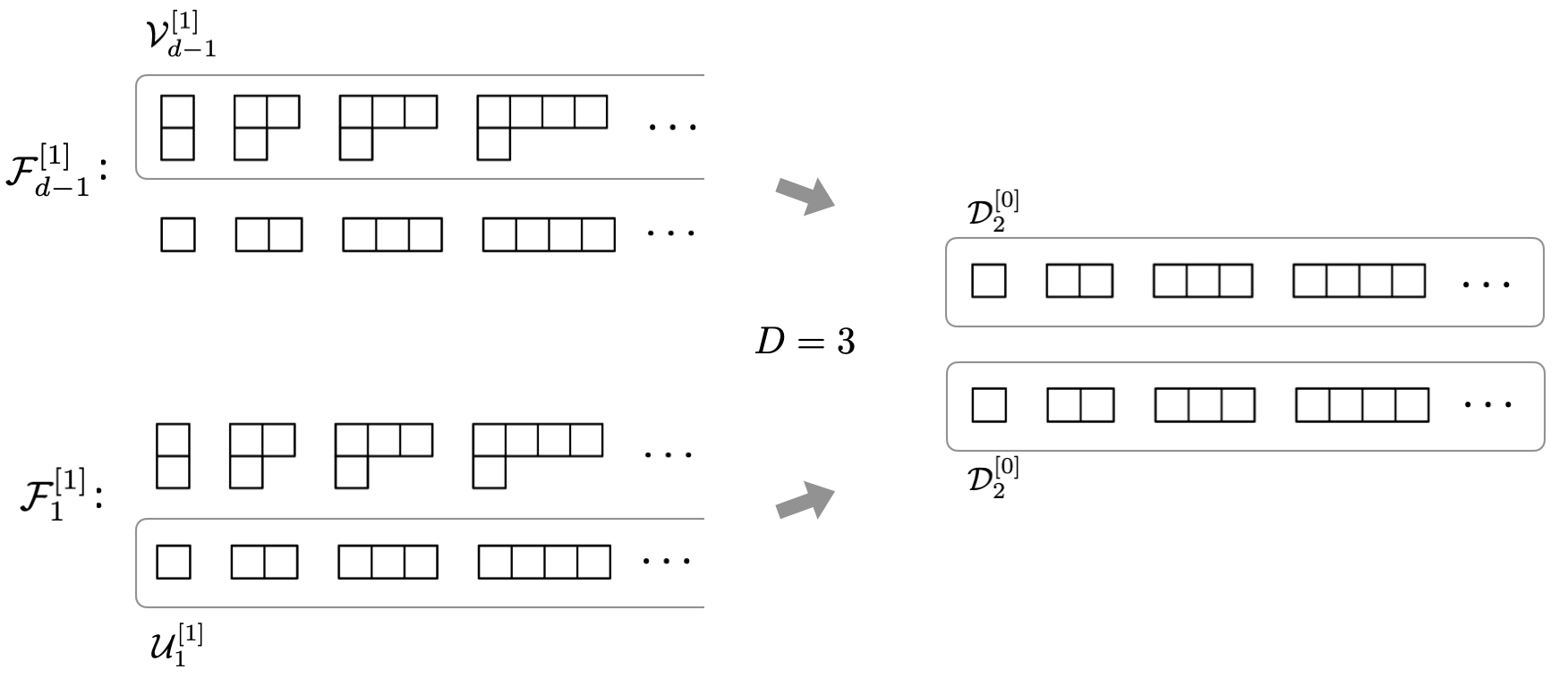,width=5.5in}}  \label{vectorsocontent13}\, \ee

\subsubsection*{$D=4$:}

In $D=4$, the $\frak{so}(4)$ reps that have 2 rows in their tableaux are no longer irreducible but instead split into two chiral parts that are self-dual and anti-self-dual upon contracting the indices in their first column with the four dimensional epsilon tensor $\epsilon_{IJKL}$,
\be [l_1,l_2]= [l_1,l_2]_+\oplus [l_1,l_2]_-\, .\label{sod4splite}\ee
Among the dS$_4$ vector reps, the only one this reducibility affects is the massless vector rep ${\cal V}^{[1]}_2$, which is the only one that has exclusively two row tableaux among its $\frak{so}(4)$ content.  Splitting each of the $\frak{so}(4)$ reps leads to a chiral splitting of the massless vector rep,
\be {\cal V}^{[1]}_2 ={\cal V}^{[1]_+}_2\oplus {\cal V}^{[1]_-}_2\,,\ \ \ D=4\,,\ee
illustrated here:
\be \raisebox{-0pt}{\epsfig{file=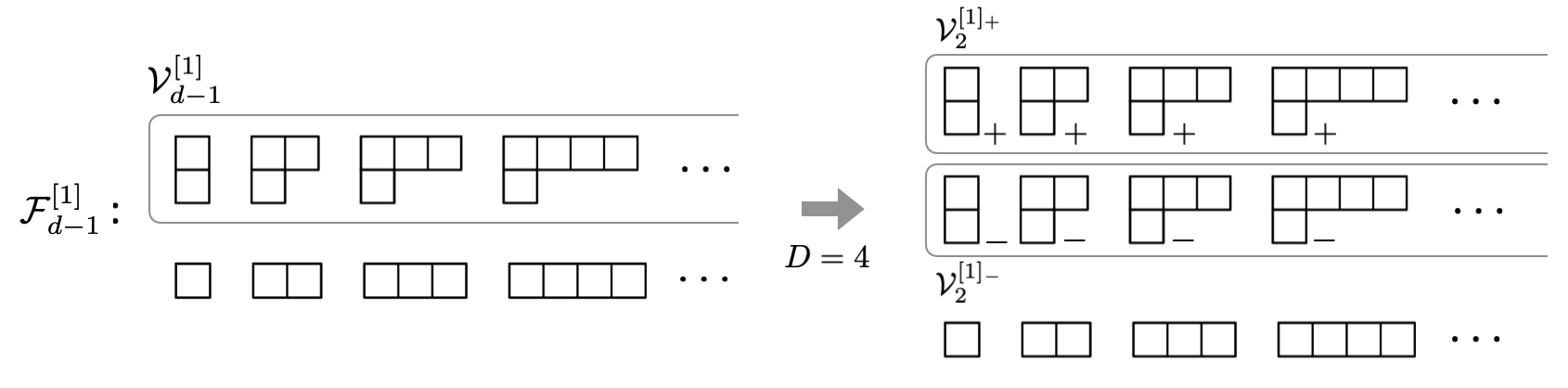,width=5.8in}}  \label{vectorsocontent14}\, \ee
The reps ${\cal V}^{[1]_\pm}_2$ represent the two independent helicities of a massless photon on dS$_4$ (just as the flat space massless photon has two independent Lorentz invariant helicities).  As we will see in section \ref{unitarylistsection}, these are accounted for among the so-called discrete series reps.

\subsection{Spin 2 representations\label{spin2sec}}

We next turn to the spin 2 reps, carried by rank 2 symmetric traceless tensor fields.  We use the mass $m$ that vanishes when the field acquires its largest gauge invariance, which from \eqref{spinsoffsete} is related to the mass $\tilde m$ in the dS$_D$ field's Klein-Gordon equation by $\tilde m^2=m^2 +2H^2$.   From \eqref{deltamregexe}, \eqref{dscfttmassrelatione}, the relations between $m$ and $\Delta$ for the late time behavior $\sim e^{-\left( \Delta-2\right) {H t}}$ of \eqref{ssassymprosole} are,
\be {m^2\over H^2}=\Delta(d-\Delta)\,, \ \ \  \Delta_\pm={d\over 2}\pm\sqrt{{d^2\over 4}-{m^2\over H^2}}\,. \label{tensmassretlationere}\ee

The vector space carrying the tensor $\frak{so}(1,D)$ reps will be the space of square integrable complex traceless symmetric rank-2 tensor fields $\phi_{ij}$ on the sphere ${\mathbb S}^d$, transforming under $\frak{so}(1,D)$ as in \eqref{latetimesheactiont3} with $r=2$.
Call this space ${\cal F}^{[2]}_\Delta$,
\be {\cal F}^{[2]}_\Delta:\ \ {\rm complex\ rank\ 2\ symmetric\ traceless\ tensor\ fields\ on\ } {\mathbb S}^d \,.\ee

To break this up into its $\frak{so}(D)$ content, we first perform the SVT decomposition of a traceless symmetric tensor into a transverse traceless symmetric tensor $\chi_{ij}$, transverse vector $\chi_i$, and scalar $\chi$ \cite{York:1973ia,York:1974psa} (see appendix D of \cite{Hinterbichler:2013kwa} for a more detailed explanation of this decomposition),
\be \phi_{ij}=\chi_{ij}+\nabla_{(i} \chi_{j)}+\nabla_{(i}\nabla_{j)_T}\chi\, ,\ \ \ \nabla^j\chi_{ij}=\chi^{i}_{\ i}=\nabla^i\chi_i=0\,.\label{SVTdedcompee}\ee

Each of these parts is in turn split up into its constituent spherical harmonics.  For the scalar part $\chi$, we use the scalar spherical harmonics \eqref{scalarphserarmoe}, but the $l=0,1$ modes will be missing because these are annihilated in the combination $\nabla_{(i}\nabla_{j)_T}\chi$ that appears in \eqref{SVTdedcompee}, so we only have the $\frak{so}(D)$ reps $[l]$ with $l\geq 2$.  For the transverse vector part $\chi_i$, we expand using the vector spherical harmonics in \eqref{vecotrspherharmonicese}, but the $l=1$ modes do not appear because these are the Killing vectors on the sphere, which vanish in the combination $\nabla_{(i} \chi_{j)_T}$ (which is nothing but the conformal Killing equation) appearing in \eqref{SVTdedcompee}, thus we only have the $\frak{so}(D)$ reps $[l,1]$ with $l\geq 2$.   The transverse traceless part $\chi_{ij}$ can be expanded in transverse traceless tensor spherical harmonics on ${\mathbb S}^d$, which take the form \cite{rubin1984eigenvalues,rubin1985symmetric,Higuchi:1986wu}
\be Y^{I_1\ldots I_l,J_1J_2}_{l,ij}(\hat X)= \partial_i \hat X^{[J_1}\, \hat X^{I_1]} \partial_j \hat X^{[J_2}\, \hat X^{I_2]}  \hat X^{I_3}\cdots\hat X^{I_l}-{\rm traces}\, ,\ \ \ l=2,3,4,\ldots \ . \label{tensorspherharmonicese}\ee
These have the index symmetries of traceless $[l,2]$ tableaux.  They form a basis of the space of transverse traceless rank 2 symmetric tensors on ${\mathbb S}^d$. The natural Laplacian on this space is the Lichnerowicz Laplacian $\Delta_L$, given by \cite{Lichnerowicz1961},
\be \Delta_L=-\nabla_\Omega^2 +2d\, , \ee 
and the transverse traceless harmonics are eigenfunctions of it, with the following eigenvalues \cite{rubin1984eigenvalues,rubin1985symmetric,Higuchi:1986wu},
\be \Delta_LY^{I_1\ldots I_l,J_1J_2}_{l,ij}=\left[ l (l + d -1) + 2 (d -1) \right] Y^{I_1\ldots I_l,J_1J_2}_{l,ij}.\ee

We can illustrate all the $\frak{so}(D)$ reps present in ${\cal F}^{[2]}_\Delta$ as follows:
\be \raisebox{-0pt}{\epsfig{file=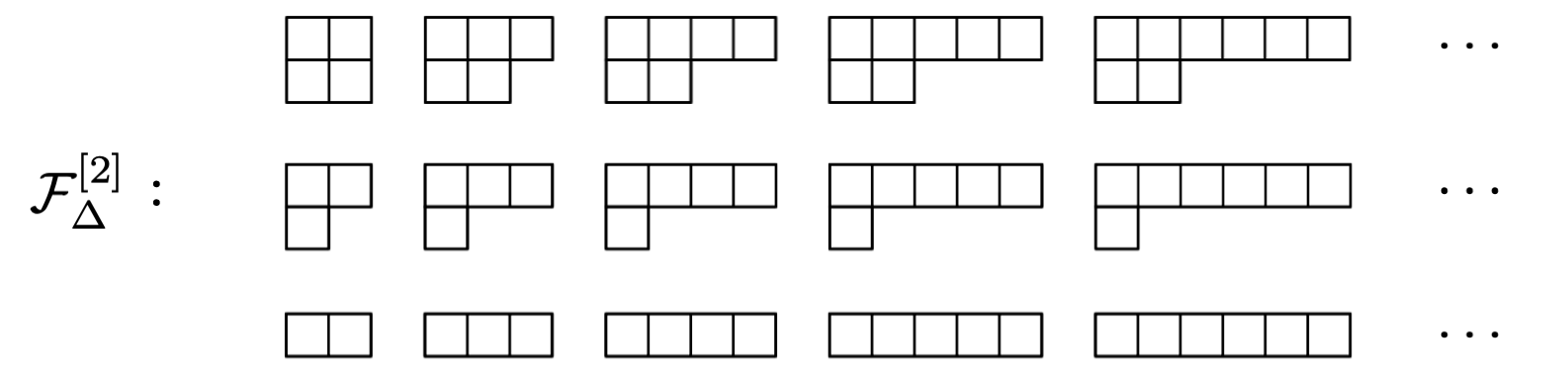,width=4.6in}}  \label{tensorsocontent0}\, \ee
The top line contains the tensor harmonics \eqref{tensorspherharmonicese} making up the transverse traceless tensor part $\chi_{ij}$ in the decomposition \eqref{SVTdedcompee}, the middle line contains the vector harmonics making up the transverse vector part $\chi_{i}$, and the bottom line contains the scalar harmonics making up the scalar part $\chi$.

For generic $\Delta$, the boost generators ${\cal K}^I$ will link the lattice of $\frak{so}(D)$ reps shown in \eqref{tensorsocontent0} through nearest neighbor interactions, as illustrated with the arrows here:
\be \raisebox{-0pt}{\epsfig{file=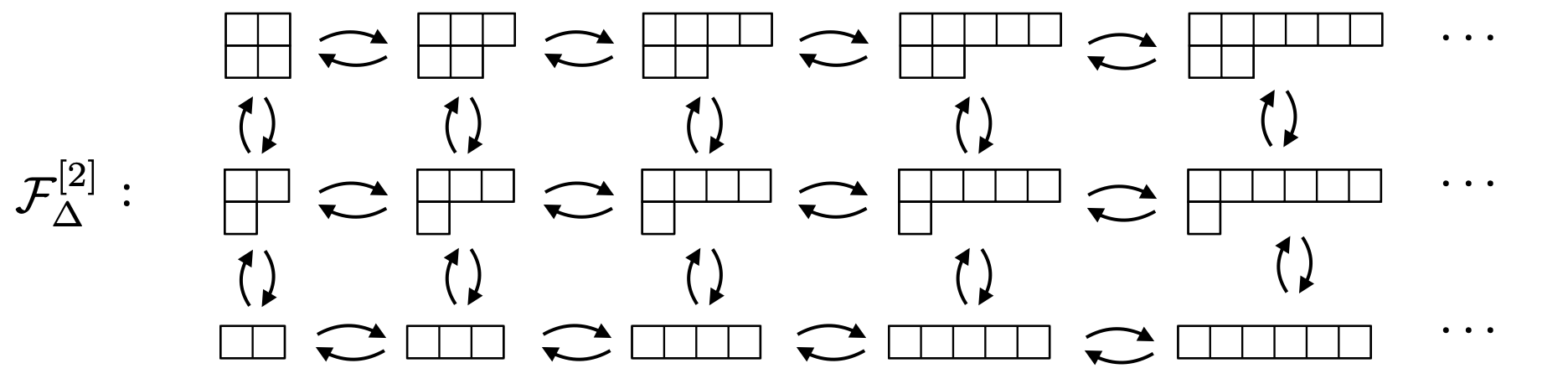,width=4.8in}}  \label{tensorsocontent}\, \ee
The maps represented by the arrows are defined as in the explanation right below \eqref{vectorsocontent2}.

\textbf{Reducible cases:} There are discrete values of $\Delta$ at which these arrows break and we get sub-reps, making the rep ${\cal F}_{\Delta}^{[2]}$ reducible but not decomposable.  These values are as follows:
\begin{itemize}

\item  
Shift symmetric points: 
\be \Delta=d+k+2\, ,\ \  k=0,1,2,\ldots \ \ .\ee  
The columns to the right of the $(k+1)$-th column in \eqref{tensorsocontent} split off into an infinite dimensional sub-rep we call
\be {\cal D}^{[2]}_{d+2+k}\,,\ee
as illustrated here for the case $k=1$:
\be \raisebox{-0pt}{\epsfig{file=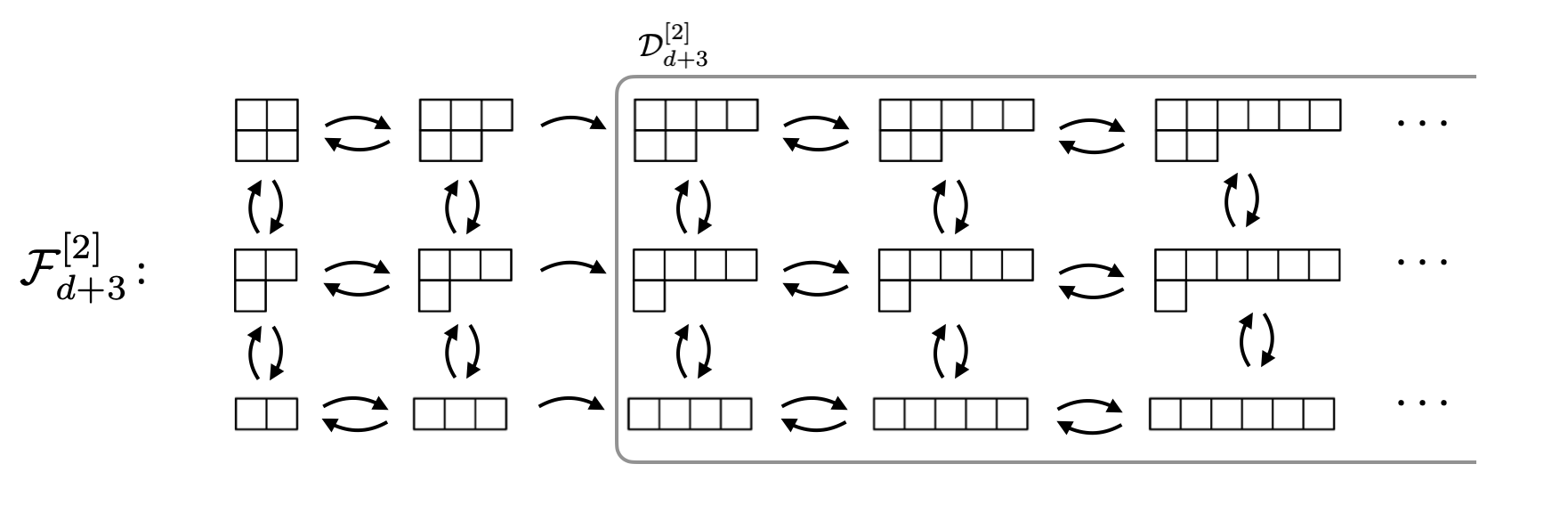,width=4.2in}}  \label{tensorsocontent2}\, \ee
These represent the physical modes of the level $k$ shift symmetric spin 2 fields.

\item 
Finite points:
\be \Delta=-k-2\, ,\ \  k=0,1,2,\ldots\ \ .\ee 
The first  $(k+1)$ columns in \eqref{tensorsocontent} split off into a finite dimensional sub-rep we call 
\be {\cal S}^{[2]}_{-k-2}\, ,\ee
as illustrated here for the case $k=1$:
\be \raisebox{-0pt}{\epsfig{file=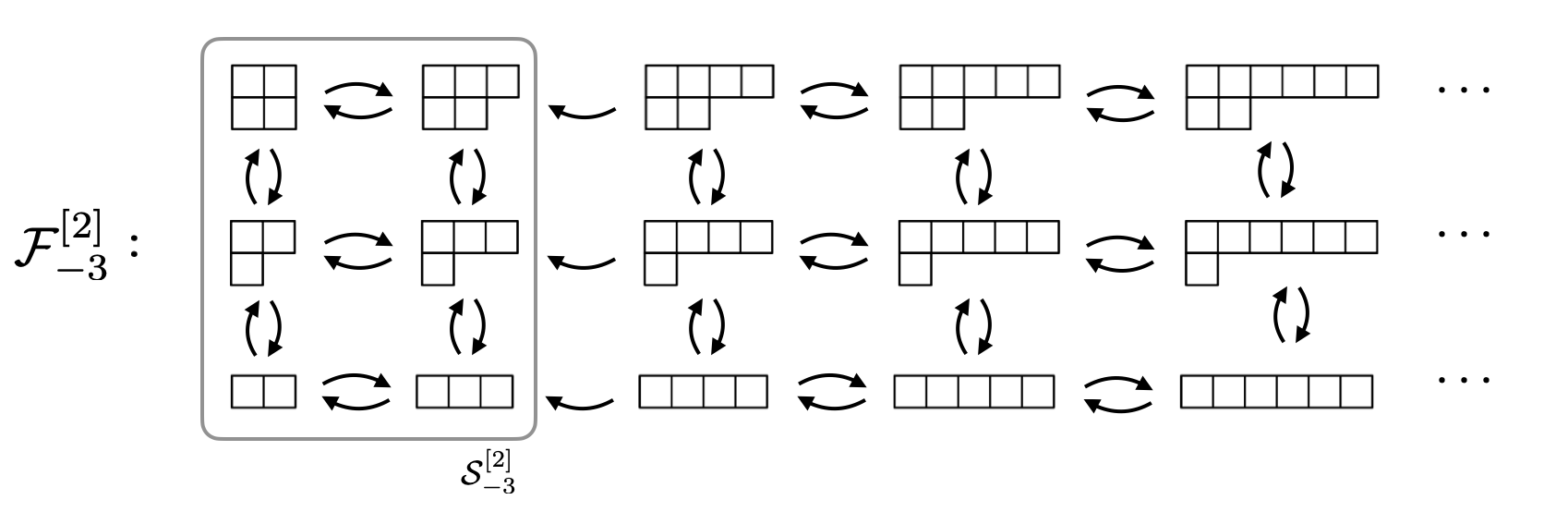,width=4.2in}}  \label{tensorsocontent3}\, \ee
This is the finite dimensional rep corresponding to the $[k+2,2]$ tensor rep of $\frak{so}(1,D)$.  When it is branched to $\frak{so}(D)$ using the branching rules  in appendix \ref{branchingappendix}, it gives all the $\frak{so}(D)$ tensors present in ${\cal S}^{[2]}_{-k-2}$.  Note that the $[k+2,2]$ tableau appears in the upper right of the ${\cal S}^{[2]}_{-k-2}$ sub-rep.   

This corresponds to the shift symmetries of the level $k$ shift symmetric field; indeed the shift symmetries are parametrized by a $[k+2,2]$ tensor in the dS$_D$ embedding space \cite{Bonifacio:2018zex}.

\item 
Massless point:
\be \Delta=d\,.\ee 
The first row of \eqref{tensorsocontent} splits off into the infinite dimensional sub-rep we call
\be {\cal V}^{[2]}_{d}\, ,\ee
as illustrated here:
\be \raisebox{-0pt}{\epsfig{file=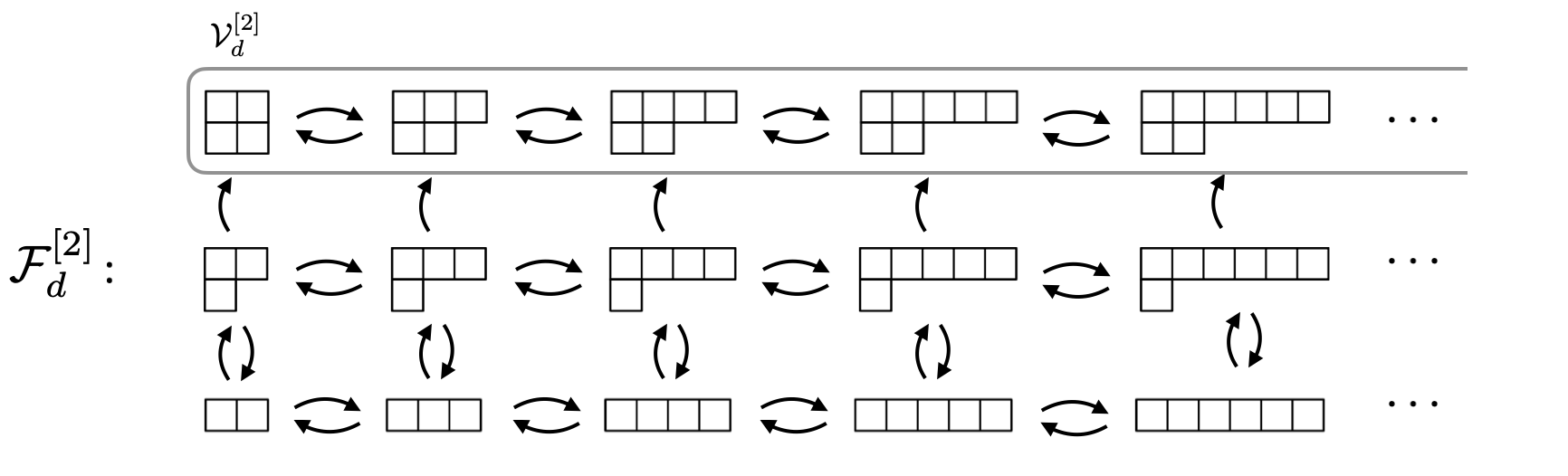,width=4.2in}}  \label{tensorsocontent4}\, \ee
This corresponds to the physical modes of the massless graviton.

\item
PM point:
\be \Delta=d-1\, .\ee
This is a new feature as compared to the vector case in section \ref{vectorsection}.
At this value, the first two rows of \eqref{tensorsocontent} split off into the infinite dimensional sub-rep we call
\be {\cal V}^{[2]}_{d-1}\, ,\ee
as illustrated here:
\be \raisebox{-0pt}{\epsfig{file=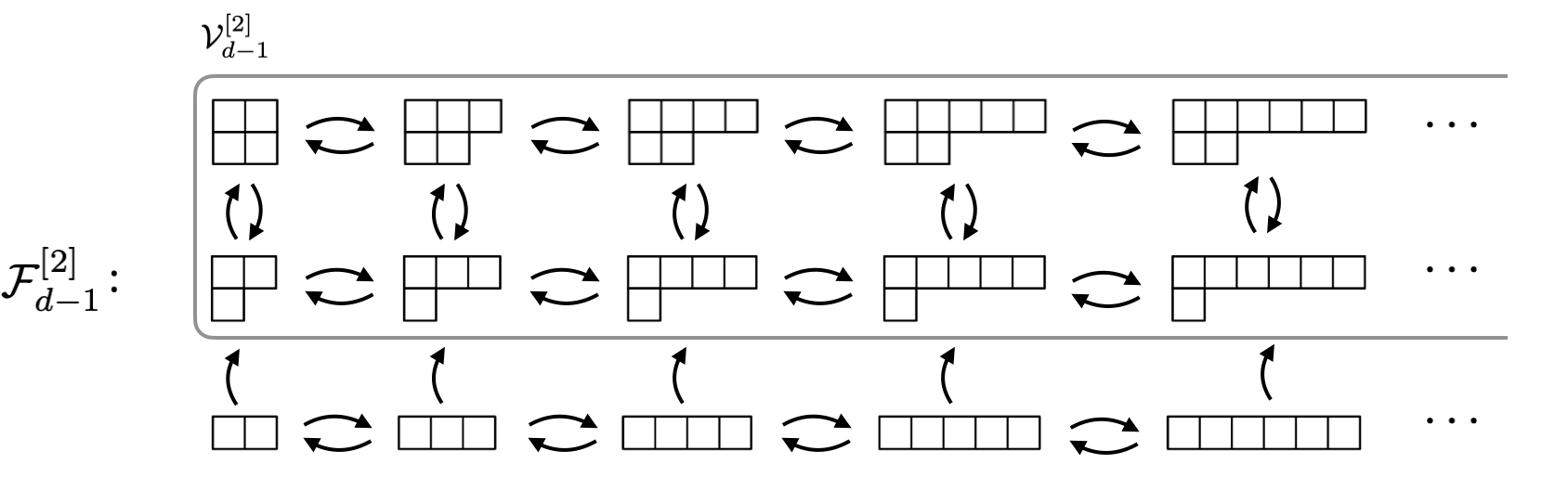,width=4.2in}}  \label{tensorsocontent6}\, \ee
This corresponds to the physical modes of the partially massless graviton.

\item 
Massless gauge point:
\be \Delta=0\,.\ee 
The last two rows of \eqref{tensorsocontent} split off into the infinite dimensional sub-rep we call
\be {\cal U}^{[2]}_{0}\, ,\ee
as illustrated here:
\be \raisebox{-0pt}{\epsfig{file=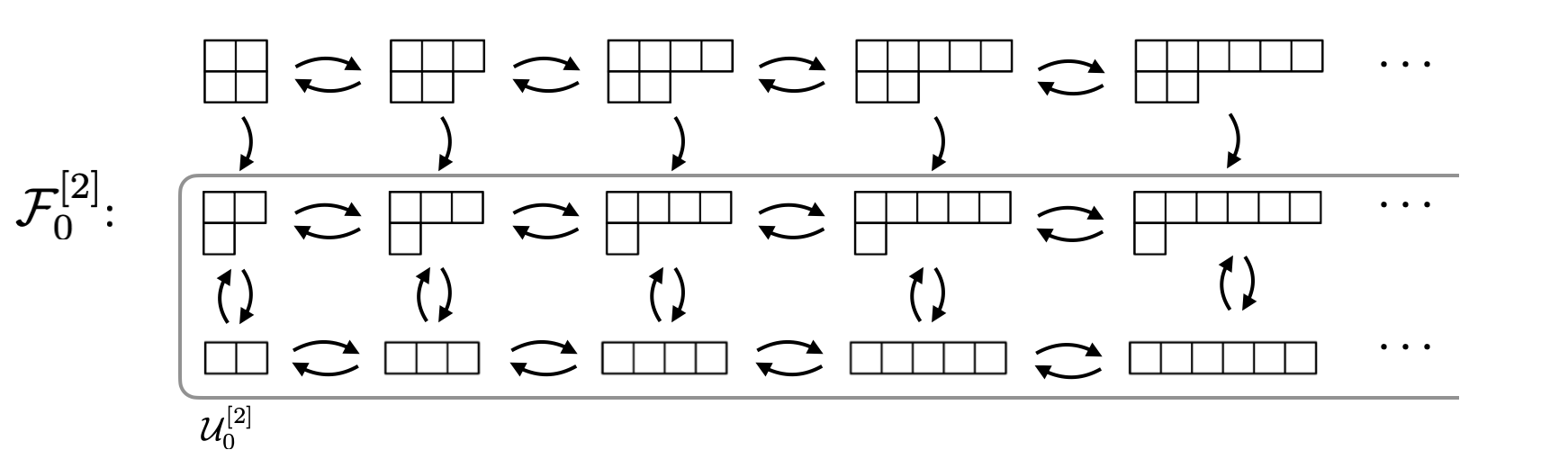,width=4.2in}}  \label{tensorsocontent5}\, \ee
This corresponds to the gauge modes of the massless graviton.  

\item
PM gauge point:
\be \Delta=1\, .\ee
The last  row of \eqref{tensorsocontent} splits off into the infinite dimensional sub-rep we call
\be {\cal U}^{[2]}_{1}\, ,\ee
as illustrated here:
\be \raisebox{-0pt}{\epsfig{file=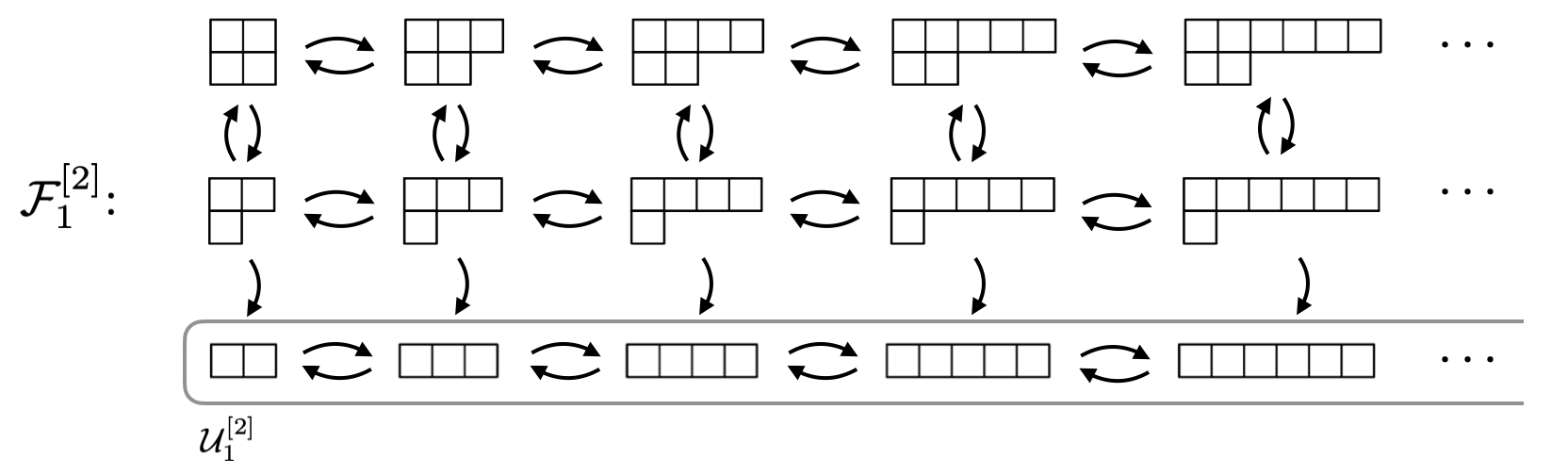,width=4.8in}}  \label{tensorsocontent7}\, \ee
This corresponds to the gauge modes of the partially massless graviton.  

\end{itemize}

\textbf{Equivalences:} There is a shadow transform intertwining operator that connects the $\Delta$ and $\bar\Delta\equiv d-\Delta$ reps, 
\be S_\Delta^{[2]} :\ {\cal F}_{\Delta}^{[2]}\rightarrow {\cal F}_{\bar\Delta}^{[2]}\,.\label{tensorinterfinte}\ee
It commutes with the $\frak{so}(D)$ rotations and satisfies $\delta_{{\cal K}^I_{\bar\Delta}}S_\Delta^{[2]}=S_\Delta^{[2]} \delta_{{\cal K}^I_{\Delta}}$.
For generic $\Delta$, this operator is invertible and the reps at $\Delta$ and $\bar\Delta$ are equivalent to each other,
\be {\cal F}_{\Delta}^{[2]}\simeq {\cal F}_{\bar\Delta}^{[2]}\,.\label{tensorequifve}\ee
  But for the special values of $\Delta$ given above where ${\cal F}_{\Delta}^{[2]}$ develops a sub-rep, $S_\Delta^{[2]}$ develops a kernel which is always precisely the sub-rep.

The shift symmetric and finite points are linked to each other via the maps $S^{[2]}_{-k-2}$, ${S}^{[2]}_{d+k+2}$, where the kernel and image of each are the sub-reps ${\cal S}^{[2]}_{-k-2}$, ${\cal D}^{[2]}_{d+k+2}$ respectively, which induces the isomorphisms
\be {\cal S}^{[2]}_{-k-2}\simeq {\cal F}^{[2]}_{d+k+2}/{\cal D}^{[2]}_{d+2+k}\,,\ \ \  {\cal D}^{[2]}_{d+k+2}\simeq {\cal F}^{[2]}_{-k-2}/{\cal S}^{[2]}_{-k-2}\,.\ee
This is illustrated here for the case $k=2$:
\be \raisebox{-0pt}{\epsfig{file=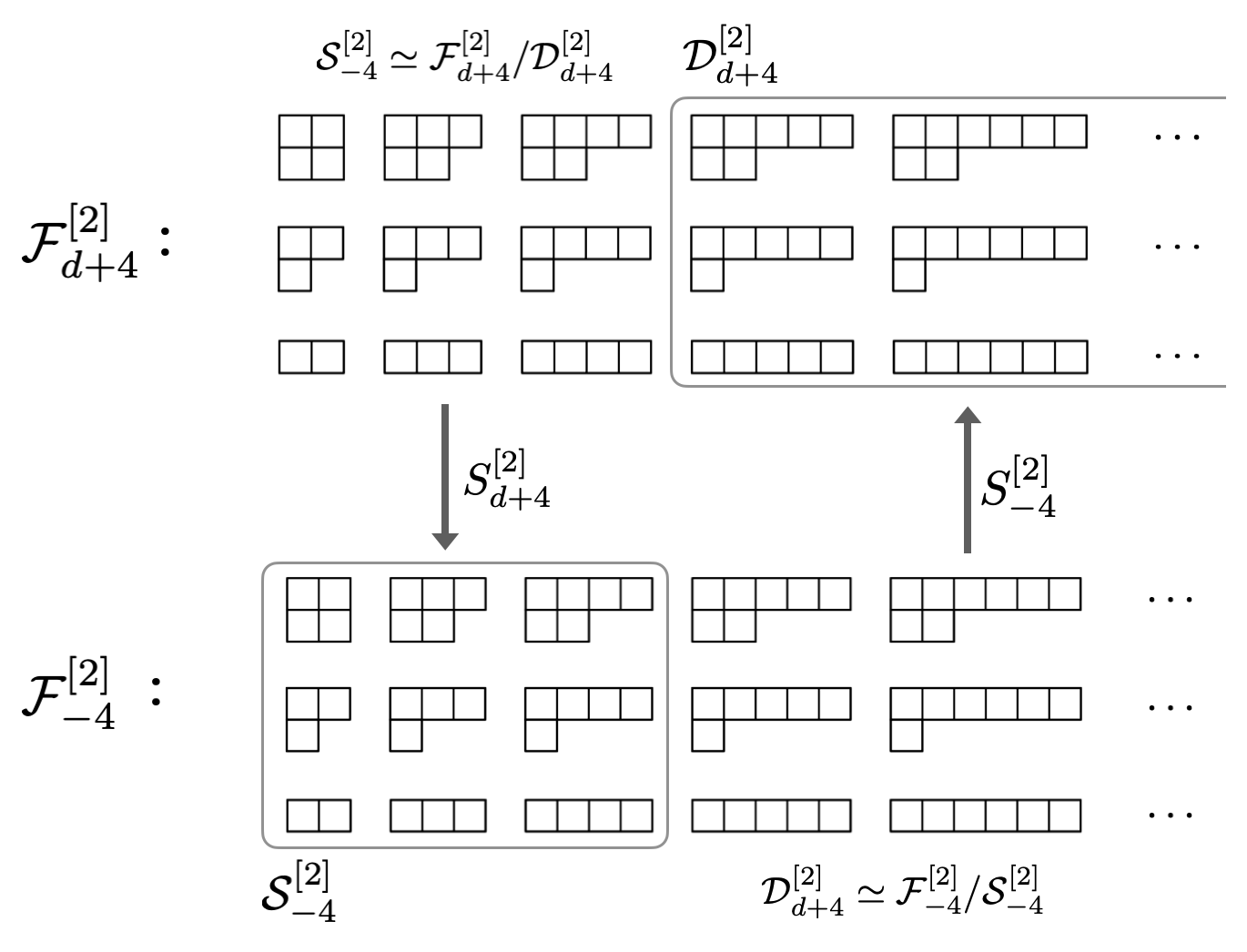,width=4.7in}}  \label{tensorsocontent7-2}\, \ee

The massless and massless gauge points are linked to each other via the maps $S^{[2]}_{0}$, ${S}^{[2]}_{d}$, where the kernel and image of each are the sub-reps ${\cal U}^{[2]}_{0}$, ${\cal V}^{[2]}_{d}$ respectively, which induces the isomorphisms
\be {\cal U}^{[2]}_{0}\simeq {\cal F}^{[2]}_{d}/{\cal V}^{[2]}_{d}\,,\ \ \  {\cal V}^{[2]}_{d}\simeq {\cal F}^{[2]}_{0}/{\cal U}^{[2]}_{0}\,.\ee
This is illustrated here
\be \raisebox{-0pt}{\epsfig{file=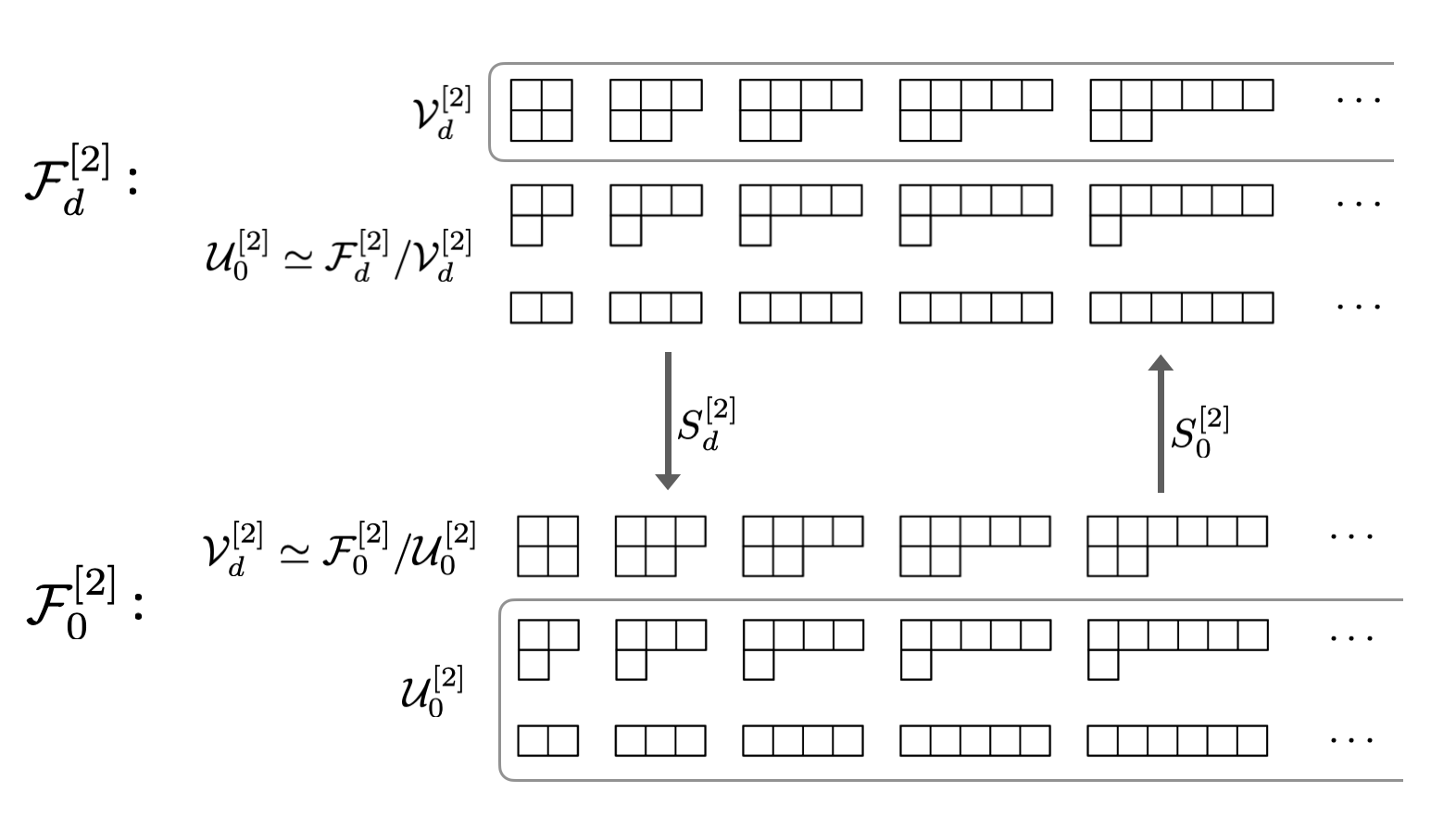,width=5.0in}}  \label{tensorsocontent7-3}\, \ee

The partially massless and partially massless gauge points are linked to each other via the maps $S^{[2]}_{1}$, ${S}^{[2]}_{d-1}$, where the kernel and image of each are the sub-reps ${\cal U}^{[2]}_{1}$, ${\cal V}^{[2]}_{d-1}$ respectively, which induces the isomorphisms
\be {\cal U}^{[2]}_{1}\simeq {\cal F}^{[2]}_{d-1}/{\cal V}^{[2]}_{d-1}\,,\ \ \  {\cal V}^{[2]}_{d-1}\simeq {\cal F}^{[2]}_{1}/{\cal U}^{[2]}_{1}\,.\ee
This is illustrated here
\be \raisebox{-0pt}{\epsfig{file=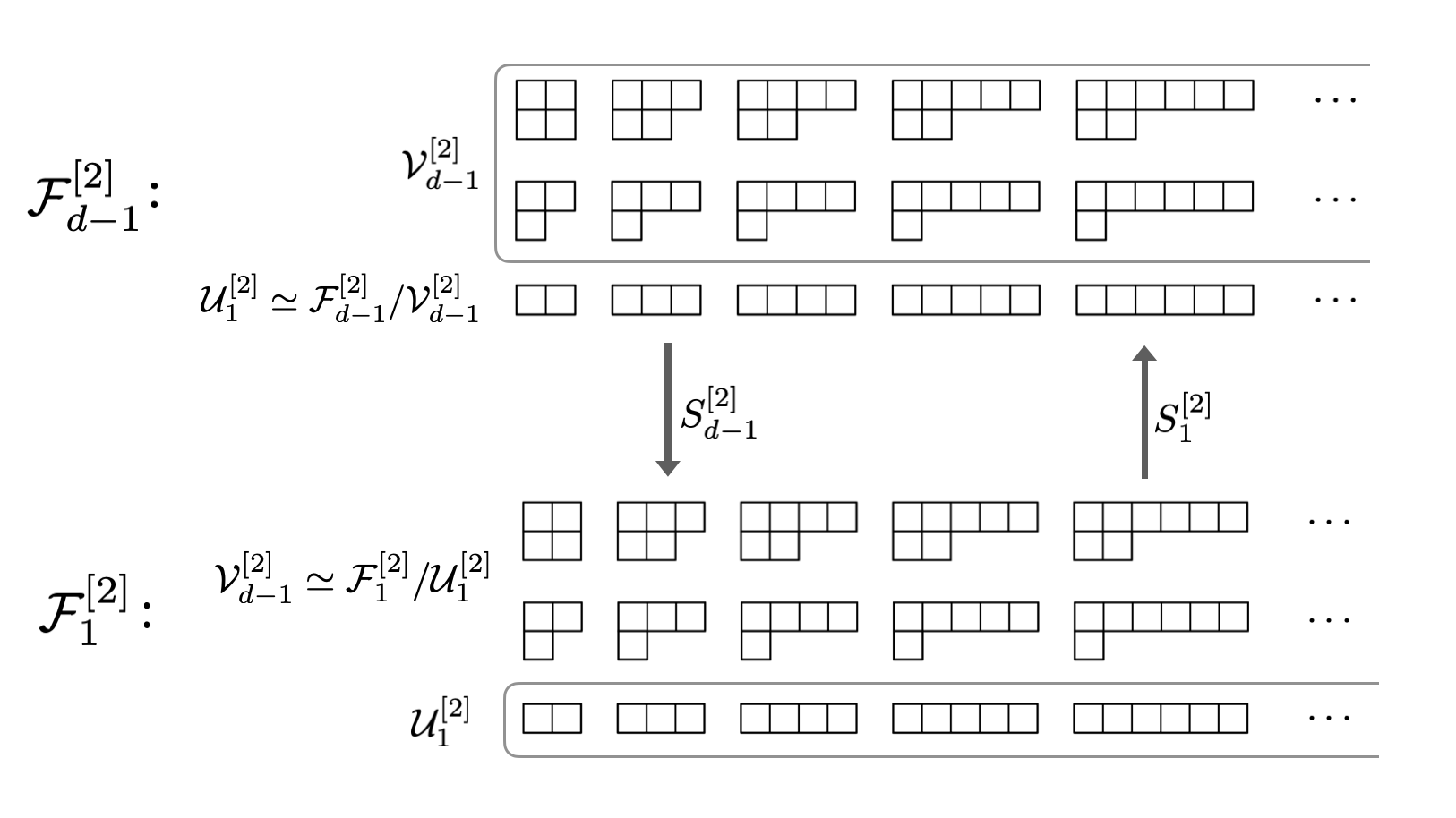,width=5.0in}}  \label{tensorsocontent7-4}\, \ee

There are two additional maps of interest in the massless case that connect vector and tensor reps.  The first is the symmetric gradient, which takes vectors with $\Delta=-1$ to traceless symmetric tensors, adding one unit of $\Delta$ in the process,
\be {\rm grad}: \  {\cal F}^{[1]}_{-1} \rightarrow {\cal F}^{[2]}_{0} \, ,\ \ \phi_i\rightarrow \nabla_{(i} \phi_{j)_T}\,.
\ee
The kernel of this gradient is the space of conformal Killing vectors, which is precisely the first column of \eqref{vectorsocontent}, and which is the sub-rep ${\cal S}^{[1]}_{-1}$.   The image is the lower two rows of \eqref{tensorsocontent0}, which is the sub-rep ${\cal U}^{[2]}_{0}$.

The other map of interest in the massless case is the divergence, which takes traceless symmetric tensors with $\Delta=d$ to vectors, adding one unit of $\Delta$ in the process,
\be {\rm div}: \  {\cal F}^{[2]}_{d} \rightarrow {\cal F}^{[1]}_{d+1} \, ,\ \ \phi_{ij} \rightarrow \nabla^j\phi_{ij}\,.
\ee
Its kernel is the space of transverse traceless tensors, which is precisely the top row of \eqref{tensorsocontent0}, and is the sub-rep ${\cal V}^{[2]}_{d}$.  Its image is all but the first column of \eqref{vectorsocontent}, which is the sub-rep ${\cal D}^{[1]}_{d+1}$.

The four spaces ${\cal F}^{[1]}_{d+1}$, ${\cal F}^{[1]}_{-1}$, ${\cal F}^{[2]}_{d}$, ${\cal F}^{[2]}_{0}$ are joined together using the grad and div maps and the intertwiners $S^{[2]}_{d}$, ${S}^{[2]}_{0}$, $S^{[1]}_{d+1}$, ${S}^{[1]}_{-1}$ to make the following commutative diagram:
\be \raisebox{-0pt}{\epsfig{file=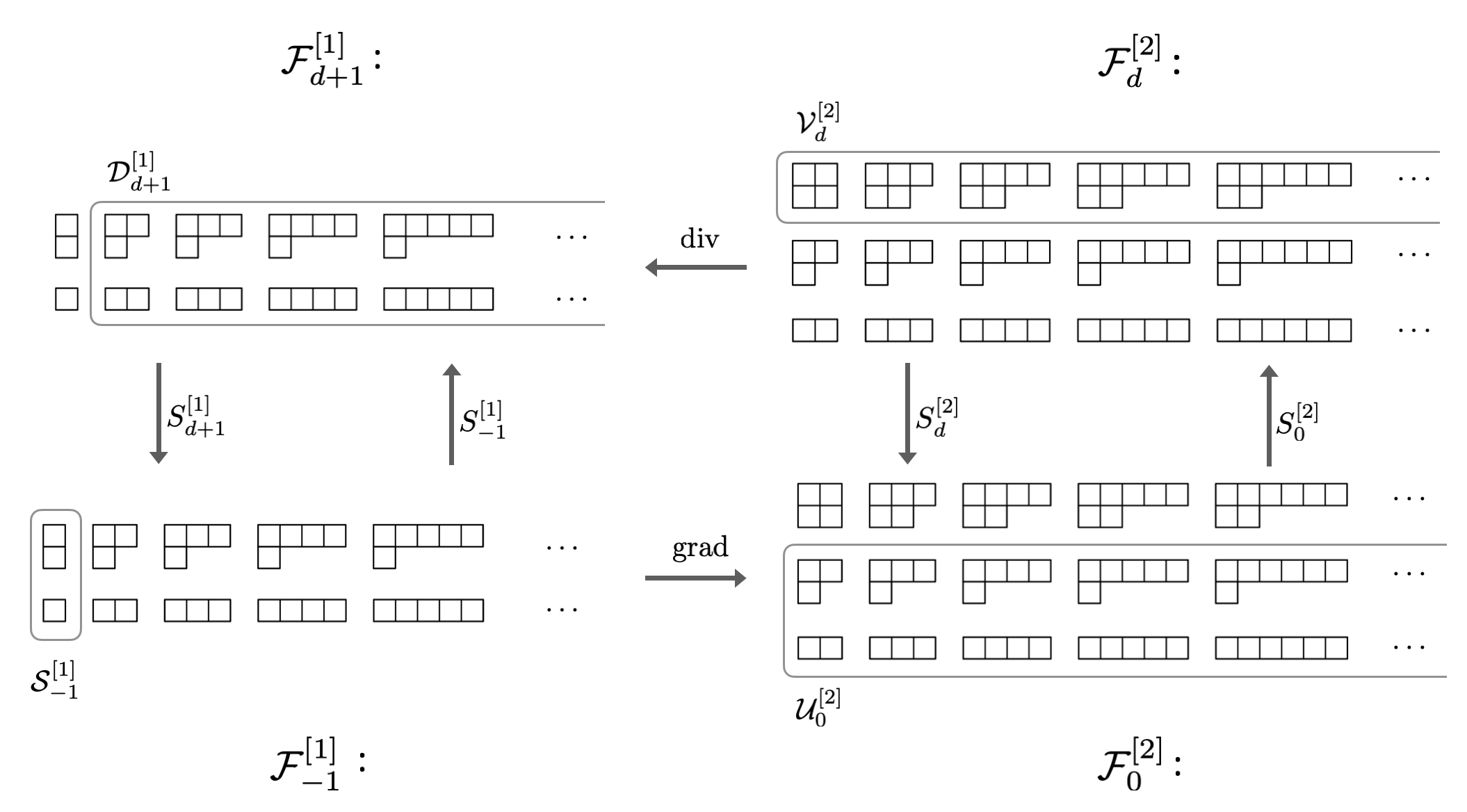,width=6.1in}}  \label{tensorsocontent8}\, \ee
The grey squares enclose the various sub-reps; these are simultaneously the kernel of every outgoing map, and the image of every ingoing map.  Going through any two arrows consecutively gives zero.  From this picture, we get the isomorphism
\be { \cal D}^{[1]}_{d+1}\simeq { \cal U}^{[2]}_0\, .\label{masslstovss32isnee}\ee
This expresses the fact that the gauge modes of a massless graviton on dS$_D$ are precisely those of a $k=0$ shift symmetric vector \cite{DeRham:2018axr}.

There are two additional maps of interest in the partially massless case that link tensor and scalar reps.  The first is the double gradient, which takes scalars with $\Delta=-1$ to traceless symmetric tensors, adding two units of $\Delta$ in the process,
\be {\rm grad}^2: \  {\cal F}^{[0]}_{-1} \rightarrow {\cal F}^{[2]}_{1} \, ,\ \ \phi\rightarrow \nabla_{(i} \nabla_{j)_T}\phi \,.
\ee
The kernel of this double gradient is the first two entries of \eqref{scalarsocontent}, which is the sub-rep ${\cal S}^{[0]}_{-1}$.   The image is the lower row of \eqref{tensorsocontent0}, which is the sub-rep ${\cal U}^{[2]}_{1}$.

The other map of interest in the partially massless case is the double divergence, which takes symmetric traceless tensors with $\Delta=d-1$ to scalars and raises the value of $\Delta$ by two in the process,
\be {\rm div}^2: \  {\cal F}^{[2]}_{d-1} \rightarrow {\cal F}^{[0]}_{d+1} \, ,\ \ \phi_{ij} \rightarrow \nabla^i\nabla^j\phi_{ij}\,.
\ee
Its kernel is the space of doubly conserved traceless tensors, which is precisely the top two rows of \eqref{tensorsocontent0}, and is the sub-rep ${\cal V}^{[2]}_{d-1}$.  Its image is all but the first two entries of \eqref{scalarsocontent}, which is the sub-rep $ {\cal D}^{[0]}_{d+1}$.
The four spaces ${\cal F}^{[0]}_{d+1}$, ${\cal F}^{[0]}_{-1}$, ${\cal F}^{[2]}_{d-1}$, ${\cal F}^{[2]}_{1}$ can be joined together using the grad$^2$ and div$^2$ maps and the intertwiners $S^{[2]}_{d-1}$, ${S}^{[2]}_{1}$, $S^{[0]}_{d+1}$, ${S}^{[0]}_{-1}$ to make the following commutative diagram:
\be \raisebox{-0pt}{\epsfig{file=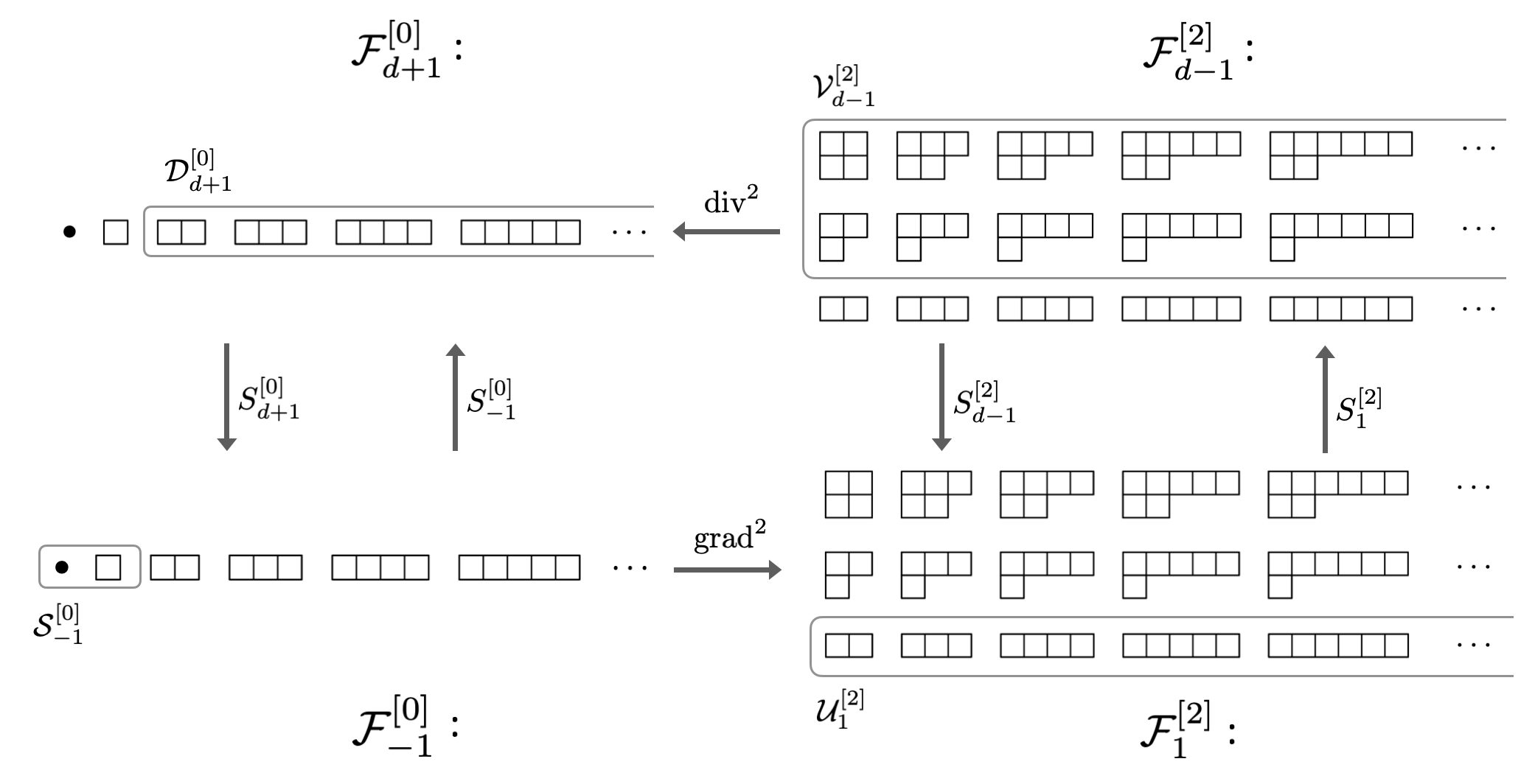,width=6.1in}}  \label{tensorsocontent9}\, \ee
The grey squares enclose the various sub-reps; these are simultaneously the kernel of every outgoing map, and the image of every ingoing map, and going through any two arrows consecutively gives zero.  From this picture, we get the isomorphism
\be { \cal D}^{[0]}_{d+1}\simeq { \cal U}^{[2]}_1\, .\label{tensaddisomre}\ee
This expresses the fact that the gauge modes of a partially massless graviton are precisely those of a $k=1$ shift symmetric scalar \cite{DeRham:2018axr}.

\textbf{Unitarity:} There is a diffeomorphism and Weyl invariant bilinear form pairing the two spaces ${\cal F}_\Delta^{[2]}$ and ${\cal F}_{\bar \Delta}^{[2]}$,
\be (\phi_1,\phi_2)\equiv \int d^d\Omega\, \phi_1^{ij}(\hat X)\phi_{2\, ij}(\hat X)\,,\ \ \ \  \phi_{1\, ij}\in {\cal F}_{\bar \Delta}^{[2]},\ \ \phi_{2\, ij}\in {\cal F}_{\Delta}^{[2]}\, . \label{bilintearformvee}\ee

In the case where $\Delta^\ast=\bar\Delta$ we can use \eqref{bilintearformvee} to form a manifestly positive definite, invariant inner product on ${\cal F}_\Delta^{[2]}$ via:
\be \la \phi_1 | \phi_2 \ra\equiv (\phi_1^{\ast},\phi_2),\ \ \ \Delta^\ast=\bar\Delta\, , \ \ \phi_{1\, ij},\phi_{2\, ij}\in {\cal F}_\Delta^{[2]} \, .\label{innerprotdvprise} \ee
This gives us the spin 2 principal series reps, which are all unitary,
\be {\rm spin\ 2 \ principal\ series:}\  \Delta={d\over 2}+i\nu,\ \ \ \nu\in {\mathbb R}\,.\ee

In the case where $\Delta$ is real, we use $S^{[2]}_{\Delta}$ to move a state from ${\cal F}_{ \Delta}^{[2]}$ to ${\cal F}_{\bar \Delta}^{[2]}$ and form an inner product on ${\cal F}_{ \Delta}^{[2]}$ as follows,
\be \la \phi_1 | \phi_2 \ra\equiv (S^{[2]}_{\Delta}\phi_1^\ast,\phi_2)\, ,\ \ \ \Delta^\ast=\Delta\, ,  \ \ \phi_{1\, ij},\phi_{2\, ij}\in {\cal F}_\Delta^{[2]} \,  .\label{innerprodtpriscsvdee} \ee 
The positivity of this inner product is now equivalent to whether the matrix elements of $S^{[2]}_{\Delta}$ are positive for all the different $\frak{so}(D)$ reps.  Away from the discrete special cases described above, it turns out that they are all positive only in the range $1<\Delta<d-1$, which gives the spin 2 complementary series,
\be {\rm spin\ 2 \ complementary\ series:}\  1<\Delta<d-1\,, \label{tensorcompnle}\ee
where, as in the scalar and vector cases, we do not yet commit the point $\Delta=d/2$ to either series.

Now turn to the cases of $\Delta$ where the reps become reducible. For the finite points $\Delta=-k-2$, $k=0,1,2,\ldots$, we use $S^{[2]}_{-k-2}$ in the inner product, and this has the finite dimensional kernel consisting of the sub-rep ${\cal S}^{[2]}_{-k-2}$.   All of these states will have zero norm in the inner product.  Apart from these null states, other states also have negative norm, so even once these null states are factored out, we are left with a non-unitary rep.  The rep obtained after this factoring is ${\cal D}^{[2]}_{d+k+2}$, realized through the quotient ${\cal D}^{[2]}_{d+k+2}\simeq {\cal F}^{[2]}_{-k-2}/{\cal S}^{[2]}_{-k-2}$, so these reps, corresponding to the shift symmetric spin 2 fields, are non-unitary.

For the shift symmetric points $\Delta=d+k+2$, $k=0,1,2,\ldots$, we use $S^{[2]}_{d+k+2}$ in the inner product, and this has the infinite dimensional kernel consisting of the states of the sub-rep ${\cal D}^{[2]}_{d+k+2}$.   All of these states will therefore be null in the inner product.  Apart from these null states, some of the norms are always still negative, so if the null states are factored out, we will be left with a finite dimensional non-unitary rep.  This is the rep ${\cal S}^{[2]}_{-k-2}$, realized through the quotient ${\cal S}^{[2]}_{-k-2}\simeq  {\cal F}^{[2]}_{d+k+2}/{\cal D}^{[2]}_{d+k+2}$.  

For the massless gauge point $\Delta=0$, we use $S^{[2]}_{0}$ in the inner product, and this has as kernel the sub-rep ${\cal U}^{[2]}_{0}$.   All of these states will have zero norm in the inner product.  Apart from these null states, the other states all have the same sign norm, so  once these null states are factored out, we will be left with a rep that can be made unitary by adjusting the overall sign if necessary, which is nothing but ${\cal V}^{[2]}_{d}$, realized through the quotient ${\cal V}^{[2]}_{d}\simeq {\cal F}^{[2]}_{0}/{\cal U}^{[2]}_{0}$.  These correspond to the physical states of the massless graviton, which is unitary on dS.

For the massless point $\Delta=d$, we use $S^{[2]}_{d}$ in the inner product, and this has as kernel the sub-rep ${\cal V}^{[2]}_{d}$.   All of these states will have zero norm in the inner product.  Apart from these null states, the other states do not all have positive norm, so  once these null states are factored out, we will be left with a non-unitary rep, which is nothing but ${\cal U}^{[2]}_{0}\simeq {\cal D}^{[1]}_{d+1} $, realized through the quotient ${\cal U}^{[2]}_{0}\simeq {\cal F}^{[2]}_{d}/{\cal V}^{[2]}_{d}$.  This corresponds to the longitudinal gauge modes of the massless graviton, which is equivalent to the $k=0$ shift symmetric vector rep.

For the PM gauge point $\Delta=1$, we use $S^{[2]}_{1}$ in the inner product, and this has as kernel the sub-rep ${\cal U}^{[2]}_{1}$.   All of these states will have zero norm in the inner product.  Apart from these null states, the other states all have the same sign norm, so  once these null states are factored out, we will be left with a rep that can be made unitary by adjusting the overall sign if necessary.  This rep is ${\cal V}^{[2]}_{d-1}$, realized through the quotient ${\cal V}^{[2]}_{d-1}\simeq {\cal F}^{[2]}_{1}/{\cal U}^{[2]}_{1}$.  These correspond to the physical states of the partially massless graviton, which is unitary on dS.

For the PM point $\Delta=d-1$, we use $S^{[2]}_{d-1}$ in the inner product, and this has as kernel the sub-rep ${\cal V}^{[2]}_{d-1}$.   All of these states will have zero norm in the inner product.  Apart from these null states, the other states all have the same sign norm, so  once these null states are factored out, we will be left with a rep that can be made unitary, which is ${\cal U}^{[2]}_{1}\simeq {\cal D}^{[0]}_{d+1} $, realized through the quotient ${\cal U}^{[2]}_{1}\simeq {\cal F}^{[2]}_{d-1}/{\cal V}^{[2]}_{d-1}$.  This corresponds to the longitudinal gauge modes of the partially massless graviton, which is equivalent to the shift symmetric $k=1$ scalar rep.

For all but the unitary cases discussed here, there is no other way to construct an invariant inner product on ${\cal F}_\Delta^{[2]}$, so all the other reps in the complex $\Delta$ plane are non-unitary.  

\textbf{Summary:} The tensor reps are summarized here:
\be \raisebox{-40pt}{\epsfig{file=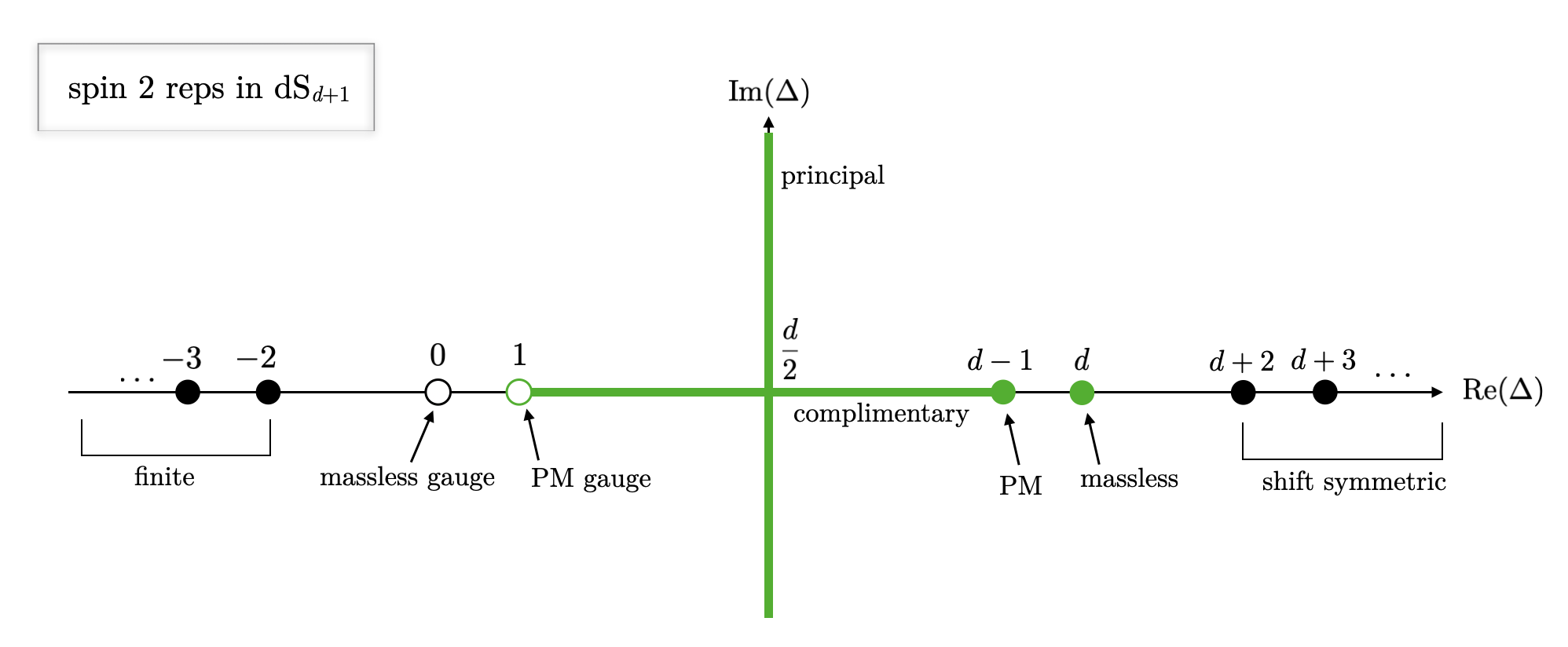,width=6.4in}}  \label{dsrepstensor}\, \ee
Each point in the plane is a rep, points in green are unitary reps.  The dots are the special points where the rep becomes reducible. The dots at $\Delta=1$ and $\Delta=0$ are empty to indicate that these are equivalent, due to  \eqref{masslstovss32isnee}, \eqref{tensaddisomre}, to a scalar rep and a vector rep respectively, and thus are already accounted for among those.
  The equivalence \eqref{tensorequifve} of the non-special reps is given by reflecting through the point $\Delta=d/2$, and due to the breakdown of the intertwiner map \eqref{tensorinterfinte}, the special reps with dots are inequivalent despite the reflection.  Other than this equivalence, all the reps are distinct.

The value of the quadratic Casimir operator ${\cal C}_2$ of section \ref{casimirsec} on the tensor reps is given by
\be {\cal C}_2= \Delta(\Delta-d)+2d = -{m^2\over H^2}+2(D-1)\,. \ee

We have the following correspondence between tensor fields on dS$_{D}$ and the spin 2 unitary reps:
\bea \begin{cases}
 {\rm principal:\ }\quad  \Delta={d\over 2}+i\nu \, , \ \ \nu\in {\mathbb R}\,, & {\rm heavy \ tensors:\ } {m^2\over H^2}\geq {(D-1)^2\over 4}\, , \\
 {\rm complementary:\ } \quad 1<\Delta<d-1 \, , & {\rm light \ tensors:\ } D-2 < {m^2\over H^2}\leq {(D-1)^2\over 4}\, , \\
  {\rm massless:\ } \quad \Delta=d \, , &  {\rm massless \ tensor:\ } m^2=0\, , \\
    {\rm PM:\ } \quad \Delta=d-1 \, , &  {\rm PM \ tensor:\ } {m^2\over H^2}=D-2\, . \\
\end{cases} \nn
\eea

The spin 2 tensor reps have some additional subtleties that occur in the lower dimensions $D=3,4$, which we turn to next (they are trivial in $D=2$, as discussed in section \ref{D2section}).

\subsubsection*{$D=3$:}

Consider the content of the spin 2 rep for generic $\Delta$ depicted in \eqref{tensorsocontent0}.  For $D=3$, these are $\frak{so}(3)$ tableaux, and those in the top row all become trivial.  For the middle row, we can use the three dimensional epsilon tensor $\epsilon_{IJK}$ to dualize all the reps into single row reps.  Once this is done, the middle row looks identical to the bottom row.  
Then, after taking a suitable linear combination of the middle and bottom rows, it can be seen that they split apart into two separate irreducible reps, which we call ${\cal F}^{[2]_\pm}_\Delta$, 
\be {\cal F}^{[2]}_\Delta={\cal F}^{[2]_+}_\Delta\oplus {\cal F}^{[2]_-}_\Delta,\ \ D=3\, .\label{d3tenssplite}\ee
This splitting is illustrated here:
\be \raisebox{-0pt}{\epsfig{file=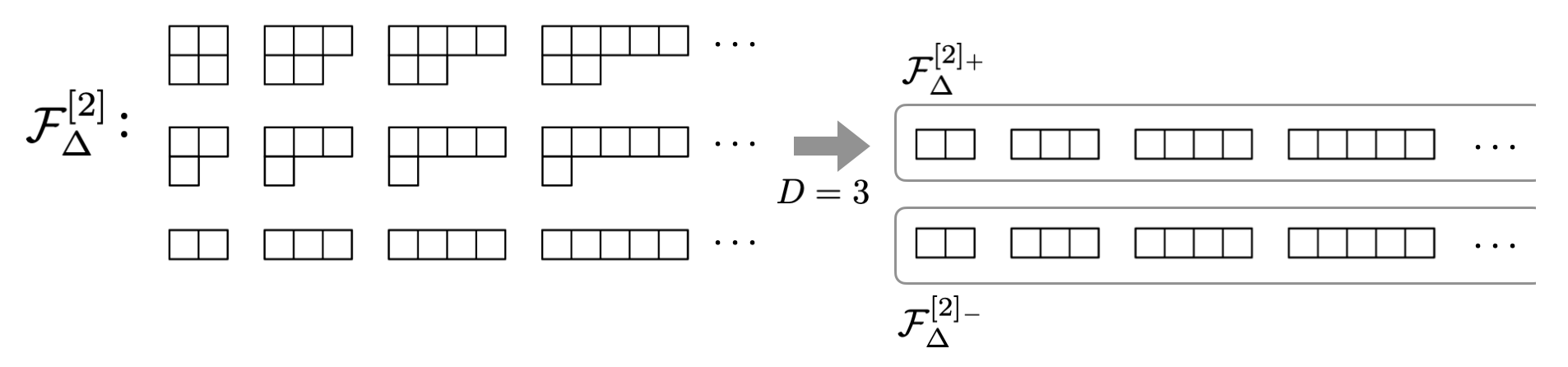,width=5.5in}}  \label{tensorsocontent10}\, \ee
The arrow indicates dualizing the middle row as well as taking appropriate linear combinations of the middle and bottom rows so that they separate under the action of the boosts ${\cal K}^I$.  Both ${\cal F}^{[2]\pm}_\Delta$ contain the same $\frak{so}(3)$ content $[2],[3],[4],\ldots$, but they form distinct $\frak{so}(1,3)$ reps.

In $D=3$, the representation space is the space of tensors on the 2-sphere.  The split \eqref{d3tenssplite} is equivalent to the splitting of this space into (imaginary) self-dual and anti-self-dual parts with respect to the $d=2$ epsilon tensor $\epsilon_{ij}$ on the sphere, acting on one of the indices of the tensor.  This reflects the fact that massive spin 2 fields on dS$_3$ come in two independent chiralities (just as on flat space, where the massive little group is $U(1)$ and massive tensor fields have independent positive and negative helicities $\pm 2$ with respect to it).  These are realized field theoretically by gravitational Chern-Simons terms that split the masses between the two chiralities \cite{Deser:1981wh}.  

The same splitting also occurs for the shortened reps ${\cal D}_{k+4}^{[2]}$ corresponding to the shift symmetric fields,
\be{\cal D}_{k+4}^{[2]}={\cal D}_{k+4}^{[2]_+}\oplus {\cal D}_{k+4}^{[2]_-}\, ,\ \ D=3\,,\label{d3tvecsplite2}\ee
illustrated here for $k=2$,
\be \raisebox{-0pt}{\epsfig{file=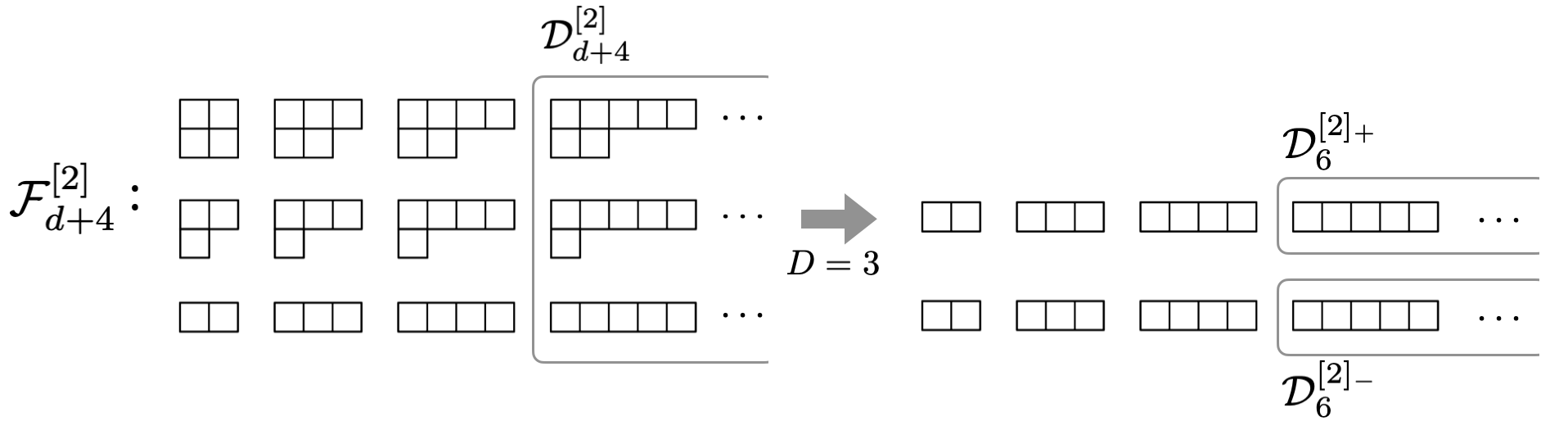,width=5.5in}}  \label{tensorsocontent11}\, \ee
And it occurs for the finite dimensional reps ${\cal S}_{-k-2}^{[2]}$ corresponding to the shift symmetries,
\be {\cal S}_{-k-2}^{[2]}={\cal S}_{-k-2}^{[2]_+} \oplus  {\cal S}_{-k-2}^{[2]_-} \, ,\ \ D=3\,,\label{d3vtecsplite2}\ee
illustrated here for $k=2$,
\be \raisebox{-0pt}{\epsfig{file=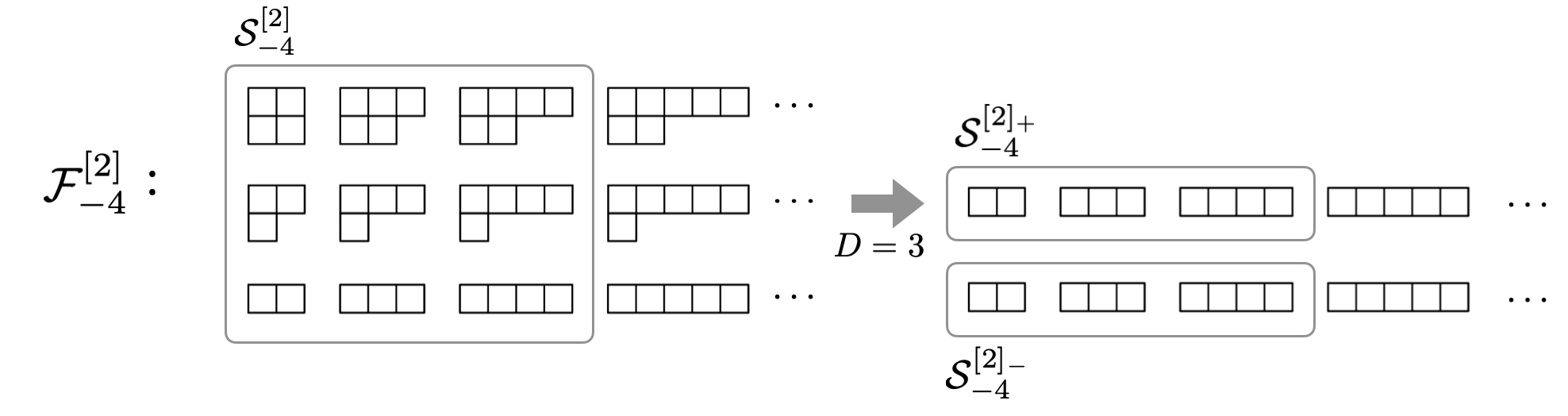,width=5.5in}}  \label{tensorsocontent12}\, \ee
These are the finite dimensional chiral reps $[k+2,2]_\pm$ of $\frak{so}(1,3)$, which when branched to $\frak{so}(3)$ using \eqref{evenDbranchinge} each give the reps $[2],[3],\ldots,[k+2]$.

The shadow transform \eqref{tensorinterfinte} that relates $\Delta$ and $\bar \Delta=2-\Delta$ flips the helicity, so we have
\be {\cal F}^{[2]_\pm}_\Delta\simeq {\cal F}^{[2]_\mp}_{\bar \Delta}\,, \ \ D=3\,.\ee
This means that at $\Delta=d/2=1$, the two reps are the same, ${\cal F}^{[2]_+}_1\simeq {\cal F}^{[2]_-}_1$.  At the shift symmetric and finite points, we have
\be  {\cal D}^{[2]_\pm }_{k+4} \simeq {\cal F}^{[2]_\mp}_{-k-2}/{\cal S}^{[2]_\mp}_{-k-2}\,, \ \ {\cal S}^{[2]_\pm }_{-k-2} \simeq {\cal F}^{[2]_\mp}_{k+4}/{\cal D}^{[2]_\mp}_{k+4} \,,\ \ D=3\,.\ee

In $D=3$, the range \eqref{tensorcompnle} degenerates and there is no complementary series for the tensor.  The point $\Delta=d-1$, which would have been where the complementary series ends at the partially massless rep, is also the point $\Delta=d/2$ which would be where the complementary series begins at its intersection with the principal series.  So ${\cal F}^{[2]}_1$ actually carries the partially massless tensor.

Looking at the partially massless rep in \eqref{tensorsocontent6}, the top row trivializes and the middle row can be dualized to obtain a rep which is identical to that of the $k=1$ shift symmetric scalar, ${\cal D}^{[0]}_{3}$, and we have the isomorphism ${\cal V}^{[2]}_1\simeq {\cal D}^{[0]}_{3}$.  This reflects the fact that a partially massless spin 2 field in $D=3$ is dual to a $k=1$ shift symmetric scalar, with the scalar's shift symmetry gauged \cite{Hinterbichler:2024vyv}.  From \eqref{tensaddisomre}, the rep ${\cal U}^{[2]}_1$ representing the PM gauge modes is also isomorphic to ${\cal D}^{[0]}_{3}$, so in $D=3$ we see that the physical and gauge modes of a partially massless spin 2 both carry the same rep, reflecting the equivalence ${\cal F}^{[2]_+}_1\simeq {\cal F}^{[2]_-}_1$.   Note that the point in \eqref{tensorsocontent6} is the same as the point in \eqref{tensorsocontent7} when $D=3$, and so in this case the massless vector modes and gauge modes split off into a direct sum rather than the factorizations and invariant subspaces that occur in general $D$.  Altogether, we have the equivalences
\be {\cal F}_1^{[2]_+}\simeq {\cal F}_1^{[2]_-} \simeq {\cal V}_1^{[2]}\simeq {\cal U}_1^{[2]} \simeq {\cal D}_3^{[0]}\,,\ \ \ D=3\,.\ee  
This degeneration and splitting of the PM rep in $D=3$ is illustrated here
\be \raisebox{-0pt}{\epsfig{file=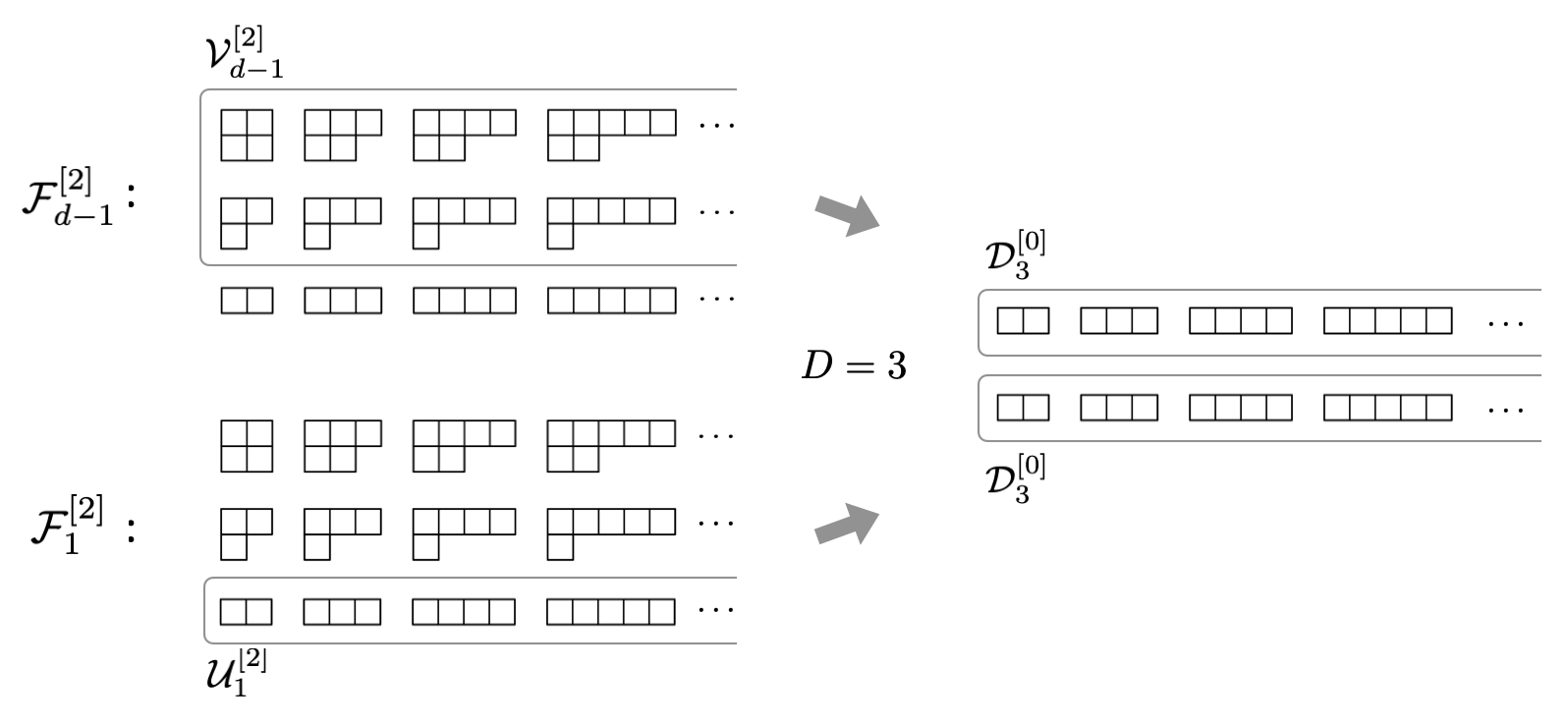,width=5.5in}}  \label{tensorsocontent13}\, \ee

Looking at the massless rep in \eqref{tensorsocontent4}, all of the $\frak{so}(3)$ reps in the top row that make up ${\cal V}^{[2]}_{2}$ are empty, and so this rep is trivial in $D=3$.  This reflects the fact that the massless graviton has no propagating degrees of freedom in $D=3$.

\subsubsection*{$D=4$:}

In $D=4$, the $\frak{so}(4)$ reps that have 2 rows in their tableaux split into chiral pairs as in \eqref{sod4splite}, and this affects the partially massless and massless spin 2 reps, the only ones with only 2-row tableaux among their 
$\frak{so}(4)$ content.  The massless reps split into two chiralities,
\be {\cal V}^{[2]}_3 ={\cal V}^{[2]_+}_3\oplus {\cal V}^{[2]_-}_3\,,\ \ \ D=4\,, \ee
illustrated here:
\be \raisebox{-0pt}{\epsfig{file=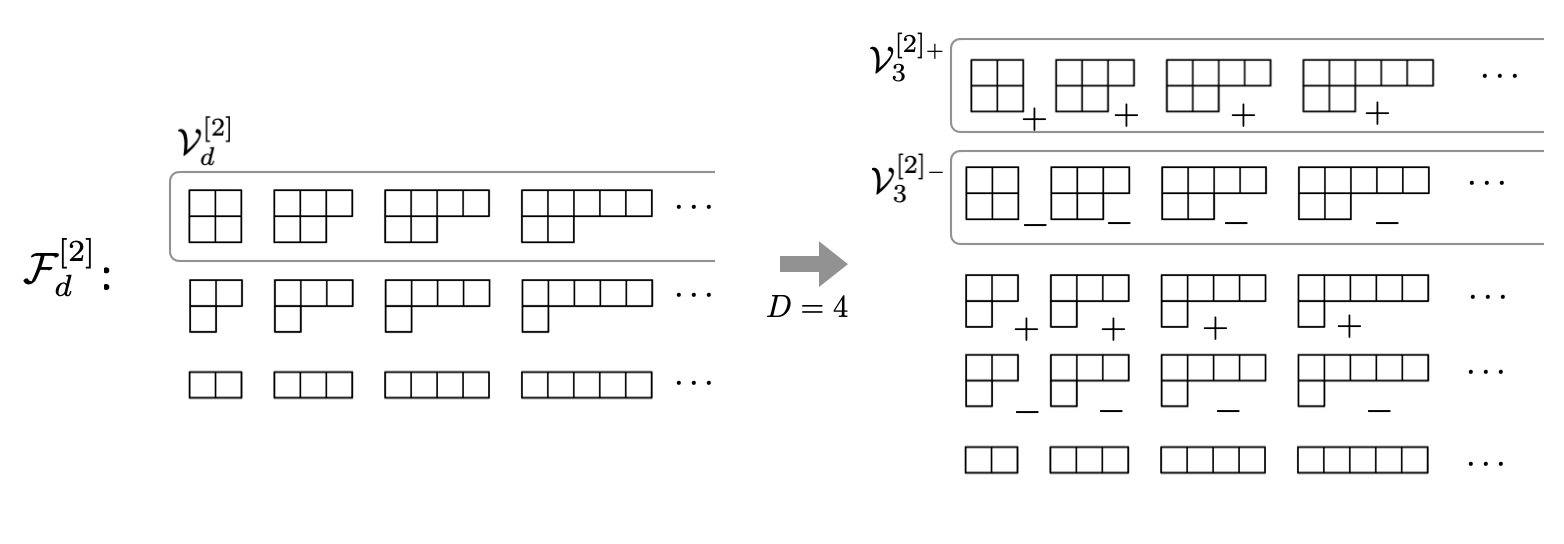,width=5.5in}}  \label{tensorsocontent14}\, \ee
These two reps represent the two independent chiralities of a massless graviton on dS$_4$ (just as the flat space massless graviton has two independent Lorentz invariant polarizations).

The same happens with the PM rep 
\be {\cal V}^{[2]}_2 ={\cal V}^{[2]_+}_2\oplus {\cal V}^{[2]_-}_2\,,\ee
illustrated here:
\be \raisebox{-0pt}{\epsfig{file=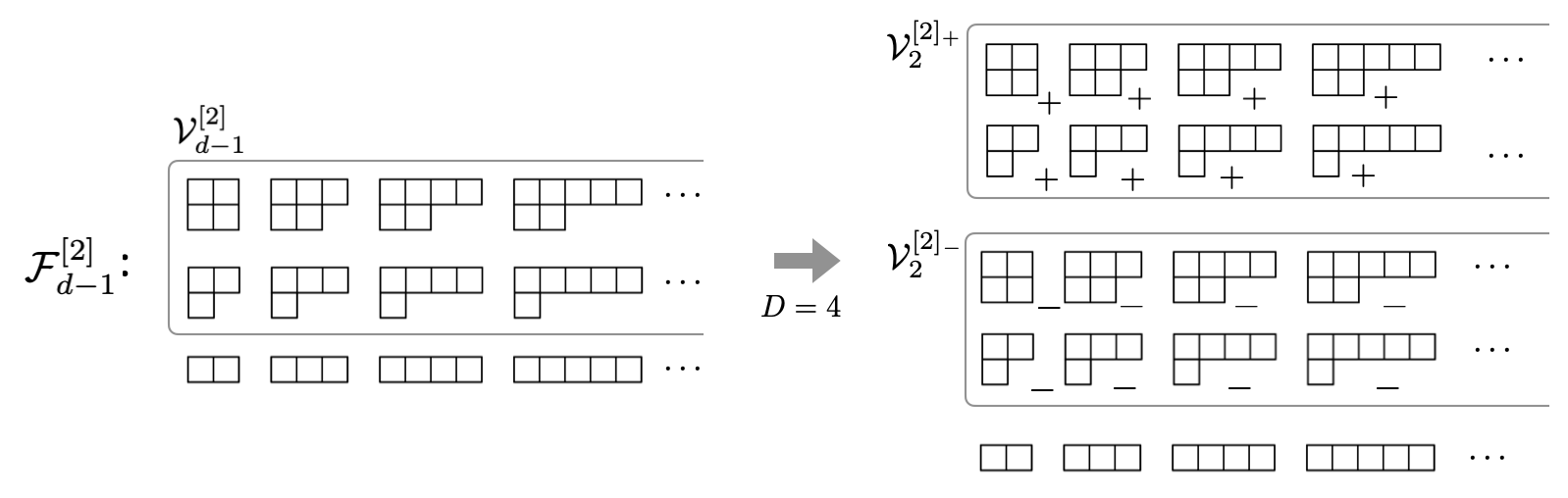,width=5.5in}}  \label{tensorsocontent15}\, \ee
This reflects the fact that the PM graviton in $D=4$, like the massless graviton, splits into two independent chiralities (this is the root of the duality invariance of the PM graviton in $D=4$ \cite{Deser:2013xb,Hinterbichler:2014xga}).

As we will see in section \ref{unitarylistsection}, the massless and partially massless reps in $D=4$ are accounted for among the so-called discrete series reps.

\subsection{Spin $s$ representations\label{spinssec}}

We now turn to the general symmetric traceless tensor reps, those corresponding to a particle with integer spin $s$.  We restrict to $s\geq 1$ in this section (the cases $s=1$ and $s=2$ reduce to sections \eqref{vectorsection} and \eqref{spin2sec} respectively).   We use the mass $m$ that vanishes when the field acquires its largest gauge invariance, which is related to the mass $\tilde m$ in the Klein-Gordon equation by  $\tilde m^2=m^2 +\left[s+D-2-(s-1)(s+D-4)\right]H^2$.  The relations \eqref{deltamregexe}, \eqref{dscfttmassrelatione} between $\Delta$ and $m^2$ in the late time behavior $\sim e^{-\left( \Delta_\pm-s\right) {H t}}$ of \eqref{ssassymprosole} for the spin $s$ field then reads
\be {m^2\over H^2}=-\left(\Delta+s-2\right)\left(\Delta-s-d+2\right)\,,\ \   \Delta_\pm= {d\over 2}\pm\sqrt{{\left({d\over 2}+s-2\right)^2 }-{m^2\over H^2}}\,. \label{tensspsssmassretlatiospsnere}\ee

The representation space will be the space of square integrable complex traceless rank $s$ symmetric tensor fields on ${\mathbb S}^d$, transforming under $\frak{so}(1,D)$ as in \eqref{latetimesheactiont3} with $r=s$.  Call this space ${\cal F}^{[s]}_\Delta$,
\be {\cal F}^{[s]}_\Delta:\ \ {\rm complex\ symmetric\ traceless\ rank\ } s {\rm \ tensor\ fields\ on\ } {\mathbb S}^d \,.\ee

To proceed with decomposing this space into $\frak{so}(D)$ reps, we first need the SVT-like decomposition of a rank $s$ traceless symmetric tensor.  On the sphere (or any locally maximally symmetric space \cite{Bonifacio:2021msa}), this decomposition breaks the field into symmetric and transverse traceless pieces $\chi_{i_1\ldots i_{u}}$ of all ranks $u=0,1,\ldots,s$:
\be \phi_{i_1\ldots i_s}=\chi_{i_1\ldots i_s}+\nabla_{(i_1} \chi_{i_2\ldots i_s)_T}+\nabla_{(i_1}\nabla_{i_2} \chi_{i_3\ldots i_s)_T}+\cdots+\nabla_{(i_1} \ldots \nabla_{i_s)_T}\chi\, .\label{SVTdedcompeese}\ee 
Each of the $\chi_{i_1\ldots i_{u}}$  are transverse and traceless,
\bea &&\nabla^{i_1}\chi_{i_1i_2\ldots i_{u}}=0\, , \ \ \ u=1,2,\ldots, s \ \ \ , \\
&& \chi^i_{\ i i_3\ldots i_{u}}=0,\ \ \ \ \  \ u=2,3,\ldots, s\, \ \ . 
\eea
Each of them can in turn be broken up into tensor spherical harmonics on ${\mathbb S}^d$, which take the form \cite{rubin1984eigenvalues,rubin1985symmetric,Higuchi:1986wu}
\be Y^{I_1\ldots I_l,J_1\ldots J_{u}}_{l,i_1\ldots i_{u}}(\hat X)= \partial_{i_1} \hat X^{[J_1}\, \hat X^{I_1]}\cdots  \partial_{i_{u}} \hat X^{[J_{u}}\, \hat X^{I_{u}]}  \hat X^{I_{{u}+1}}\cdots\hat X^{I_l}-{\rm traces}\, ,\ \ \ l={u},{u}+1,{u}+2,\ldots \ .\label{tensorspherharmonicese2}\ee
These transform in the $\frak{so}(D)$ rep given by a traceless $[l,{u}]$ tableau.  They form a basis of the space of transverse traceless symmetric rank ${u}$ tensors on ${\mathbb S}^d$.  The natural Laplacian on this space is 
\be \Delta_L=-\nabla^2_\Omega+{u} ({u}+d-2)\, ,\label{spinslaplacianue}\ee 
which is the rank ${u}$ version of the Lichnerowicz Laplacian, and the transverse harmonics \eqref{tensorspherharmonicese2} are eigenfunctions of it, with the following eigenvalues \cite{rubin1984eigenvalues,rubin1985symmetric,Higuchi:1986wu},
\be \Delta_L Y^{I_1\ldots I_l,J_1\ldots J_{u}}_{l,i_1\ldots i_{u}}=   \left[ l(l+d-1)+{u}({u}+d-3)\right] Y^{I_1\ldots I_l,J_1\ldots J_{u}}_{l,i_1\ldots i_{u}} \, .\ee

The harmonics in $\chi_{i_1\ldots i_{u}}$ with $l<s$ will not appear in the decomposition \eqref{SVTdedcompeese}, because they are generalized conformal Killing tensors that are annihilated by the derivative combination $\nabla_{(i_{{u}+1}}\cdots \nabla_{i_s} \chi_{i_1\ldots i_{u})_T}$ that appears in  \eqref{SVTdedcompeese} (see appendix B of \cite{Brust:2016gjy} for a discussion of these generalized conformal Killing tensors). 

The $\frak{so}(D)$ rep content of the space ${\cal F}^{[s]}_\Delta$ can be illustrated as a lattice with $s+1$ rows and an infinite number of columns labelled by $l=s,s+1,s+2,\ldots$ as illustrated here:
\be \raisebox{-0pt}{\epsfig{file=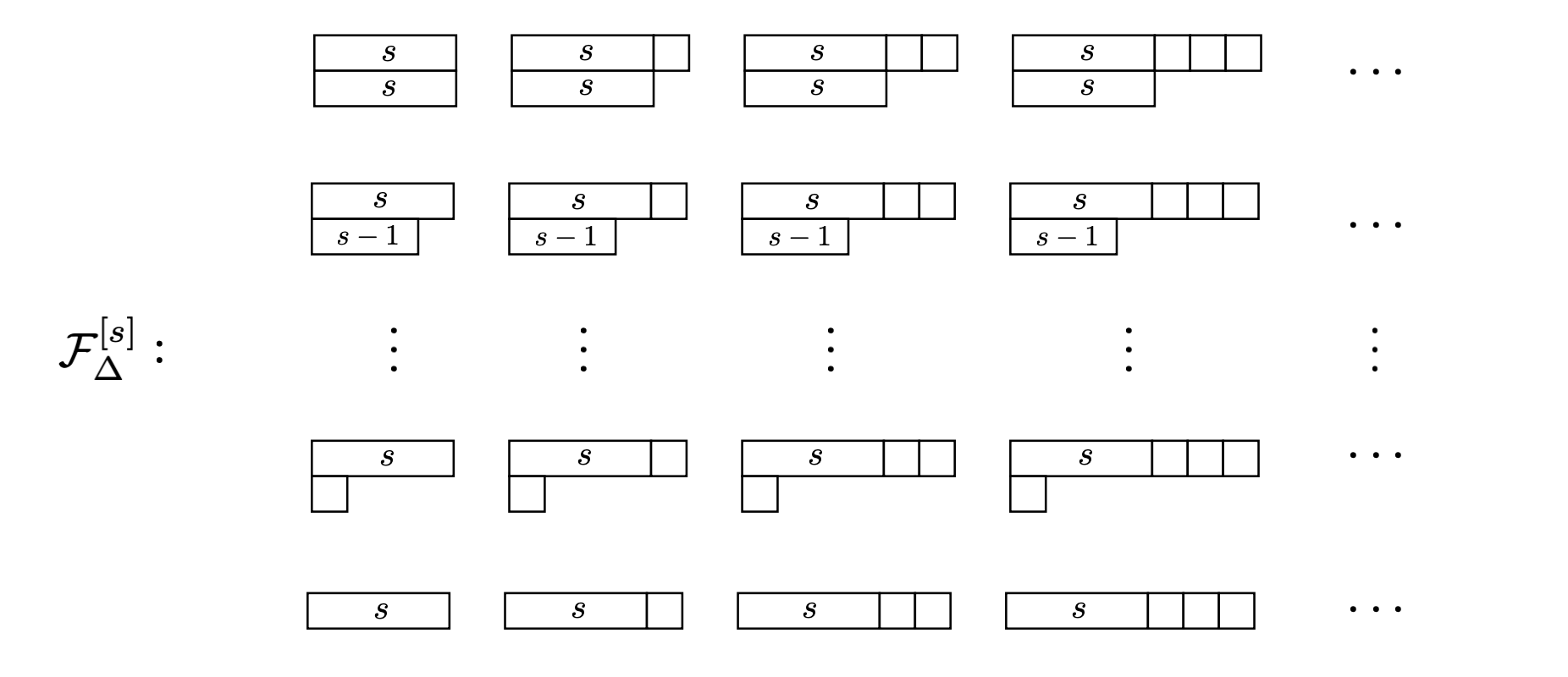,width=5.5in}}  \label{spinssocontent}\, \ee
The first row contains the harmonics in the highest rank part $\chi_{i_1\ldots i_s}$ of the SVT decomposition \eqref{SVTdedcompeese}, the second row contains the harmonics in the next highest rank part $\chi_{i_1\ldots i_{s-1}}$, and so on down to the last row which contains the harmonics of the scalar part $\chi$.
For generic $\Delta$, this lattice of $\frak{so}(D)$ reps will be all be linked together by the action of the boost generators ${\cal K}^I$ through two-way nearest neighbor interactions analogous to those displayed in \eqref{tensorsocontent} (to avoid unwieldy pictures we will suppress these arrows from now on), forming an irreducible rep of $\frak{so}(1,D)$.

\textbf{Reducible cases:} There are discrete values of $\Delta$ at which these arrows break and we get sub-reps, making the rep ${\cal F}_{\Delta}^{[s]}$ reducible but not decomposable.  These values are as follows:
\begin{itemize}

\item  
Shift symmetric points: 
\be \Delta=d+s+k\, ,\ \  k=0,1,2,\ldots \ \ .\ee 
At these values the columns to the right of the $(k+1)$-th column in \eqref{spinssocontent} split off into the infinite dimensional sub-reps 
\be {\cal D}^{[s]}_{d+s+k}\, ,\ \ \ k=0,1,2,\ldots\,.\ee
This is illustrated here for the case where $k=1$,
\be \raisebox{-0pt}{\epsfig{file=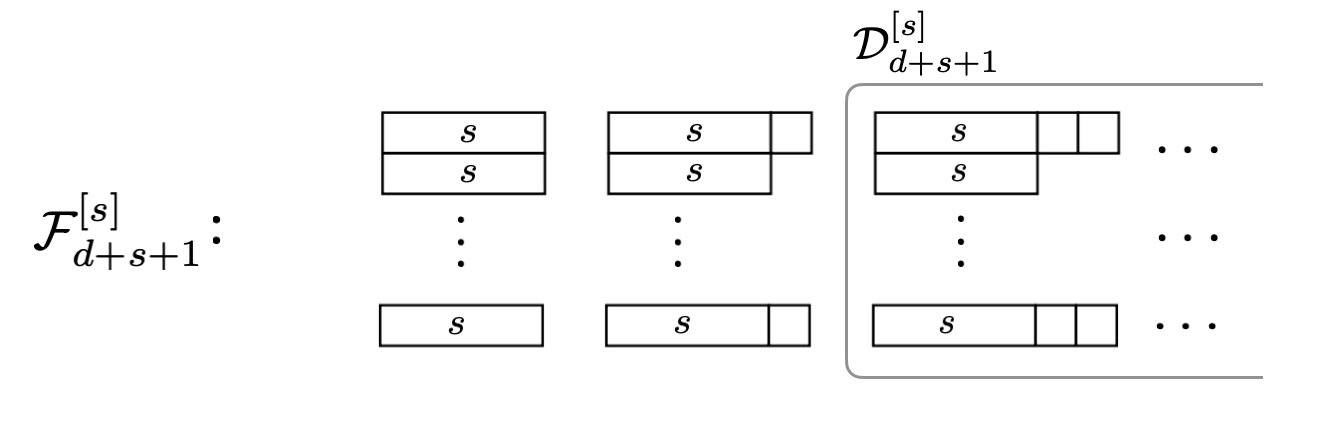,width=4.7in}}  \label{spinssocontent2}\, \ee
These represent the physical modes of the level $k$ shift symmetric spin $s$ fields.

\item
Finite points:
\be \Delta=-s-k\, ,\ \  k=0,1,2,\ldots\ \ .\ee 
Here the first  $(k+1)$ columns in \eqref{spinssocontent} split off into the finite dimensional sub-rep
\be {\cal S}^{[s]}_{-s-k}\, ,\ \ \ k=0,1,2,\ldots\,.\ee
This is illustrated here for the case where $k=1$,
\be \raisebox{-0pt}{\epsfig{file=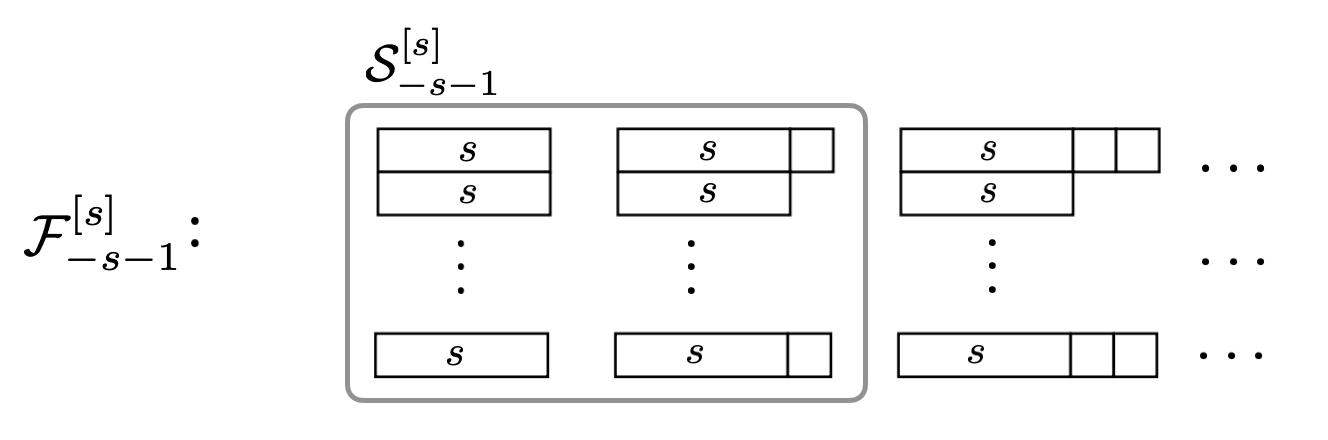,width=4.0in}}  \label{spinssocontent3}\, \ee
These are the finite dimensional reps corresponding to the $[s+k,s]$ tensor reps of $\frak{so}(1,D)$.  When it is branched to $\frak{so}(D)$ using the branching rules in appendix \ref{branchingappendix}, it gives all the $\frak{so}(D)$ tensors present in ${\cal S}^{[s]}_{-s-k}$.  Note that the $[s+k,s]$ tableau appears in the upper right of the ${\cal S}^{[s]}_{-s-k}$ sub-rep.   

These correspond to the shift symmetries of the level $k$ shift symmetric spin $s$ field; indeed the shift symmetries are parametrized by a $[k+s,s]$ tensor in the dS$_D$'s embedding space \cite{Bonifacio:2018zex}.

\item 
PM points:
\be \Delta=d+s-t-1,\ \ \ t=1,2,\ldots,s\,.\ee
There are now $s$ different points, labelled by $t$, corresponding to the different depths \cite{Deser:2001pe} of partially massless fields.\footnote{We use the convention where $t$ is the number of derivatives in the PM gauge transformation, so $t=1$ is the massless case.  This differs from the conventions in some of our other papers such as \cite{Hinterbichler:2016fgl,Hinterbichler:2022vcc,Baumann:2025tkm}: $t_{\rm there}=s-t_{\rm here}$.}
At these values, the top $t$ rows in \eqref{spinssocontent} split off into the infinite dimensional sub-rep 
\be {\cal V}^{[s]}_{d+s-t-1}\, ,\ \ \ t=1,2,\ldots,s\,.\ee
These correspond to the physical modes of the depth $t$ partially massless spin $s$ field.   This is illustrated here:
\be \raisebox{-0pt}{\epsfig{file=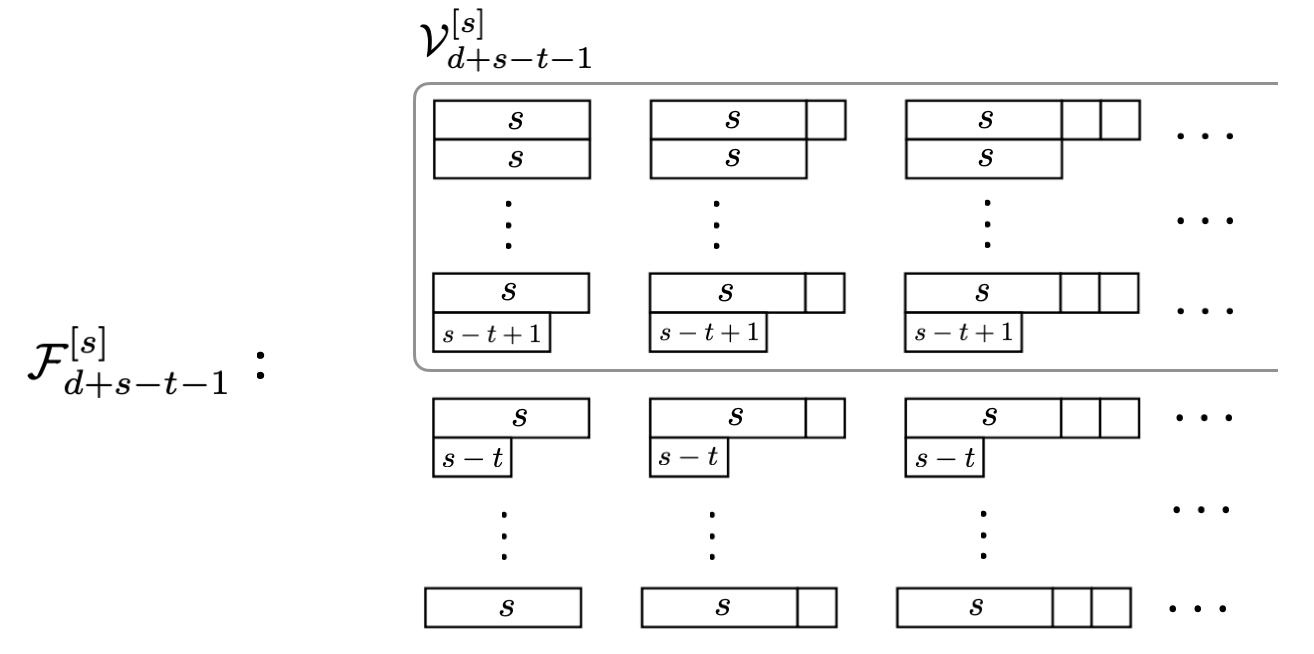,width=4.3in}}  \label{spinssocontent4}\, \ee

\item 
Gauge points:
\be \Delta=-s+t+1\ \ \ t=1,2,\ldots,s\, . \ee
The last  $s-t+1$ rows in \eqref{spinssocontent} split off into the infinite dimensional sub-rep
\be {\cal U}^{[s]}_{-s+t+1}\, ,\ \ \ t=1,2,\ldots,s\,.\ee
This is illustrated here:
\be \raisebox{-0pt}{\epsfig{file=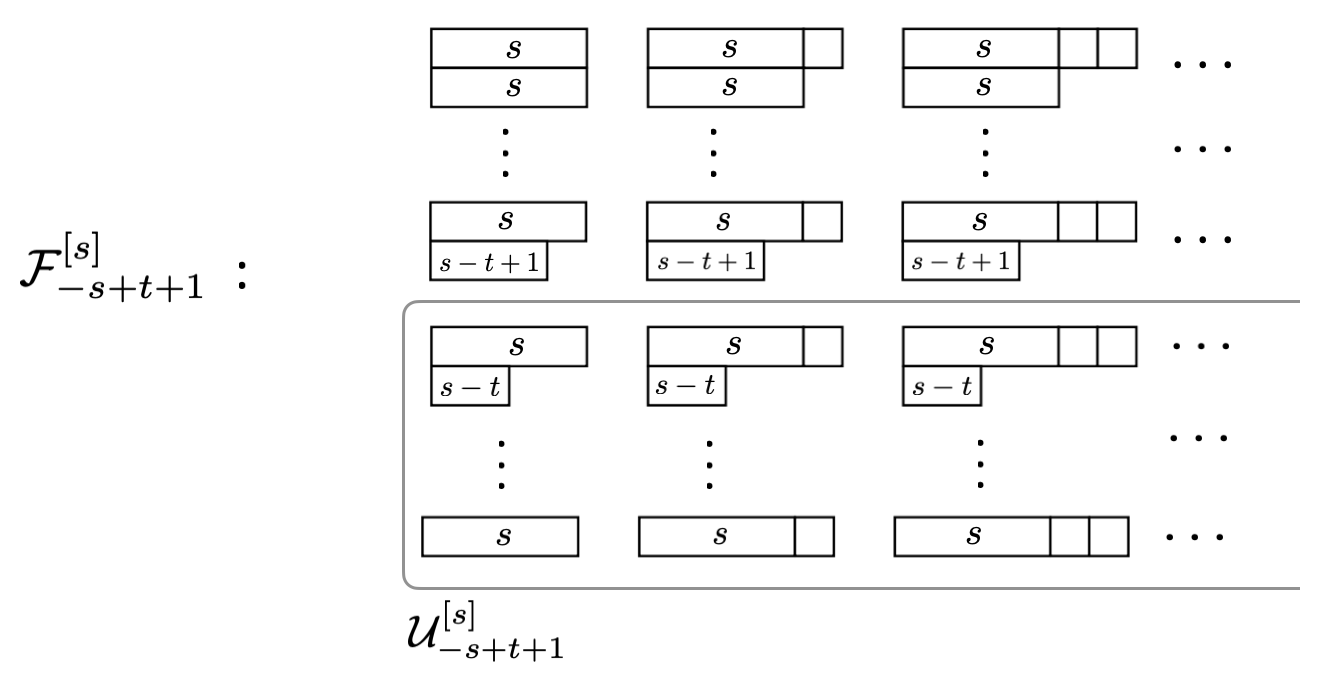,width=4.0in}}  \label{spinssocontent5}\, \ee
These correspond to the gauge modes of the partially massless fields.

\end{itemize}
Note that we have now taken the terminology ``partially massless'' to also encompass the massless case: the case $t=1$ corresponds to the massless field.  We will use this definition of ``partially massless'' going forward.

\textbf{Equivalences:} There is a shadow transform intertwining operator that connects the $\Delta$ and $\bar\Delta\equiv d-\Delta$ reps, 
\be S_\Delta^{[s]} :\ {\cal F}_{\Delta}^{[s]}\rightarrow {\cal F}_{\bar\Delta}^{[s]}\,.\label{spinsintertwineree}\ee
It commutes with the $\frak{so}(D)$ rotations and satisfies $\delta_{{\cal K}^I_{\bar\Delta}}S_\Delta^{[s]}=S_\Delta^{[s]} \delta_{{\cal K}^I_{\Delta}}$.  For generic $\Delta$, it is invertible and the $\Delta$ and $\bar\Delta$ reps are equivalent to each other,
\be {\cal F}_{\Delta}^{[s]}\simeq {\cal F}_{\bar\Delta}^{[s]}\,.\label{spinsequivljee}\ee
  But for the special values of $\Delta$ given above where ${\cal F}_{\Delta}^{[s]}$ develops a sub-rep, $S_\Delta^{[s]}$ develops a kernel which is always precisely the sub-rep.

The shift symmetric and finite points are linked to each other via the maps $S^{[s]}_{-s-k}$, ${S}^{[s]}_{d+s+k}$, where the kernel and image of each are the sub-reps ${\cal S}^{[s]}_{-s-k}$, ${\cal D}^{[s]}_{d+s+k}$, which induces the isomorphisms
\be {\cal S}^{[s]}_{-s-k}\simeq {\cal F}^{[s]}_{d+s+k}/{\cal D}^{[s]}_{d+s+k}\,,\ \ \  {\cal D}^{[s]}_{d+s+k}\simeq {\cal F}^{[s]}_{-s-k}/{\cal S}^{[s]}_{-s-k}\,.\ee
This is illustrated here for the case $k=1$,
\be \raisebox{-0pt}{\epsfig{file=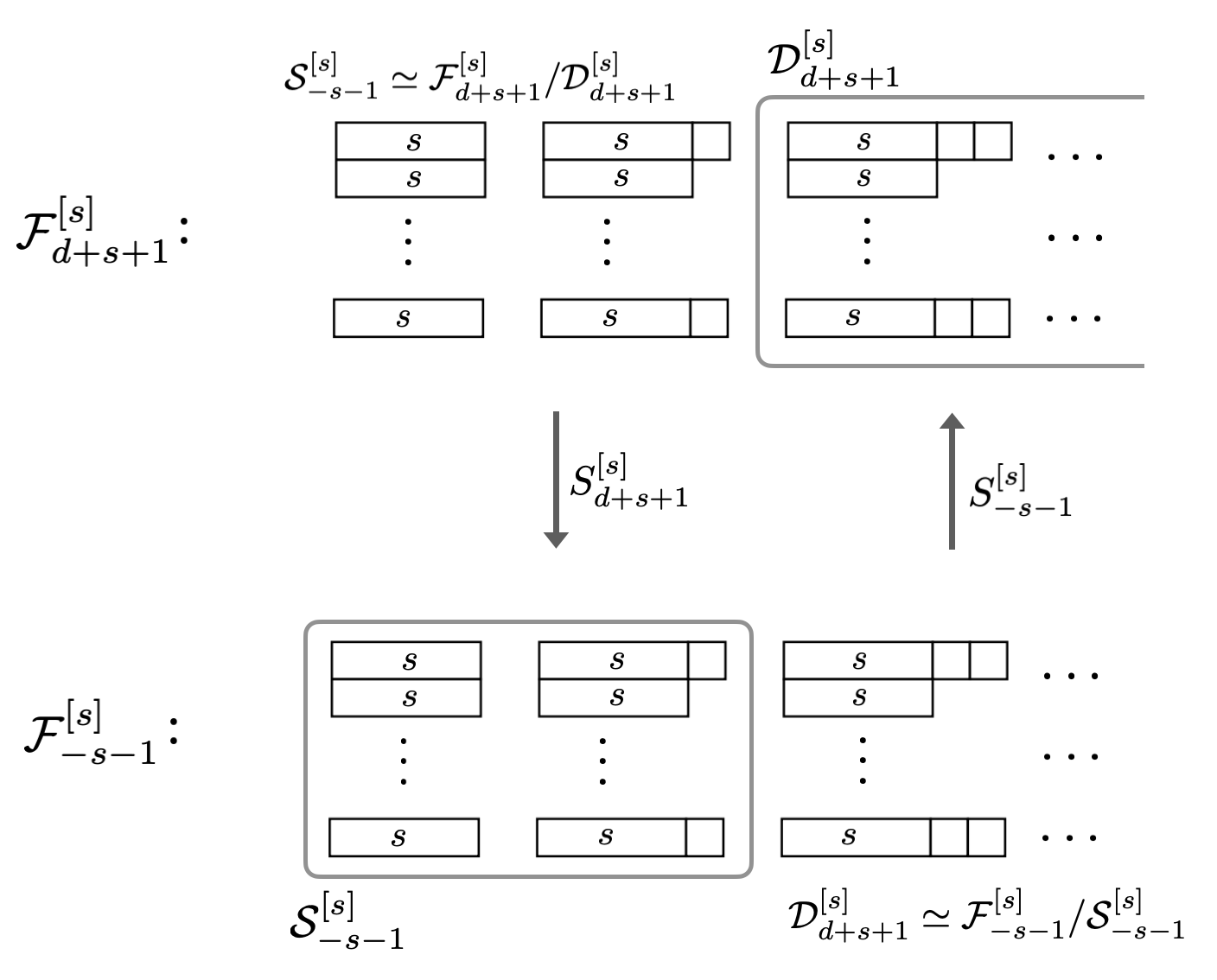,width=4.3in}}  \label{spinssocontent6}\, \ee

The PM points and their gauge mode points are are linked to each other via the maps $S^{[s]}_{1-s+t}$, ${S}^{[s]}_{d-1+s-t}$, where the kernel and image of each are the sub-reps ${\cal S}^{[s]}_{1-s+t}$, ${\cal D}^{[s]}_{d-1+s-t}$, which induces the isomorphisms
\be {\cal U}^{[s]}_{1-s+t}\simeq {\cal F}^{[s]}_{d-1+s-t}/{\cal V}^{[s]}_{d-1+s-t}\,,\ \ \  {\cal V}^{[s]}_{d-1+s-t}\simeq {\cal F}^{[s]}_{1-s+t}/{\cal U}^{[s]}_{1-s+t}\,.\label{PMgagspindslinkeee}\ee
This is illustrated here:
\be \raisebox{-0pt}{\epsfig{file=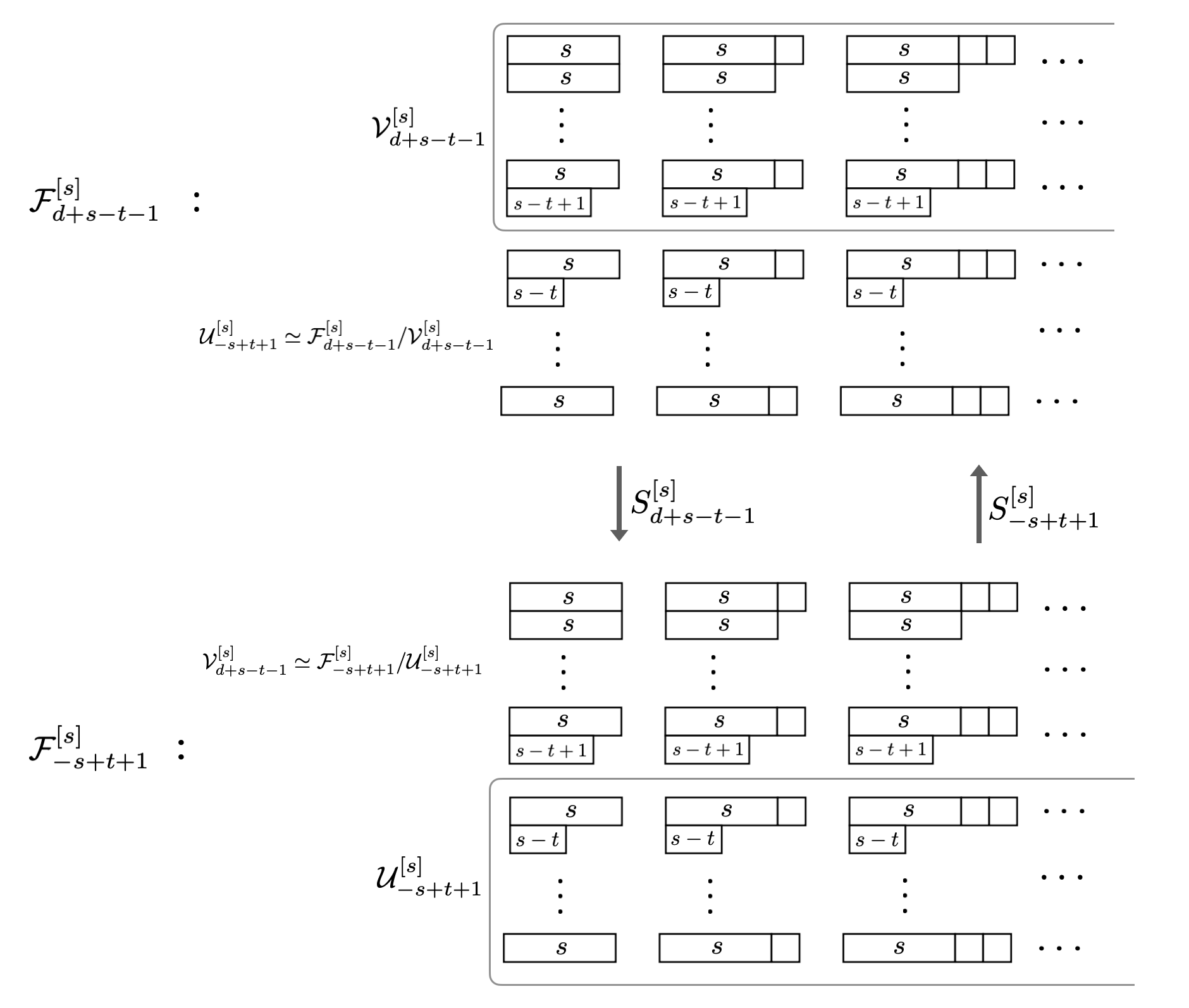,width=5.0in}}  \label{spinssocontent7}\, \ee

There are two other maps of interest that map between reps with different values of $s$ and that are important for the PM cases.  The first is the order $t$ symmetrized gradient, which takes spin $s-t$ tensors with $\Delta=1-s$ to spin $s$ tensors, adding $t$ units of $\Delta$ in the process,
\be {\rm grad}^{t}: \  {\cal F}^{[s-t]}_{1-s} \rightarrow {\cal F}^{[s]}_{1-s+t} \, ,\ \ \phi_{i_1\ldots i_{s-t}} \rightarrow \nabla_{(i_{s-t+1}}\cdots \nabla_{ i_s} \phi_{i_1\ldots i_{s-t})_{T}}\,.
\ee
The kernel of ${\rm grad}^{t}$ is the space ${\cal S}^{[s-t]}_{1-s}$ and the image is the space ${\cal U}^{[s]}_{1-s+t}$ (the kernel is the space of generalized conformal Killing vectors).
The other map of interest is the order $t$ divergence, which takes spin $s$ tensors with $\Delta=d-1+s-t$ to spin $s-t$ tensors, adding $t$ units of $\Delta$ in the process,
\be {\rm div}^{t}: \  {\cal F}^{[s]}_{d-1+s-t} \rightarrow {\cal F}^{[s-t]}_{d-1+s} \, ,\ \ \phi_{i_1\ldots i_{s}} \rightarrow \nabla^{i_1}\cdots \nabla^{i_{t}} \phi_{i_1\ldots i_t i_{t+1}\ldots i_{s}}\,.
\ee
The kernel of ${\rm div}^{t}$ is the space ${\cal V}^{[s]}_{d-1+s-t}$ (this is the space of multiply conserved currents) and the image is the space ${\cal D}^{[s-t]}_{d-1+s}$.  

The four spaces ${\cal F}^{[s]}_{d+s-t-1}$, ${\cal F}^{[s]}_{-s+t+1}$, ${\cal F}^{[s-t]}_{d+s-1}$, ${\cal F}^{[s-t]}_{-s+1}$ are joined together into the commutative diagram illustrated here:
\be \raisebox{-0pt}{\epsfig{file=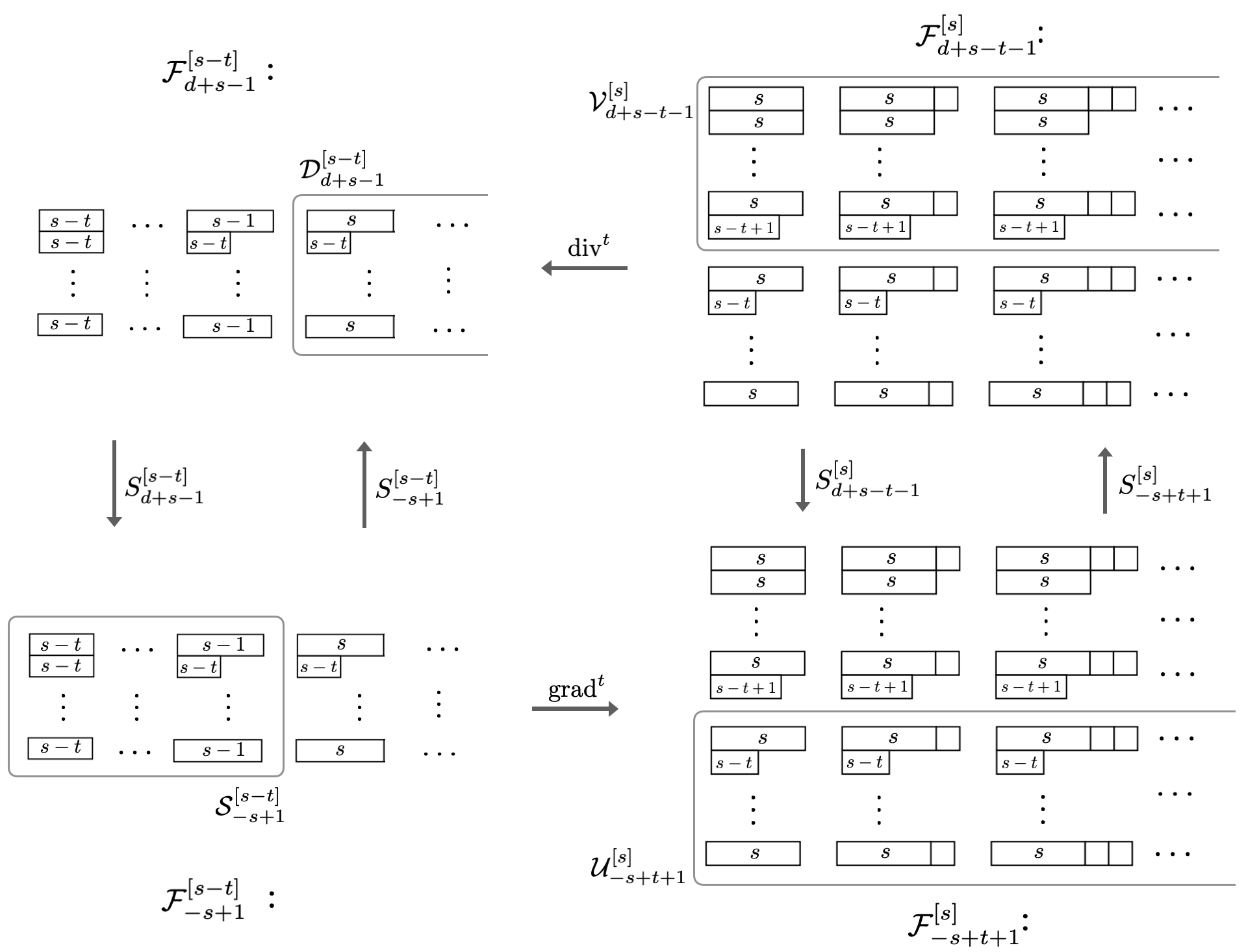,width=6.2in}}  \label{spinssocontent10}\, \ee
The grey rectangles enclose the various sub-reps; these are simultaneously the kernel of every outgoing map, and the image of every ingoing map, and going through any two arrows consecutively gives zero.  
From this, we get the isomorphisms
\be { \cal D}^{[s-t]}_{d+s-1}\simeq { \cal U}^{[s]}_{-s+t+1}\, .\label{spinsudisoeee}\ee
This expresses the statement that the gauge modes of a partially massless spin $s$, depth $t$, field are precisely those of a $k=t-1$ shift symmetric spin $s-t$ field \cite{Bonifacio:2018zex}.

\textbf{Unitarity:} There is a diffeomorphism and Weyl invariant bilinear form pairing the two spaces ${\cal F}_\Delta^{[s]}$ and ${\cal F}_{\bar \Delta}^{[s]}$,
\be (\phi_1,\phi_2)\equiv \int d^d\Omega\, \phi_1^{i_1\ldots i_s}(\hat X)\phi_{2\,  i_1\ldots i_s}(\hat X)\,,\ \ \ \  \phi_{1\,  i_1\ldots i_s}\in {\cal F}_{\bar\Delta}^{[s]},\ \ \phi_{2\, i_1\ldots i_s}\in {\cal F}_{\Delta}^{[s]}\, . \label{bilintsearformvee}\ee

In the case where $\Delta^\ast=\bar\Delta$ we can use this to form a manifestly positive definite, invariant inner product on ${\cal F}_\Delta^{[2]}$ via:
\be \la \phi_1 | \phi_2 \ra\equiv (\phi_1^{\ast},\phi_2),\ \ \ \Delta^\ast=\bar\Delta\,, \ \  \phi_{1\, i_1\ldots i_s}, \phi_{2\, i_1\ldots i_s}\in {\cal F}_{\Delta}^{[s]}\,. \label{innerprsotdvprise} \ee
This gives us the spin 2 principal series reps, which are all unitary
\be {\rm spin\ } s {\rm \ principal\ series:}\  \Delta={d\over 2}+i\nu\, ,\ \ \ \nu\in {\mathbb R}\,.\ee

In the case where $\Delta$ is real, we use $S^{[s]}_{\Delta}$ to move a state from ${\cal F}_{ \Delta}^{[s]}$ to ${\cal F}_{\bar \Delta}^{[s]}$ and form an inner product on ${\cal F}_{ \Delta}^{[s]}$ as follows,
\be \la \phi_1 | \phi_2 \ra\equiv (S^{[s]}_{\Delta}\phi_1^\ast,\phi_2)\, ,\ \ \ \Delta^\ast=\Delta\, , \ \  \phi_{1\, i_1\ldots i_s}, \phi_{2\, i_1\ldots i_s}\in {\cal F}_{\Delta}^{[s]}\,.  \label{innerprodtprisscsvdee} \ee 
The positivity of this inner product is now equivalent to whether the matrix elements of $S^{[s]}_{\Delta}$ are positive for all the different $\frak{so}(D)$ reps in ${\cal F}^{[s]}_\Delta$.  Away from the discrete special cases described above, it turns out that they are all positive only in the range $1<\Delta<d-1$, which gives the spin $s$ complementary series,
\be {\rm spin\ } s {\rm \ complementary\ series:}\  1<\Delta<d-1\,, \label{spinscomgrangee}\ee
where we have remained ambiguous as to which series $\Delta=d/2$ belongs to.

Now turn to the cases of $\Delta$ where the reps become reducible. For the finite points $\Delta=-s-k$, $k=0,1,2,\ldots$, we use $S^{[s]}_{-s-k}$ in the inner product, and this has the finite dimensional kernel consisting of the sub-rep ${\cal S}^{[s]}_{-s-k}$.   All of these states will have zero norm in the inner product.  Apart from these null states, other states also have negative norm, so even once these null states are factored out, we will be left with non-unitary reps.  The rep obtained after this factoring is nothing but ${\cal D}^{[s]}_{d+s+k}$, realized through the quotient ${\cal D}^{[s]}_{d+s+k}\simeq {\cal F}^{[s]}_{-s-k}/{\cal S}^{[s]}_{-s-k}$, so these reps, corresponding to the shift symmetric spin $s$ fields, are non-unitary.

For the shift symmetric points $\Delta=d+s+k$, $k=0,1,2,\ldots$, we use $S^{[s]}_{d+s+k}$ in the inner product, and this has the infinite dimensional kernel consisting of the states of the sub-rep ${\cal D}^{[s]}_{d+s+k}$.   All of these states will therefore be null in the inner product.  Apart from these null states, some of the norms are always still negative, so if the null states are factored out, we will be left with a finite dimensional non-unitary rep.  This is the rep ${\cal S}^{[s]}_{-s-k}$, realized through the quotient ${\cal S}^{[s]}_{-s-k}\simeq  {\cal F}^{[s]}_{d+s+k}/{\cal D}^{[s]}_{d+s+k}$.  

For the gauge points $\Delta=-s+t+1$, we use $S^{[s]}_{-s+t+1}$ in the inner product, and this has as kernel the sub-rep ${\cal U}^{[s]}_{-s+t+1}$.   All of these states will have zero norm in the inner product.  Apart from these null states, the other states all have positive norm, so  once these null states are factored out, we will be left with a unitary rep, which is nothing but ${\cal V}^{[s]}_{d+s-t-1}$, realized through the quotient ${\cal V}^{[s]}_{d+s-t-1}\simeq {\cal F}^{[s]}_{-s+t+1}/{\cal U}^{[s]}_{-s+t+1}$.  These correspond to the physical states of the partially massless spin $s$ fields, which are all unitary on dS.

For the PM points $\Delta=d+s-t-1$, we use $S^{[s]}_{d+s-t-1}$ in the inner product, and this has as kernel the sub-rep ${\cal V}^{[s]}_{d+s-t-1}$.   All of these states will have zero norm in the inner product.  Apart from these null states, the other states do not have vanishing norm, so once these null states are factored out, we will be left with a rep which is nothing but ${\cal U}^{[s]}_{-s+t+1}$, realized through the quotient ${\cal U}^{[s]}_{-s+t+1}\simeq {\cal F}^{[s]}_{d+s-t-1}/{\cal V}^{[s]}_{d+s-t-1}$.  These correspond to the longitudinal gauge modes of the partially massless fields, which are equivalent to the shift symmetric $k=t-1$ spin $s-t$ field due to \eqref{spinsudisoeee}.  For $t<s$, these will all be non-unitary, whereas for $t=s$ they are unitary, since ${\cal U}^{[s]}_{1}$ is equivalent to the unitary shift symmetric scalar rep ${\cal D}^{[0]}_{d+s-1}$.

For all but the unitary cases discussed here, there is no other way to construct an invariant inner product on ${\cal F}_\Delta^{[s]}$, so all the other reps in the complex $\Delta$ plane are non-unitary.  

\textbf{Summary:}  The spin $s$ reps are summarized here:
\be \raisebox{-40pt}{\epsfig{file=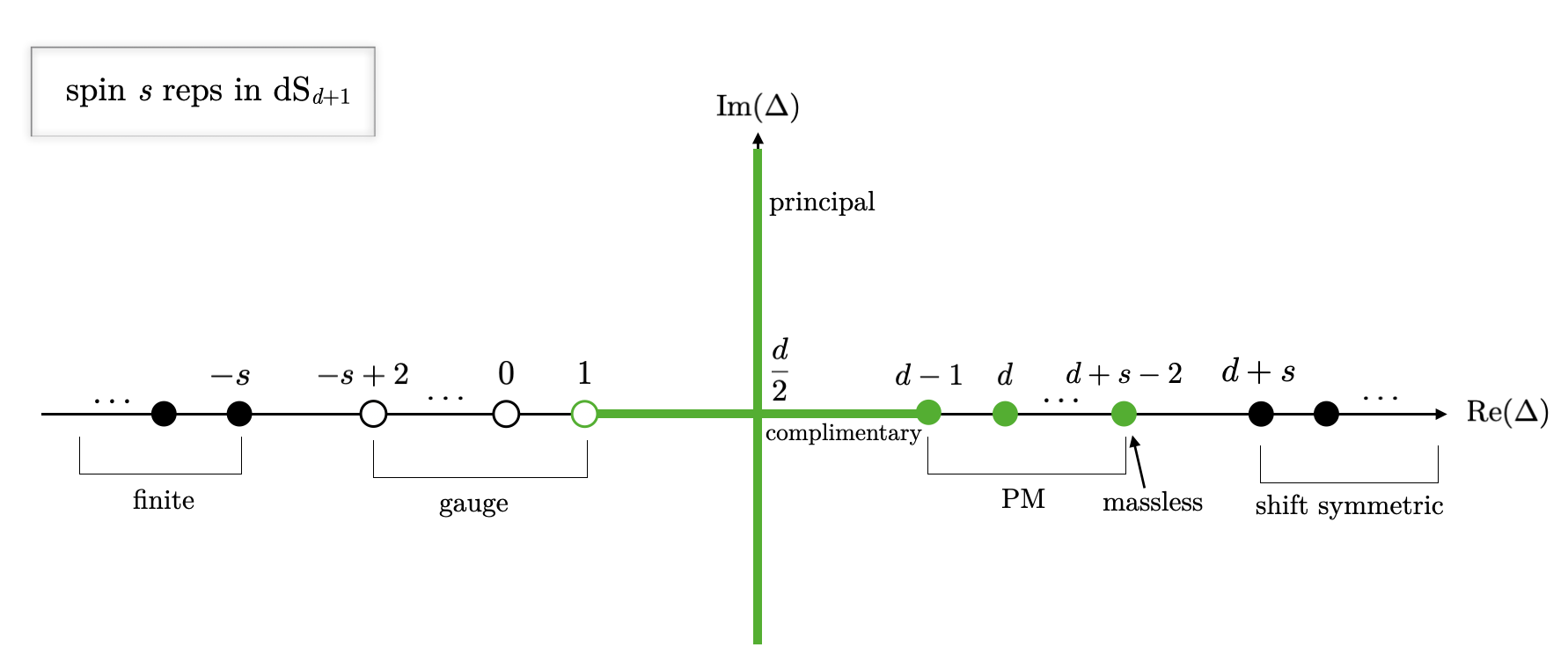,width=6.4in}}  \label{dsrepsspins}\, \ee
 Points in green are unitary reps.  The dots are the special points where the rep becomes reducible. The dots representing the gauge points at $\Delta=-s+2,-s+3,\ldots,1$ are empty to indicate that these are equivalent through \eqref{spinsudisoeee} to shift symmetric spin $0,1,\ldots,s-1$ reps respectively, and thus already accounted for among the lower spin reps.
  The equivalence \eqref{spinsequivljee} of the non-special reps is given by reflecting through the point $\Delta=d/2$, and due to the breakdown of the intertwiner map \eqref{spinsintertwineree}, the special reps with dots are inequivalent despite the reflection.  Other than this equivalence, all the reps are distinct. 

The value of the quadratic Casimir operator ${\cal C}_2$ of section \ref{casimirsec} on the spin $s$ reps is given by
\be {\cal C}_2= \Delta(\Delta-d)+s(s+d-2) = -{m^2\over H^2}+2(s-1)(D+s-3)\, .\ee

We have the following correspondence between spin $s$ fields on dS$_{d+1}$ and spin $s$ unitary reps of $\frak{so}(1,D)$:
\bea \begin{cases}
 {\rm principal:\ }\quad  \Delta={d\over 2}+i\nu \, , \ \ \nu\in {\mathbb R}\,, & {\rm heavy:\ } {m^2\over H^2 }\geq {\left(D+2s-5\right)^2\over 4}\, , \\
 {\rm complementary:\ } \quad 1<\Delta<d-1 \, , & {\rm light:\  } (s+D-4)(s-1) < {m^2\over H^2 }\leq  {\left(D+2s-5\right)^2\over 4}\, , \\
    {\rm PM:\ } \quad \Delta=d-1+s-t \, , \ \ t=1,2,\ldots,s\,,  &  {\rm PM:\ } {m^2\over H^2 }=(2s+D-4-t)(t-1)\, . \\
\end{cases} \nn
\eea

The spin $s$ reps have additional subtleties that occur in the lower dimensions $D=3,4$, which we turn to next ($D=2$ is discussed in section \ref{D2section}).

\subsubsection*{$D=3$:}

Consider the content of the spin $s$ rep for generic $\Delta$ depicted in \eqref{spinssocontent}.  In $D=3$ these are $\frak{so}(3)$ tableaux, and all the rows except for the last two become trivial.  We can use the three dimensional epsilon tensor $\epsilon_{IJK}$ to dualize all the reps in the second to last row into single row reps, and once this is done, it looks identical to the last row.  
Then, after taking a suitable linear combination of the bottom two rows, it can be seen that they split apart into two separate irreducible reps, which we call ${\cal F}^{[s]_\pm}_\Delta$, 
\be {\cal F}^{[s]}_\Delta={\cal F}^{[s]_+}_\Delta\oplus {\cal F}^{[s]_-}_\Delta,\ \ D=3\, .\label{d3tenssssplite}\ee
Both ${\cal F}^{[s]\pm}_\Delta$ contain the same $\frak{so}(3)$ content $[s],[s+1],[s+2],\ldots$, but they form distinct $\frak{so}(1,3)$ reps.

In $D=3$, the rep space is the space of spin $s$ tensors on the 2-sphere.  The split \eqref{d3tenssssplite} is equivalent to the statement that this space can be broken up into (imaginary) self-dual and anti-self-dual parts with respect to the $d=2$ epsilon tensor $\epsilon_{ij}$ on the 2-sphere, acting on one of the indices of the tensor.  This reflects the fact that all the massive spin $s$ fields on dS$_3$ come in two independent chiralities (just as they do on flat space).  

The same splitting also occurs for the shortened shift symmetric reps ${\cal D}_{s+k+2}^{[s]}$ corresponding to the shift symmetric fields,
\be{\cal D}_{s+k+2}^{[s]}={\cal D}_{s+k+2}^{[s]_+}\oplus {\cal D}_{s+k+2}^{[s]_-}\, ,\ \ D=3\, .\label{d3tvecssplite2}\ee
Each of these contains the $\frak{so}(3)$ tensors $[s+k+1],[s+k+2],[s+k+3],\ldots$.

And it occurs for the finite dimensional reps ${\cal S}_{-s-k}^{[2]}$ corresponding to the shift symmetries,
\be {\cal S}_{-s-k}^{[s]}={\cal S}_{-s-k}^{[s]_+} \oplus  {\cal S}_{-s-k}^{[s]_-} \, ,\ \ D=3\, .\label{d3vtecssplidte2}\ee
Each of these is finite dimensional, containing the $\frak{so}(3)$ tensors $[s],[s+1],\ldots,[s+k].$
These are the finite dimensional chiral reps $[s+k,s]_\pm$ of $\frak{so}(1,3)$, which when branched to $\frak{so}(3)$ using \eqref{evenDbranchinge} each give the reps $[s],[s+1],\ldots,[s+k].$

The shadow transform that relates $\Delta$ and $\bar \Delta$ flips the helicity, so away from the points with shortened reps we have
\be {\cal F}^{[s]_\pm}_\Delta\simeq {\cal F}^{[s]_\mp}_{\bar \Delta}\,, \ \ D=3\,.\ee
This means that at $\Delta=d/2=1$, the two reps are the same, ${\cal F}^{[s]_+}_1\simeq {\cal F}^{[s]_-}_1$.  At the shift symmetric and finite points, we have
\be  {\cal D}^{[s]_\pm }_{s+k+2} \simeq {\cal F}^{[s]_\mp}_{-s-k}/{\cal S}^{[s]_\mp}_{-s-k}\,, \ \ {\cal S}^{[s]_\pm }_{-s-k} \simeq {\cal F}^{[s]_\mp}_{s+k+2}/{\cal D}^{[s]_\mp}_{s+k+2} \,,\ \ D=3\,.\ee

In $D=3$, the range \eqref{spinscomgrangee} degenerates and there is no complementary series for any of the symmetric tensors.  The point $\Delta=d-1$, which would have been where the complementary series ends at the $t=s$ PM rep, is also the point $\Delta=d/2$ that would be where the complementary series begins at its intersection with the principal series.  So ${\cal F}^{[s]}_1$ actually carries the $t=s$ PM tensor.

The $t=s$ PM rep contains everything except for the last row of \eqref{spinssocontent}.  In $D=3$ these all trivialize except for the second to last row of \eqref{spinssocontent}, which can be dualized to obtain a rep which is identical to that of the $k=s-1$ shift symmetric scalar ${\cal D}^{[0]}_{s+1}$, and we have the isomorphism ${\cal V}^{[s]}_1\simeq {\cal D}^{[0]}_{s+1}$.
This reflects the fact that a depth $t=s$ PM spin $s$ field in $D=3$ is dual to a $k=s-1$ shift symmetric scalar, with the scalar's shift symmetry gauged \cite{Hinterbichler:2024vyv}.  From \eqref{spinsudisoeee}, the rep ${\cal U}^{[s]}_1$ representing the gauge modes is also isomorphic to ${\cal D}^{[0]}_{s+1}$, so in $D=3$ we see that the physical and gauge modes of the PM spin $s$ both carry the same rep, reflecting the equivalence ${\cal F}^{[s]+}_1\simeq {\cal F}^{[s]-}_1$.   Note that the point in \eqref{spinssocontent4} is the same as the point in \eqref{spinssocontent5} when $D=3$ and $t=s$, and so in this case the PM massless modes and gauge modes split off into a direct sum rather than the factorizations and invariant subspaces that occur in general $D$.  In summary, we have the equivalences
\be {\cal F}_1^{[s]_+}\simeq {\cal F}_1^{[s]_-} \simeq {\cal V}_1^{[s]}\simeq {\cal U}_1^{[s]} \simeq {\cal D}_{s+1}^{[0]}\,,\ \ \ D=3\,.\label{d3sphpfeqsce}\ee

For any of the other PM points $t=1,2,\ldots,s-1$ (including the massless point at $t=1$), all of the $\frak{so}(3)$ reps in the rows that make up ${\cal V}^{[s]}_{s-t+1}$ are empty, and so these reps are all trivial in $D=3$.  This reflects the fact that all of the PM higher spins other than the maximal depth one at $t=s$, including the massless ones, have no propagating degrees of freedom on dS$_3$.

\subsubsection*{$D=4$:}

In $D=4$, the $\frak{so}(4)$ reps that have 2 rows in their tableaux split as in \eqref{sod4splite}, and this affects all of the PM spin $s$ reps, which are the only ones with only 2-row tableaux among their  $\frak{so}(4)$ content.  These reps split into two chiralities,
\be {\cal V}^{[s]}_{s-t+2} ={\cal V}^{[s]_+}_{s-t+2}\oplus {\cal V}^{[s]_-}_{s-t+2}\,,\ \ D=4\,,\label{chirald4splitpne}\ee
where ${\cal V}^{[s]_+}_{s-t+2}$ contains only the plus chirality $\frak{so}(4)$ reps and ${\cal V}^{[s]_-}_{s-t+2}$ contains only the minus chirality $\frak{so}(4)$ reps (this can be illustrated with pictures analogous to \eqref{tensorsocontent14} or \eqref{tensorsocontent15}).
This reflects the fact that all the PM higher spins fields in $D=4$ split into two independent chiralities (this is the root of the duality invariance of all the spin $s$ PM modes in $D=4$ \cite{Deser:2004xt,Deser:2013xb,Hinterbichler:2016fgl}).  As we will see in section \ref{unitarylistsection}, the PM reps in $D=4$ are all accounted for among the so-called discrete series reps.

\subsection{$p$-form representations\label{pformsec}}

We now turn to the simplest class of reps beyond the symmetric tensors, the rank $p$ fully anti-symmetric tensors, i.e. $p$-forms (we restrict to $p\geq 1$ in this section; the case $p=1$ reduces to section \ref{vectorsection}).    We use the mass $m$ that vanishes when the field acquires its largest gauge invariance, which is related to the mass $\tilde m$ in the Klein-Gordon equation by $\tilde m^2=m^2 +p(D-p)H^2$.  The relation between $\Delta$ and $m^2$ coming from \eqref{deltamregexe}, \eqref{dscfttmassrelatione} in the late time behavior $\sim e^{-\left( \Delta_\pm-p\right) {H t}}$ of \eqref{ssassymprosole} then reads
\be {m^2\over H^2}=(\Delta-p)(d-\Delta-p) \,,\ \ \Delta_\pm ={d\over 2}\pm \sqrt{\left({d\over 2}-p\right)^2-{m^2\over H^2}}\,.  \label{pformdggere}\ee

The representation space corresponding to a $p$-form rep consists of the space of square integrable $p$-forms $\phi_{i_1\ldots i_p}$ on ${\mathbb S}^d$, transforming under $\frak{so}(1,D)$ as in \eqref{latetimesheactiont3} with $r=p$.  Call this space ${\cal F}^{[1^p]}_\Delta$,
\be {\cal F}^{[1^p]}_\Delta:\ \ {\rm complex\ } p \text{-forms\ on\ } {\mathbb S}^d \,.\ee
Here the label $[1^p]\equiv \underset{p\ {\rm rows}}{\underbrace{[1,\ldots,1]}}$ indicates the Young tableau with a single column of $p$ rows, indicating the totally anti-symmetric index symmetry of the field.

To decompose the $p$-form field, we use the Hodge decomposition, which breaks up a $p$-form into an exact, co-exact (i.e. transverse) and harmonic form.  The $d$-sphere with $d\geq 2$ has no harmonic $p$-forms for $p\geq 1$, so the decomposition becomes
\be \phi_{i_1\ldots i_p}=\chi_{i_1\ldots i_p}+\nabla_{[i_1}\chi_{i_2 \ldots i_p]}\, ,\ \ \ \nabla^{i_1} \chi_{i_1i_2\ldots i_p}=\nabla^{i_1} \chi_{i_1i_2\ldots i_{p-1}}=0\, .\label{hodgedecpe} \ee
The co-exact form is the transverse $p$-form $\chi_{i_1\ldots i_p}$, and the exact form is proportional to the exterior derivative of the $(p-1)$-form $\chi_{i_1\ldots i_{p-1}}$.  This $(p-1)$-form is itself taken to be co-exact, i.e. transverse, because any exact or harmonic part would be annihilated by the exterior derivative.  

A transverse $q$-form can be split up into its constituent transverse $q$-form spherical harmonics on ${\mathbb S}^d$, which take the form
\be Y^{I_1\ldots I_{l},J_1\ldots J_q}_{l,i_1\ldots i_{q}}(\hat X)=  \partial_{i_1}  \hat X^{[J_1}\cdots \partial_{i_{q}} \hat X^{J_{q} } \hat X^{I_1]} \hat X^{I_2}\cdots \hat X^{I_l}-{\rm traces}\, ,\ \ \ l=1,2,3,\ldots \ . \label{tensorspherharmonicese3}\ee
These transform under $\frak{so}(D)$ as tensors with the symmetries of traceless $[l,1^{q}]$ tableaux.  They form a basis of the space of transverse ${q}$-forms on ${\mathbb S}^d$.  The natural Laplacian on the space of ${q}$-forms is the Hodge Laplacian, given by 
\be \Delta_H=-\nabla^2_\Omega+{q}(d-{q})\, ,\ee 
and the transverse vector harmonics are eigenfunctions of it, with the following eigenvalues \cite{Folland1989},
\be \Delta_H Y^{I_1\ldots I_{l},J_1\ldots J_q}_{l,i_1\ldots i_{q}}=(l+ d-{q}-1)(l+ {q})   Y^{I_1\ldots I_{l},J_1\ldots J_q}_{l,i_1\ldots i_{q}}\, .\ee

Using this spherical harmonic expansion on both the $p$-form and the $(p-1)$-form in \eqref{hodgedecpe},
the $\frak{so}(D)$ content of ${\cal F}^{[1^p]}_\Delta$ can be illustrated as follows:
\be \raisebox{-0pt}{\epsfig{file=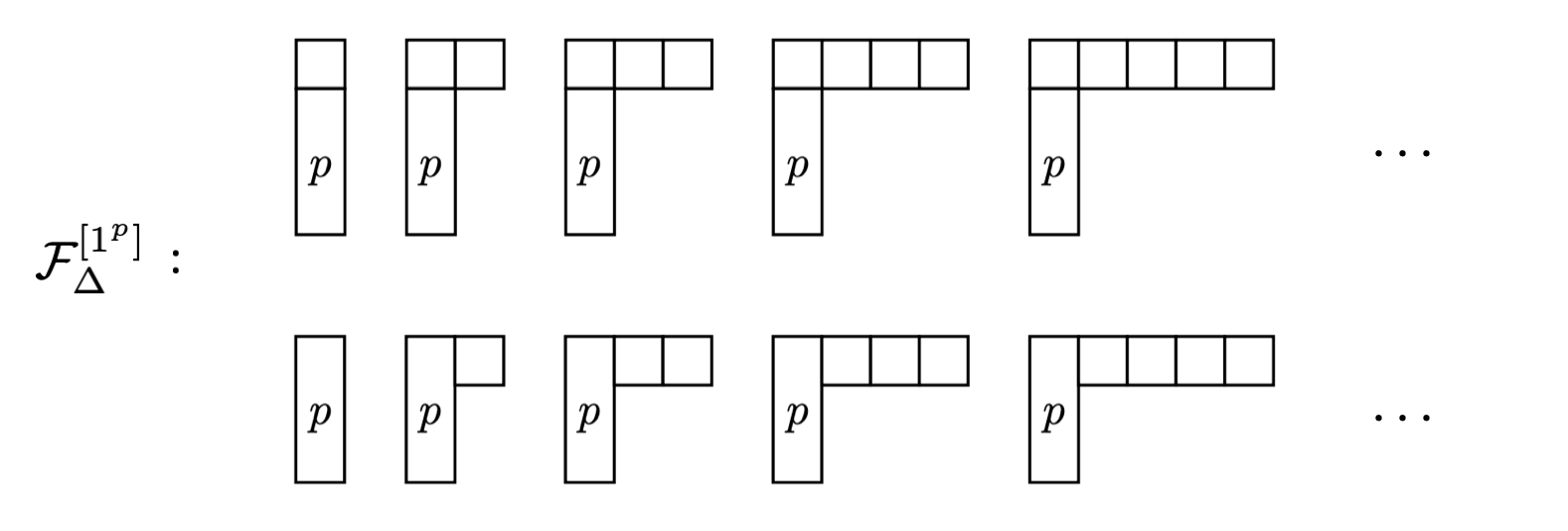,width=5.5in}}  \label{pformsocontent}\, \ee
The top line consists of the harmonics in the transverse $p$-form $\chi_{i_1\ldots i_p}$, and the bottom line consists of the harmonics in the transverse $(p-1)$-form $\chi_{i_1\ldots i_{p-1}}$. 

For generic $\Delta$, the boost generators ${\cal K}^I$ will link together the lattice of $\frak{so}(D)$ reps shown in \eqref{pformsocontent} with nearest neighbor interactions, resulting in a picture just like \eqref{vectorsocontent2}, with the maps represented by the arrows defined as in the explanation right below \eqref{vectorsocontent2}.

\textbf{Reducible cases:} There are discrete values of $\Delta$ at which these arrows break and we get sub-reps, making the rep ${\cal F}_{\Delta}^{[1^p]}$ reducible but not decomposable.  These values are as follows:
\begin{itemize}

\item
Shift symmetric points: 
\be \Delta=d+k+1\, ,\ \ \ k=0,1,2,\ldots\ . \ee
The $(k+2)$-th and higher columns of \eqref{pformsocontent} form a sub-rep we call
\be {\cal D}^{[1^p]}_{d+k+1}\,,\ee
illustrated here for the $k=2$ case,
\be \raisebox{-0pt}{\epsfig{file=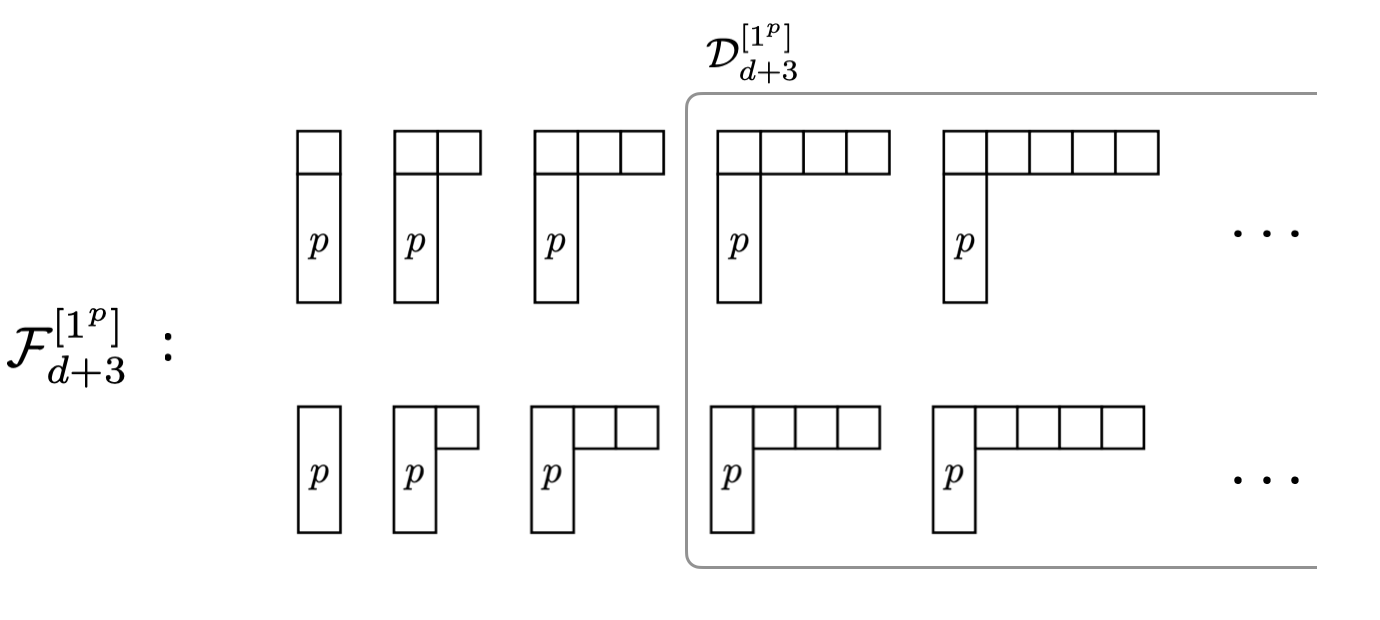,width=5.0in}}  \label{pformsocontent2}\, \ee
These represent the physical modes of the level $k$ shift symmetric $p$-form fields \cite{Hinterbichler:2022vcc}.

\item
Finite points:
\be \Delta=-k-1\, ,\ \ \ k=0,1,2,\ldots \,.\ee 
The first $k+1$ columns of \eqref{pformsocontent} form a finite dimensional sub-rep we call
\be {\cal S}^{[1^p]}_{-k-1}\,,\ee
illustrated here for the $k=2$ case,
\be \raisebox{-0pt}{\epsfig{file=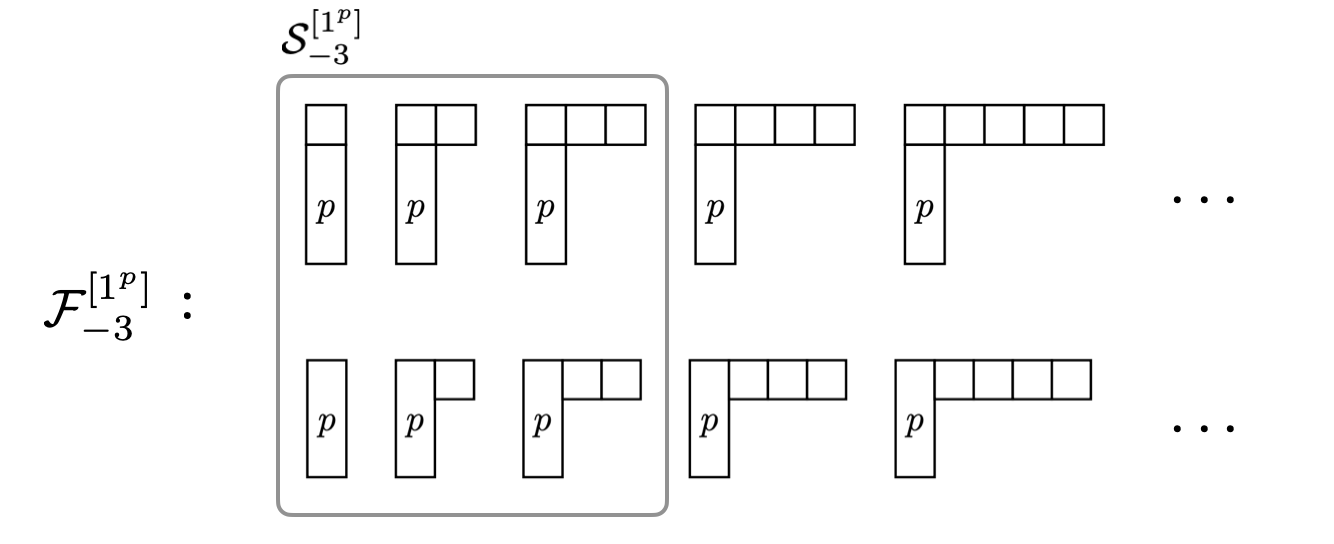,width=5.0in}}  \label{pformsocontent3}\, \ee
This is nothing but the $[k+1,1^p]$ tensor rep of $\frak{so}(1,D)$, which upon branching to $\frak{so}(D)$ using the branching rules in appendix \ref{branchingappendix} gives precisely the reps in ${\cal S}^{[1^p]}_{-k-1}$.  (Note that the tableau for this $\frak{so}(1,D)$ rep appears in the upper right corner of the subrep.)  This rep corresponds to the shift symmetries of the level $k$ shift symmetric $p$-form field \cite{Hinterbichler:2022vcc}.

\item
Massless point:
\be \Delta=d-p\,.\ee
The first row of 
\eqref{pformsocontent} forms a sub-rep we call
\be {\cal V}^{[1^p]}_{d-p}\, ,\ee
illustrated here
\be \raisebox{-0pt}{\epsfig{file=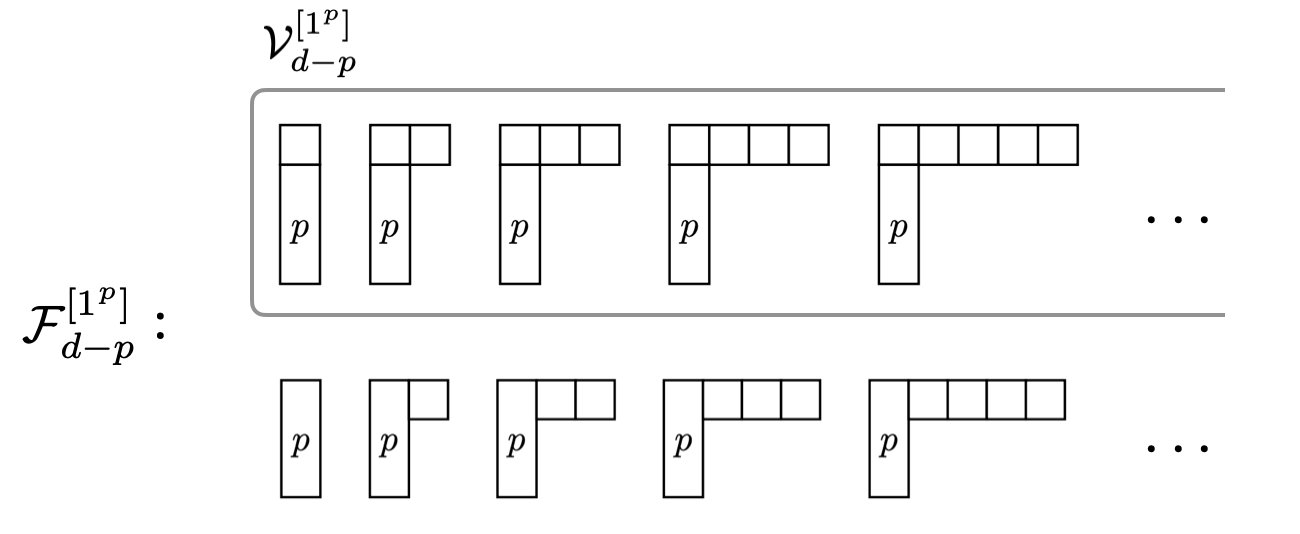,width=5.1in}}  \label{pformsocontent4}\, \ee
This rep represents the physical modes of the massless ($m=0$) $p$-form field.

\item
Gauge point:
\be \Delta=p\, .\ \ \ \ee 
The final row of \eqref{pformsocontent} forms a finite dimensional sub-rep we call
\be {\cal U}^{[1^p]}_{p}\, ,\ee
illustrated here
\be \raisebox{-0pt}{\epsfig{file=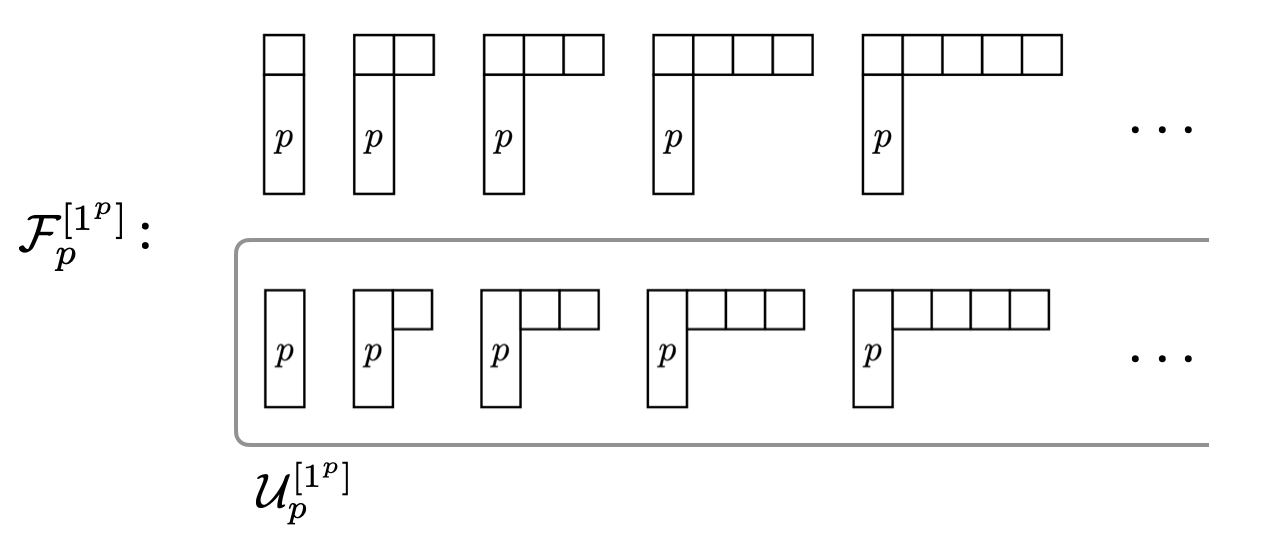,width=5.1in}}  \label{pformsocontent5}\, \ee
This rep represents the gauge modes of the massless $p$-form field.

\end{itemize}

\textbf{Equivalences:} There is a shadow transform intertwining operator that connects the $\Delta$ and $\bar\Delta\equiv d-\Delta$ reps, 
\be S_\Delta^{[1^p]} :\ {\cal F}_{\Delta}^{[1^p]}\rightarrow {\cal F}_{\bar\Delta}^{[1^p]}\,.\label{pformshwadee}\ee
It commutes with the $\frak{so}(D)$ rotations and satisfies $\delta_{{\cal K}^I_{\bar\Delta}}S_\Delta^{[1^p]}=S_\Delta^{[1^p]} \delta_{{\cal K}^I_{\Delta}}$.  For generic $\Delta$, this operator is invertible and the $\Delta$ and $\bar\Delta$ reps are equivalent to each other,
\be {\cal F}_{\Delta}^{[1^p]}\simeq {\cal F}_{\bar\Delta}^{[1^p]}\,.\label{pformeejfee2}\ee
  But for the special values of $\Delta$ indicated above where ${\cal F}_{\Delta}^{[1^p]}$ develops a sub-rep, $S_\Delta^{[1^p]}$ develops a kernel which is always precisely the sub-rep.

The shift symmetric and finite points are linked to each other via the maps $S^{[1^p]}_{-k-1}$, ${S}^{[1^p]}_{d+k+1}$, where the kernel and image of each are the sub-reps ${\cal S}^{[1^p]}_{-k-1}$, ${\cal D}^{[1^p]}_{d+k+1}$, which induces the isomorphisms
\be {\cal S}^{[1^p]}_{-k-1}\simeq {\cal F}^{[1^p]}_{d+k+1}/{\cal D}^{[1^p]}_{d+k+1}\,,\ \ \  {\cal D}^{[1^p]}_{d+k+1}\simeq {\cal F}^{[1^p]}_{-k-1}/{\cal S}^{[1^p]}_{-k-1} \,.\ee
This is illustrated here in the $k=1$ case,
\be \raisebox{-0pt}{\epsfig{file=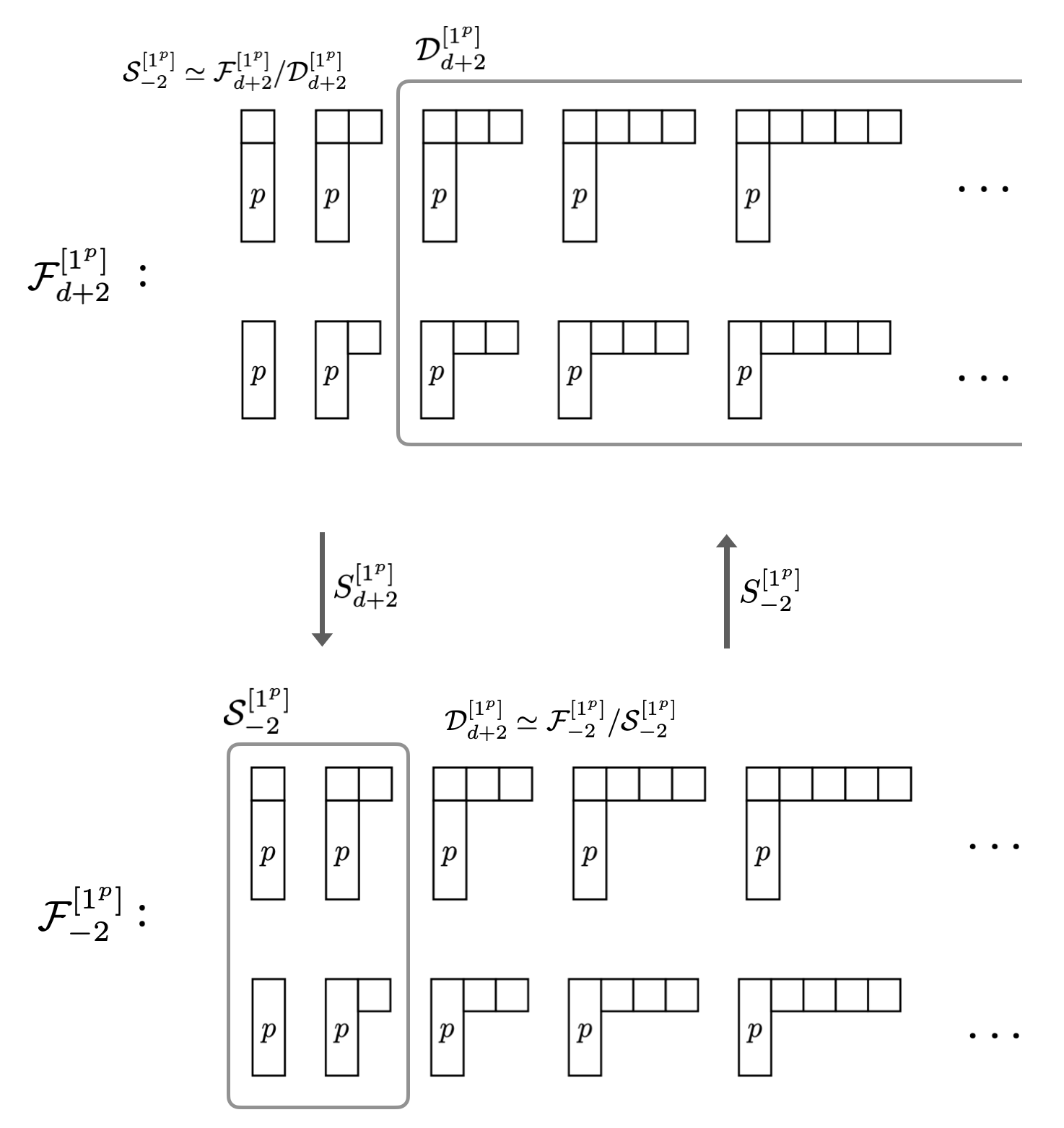,width=4.7in}}  \label{pformsocontent5-2}\, \ee

The massless and gauge points are linked to each other via the maps $S^{[1^p]}_{p}$, ${S}^{[1^p]}_{d-p}$, where the kernel and image of each are the sub-reps ${\cal U}^{[1^p]}_{p}$, ${\cal V}^{[1^p]}_{d-p}$, which induces the isomorphisms
\be {\cal U}^{[1^p]}_{p}\simeq {\cal F}^{[1^p]}_{d-p}/{\cal V}^{[1^p]}_{d-p}\,,\ \ \  {\cal V}^{[1^p]}_{d-p}\simeq {\cal F}^{[1^p]}_{p}/{\cal U}^{[1^p]}_{p}\,.\ee
This is illustrated here:
\be \raisebox{-0pt}{\epsfig{file=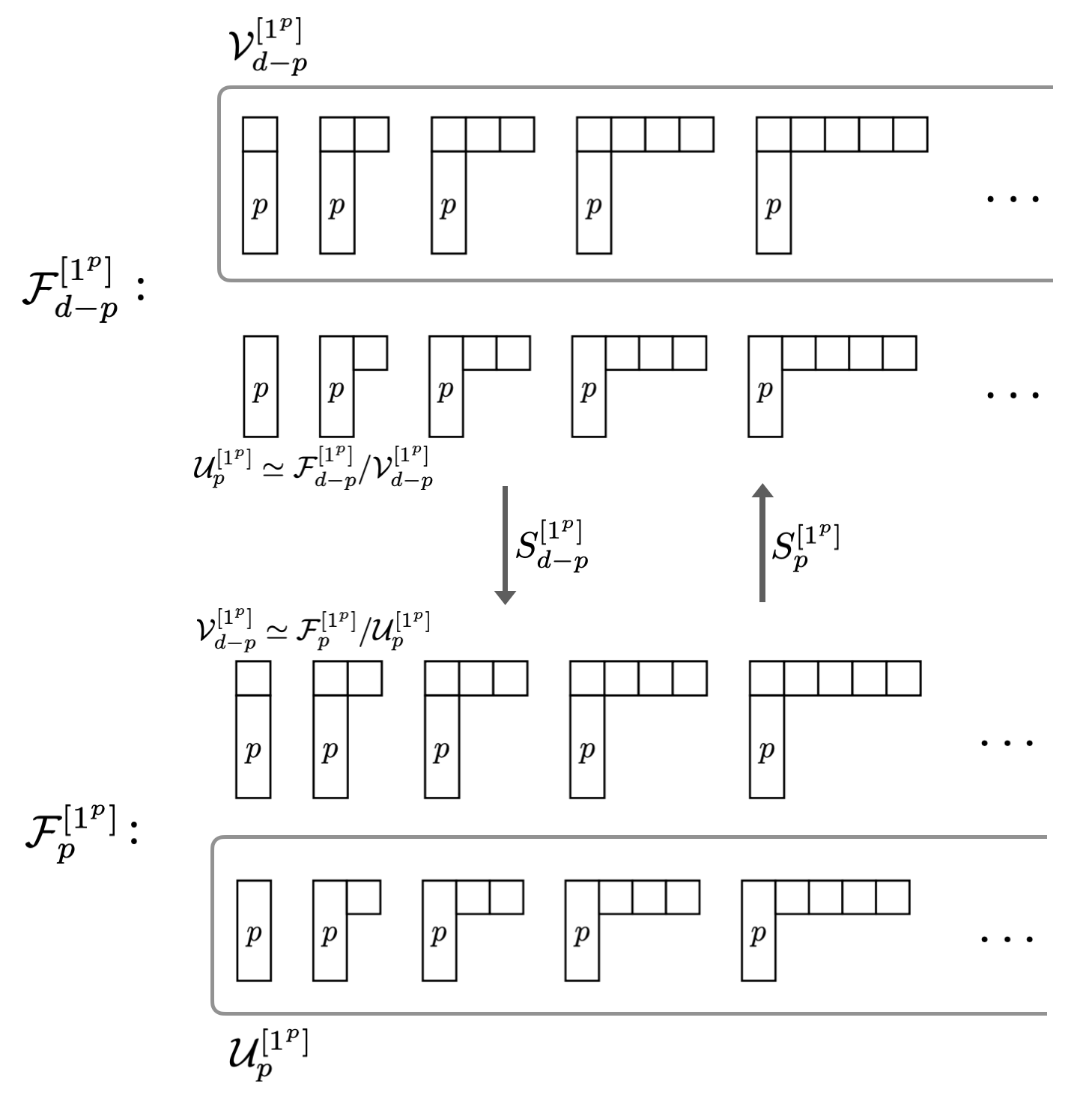,width=4.7in}}  \label{pformsocontent5-3}\, \ee

The other maps that will be of interest for the massless points are the de Rham exterior derivative operator $\rd$ and the co-exterior derivative operator\footnote{The usual exterior and co-exterior derivatives are defined as $(q+1)\nabla_{[i_{1}}\phi_{i_2\ldots i_{q+1}]}$ and $-\nabla^{i_1}\phi_{i_1\ldots i_{q}}$, with different overall normalizations than what we have used here.} $\rd^\dag$.  For $q=0,1,\ldots,p-1$, the exterior derivative takes $q$ forms with $\Delta=q$ to $(q+1)$-forms, raising the degree of $\Delta$ by one in the process,
\be \rd: \  {\cal F}^{[1^{q}]}_{q} \rightarrow {\cal F}^{[1^{q+1}]}_{q+1} \, ,\ \ \phi_{i_1\ldots i_{q}}\rightarrow \nabla_{[i_1}\phi_{i_2\ldots i_{q+1}]}\,.
\ee
For $q=1,2,\ldots,p$, the co-exterior derivative $\rd^\dag$ takes $q$-forms with $\Delta=d-q$ into $(q-1)$-forms, raising the degree of $\Delta$ by one in the process,
\be \rd^\dag: \  {\cal F}^{[1^{q}]}_{d-q} \rightarrow {\cal F}^{[1^{q-1}]}_{d-q+1} \, ,\ \phi_{i_1\ldots i_{q}}\rightarrow \nabla^{i_1}\phi_{i_1i_2\ldots i_{q}} \,.
\ee

For $p$ forms with $p>1$ the gauge symmetry is reducible.  This is reflected in the $\rd$ and $\rd^\dag$ operator linking together more than two ranks worth of reps, through the de Rham complex.  The spaces ${\cal F}^{[1^p]}_{p}$, ${\cal F}^{[1^{p-1}]}_{p-1}$, and so on down to the scalar space ${\cal F}^{[0]}_{0}$ are linked by $\rd$, and the spaces ${\cal F}^{[1^p]}_{d-p}$, ${\cal F}^{[1^{p-1}]}_{d-p+1}$, and so on down to the scalar space ${\cal F}^{[0]}_{d}$, are linked by $\rd^\dag$.  The spaces ${\cal F}^{[1^p]}_{p},\ldots,{\cal F}^{[1]}_{1},{\cal F}^{[0]}_{0}$ have as invariant subspaces ${\cal U}^{[1^p]}_{p},\ldots,{\cal U}^{[1]}_{1},{\cal S}^{[0]}_{0}$, and the spaces ${\cal F}^{[1^p]}_{d-p},\ldots,{\cal F}^{[1]}_{d-1},{\cal F}^{[0]}_{d}$ have as invariant subspaces ${\cal V}^{[1^p]}_{d-p},\ldots,{\cal V}^{[1]}_{d-1},{\cal D}^{[0]}_{d}$.

The pair ${\cal F}^{[1^p]}_{d-p}$ and ${\cal F}^{[1^p]}_{p}$ will in turn be linked together by $S^{[1^p]}_{d-p}$, $S^{[1^p]}_{p}$, the pair ${\cal F}^{[1^{p-1}]}_{d-p+1}$ and ${\cal F}^{[1^{p-1}]}_{p-1}$ will be linked together by $S^{[1^{p-1}]}_{d-p+1}$, $S^{[1^{p-1}]}_{p-1}$ and so on down to the pair ${\cal F}^{[0]}_{d}$ and ${\cal F}^{[0]}_{0}$ which are linked together by $S^{[0]}_{d}$, $S^{[0]}_{0}$.  This forms a large commutative diagram interlinking all of these reps, illustrated here:
\be \raisebox{-0pt}{\epsfig{file=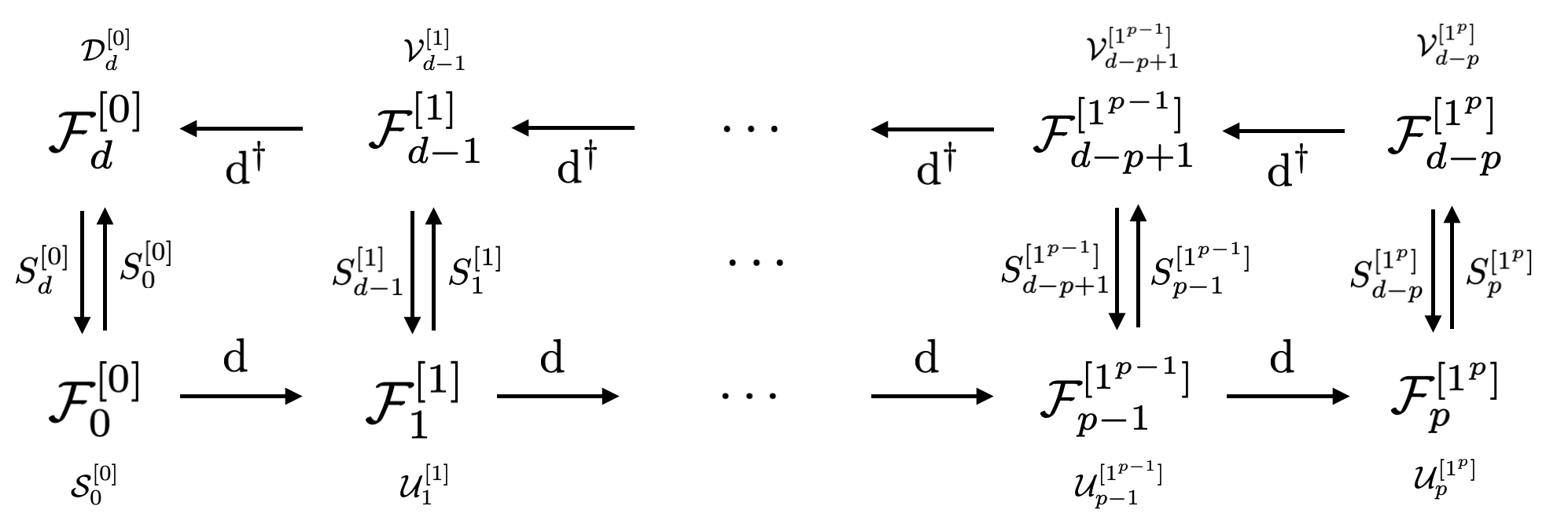,width=5.5in}}  \label{dsrepspformcomplex2}\, \ee
Each rep is shown, along with its sub-rep in smaller font.  Each sub-rep is simultaneously the kernel of all the outgoing maps and the image of all the ingoing maps, and going through any two arrows consecutively gives zero.
Here is an illustration of the same diagram for $p=2$, now showing the full $\frak{so}(D)$ content of each rep, with grey rectangles enclosing the various sub-reps:
\be \raisebox{-0pt}{\epsfig{file=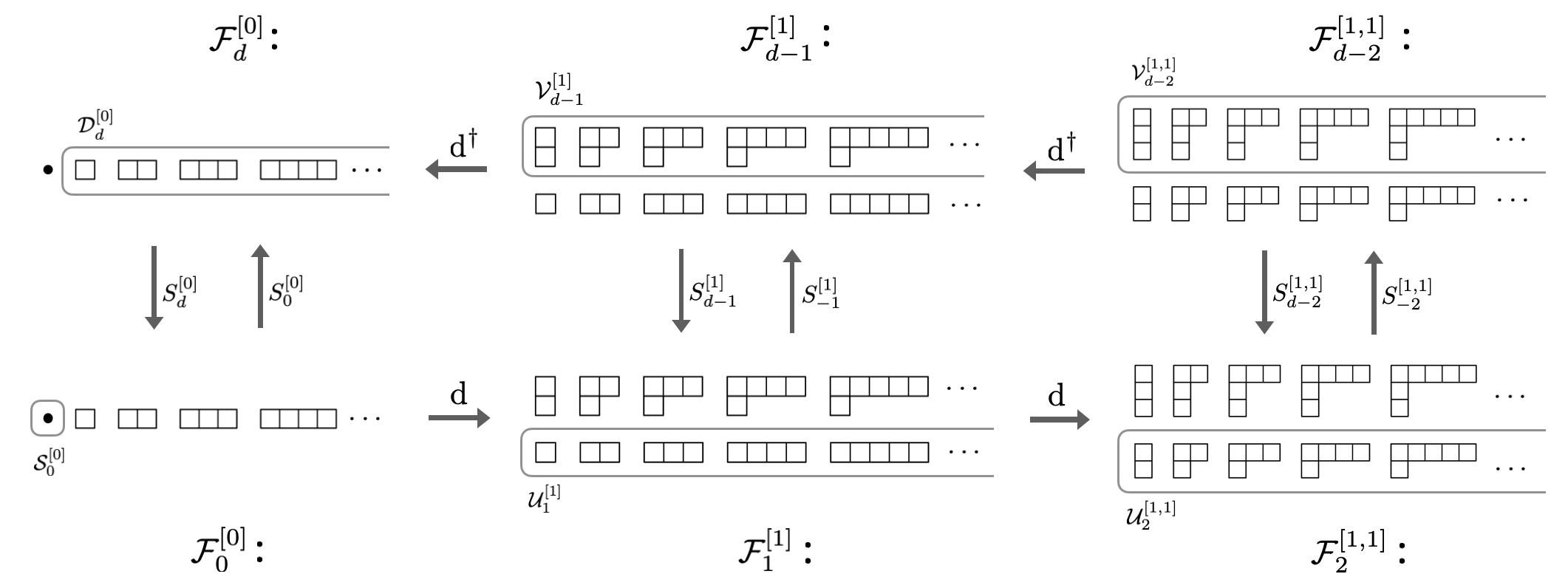,width=7.0in}}  \label{pformsocontent6}\, \ee

From this commutative diagram for the massless $p$-form we see the sequence of isomorphisms
\be { \cal D}^{[0]}_d\simeq { \cal U}^{[1]}_1\,,\ \ { \cal V}^{[1]}_{d-1}\simeq { \cal U}^{[1,1]}_2\,,\ \ \ { \cal V}^{[1,1]}_{d-2}\simeq { \cal U}^{[1,1,1]}_3\,,\  \ldots \ ,\  { \cal V}^{[1^{p-1}]}_{d-p+1}\simeq { \cal U}^{[1^p]}_{p}\,   . \label{pformuveqveuveee}\ee
This expresses the fact that a massless $p$-form has as its gauge modes a massless $(p-1)$-form, its gauge-for-gauge modes a massless $(p-2)$-form, and so on all the way down, ending with a massless scalar.

\textbf{Unitarity:}  There is a diffeomorphism and Weyl invariant bilinear form pairing the two spaces ${\cal F}_\Delta^{[1^p]}$ and ${\cal F}_{\bar \Delta}^{[1^p]}$,
\be (\phi_1,\phi_2)\equiv {1\over p!} \int d^d\Omega\, \phi_1^{i_1\ldots i_p}(\hat X)\phi_{2\,  i_1\ldots i_p}(\hat X)\,,\ \ \ \  \phi_{1\,  i_1\ldots i_p}\in {\cal F}_{\bar\Delta}^{[1^p]},\ \ \phi_{2\, i_1\ldots i_p}\in {\cal F}_{\Delta}^{[1^p]}\, .\label{bilinteaprformvee}\ee

In the case where $\Delta^\ast=\bar\Delta$ we can use this to form a manifestly positive definite, invariant inner product on ${\cal F}_\Delta^{[1^p]}$ via:
\be \la \phi_1 | \phi_2 \ra\equiv (\phi_1^{\ast},\phi_2)\, ,\ \ \ \Delta^\ast=\bar\Delta\,, \ \  \phi_{1\, i_1\ldots i_p}, \phi_{2\, i_1\ldots i_p}\in {\cal F}_{\Delta}^{[1^p]}\,. \label{innerprotdpvprise} \ee
This gives us the $p$-form principal series reps, which are all unitary
\be p\text{-form \ principal\ series:}\  \Delta={d\over 2}+i\nu\, ,\ \ \ \nu\in {\mathbb R}\,.\ee

In the case where $\Delta$ is real, we use $S^{[1^p]}_{\Delta}$ to move a state from ${\cal F}_{ \Delta}^{[1^p]}$ to ${\cal F}_{\bar \Delta}^{[1^p]}$ and form an inner product on ${\cal F}_{ \Delta}^{[1^p]}$ as follows,
\be \la \phi_1 | \phi_2 \ra\equiv (S^{[1^p]}_{\Delta}\phi_1^\ast,\phi_2)\, ,\ \ \ \Delta^\ast=\Delta\, , \ \  \phi_{1\, i_1\ldots i_p}, \phi_{2\, i_1\ldots i_p}\in {\cal F}_{\Delta}^{[1^p]}\,. \label{innerprodtprispcsvdee} \ee 
The positivity of this inner product is now equivalent to whether the matrix elements of $S^{[1^p]}_{\Delta}$ are positive for all the different $\frak{so}(D)$ reps.  Away from the discrete special cases described above, it turns out that they are all positive only in the range $p<\Delta<d-p$, which gives the $p$-form complementary series,
\be p \text{-form \ complementary\ series:}\  p<\Delta<d-p\,, \label{comprangepfe}\ee
where the point $\Delta=d/2$ has not been committed to either series.

For the cases of $\Delta$ where the reps become reducible, we have the following.  For the finite points $\Delta=-k-1$, $k=0,1,2,\ldots$, we use $S^{[1^p]}_{-k-1}$ in the inner product, and this has the finite dimensional kernel consisting of the sub-rep ${\cal S}^{[1^p]}_{-k-1}$.   All of these states will have zero norm in the inner product.  Apart from these null states, other states also have negative norm, so even once these null states are factored out, we will be left with non-unitary reps.  The rep obtained after this factoring is nothing but ${\cal D}^{[1^p]}_{d+k+1}$, realized through the quotient ${\cal D}^{[1^p]}_{d+k+1}\simeq {\cal F}^{[1^p]}_{-k-1}/{\cal S}^{[1^p]}_{-k-1}$, so these reps, corresponding to the shift symmetric $p$-form fields, are non-unitary reps.

For the shift symmetric points $\Delta=d+k+1$, $k=0,1,2,\ldots$, we use $S^{[1^p]}_{d+k+1}$ in the inner product, and this has the infinite dimensional kernel consisting of the states of the sub-rep ${\cal D}^{[1^p]}_{d+k+1}$.   All of these states will therefore be null in the inner product.  Apart from these null states, some of the norms are always still negative, so if the null states are factored out, we will be left with a finite dimensional non-unitary rep.  This is the rep ${\cal S}^{[1^p]}_{-k-1}$, realized through the quotient ${\cal S}^{[1^p]}_{-k-1}\simeq  {\cal F}^{[1^p]}_{d+k+1}/{\cal D}^{[1^p]}_{d+k+1}$.  

For the gauge point $\Delta=p$, we use $S^{[1^p]}_{p}$ in the inner product, and this has as kernel the sub-rep ${\cal U}^{[1^p]}_{p}$.   All of these states will have zero norm in the inner product.  Apart from these null states, the other states can be taken to have positive norm by adjusting an overall sign if necessary, so once the null states are factored out, we will be left with a unitary rep, which is nothing but ${\cal V}^{[1^p]}_{d-p}$, realized through the quotient ${\cal V}^{[1^p]}_{d-p}\simeq {\cal F}^{[1^p]}_{p}/{\cal U}^{[1^p]}_{p}$.  These correspond to the physical states of the massless $p$-form, which is unitary on dS.

For the massless point $\Delta=d-p$, we use $S^{[1^p]}_{d-p}$ in the inner product, and this has as kernel the sub-rep ${\cal V}^{[1^p]}_{d-p}$.   All of these states will have zero norm in the inner product.  Apart from these null states, the other states can all be taken to have positive norm, so  once the null states are factored out, we will be left with a unitary rep, which is nothing but ${\cal U}^{[1^p]}_{p}\simeq {\cal V}^{[1^{p-1}]}_{d-p+1}$ (${\cal U}^{[1]}_1\simeq {\cal D}^{[0]}_d$ for $p=1$), realized through the quotient ${\cal U}^{[1^p]}_{p}\simeq {\cal F}^{[1^p]}_{d-p}/{\cal V}^{[1^p]}_{d-p}$.  These correspond to the longitudinal gauge modes of a $p$-form, which are equivalent to the physical modes of a massless $(p-1)$-form.

For all but the unitary cases discussed here, there is no other way to construct an invariant inner product on ${\cal F}_\Delta^{[1^p]}$, so all the other reps in the complex $\Delta$ plane are non-unitary.  

\textbf{Summary:} The $p$-form reps are summarized here:
\be \raisebox{-40pt}{\epsfig{file=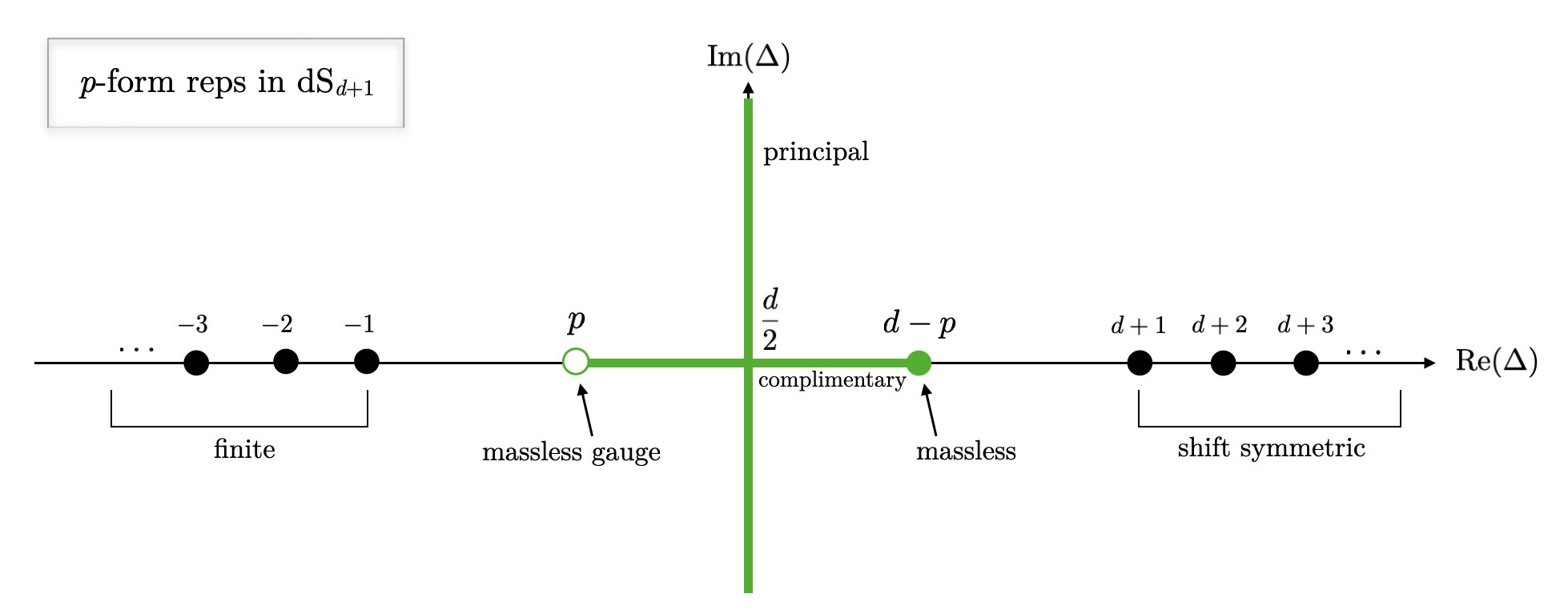,width=6.4in}}  \label{dsrepspform}\, \ee
Points in green are unitary reps.  The dots are the special points where the rep becomes reducible. The dot at $\Delta=p$ is the gauge point, it is empty to indicate that this is equivalent to a $(p-1)$-form rep through \eqref{pformuveqveuveee}, and thus is already accounted for among those.  The equivalence \eqref{pformeejfee2} of the non-special reps is given by reflecting through the point $\Delta=d/2$, and due to the breakdown of the intertwiner map \eqref{pformshwadee}, the special reps with dots are inequivalent despite the reflection.  Other than this equivalence, all the reps are distinct.

The value of the quadratic Casimir operator ${\cal C}_2$ of section \ref{casimirsec} on the $p$-form reps is given by
\be {\cal C}_2= (\Delta-p)(\Delta-d+p)= -{m^2\over H^2}\,. \ee

We have the following correspondence between $p$-form fields on dS$_{D}$ and $p$-form unitary reps of $\frak{so}(1,D)$:
\bea \begin{cases}
 {\rm principal :\ }\quad  \Delta={d\over 2}+i\nu \, , \ \ \nu\in {\mathbb R}\,, & {\rm heavy \ } p \text{ -forms:\ }  {m^2\over H^2 }\geq {\left(D-1-2p\right)^2\over 4}\,, \\
 {\rm complementary:\ } \quad p<\Delta<d-p \, ,  & {\rm light \ } p \text{ -forms:\ }  0 < {m^2\over H^2 }\leq  {\left(D-1-2p\right)^2\over 4} \, ,\\
  {\rm massless:\ } \quad \Delta=d-p \, , &  {\rm massless \ } p \text{ -forms:\ } m^2=0\, . \\
 \end{cases} \nn
\eea

\subsubsection*{Duality equivalences and splittings:}

In a given $d$, the space of $p$-form fields on ${\mathbb S}^d$ is equivalent to the space of $d-p$ form fields, since we can Hodge dualize using the $d$-dimensional volume form $\epsilon_{i_1\ldots i_d}$ on the sphere.  For the dS$_D$ reps we thus have the equivalence 
\be {\cal F}^{[1^p]}_\Delta \simeq {\cal F}^{[1^{d-p}]}_{  \Delta}\,.\label{mainpformdualee}\ee
From the point of view of the $\frak{so}(D)$ content of the rep, there is the $D$ dimensional epsilon tensor $\epsilon_{I_1\ldots I_D}$ that can be used to dualize each $\frak{so}(D)$ by contracting it with the first column of the tableau.  
Doing this interchanges the two rows in \eqref{pformsocontent} and gives back the $\frak{so}(D)$ content of a $(D-p-1)$-form, as illustrated here:
\be \raisebox{-0pt}{\epsfig{file=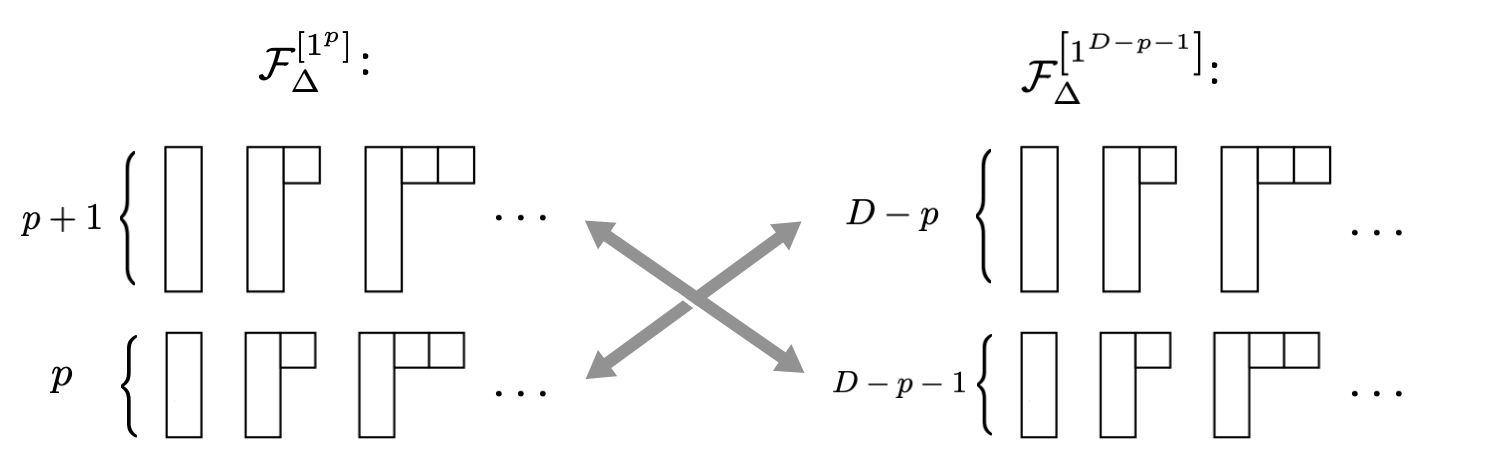,width=5.5in}}  \label{pformsocontent7}\, \ee
This is the statement that a generic massive $p$-form field is dual to a massive $(D-p-1)$-form field (with the same mass, as can be seen from \eqref{pformdggere}). Note that if $p=d$, then the $d$-form is dual to a scalar rep, and if $p\geq d+1$ then the rep is empty or trivial: this can be seen from the fact that all the $\frak{so}(D)$ reps appearing in \eqref{pformsocontent} are empty, with the exception of the single $[1^D]$ tableau in the lower left corner when $p=D$, which is equivalent to the trivial $\frak{so}(D)$ rep, so in the case $p=D$ the $\frak{so}(D)$ rep is equivalent to the trivial rep.

These equivalences also relate the shift symmetric reps \cite{Hinterbichler:2024vyv}.  The level $k$ shift symmetric $p$-form rep is equivalent to the level $k$ shift symmetric $(D-p-1)$-form rep when $p=1,\ldots,D-2$, and the level $k$ shift symmetric $(D-1)$-form is dual to the level $k+1$ shift symmetric scalar,
\be {\cal D}^{[1^p]}_{d+k+1}\simeq {\cal D}^{[1^{d-p}]}_{d+k+1} \,. \ee

The corresponding finite reps are also related, 
\be {\cal S}^{[1^p]}_{-k-1}\simeq {\cal S}^{[1^{d-p}]}_{-k-1} \,, \ee
which is the statement that the $[k+1,1^p]$ tensor rep of $\frak{so}(1,D)$ is equivalent to the $[k+1,1^{d-p}]$ tensor rep, obtained by dualizing the first column of the $\frak{so}(1,D)$ tensor's tableau using the $(D+1)$-dimensional epsilon symbol.

Consider now the massless case $\Delta=d-p$, where the top row of \eqref{pformsocontent} forms the massless rep ${\cal V}^{[1^p]}_{d-p}$.  Under the dualization \eqref{mainpformdualee}, ${\cal V}^{[1^p]}_{d-p}$ maps to ${\cal U}^{[1^{d-p}]}_{d-p}$, which by \eqref{pformuveqveuveee} is equivalent to the massless $(D-p-2)$-form rep ${\cal V}^{[1^{d-p-1}]}_{p+1}$, so we have the equivalence
\be {\cal V}^{[1^p]}_{d-p}\simeq {\cal V}^{[1^{d-p-1}]}_{p+1}\,, \ \ 1\leq p\leq D-3\,.\ee
This reflects the well-known fact that a massless $p$-form field is equivalent to a massless $D-p-2$ form field in $D$ dimensions.
If $p=D-2$, then the form is dual to a massless scalar (the $k=0$ shift symmetric scalar), 
\be {\cal D}_d^{[0]} \simeq {\cal V}^{[1^{d-1}]}_1 \,.\ee
If $p\geq D-1$ then the rep is empty or trivial: this can be seen from the fact that all the $\frak{so}(D)$ reps appearing in the top line of \eqref{pformsocontent} are empty, with the exception of the single $[1^D]$ tableau in the lower left corner when $p=D-1$, which is equivalent to the scalar $\frak{so}(D)$ rep, so in the case $p=D-1$ the $\frak{so}(D)$ rep is equivalent to the trivial rep.
Due to these equivalences, we only need to worry about the massless $p$-form reps with $p\leq  {{d-1}\over 2}$.  

Taking into account all the dualities, we only need to worry about the $p$-form reps with $p\leq {d\over 2}$, since all others will be either trivial or equivalent to one of these.

\subsubsection*{$D$ odd:}

Let $D=2n+1$, $d=2n$, $n=1,2,3,\ldots$.  There is a further splitting that occurs when $p=n$, which is the case where $p$-forms on the $d$-sphere can be split into (imaginary for $n$ odd, real for $n$ even) self-dual and anti-self-dual parts.  In this case, the $\frak{so}(D)$ reps in the top row of \eqref{pformsocontent} can be dualized and they become identical to those of the bottom row.  By taking a linear combination of the two rows, it can then be seen that they split into two separate irreps that we call ${\cal F}^{[1^{d\over 2}]_\pm}_\Delta$,
\be {\cal F}^{[1^{d\over 2}]}_\Delta={\cal F}^{[1^{d\over 2}]_+}_\Delta\oplus {\cal F}^{[1^{d\over 2}]_-}_\Delta\, , \ \ \ D \ {\rm odd}\,,\ee
as illustrated here:
\be \raisebox{-40pt}{\epsfig{file=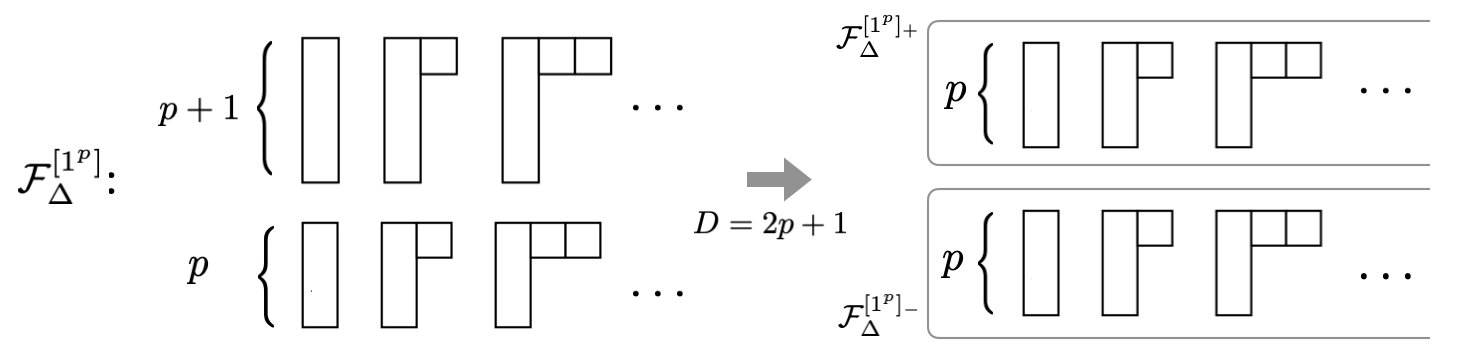,width=5.5in}}  \label{dsrepspformdual2}\, \ee
This is the statement that in odd $D$ a massive ${d\over 2}$-form field splits up into chiral halves (just as it does on flat space, where the massive little group rep $[1^{d\over 2}]$ splits into two irreducible parts $[1^{d\over 2}]_\pm$).

The same splitting also occurs for the shift symmetric reps,
\be {\cal D}^{[1^{d\over 2}]}_{d+k+1}={\cal D}^{[1^{d\over 2}]_+}_{d+k+1}\oplus {\cal D}^{[1^{d\over 2}]_-}_{d+k+1} \, ,\ \ \ D \ {\rm odd}\,,\ee
and the finite dimensional reps
\be {\cal S}^{[1^{d\over 2}]}_{-k-1}={\cal S}^{[1^{d\over 2}]_+}_{-k-1}\oplus {\cal S}^{[1^{d\over 2}]_-}_{-k-1} \ , \ \ \ D \ {\rm odd}\,,\ee
which is the statement that the $[k+1,1^{{d\over 2}}]$ tensor rep of $\frak{so}(1,D)$ splits into two irreducible pieces $[k+1,1^{{d\over 2}}]_\pm$, the self-dual and anti-self-dual parts with respect to the $(D+1)$-dimensional $\frak{so}(1,D)$ epsilon tensor.

The shadow transform that relates $\Delta$ and $\bar \Delta$ flips the chirality and we have the equivalence
\be {\cal F}^{[1^{d\over 2}]_+}_\Delta \simeq {\cal F}^{[1^{d\over 2}]_-}_{\bar \Delta} \,, \ \ \ D \ {\rm odd}\,. \label{Fhdbpfordeme}\ee
At the shift symmetric and finite points, this flip in chirality tells us that we have
\be  {\cal D}^{[1^{d\over 2}]_\pm }_{d+k+1} \simeq {\cal F}^{[1^{d\over 2}]_\mp}_{-k-1}/{\cal S}^{[1^{d\over 2}]_\mp}_{-k-1}\,, \ \ {\cal S}^{[1^{d\over 2}]_\pm }_{-k-1} \simeq {\cal F}^{[1^{d\over 2}]_\mp}_{d+k+1}/{\cal D}^{[1^{d\over 2}]_\mp}_{d+k+1} \,,\ \ D\ {\rm odd}\,.\ee 

The ${d\over 2}$-form fields have no complementary series: the point $\Delta={d\over 2}$ where the would-be complementary series intersects the principal series is also the massless point where $\Delta=d-p$.  In this case \eqref{Fhdbpfordeme} becomes ${\cal F}^{[1^{d\over 2}]_+}_{d\over 2} \simeq {\cal F}^{[1^{d\over 2}]_-}_{d\over 2}$ so that there is only one inequivalent rep: both are equivalent to the massless rep ${\cal V}^{[1^{{d\over 2}}]}_{{d\over 2}}$, which is in turn equivalent to the gauge rep ${\cal U}^{[1^{d\over 2}]}_{d\over 2}$, and to the massless $\left({d\over 2}-1\right)$-form via \eqref{pformuveqveuveee}.  In summary, we have 
\be {\cal F}^{[1^{d\over 2}]_+}_{d\over 2} \simeq {\cal F}^{[1^{d\over 2}]_-}_{d\over 2}\simeq {\cal V}^{[1^{d\over 2}]}_{d\over 2}\simeq {\cal U}^{[1^{d\over 2}]}_{{d\over 2}}\simeq {\cal V}^{[1^{{d\over2}-1}]}_{{d\over 2}+1} \,, \  \ D\ {\rm odd}\,. \ee
In this case the rep ${\cal F}^{[1^{d\over 2}]}_{{d\over 2}}$ is decomposable into the gauge modes and the physical modes, which live in equivalent reps.

\subsubsection*{$D$ even:}

Let $D=2n$, $d=2n-1$, $n=2,3,\ldots$.  As illustrated in \eqref{pformsocontent4}, the massless $\left(n-1\right)$-form has $\frak{so}(D)$ content consisting only of tableaux whose first column has height $n$.  These reps are not irreducible: they can be split into self-dual and anti-self-dual parts with respect to the $D$-dimensional epsilon tensor acting on the first column,
\be [l,1^{d-1\over 2}]=[l,1^{d-1\over 2}]_+\oplus [l,1^{d-1\over 2}]_-\, , \ \ \ D \ {\rm even}\,.\ee
This causes the rep ${\cal V}^{[1^{d-1\over 2}]}_{{d+1\over 2}}$ to split into two pieces that we call ${\cal V}^{[1^{d-1\over 2}]_\pm }_{{d+1\over 2}}$,
\be {\cal V}^{[1^{d-1\over 2}]}_{{d+1\over 2}}={\cal V}^{[1^{d-1\over 2}]_+}_{{d+1\over 2}}\oplus {\cal V}^{[1^{d-1\over 2}]_-}_{{d+1\over 2}}\, , \ \ \ D \ {\rm even}\,. \ee
These represent the self-dual and anti-self-dual massless $\left({D\over 2}-1\right)$-forms in even $D$, and they are unitary.  As we will see in section \ref{unitarylistsection}, these self-dual and anti-self-dual massless forms in even $D$ are accounted for among the discrete series reps.

\subsection{Mixed symmetry representations\label{mixsymsec}}

We now turn to the most general kind of bosonic rep, one carried by a mixed symmetry tensor with indices in the $p$-row tableau $[s_1,\ldots, s_p]$.  The relation between $\Delta$ in the late time behavior \eqref{ssassymprosole} and the mass $\tilde m$ in the Klein-Gordon equation \eqref{KGequationfre} is as in \eqref{deltamregexe}, \eqref{dscfttmassrelatione} with $r=\sum_{i=1}^p s_i$.

The vector space carrying the rep will be the space of square integrable complex fully traceless tensor fields on the sphere ${\mathbb S}^d$ with the index symmetry of a $[s_1,\ldots, s_p]$ tableau, transforming under $\frak{so}(1,D)$ as in \eqref{latetimesheactiont3} with $r=\sum_{i=1}^p s_i$.  Call this space ${\cal F}^{[s_1,\ldots, s_p]}_\Delta$,
\be {\cal F}^{[s_1,\ldots, s_p]}_\Delta:\ \ {\rm complex\ traceless\ tensor\ fields\ of\ type\ } [s_1,\ldots, s_p] {\rm \ on\ } {\mathbb S}^d \,.\ee

To break up this space into its $\frak{so}(D)$ content, we need the generalization of the SVT decomposition for symmetric tensors, and the Hodge decomposition for $p$-forms, to tensors of general mixed symmetry type.  To obtain this generalization we proceed as follows.  Call the $\frak{so}(d)$ tensor rep $[s_1,\ldots, s_p]$ the parent field rep.  We first branch the parent field rep down one dimension using the branching rules in appendix \ref{branchingappendix}.   In terms of tableaux (i.e. for large enough $D$), this rule amounts to the following: we first write the tableau of the parent field rep, then we list all possible ways of removing a single box while remaining a legal tableau.  Then we list all possible ways of removing two boxes from the parent field rep while remaining a tableau, with the restriction that the two boxes cannot be taken from the same column.  Then we continue, removing three boxes, then four boxes, and so on, where at each stage any two boxes cannot be taken from the same column.  At some point we will not be able to continue without removing more than one box from a column, and then we are done.  The resulting list of tableaux gives the index symmetries of fields appearing in the SVT decomposition, all of which are transverse in all indices and fully traceless; we call these the TT fields.  The number of derivatives appearing on each resulting TT field is equal to the number of boxes removed as compared to the parent field rep, and these derivatives are then projected according to the symmetries of the parent field rep, with the derivatives placed into the removed boxes.  

As an example, consider the $[2,1]$ field $\phi_{ij,k}$.  To find the TT fields we first write the parent field rep $[2,1]$, the reps $[2]$, $[1,1]$ with one box removed, and the rep $[1]$ with two boxes removed.  We cannot remove three boxes without removing more than one in a column, so we are done.   The branching is thus
\be \raisebox{1.15ex}{\yng(2,1)}\underset{\frak{so}(d)\rightarrow \frak{so}(d-1)}{\Rightarrow}\raisebox{1.15ex}{\yng(2,1)}\oplus \raisebox{0.15ex}{\yng(2)}\oplus\raisebox{1.15ex}{\yng(1,1)}\oplus\raisebox{0.15ex}{\yng(1)}\ . \label{hookmassmsdece} \ee
 Call the TT fields on the right hand side $\chi_{ij,k},\ \chi_{ij},\ \chi_{i,j},\ \chi_i$, respectively.
Among the resulting fields, the $[2,1]$ will appear in the decomposition with no derivatives, the $[1,1]$ and $[2]$ will appear with one derivative (since they have one fewer box than the parent field rep $[2,1]$), and the $[1]$ will have two derivatives (since it has two fewer boxes than the parent field rep $[2,1]$).   Each field, with its derivatives, is then projected onto the $[2,1]$, with the removed boxes corresponding to the derivatives: for the $[2]$ we must project ${\cal Y}^T_{\resizebox{.5cm}{!}{\gyoung(ik,j)}}(\nabla_j \chi_{ik})\propto \nabla_{[i} \chi_{j]k}$, for the $[1,1]$ we must project ${\cal Y}^T_{\resizebox{.5cm}{!}{\gyoung(ik,j)}}(\nabla_k \chi_{i,j}) \propto \nabla_k \chi_{i,j} - {1\over 2} \nabla_{[i}  \chi_{j],k}$, and for the $[1]$ we must project ${\cal Y}^T_{\resizebox{.5cm}{!}{\gyoung(ik,j)}}(\nabla_j \nabla_k \chi_{i})\propto\nabla_k\nabla_{[i}\chi_{j]} -{1\over (d-1)}g_{k[i}\left(\nabla^2-(d-1)\right) \chi_{j]}$.
The full TT decomposition can thus be written as
\be \phi_{ij,k}=\chi_{ij,k}+\nabla_{[i} \chi_{j]k}+\nabla_k \chi_{i,j} - {1\over 2} \nabla_{[i}  \chi_{j],k}+ \nabla_k\nabla_{[i}\chi_{j]} -{1\over (d-1)}g_{k[i}\left(\nabla^2-(d-1)\right) \chi_{j]} \,.\label{21exdecomde}\ee

Each TT field is then expanded in spherical harmonics.  The spherical harmonics for a general TT field of type $[u_1,\ldots, u_{q}]$ take the form
\be Y^{I_1\ldots I_l,J_1J_2\ldots}_{l,i_1i_2\ldots }(\hat X)={\cal Y}^T_{[u_1+l,u_1,\ldots,u_{q}]} \partial_{i_1}  \hat X^{J_1}\partial_{i_2}  \hat X^{J_2 }\cdots \hat X^{I_1} \cdots \hat X^{I_l}\,, \ \ l=u_1,u_1+1,\ldots\ . \label{mixsesspjsjre}\ee
They transform under $\frak{so}(D)$ as tensors in the tableaux $[l,u_1,u_2,\ldots, u_{q}]$.  The lowest $l$ harmonic has a tableau $[u_1,u_1,u_2,\ldots, u_{q}]$; we can think of this as being obtained by ``capping'' the original tableau, i.e. adding a row to the top with the same length as the top row.  The higher $l$'s are then given by adding further boxes to the new top row.  

The spherical harmonics \eqref{mixsesspjsjre} form a basis of the space of TT fields of type $[u_1,\ldots, u_{q}]$ on ${\mathbb S}^d$. 
The natural Laplacian on this space is given by the action of the quadratic Casimir operator $C_2=-{1\over 2}{\cal L}_{{\cal M}_{IJ}} {\cal L}_{{\cal M}^{IJ}}$ of the rotation algebra $\frak{so}(D)$ of ${\mathbb S}^d$, which is given in terms of the bare Laplacian by 
\be C_2= -\nabla_\Omega^2+\sum_{i=1}^{q} u_i(u_i+d-2i)\, .\label{mixedcasimire}\ee  
The sum on the right hand side is the value of the quadratic Casimir operator in \eqref{quadcasmideweaee} for the $\frak{so}(d)$ rep $[u_1,\ldots, u_{q}]$ in which the field indices live.
The spherical harmonics \eqref{mixsesspjsjre} of the transverse traceless fields are eigenfunctions of this Casimir operator, with the following eigenvalues \cite{BRANSON1992314},
\be C_2 Y^{I_1\ldots I_l,J_1\ldots}_{l,i_1\ldots }=\left[ l(l+d-1)+ \sum_{i=1}^{q} u_i(u_i+d-1-2i)\right] Y^{I_1\ldots I_l,J_1\ldots}_{l,i_1\ldots } \,. \ee
The eigenvalue on the right hand side is the value of the quadratic Casimir operator in \eqref{quadcasmideweaee} for the $\frak{so}(D)$ rep $[l,u_1,\ldots, u_{q}]$ in which the harmonics transform.

Among the lower rank TT fields, the spherical harmonics with $l<s_1$  will vanish in the TT decomposition, because they are annihilated by the combinations of derivatives that appear in the decomposition into TT fields.  A way to visualize this rule is the following: take the tableau $[u_1,\ldots, u_{q}]$ for the TT field, add a row to the top of length $s_1$, where $s_1$ is the length of the top row of the original parent field.  This then forms the tableau $[s_1,u_1,\ldots, u_{q}]$ for the lowest $l$ harmonic appearing in the decomposition. (We can get this same content by first capping the original rep, then branching it down one dimension.) For example, for the $[2,1]$ field, the TT fields are $[2]$, $[1,1]$ and $[1]$, to each of these we add a row on top of length 2, obtaining $[2,2]$, $[2,1,1]$, $[2,1]$.  These are the tableaux for the lowest $l$ harmonic appearing in the decomposition, i.e. $l=2$ for each of them.  This means the lowest $l=1$ harmonics of the $[1,1]$ field and $[1]$ field vanish (as can be seen explicitly by plugging into \eqref{21exdecomde}).    
The $\frak{so}(D)$ content for the $[2,1]$ field can thus be visualized as follows:
\be \raisebox{-40pt}{\epsfig{file=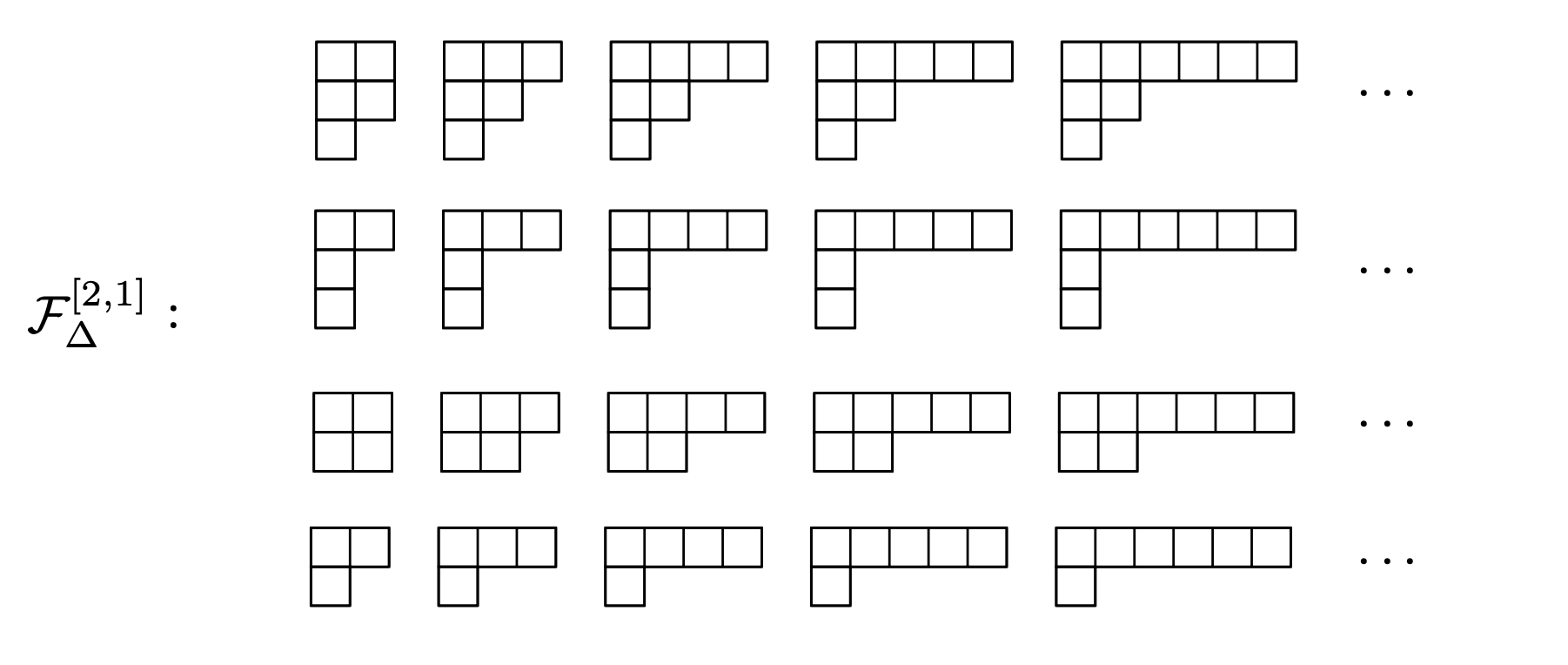,width=5.0in}}  \label{mixedcontent1}\, \ee

The boost operators ${\cal K}^I$ will connect these reps.  For a given $\frak{so}(D)$ rep $\rho$, the other $\frak{so}(D)$ reps that can be generically reached by the action of the ${\cal K}^I$ are those that appear in the product $[1]\otimes \rho$.  We can  think of the $\frak{so}(D)$ reps as sites on a lattice, and the links generated by the action of ${\cal K}^I$ as lattice interactions.  These will be nearest neighbor interactions if the lattice is arranged in the right way.  For example, the $\frak{so}(D)$ content for the $[2,1]$ field shown in \eqref{mixedcontent1} is better visualized by arranging the rows into the form of a 3 dimensional lattice as follows:
\be \raisebox{-40pt}{\epsfig{file=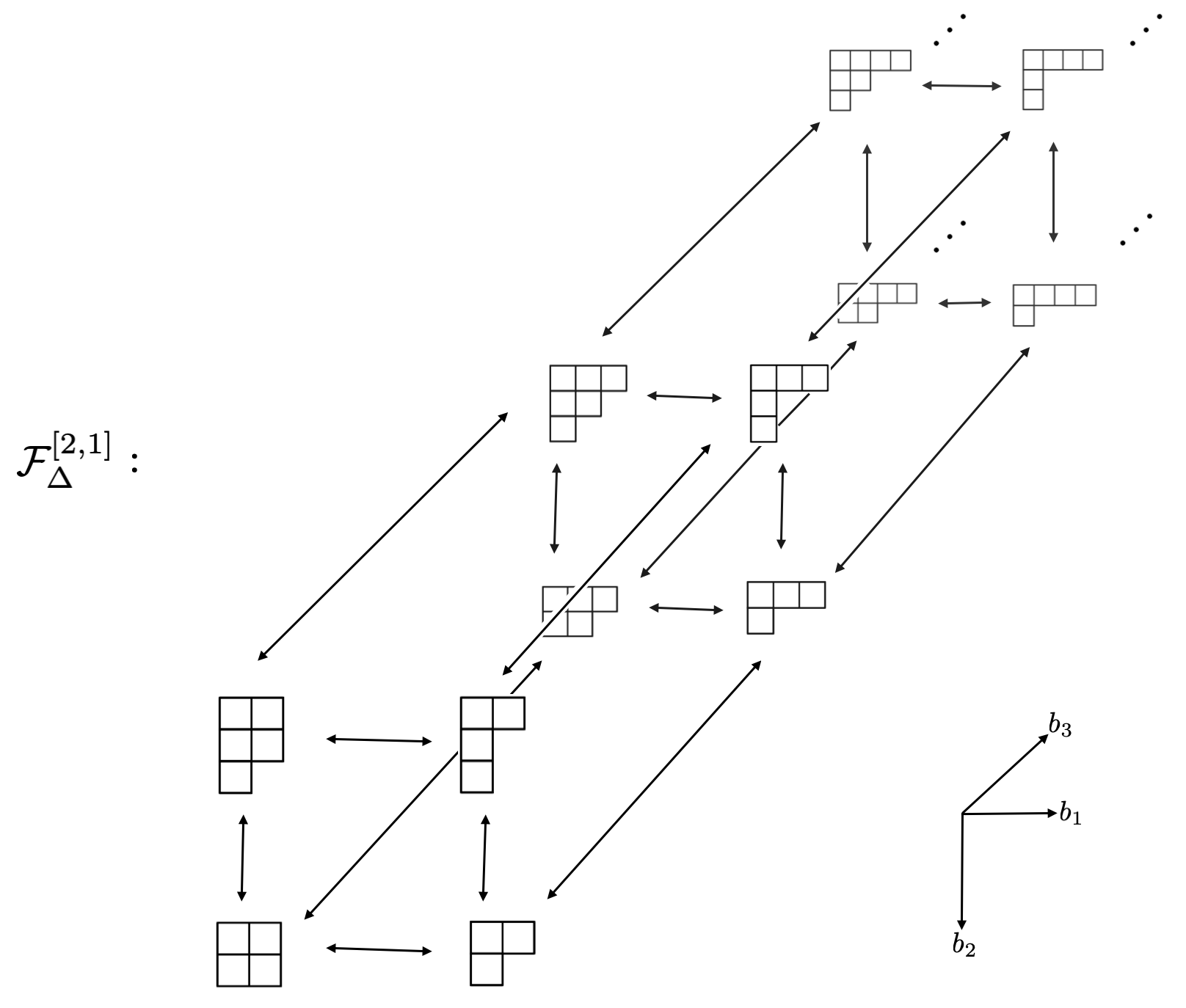,width=4.3in}}  \label{mixedcontent2}\, \ee
The lattice dimensions are labelled $b_1,b_2,b_3$, where the lattice has two sites in the first two directions and an infinite number of sites in the third, i.e. $b_1=1,2$, $b_2=1,2$, $b_3=1,2,3,\ldots$.
In this arrangement the interactions are all local nearest neighbor interactions, and the arrows representing the ${\cal K}^I$ actions are shown.

For the general ${\cal F}^{[s_1,\ldots, s_p]}_\Delta$ rep, the lattice will live in a $B+1$ dimensional space, where $B$ is the number of blocks in the tableau $[s_1,\ldots, s_p]$ (a block is the set of all the rows of a given length), so we will label the axes by $b_I$, where $I=1,2,\ldots, B+1$.  Let $T_I$ be the number of rows in the $I$-th block minus the number of rows in the $(I+1)$-th block, i.e. $T_I$ is the number of partially massless points possible from activating the $I$-th block\footnote{We assume in this section some familiarity with the classification of PM points among the mixed symmetry fields, as described in \cite{Metsaev:1995re,Metsaev:1997nj,Alkalaev:2003qv,Boulanger:2008up,Boulanger:2008kw,Skvortsov:2009zu,Skvortsov:2009nv,Basile:2016aen}.  Briefly, the PM points correspond to `activating' some number of boxes in the final row of a block in the tableau.   If it is the boxes from row $q$ that are activated, then the number of activated boxes is given by the depth parameter $t$, which can range over $1,2,\ldots, s_q-s_{q+1}$.}.  Then there are $T_I+1$ points in the $b_I$ direction of the lattice direction, so that $B_I=1,2,\ldots ,T_I+1$.  The size of the remaining $(B+1)$ direction is infinite, $b_{B+1}=1,2,3,\ldots$.  The face $b_{B+1}=1$ consists of the $\frak{so}(D)$ reps obtained from branching the capped rep $[s_1,s_1,s_2,\ldots, s_p]$: the original rep $[s_1,s_1,s_2,\ldots, s_p]$ appears in the corner $b_I=1$, $I=1,\ldots,B$, and the remaining reps obtained from the branching are arranged so that a box is removed from the $I$-th block as we move one space in the $b_I$ direction.   The faces $b_{B+1}=n+1$, $n\geq1$, are then obtained by adding $n$ boxes to the first row of each tableau in the initial face $b_{B+1}=1$.   Note that in the case of the symmetric tensors in section \ref{spinssec} and $p$-forms in section \ref{pformsec}, there is only one block and so the lattice in these cases was two dimensional.

\textbf{Reducible cases:} There are discrete values of $\Delta$ at which the arrows in the lattice break and we get sub-reps, making the rep ${\cal F}_{\Delta}^{[s_1,\ldots, s_p]}$ reducible but not decomposable.  
The sub-reps of ${\cal F}^{[s_1,\ldots, s_p]}_\Delta$  that emerge when $\Delta$ takes these values all take simple forms: they are always a grouping of faces of the lattice. 
The classes of values of $\Delta$ where ${\cal F}^{[s_1,\ldots, s_p]}_\Delta$ becomes reducible are as follows\footnote{The structure of the $\frak{so}(D)$ reps within the full $\frak{so}(1,D)$ rep mirrors exactly the structure of the Weyl module of the corresponding field.  The Weyl module is the space of all gauge invariant on-shell non-trivial local operators in the theory that are linear in the fields (which are the building blocks of the space of all local operators), along with the derivatives operators that map between them \cite{Lopatin:1987hz,Vasiliev:2001wa,Alkalaev:2003qv,Boulanger:2008up,Boulanger:2008kw,Skvortsov:2009zu,Skvortsov:2009nv,Alkalaev:2009vm,Ponomarev:2010st,Alkalaev:2011zv,Boulanger:2014vya,Boulanger:2015mka,Khabarov:2019dvi}.  These modules for all the massive, partially massless and shift symmetric fields are described in \cite{Hinterbichler:2024vyv}, and the description there mirrors precisely the description of the dS reps here. In particular, a more intricate example, the $[4,2,2,1]$ field, is described there at the end of section 2.2. }:  
\begin{itemize}

\item 
Shift symmetric points: 
\be \Delta=d+s_1+k\, ,\ \  k=0,1,2,\ldots \ \ .\ee
The sub-rep is called
\be  {\cal D}^{[s_1,\ldots, s_p]}_{d+s_1+k}\,,\ee
and is given by the harmonics with $l\geq s_1+k+1$, which corresponds to everything other than the first $k+1$ faces of the lattice of $\frak{so}(D)$ reps going in the $b_{B+1}$ direction, i.e. $b_{B+1}=k+2,k+2,\ldots$.

This represents the physical modes of the level $k$ shift symmetric mixed symmetry field \cite{Hinterbichler:2022vcc}.

\item
Finite points:
\be  \Delta=-s_1-k\, ,\ \  k=0,1,2,\ldots \ \ . \ee
The sub-rep is called
\be  {\cal S}^{[s_1,\ldots, s_p]}_{-s_1-k}\, , \ee
and is given by the harmonics with $l\leq  s_1+k$, which corresponds to the first $k+1$ faces of the lattice of $\frak{so}(D)$ reps going in the $b_{B+1}$ direction, i.e. $b_{B+1}=1,2,\ldots,k+1$. These finite dimensional reps are the tensor reps of $\frak{so}(1,D)$; the tableau for this specific tensor is the one with the most number of boxes in the ${\cal S}^{[s_1,\ldots, s_p]}_{-s_1-k}$ rep, i.e. the tableau $[s_1+k,s_1,s_2,\ldots, s_p]$ (all the tensor reps of $\frak{so}(1,D)$ are accounted for in this way).  Branching an $\frak{so}(1,D)$ tensor with this tableau to $\frak{so}(D)$ using the rules in appendix \eqref{branchingappendix} gives the $\frak{so}(D)$ tensors in ${\cal S}^{[s_1,\ldots, s_p]}_{-s_1-k}$.  

This rep corresponds to the shift symmetries of the level $k$ mixed symmetry shift symmetric field.

\item
PM points:
\be \Delta=d - q + s_{q} - t\, , \ \ t=1,2,\ldots, s_q-s_{q+1} \,. \label{dsgPMponctee}\ee
Here the $q$-th row is any row whose length is larger than the row below it, i.e. it is the bottom row of a block.
The sub-rep is called
\be  {\cal V}^{[s_1,\ldots, s_p]}_{d - q + s_{q} - t}\, . \ee
These values correspond to the physical modes of the mixed symmetry PM fields.   Recall that if the $q$-th row is being activated for a PM symmetry, then the depth $t$ indicates the number of boxes that are activated and it ranges over $1,2,\ldots, s_q-s_{q+1}$.  If the row $q$ corresponds to the $I$-th block, then this sub-rep consists of those $\frak{so}(D)$ reps with $b_{I}=1,2,\ldots, t$.

\item
Gauge points,
\be \Delta=q - s_{q} + t\,, \ \ t=1,2,\ldots, s_q-s_{q+1} \,, \ee
corresponding to the PM points \eqref{dsgPMponctee}.
The sub-rep is called
\be  {\cal U}^{[s_1,\ldots, s_p]}_{q - s_{q} + t}\,,\ee
and consists of those $\frak{so}(D)$ reps with $b_{I}= t+1,t+2,\ldots, T_I+1$.  These sub-reps correspond to the gauge modes of the corresponding PM field.

\end{itemize}

\textbf{Equivalences:} There is a shadow transform intertwining operator that connects the $\Delta$ and $\bar\Delta\equiv d-\Delta$ reps, 
\be S_\Delta^{[s_1,\ldots, s_p]} :\ {\cal F}_{\Delta}^{[s_1,\ldots, s_p]}\rightarrow {\cal F}_{\bar\Delta}^{[s_1,\ldots, s_p]}\,.\label{pformshwamixdee}\ee
It commutes with the $\frak{so}(D)$ rotations and satisfies $\delta_{{\cal K}^I_{\bar\Delta}}S_\Delta^{[s_1,\ldots, s_p]}=S_\Delta^{[s_1,\ldots, s_p]} \delta_{{\cal K}^I_{\Delta}}$.
For generic $\Delta$, this operator is invertible and the $\Delta$ and $\bar\Delta$ reps are equivalent to each other,
\be {\cal F}_{\Delta}^{[s_1,\ldots, s_p]}\simeq {\cal F}_{\bar\Delta}^{[s_1,\ldots, s_p]}\,.\label{pformeejfmixee}\ee
But for the special values of $\Delta$ indicated above where ${\cal F}_{\Delta}^{[s_1,\ldots, s_p]}$ develops a sub-rep, $S_\Delta^{[s_1,\ldots, s_p]}$ develops a kernel which is always precisely the sub-rep.

The shadow transform map \eqref{pformshwamixdee} links the shift symmetric to the finite points, giving the isomorphisms
\be {\cal D}_{d+s_1+k}^{[s_1,\ldots, s_p]} \simeq {\cal F}_{-s_1-k}^{[s_1,\ldots, s_p]}/{\cal S}_{-s_1-k}^{[s_1,\ldots, s_p]}\,, \ \ {\cal S}_{-s_1-k}^{[s_1,\ldots, s_p]} \simeq {\cal F}_{d+s_1+k}^{[s_1,\ldots, s_p]}/{\cal D}_{d+s_1+k}^{[s_1,\ldots, s_p]}\,,\ee
and it links the PM to the gauge points, giving the isomorphisms
\be {\cal V}_{d-q+s_q-t}^{[s_1,\ldots, s_p]} \simeq {\cal F}_{q-s_q+t}^{[s_1,\ldots, s_p]}/{\cal U}_{q-s_q+t}^{[s_1,\ldots, s_p]}\,, \ \ {\cal U}_{q-s_q+t}^{[s_1,\ldots, s_p]} \simeq {\cal F}_{d-q+s_q-t}^{[s_1,\ldots, s_p]}/{\cal V}_{d-q+s_q-t}^{[s_1,\ldots, s_p]}\,.\label{gesmfgixisogpmjdee}\ee

There are generalized gradient and divergence operators that move along a generalized chain complex of traceless tensors that generalize the de Rham complex of differential forms~\cite{olver1982differential,Dubois-Violette:1999iqe,Dubois-Violette:2001wjr,Bekaert:2002dt} (these generalized chain complexes are also known as BGG complexes \cite{92c50ea1-4549-3a51-8381-76d7c6eb773e,eastwood1999variations,Arnold2020ComplexesFC}, they are reviewed in section 2.1.1 of \cite{Hinterbichler:2024cxn}).   Each PM point is associated to a unique complex: the PM point in which the $q$-th row is activated by removing $t$ boxes is associated with the complex of length $q$ that we find as follows.  We first find a set of integers $r_{(a)}$, $a=1,2,\ldots,q$, by starting with $r_{(q)}=t$ and using the relation $r_{(a)}=s_{a}-s_{a+1}+1$ to fix the $r_{(a)}$ with $a=1,\ldots,q-1$.
Starting with the field's tableau, we then remove $r_{(q)}$ boxes from the $q$-th row, then $r_{(q-1)}$ boxes from the $(q-1)$-th row, and so on until at the last step we remove $r_{(1)}$ boxes from the first row.  This gives a sequence of tableaux ${\mathbb Y}_{(a)}$, $a=0,1,\ldots,q$, with ${\mathbb Y}_{(q)}$ corresponding to the PM field, ${\mathbb Y}_{(q-1)}$ its gauge parameter, ${\mathbb Y}_{(q-2)}$ the first reducibility parameter, and so on down the chain of reducibility parameters, ending with ${\mathbb Y}_{(0)}$.

The generalized gradient operators $\rd_{(a)}:{\mathbb Y}_{(a)}\rightarrow {\mathbb Y}_{(a+1)}$, $a=0,1,\ldots,q-1$, act by adding $r_{(a+1)}$ derivatives to the $(a+1)$-th row of ${\mathbb Y}_{(a)}$ and then projecting onto ${\mathbb Y}_{(a+1)}$.  They satisfy the fundamental condition $\rd_{(a+1)}\rd_{(a)}=0$.  The generalized divergence operators $\rd^\dag_{(a)}:{\mathbb Y}_{(a)}\rightarrow {\mathbb Y}_{(a-1)}$, $a=1,2,\ldots,q$, act by taking $r_{(a)}$ divergences on the indices of the $a$-th row of ${\mathbb Y}_{(a)}$ and then projecting onto ${\mathbb Y}_{(a-1)}$.  They satisfy $\rd^\dag_{(a-1)}\rd^\dag_{(a)}=0$.  

These operators act on spaces with specific weights $\Delta$ associated with the PM and gauge points.
The generalized divergence $\rd_{(a)}^\dag $ raises the value of $\Delta$ by $r_{(a)}$, which is the number of derivatives in the operator: starting with the PM point ${\mathbb Y}_{(q)}$, this generates a sequence of weights $\Delta_{(a)}$, $a=0,1,\ldots,q$ with $\Delta_{(q)}= d - q + s_{q} - t$.  The generalized gradient $\rd_{(a)}^\dag $ also raises the value of $\Delta$ by $r_{(a)}$, which is the number of derivatives in the operator: this generates a sequence of weights $\bar\Delta_{(a)}$, $a=0,1,\ldots,q$ with $\bar\Delta_{(q)}= q - s_{q} + t$ the gauge point.
Both $\rd_{(a)}$ and $\rd_{(a)}^\dag$ commute with the Casimir operator \eqref{mixedcasimire}. 

These generalized gradient and divergence operators, along with the shadow operators \eqref{pformshwamixdee}, group together PM reps, those of its gauge parameters and reducibility parameters, and their shadows, into the following two row commutative diagram:   
\be \raisebox{-40pt}{\epsfig{file=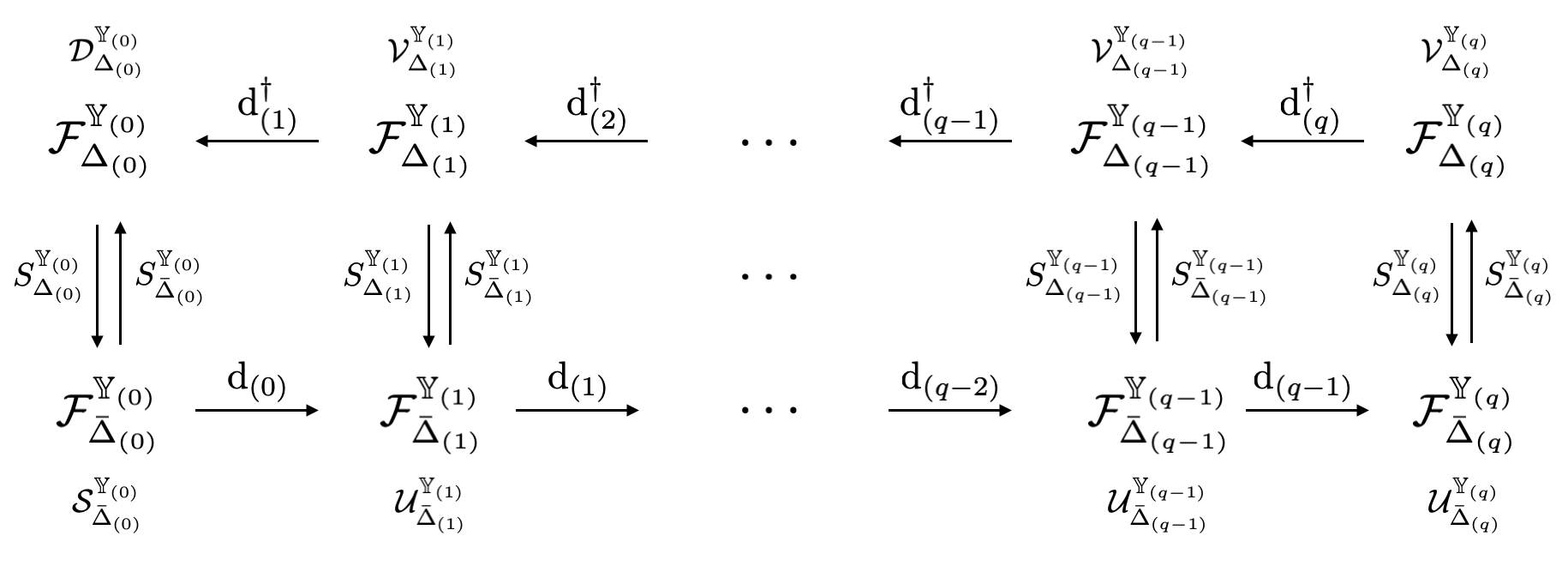,width=5.5in}}  \label{mixedcommdiagram}\, \ee
Each rep is shown, along with its sub-rep in smaller font.  Each sub-rep is simultaneously the kernel of all the outgoing maps and the image of all the ingoing maps, and going through any two arrows consecutively gives zero.  In the upper row we have the reps corresponding to the gauge parameters and reducibility parameters, which are all themselves PM reps, with the exception of the left-most one in the chain, which is a shift symmetric rep.  In the bottom row we have the corresponding gauge reps, ending with a finite rep at the left.
This diagram gives the following isomorphisms between the PM reps, the gauge mode reps, and the shift symmetric reps,
\be {\cal V}_{\Delta_{(q-1)}}^{{\mathbb Y}_{(q-1)}} \simeq {\cal U}_{\bar\Delta_{(q)}}^{{\mathbb Y}_{(q)}} \,, \ \ {\cal V}_{\Delta_{(q-2)}}^{{\mathbb Y}_{(q-2)}} \simeq {\cal U}_{\bar\Delta_{(q-1)}}^{{\mathbb Y}_{(q-1)}}  \,, \ \ldots\, , \ \ {\cal V}_{\Delta_{(1)}}^{{\mathbb Y}_{(1)}} \simeq {\cal U}_{\bar\Delta_{(2)}}^{{\mathbb Y}_{(2)}}  \,,\ \  {\cal D}_{\Delta_{(0)}}^{{\mathbb Y}_{(0)}} \simeq {\cal U}_{\bar\Delta_{(1)}}^{{\mathbb Y}_{(1)}} \,.\label{mixsyisomfsfee}\ee

As an example, consider again the $[2,1]$ field.  The PM point in which the top block is activated has $q=1$, $r_1=1$, the sequence of tableaux is ${\mathbb Y}_{(0)}=[1,1]$, ${\mathbb Y}_{(1)}=[2,1]$, the sequence of weights is $\Delta_{(0)}=d+1$, $\Delta_{(1)}=d$, $\bar\Delta_{(0)}=-1$, $\bar\Delta_{(1)}=0$, and the commutative diagram looks like this, now showing the $\frak{so}(D)$ content:
\be \raisebox{-40pt}{\epsfig{file=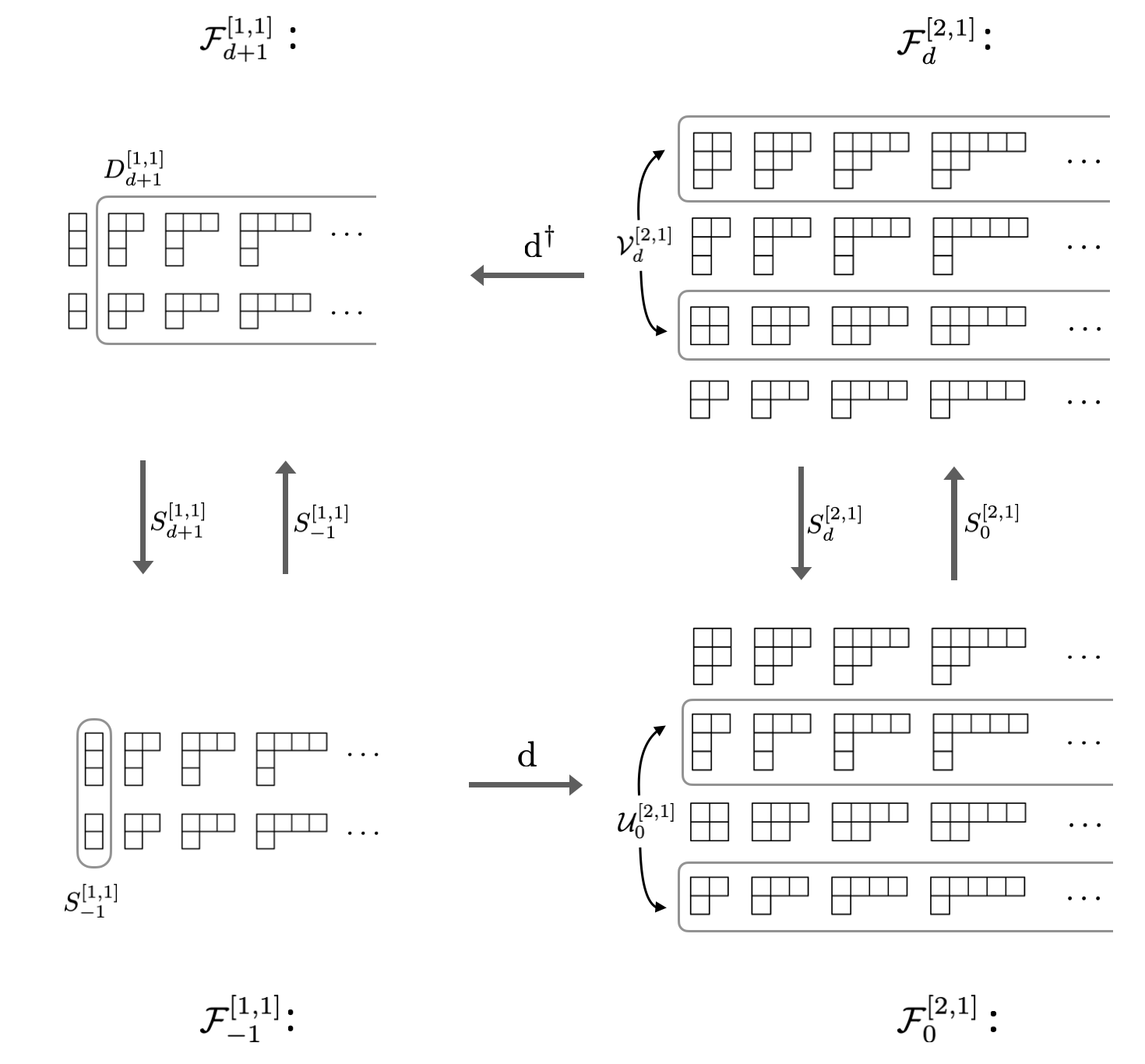,width=5.0in}}  \label{mixedcontent3}\, \ee
For clarity, we have displayed ${\cal F}_d^{[2,1]}$ and ${\cal F}_0^{[2,1]}$ as a set of four rows, but it should really be visualized as in \eqref{mixedcontent2}, where it can be seen that ${\cal V}_d^{[2,1]}$ corresponds to the left-hand face of the lattice and ${\cal U}_0^{[2,1]}$ corresponds to the right-hand face.
The diagram \eqref{mixedcontent3} illustrates the isomorphisms
\bea && {\cal D}^{[1,1]}_{d+1}\simeq {\cal F}^{[1,1]}_{-1}/{\cal S}^{[1,1]}_{-1}\, , \ \ {\cal S}^{[1,1]}_{-1}\simeq {\cal F}^{[1,1]}_{d+1}/{\cal D}^{[1,1]}_{d+1}\, , \nn\\
 && {\cal V}^{[2,1]}_{d}\simeq {\cal F}^{[2,1]}_{0}/{\cal U}^{[2,1]}_{0}\, , \ \ {\cal U}^{[2,1]}_{0}\simeq {\cal F}^{[2,1]}_{d}/{\cal V}^{[2,1]}_{d}\, , \nn\\ 
&& {\cal D}^{[1,1]}_{d+1}\simeq {\cal U}^{[2,1]}_{0}\,.  
 \eea 

The $[2,1]$ PM field in which the bottom block is activated has $q=2$, $r_2=1$, $r_1=2$, the sequence of tableaux is ${\mathbb Y}_{(0)}=[0]$, ${\mathbb Y}_{(1)}=[2]$, ${\mathbb Y}_{(2)}=[2,1]$, the sequence of weights is $\Delta_{(0)}=d+1$, $\Delta_{(1)}=d-1$, $\Delta_{(2)}=d-2$, $\bar\Delta_{(0)}=-1$, $\bar\Delta_{(1)}=1$, $\bar\Delta_{(2)}=2$, and the commutative diagram looks like this:
\be \raisebox{-40pt}{\epsfig{file=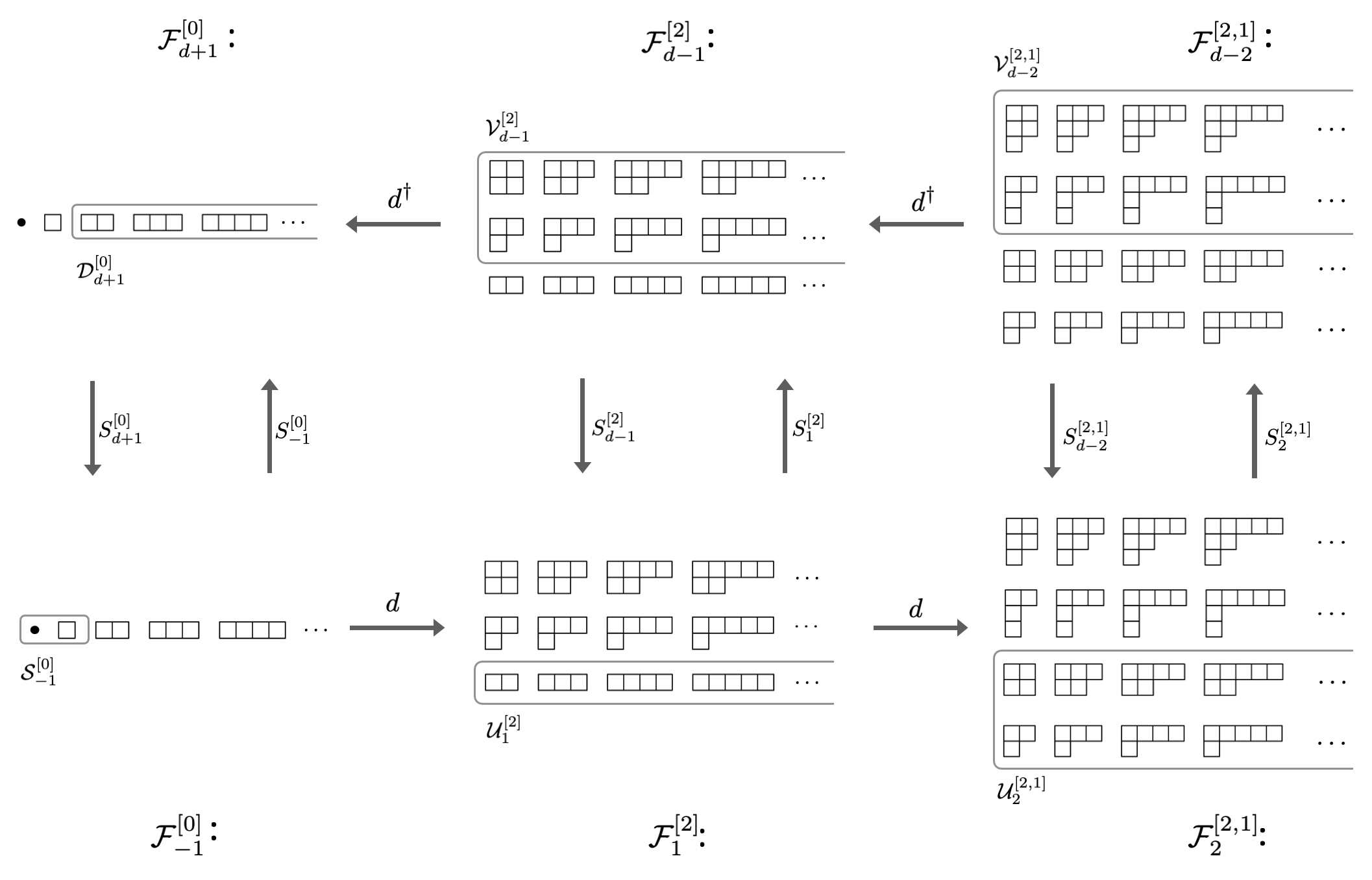,width=5.7in}}  \label{mixedcontent4}\, \ee
Again, ${\cal F}_{d-2}^{[2,1]}$, ${\cal F}_2^{[2,1]}$ should really be visualized as in \eqref{mixedcontent2}, where it can be seen that ${\cal V}_{d-2}^{[2,1]}$ corresponds to the top face of the lattice and ${\cal U}_{2}^{[2,1]}$ corresponds to the bottom face. 
The diagram \eqref{mixedcontent4} illustrates the isomorphisms 
\bea && {\cal D}^{[0]}_{d+1}\simeq {\cal F}^{[0]}_{-1}/{\cal S}^{[0]}_{-1}\, , \ \  {\cal S}^{[0]}_{-1}\simeq {\cal F}^{[0]}_{d+1}/{\cal D}^{[0]}_{d+1}\, , \nn\\
 &&{\cal V}^{[2]}_{d-1}\simeq {\cal F}^{[2]}_{1}/{\cal U}^{[2]}_{1}\, , \ \ {\cal U}^{[2]}_{1}\simeq {\cal F}^{[2]}_{d-1}/{\cal V}^{[2]}_{d-1}\, , \nn\\ 
&& {\cal V}^{[2,1]}_{d-2}\simeq {\cal F}^{[2,1]}_{2}/{\cal U}^{[2,1]}_{2}\, , \ \ {\cal U}^{[2,1]}_{2}\simeq {\cal F}^{[2,1]}_{d-2}/{\cal V}^{[2,1]}_{d-2} \, , \nn \\
&& {\cal D}^{[0]}_{d+1}\simeq {\cal U}^{[2]}_{1}\, ,\ \  {\cal V}^{[2]}_{d-1}\simeq {\cal U}^{[2,1]}_{2} \,.
 \eea

\textbf{Unitarity:}  There is a diffeomorphism and Weyl invariant bilinear form pairing the two spaces ${\cal F}_\Delta^{{[s_1,\ldots, s_p]}}$ and ${\cal F}_{\bar \Delta}^{{[s_1,\ldots, s_p]}}$,
\be (\phi_1,\phi_2)\equiv {1\over c_1!c_2!\cdots }\int d^d\Omega\, \phi_1^{i_1\ldots }(\hat X)\phi_{2\, i_1\ldots }(\hat X)\,,\ \ \ \  \phi_{1\, i_1\ldots }\in {\cal F}_{\bar\Delta}^{{[s_1,\ldots, s_p]}},\ \ \phi_{2\, i_1\ldots }\in {\cal F}_{\Delta}^{{[s_1,\ldots, s_p]}}\, ,\label{bilinteaprformveeme}\ee
where in the normalization $c_1,c_2,\ldots$ are the lengths of the columns of the tableau $[s_1,\ldots, s_p]$.

In the case where $\Delta^\ast=\bar\Delta= d-\Delta$ we can use \eqref{bilinteaprformveeme} to form a manifestly positive definite, invariant inner product on ${\cal F}_\Delta^{{[s_1,\ldots, s_p]}}$ via:
\be \la \phi_1 | \phi_2 \ra\equiv (\phi_1^{\ast},\phi_2),\ \ \ \Delta^\ast=\bar\Delta\,, \ \ \phi_{1\, i_1\ldots }, \phi_{2\, i_1\ldots }\in {\cal F}_{\Delta}^{{[s_1,\ldots, s_p]}}  \,. \label{innerprotdpvprise2} \ee
This gives us the mixed symmetry principal series reps, 
\be {\rm mixed\ symmetry \ principal\ series:}\  \Delta={d\over 2}+i\nu\, ,\ \ \ \nu\in {\mathbb R}\,,\ee
which are all unitary.

In the case where $\Delta$ is real, we use $S^{{[s_1,\ldots, s_p]}}_{\Delta}$ to move a state from ${\cal F}_{ \Delta}^{{[s_1,\ldots, s_p]}}$ to ${\cal F}_{\bar \Delta}^{{[s_1,\ldots, s_p]}}$ and form an inner product on ${\cal F}_{ \Delta}^{{[s_1,\ldots, s_p]}}$ as follows,
\be \la \phi_1 | \phi_2 \ra\equiv (S^{{[s_1,\ldots, s_p]}}_{\Delta}\phi_1^\ast,\phi_2)\, ,\ \ \ \Delta^\ast=\Delta\, , \ \ \phi_{1\, i_1\ldots }, \phi_{2\, i_1\ldots }\in {\cal F}_{\Delta}^{{[s_1,\ldots, s_p]}}  \,. \label{innerprodtprispcsvmdee} \ee 
The positivity of this inner product is now equivalent to whether the matrix elements of $S^{{[s_1,\ldots, s_p]}}_{\Delta}$ are positive for all the different $\frak{so}(D)$ reps in ${\cal F}^{[s_1,\ldots, s_p]}_\Delta$.  Away from the discrete special cases described above, it turns out that they are positive only in the range $p<\Delta<d-p$, which gives the mixed symmetry complementary series,
\be {\rm mixed\ symmetry \ complementary\ series:}\  p<\Delta<d-p\,, \label{comprangepmfe}\ee
where we have avoided committing the point $\Delta=d/2$ to either series.

Among the discrete values of $\Delta$ with shortened reps, the inner product \eqref{innerprodtprispcsvmdee} degenerates on the sub-rep.  When the sub-rep is factored out, we obtain a non-degenerate inner product on the remaining factor rep.  This non-degenerate inner product is unitary only if there are no negative norm states.  It turns out that the only unitary cases among the PM points are those in which boxes in the last row of the tableau are activated, i.e. those with $q=p$.  The end of the complementary series at $\Delta=d-p$ is the PM point in which all the boxes of the last row are activated.   None of the shift symmetric or finite points are unitary (with the exception of the scalar shift symmetric points and the one-dimensional trivial rep).  For a given gauge point, it is unitary if the PM or shift symmetric rep it is equivalent to is unitary, which is only the case when it is a gauge point for a PM field in which all the boxes in the bottom row are activated.

\textbf{Summary:} The mixed symmetry reps are summarized here:
\be \raisebox{-40pt}{\epsfig{file=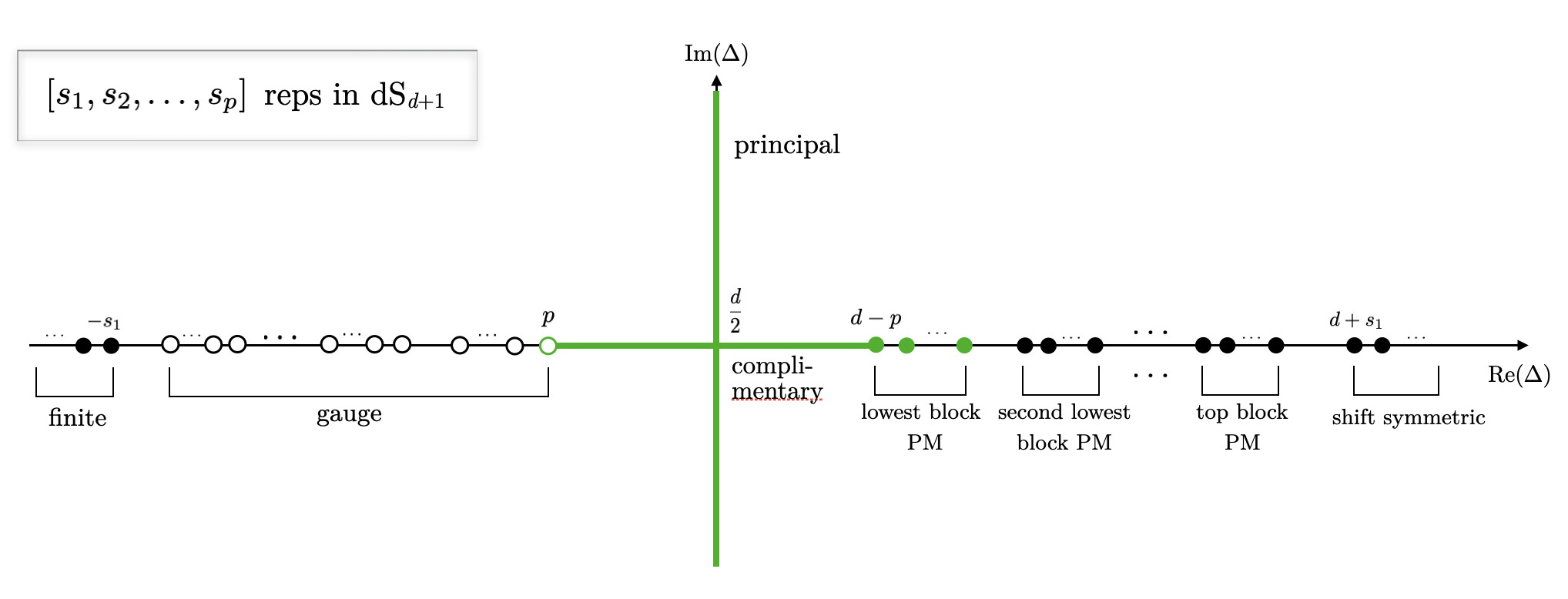,width=6.7in}}  \label{dsrepsspingen}\, \ee
 Points in green are unitary reps.  The dots are the special points where the rep becomes reducible. The hollow dots indicate that these are equivalent to a rep with a smaller tableau, and thus already accounted for there.\footnote{Note that only one of  these is green, the one corresponding to the gauge modes of the PM point in which all the boxes in the bottom row are activated, $q=p$, $t=s_p$.  This gauge point is equivalent via \eqref{mixsyisomfsfee} to the unitary PM point among $(p-1)$-row fields with $q=p-1$, $t=s_{p-1}-s_p+1$.  This fact will be important in section \ref{unitarylistsection} when accounting for the unitary reps.}
  The equivalence \eqref{pformeejfmixee} of the non-special reps is given by reflecting through the point $\Delta=d/2$, and due to the breakdown of the intertwiner map \eqref{pformshwamixdee}, the special reps with dots are inequivalent despite the reflection.  Other than this equivalence, all the reps are distinct.

The value of the quadratic Casimir operator ${\cal C}_2$ of section \ref{casimirsec} on the general mixed symmetry reps is given by \eqref{c2gnsmdee}.

We have the following correspondence between mixed symmetry $[s_1,\ldots,s_p]$ fields on dS$_{D}$ and unitary reps of $\frak{so}(1,D)$:
\bea \begin{cases}
 {\rm principal \ series:\ }  \Delta={d\over 2}+i\nu \, , \  \nu\in {\mathbb R}\, , & {\rm heavy \ fields:\ }  {\tilde m^2\over H^2 }\geq {(D-1)^2\over 4}+r \, , \\
 {\rm complementary \ series:\ }  p<\Delta<d-p  \, , & {\rm light \  fields:\ } p(D-p-1)+r < {\tilde m^2\over H^2 } \leq {(D-1)^2\over 4}+r  \,, \\
  {\rm  PM:\ }  \Delta=d-p+s_p-t \, , \ \ t=1,2,\ldots, s_p \, , &  {\rm PM :\ }{ \tilde m^2\over H^2}=(p-s_p+t)(D-p-1+s_p-t)+r\, .\\
 \end{cases} \nn
\eea
Here $r=\sum_{i=1}^p s_i $ is the total rank of the field, and the PM cases are only those in which the last row, i.e. the $p$-th row, is activated.

\subsubsection*{Equivalences and dualities}

In any given $D$, reps whose tableaux have too many rows may become trivial, and there are duality equivalences between some tableaux and others.  The detailed pattern of equivalences is precisely that which is catalogued in \cite{Hinterbichler:2024vyv} and summarized in sections 3 and 4 of that paper, so we will not repeat them in detail here.  The features that are necessary for understanding the classification of reps are detailed in the following.  

\subsubsection*{$D$ odd:}

Let $D=2n+1$, $n=1,2,3,\ldots$.  Any reps corresponding to a $p$-row tableau with $p> n$ is either trivial or related by duality to one with $p\leq n $, so we only need to consider the reps with $p\leq n$.  

Those with $n$ rows, the ${\cal F}^{[s_1,\ldots,s_n]}_\Delta$ with $s_n>0$, have an $\frak{so}(D)$ content that, after removing trivial $\frak{so}(D)$ tableaux, can be split into two identical sets, after which a change of basis will split the rep into two chiral pieces that we call ${\cal F}^{[s_1,\ldots,s_n]_\pm}_\Delta$,
\be {\cal F}^{[s_1,\ldots,s_n]}_\Delta={\cal F}^{[s_1,\ldots,s_n]_+}_\Delta\oplus {\cal F}^{[s_1,\ldots,s_n]_-}_\Delta\,,\ \ \ D \ {\rm odd}\,,\ee
This corresponds to the splitting of the tensor field on ${\mathbb S}^d$ into (imaginary for $n$ odd, real for $n$ even) self-dual and anti-self-dual parts with respect to the indices in the first column.
This split affects the shift symmetric points,
\be {\cal D}^{[s_1,\ldots,s_n]}_{}={\cal D}^{[s_1,\ldots,s_n]_+}_{}\oplus {\cal D}^{[s_1,\ldots,s_n]_-}_{}\,,  \ \ D \ {\rm odd}\,, \ee
and the finite dimensional reps
\be {\cal S}^{[s_1,\ldots,s_n]}_{}={\cal S}^{[s_1,\ldots,s_n]_+}_{}\oplus {\cal S}^{[s_1,\ldots,s_n]_-}_{}\,.  \ \ D \ {\rm odd}\,, \ee
corresponding to the chiral splitting of the $[s_1+k,s_1,\ldots, s_n]$ tensor rep of $\frak{so}(1,D)$ using the $(D+1)$-dimensional epsilon tensor acting on the first column.

The equivalence \eqref{pformeejfmixee} under the shadow transform flips the chirality, so we have
\be {\cal F}^{[s_1,\ldots,s_n]_\pm }_\Delta \simeq {\cal F}^{[s_1,\ldots,s_n]_\mp}_{\bar\Delta}\,,  \ \ D \ {\rm odd}\,.\label{mixedshadowdo2e}\ee
At the shift symmetric and finite points, this leads to
\be  {\cal D}^{[s_1,\ldots,s_n]_\pm }_{d+s_1+k} \simeq {\cal F}^{[s_1,\ldots,s_n]_\mp }_{-s_1-k}/{\cal S}^{[s_1,\ldots,s_n]_\mp }_{-s_1-k} \,, \ \   {\cal S}^{[s_1,\ldots,s_n]_\pm }_{-s_1-k} \simeq {\cal F}^{[s_1,\ldots,s_n]_\mp }_{d+s_1+k}/{\cal D}^{[s_1,\ldots,s_n]_\mp }_{d+s_1+k}  \,, \ \ D \ {\rm odd}\,.   \ee

At $\Delta=d/2$, \eqref{mixedshadowdo2e} tells us that we have a single unique rep, ${\cal F}^{[s_1,\ldots,s_n]_+}_{d\over 2} \simeq {\cal F}^{[s_1,\ldots,s_n]_-}_{d\over 2}$.
There is no complementary series for the $n$-row tableaux, and the point $\Delta=d/2$ is the PM point in which all the boxes in the bottom row are activated.  For this PM point the gauge modes carry the same rep as the physical modes, which are themselves equivalent via \eqref{mixsyisomfsfee} to a $(n-1)$-row PM field with $q=n-1$, $t=s_{n-1}-s_n+1$.  In summary, we have the equivalences,
\be {\cal F}^{[s_1,\ldots,s_n]_+}_{d\over 2} \simeq {\cal F}^{[s_1,\ldots,s_n]_-}_{d\over 2}\simeq  {\cal V}^{[s_1,\ldots,s_n]}_{d\over 2}\simeq {\cal U}^{[s_1,\ldots,s_n]}_{d\over 2}\simeq {\cal V}^{[s_1,\ldots,s_{n-1}]}_{{d\over 2}+s_n}   \,,  \ \ D \ {\rm odd}\,.\label{oddnmeqfee}\ee

There are several important features involving the PM fields.   If we have an $n$-row rep $[s_1,\ldots,s_n]$, then its unitary PM points, i.e. those with boxes in the last row activated, are all trivial except for one.  The ones that are trivial are the ones in which less than $s_n$ of the boxes in the final row are activated.  For these, all of the $\frak{so}(D)$ tableaux in the rep become empty or singlets, and the bulk field has no propagating degrees of freedom. 
The remaining PM point where all $s_n$ boxes in the final row are activated is non-trivial but from \eqref{oddnmeqfee} is equivalent to the PM rep $[s_1,\ldots,s_{n-1}]$ where $s_{n-1}-s_n+1$ boxes in the last row are activated (in the $D=3$ case, it is equivalent to the $k=s_1-1$ shift symmetric scalar, as in \eqref{d3sphpfeqsce}).    Since the only PM points that are unitary are those in which the last row is activated, this means there are no new unitary PM fields on dS$_{2n+1}$ coming from the $[s_1,\ldots,s_n]$ tableaux with $s_n>0$ (as we will see later in section \ref{unitarylistsection}, this is consistent with, and important for understanding, the fact that there are no discrete series reps in odd $D$).

\subsubsection*{$D$ even:}

Let $D=2n$, $n=2,3,4,\ldots$.  Any reps corresponding to a $p$-row tableau with $p\geq n$ is either trivial or related by equivalence to one with $p< n $, so we only need to consider the reps with $p\leq n-1$.  

There is a chiral splitting that occurs for the partially massless points of the $[s_1,\ldots,s_{n-1}]$ tableaux with $s_{n-1}>0$, where any of the boxes in the bottom row are activated.  For these cases, all the $\frak{so}(D)$ reps that occur among the rep have precisely $D/2$ rows, and they can thus be split up into self-dual and anti-self-dual parts with respect to the $D$ dimensional epsilon tensor acting on the first column.  The self-dual and anti-self-dual parts split off into separate reps that we call ${\cal V}^{[s_1,\ldots,s_{n-1}]_\pm}_{}$,
\be {\cal V}^{[s_1,\ldots,s_{n-1}]}_{}={\cal V}^{[s_1,\ldots,s_{n-1}]_+}_{}\oplus {\cal V}^{[s_1,\ldots,s_{n-1}]_-}_{}\,,\ \ D\ {\rm even}\,.\ee
As we will see later in section \ref{unitarylistsection}, these are the bosonic discrete series reps.  They include the chiral massless $({D\over 2}-1)$-forms in all even $D$, as well as the symmetric tensor PM fields in $D=4$.

\section{Fermionic Fields and Transformations\label{fermionfieldssection}}

We now turn to the fermionic reps, those carried by spaces of fermionic fields on the boundary sphere of dS$_D$.  The bosonic reps covered in section \ref{bosonsection} are those reps of the algebra $\frak{so}(1,D)$ that can be exponentiated to reps of the connected part of the group, $SO^+(1,D)$.  The fermionic reps of the algebra, on the other hand, can only be exponentiated to reps of the double cover of $SO^+(1,D)$, or to projective reps of $SO^+(1,D)$.

\subsection{Fermionic fields on dS$_D$}

To describe the fermionic reps, we will need to consider spinor fields on dS$_D$.   For $D\geq3$, dS$_D$ admits a unique spin structure, so there is no ambiguity in defining the spinors. (For $D=2$ there are two spin structures, and more general anyon-like possibilities that we save for section \ref{D2section}.) 

The general spinor field is a rank $r$ spinor-tensor $\Psi_{\mu_1\ldots \mu_{r}}$, with $r$ tensor indices in some tableau, and a Dirac spinor index (which is suppressed).\footnote{Our dS$_D$ spinor conventions are as follows: the gamma matrices $\gamma^{\mu}$ are the curved space $D$-dimensional gamma matrices satisfying the anti-commutation relations
\be 
\left\{\gamma_\mu,\gamma_\nu\right\}=2g_{\mu\nu}.\label{gammalebgeraee}
\ee
They are related to the usual flat gamma matrices $\gamma^A$ satisfying the flat space commutation relations $\left\{\gamma_A,\gamma_A\right\}=2\eta_{AB}$, with the Minkowski metric $\eta_{AB}={\rm diag}(-1,1,1,\dots)$, by contraction with the dS$_D$ vielbein $e_\mu^{\ A}$, $\gamma_\mu=e_\mu^{\ A}\gamma_A$.

The curved space covariant derivative ${\cal D}_\mu$ acting on a spinor-tensor includes a connection for the spinor index, ${\cal D}_\mu=\nabla_\mu+{1\over 8}\omega_\mu^{AB}\left[\gamma_A,\gamma_B\right]$, where $\omega_\mu^{AB}$ is the Levi-Civita spin connection and $\nabla_\mu$ is the usual tensor covariant derivative containing only the Christoffel symbols acting on the tensor indices.  This covariant derivative commutes with $\gamma^\mu$, $g_{\mu\nu}$ and $e_\mu^{\ A}$.  The Dirac operator is
\be 
\slashed{\cal D}\equiv \gamma^\mu {\cal D}_\mu\,.
\ee
}  It satisfies the Dirac equation
\be
 \left(\slashed{\cal D}+\tilde m \right) \Psi_{\mu_1\ldots\mu_{r},\ldots} =0\,, \label{bulkdirecee}
 \ee
  it is transverse in all indices (${\cal D}^{\mu_1}\Psi_{\mu_1\mu_2\ldots \mu_{r}}=0$, and similarly for all the other indices) and it is gamma-traceless in all indices ($\gamma^{\mu_1}\Psi_{\mu_1\mu_2\ldots \mu_{r}}=0$, and similarly for all the other indices).  Gamma-tracelessness also implies that it is fully traceless in all the tensor indices.  Acting on \eqref{bulkdirecee} with $\left(\slashed{\cal D}-\tilde m \right)$ and using gamma-tracelessness of the field shows that the field also satisfies the Klein-Gordon equation,
 \be \left({\cal D}^2 -H^2\left[{D(D-1)\over 4}+r\right]-\tilde m^2\right) \Psi_{\mu_1\ldots \mu_{r}} =0\,.\ee
Note that the mass $\tilde m$ for the fermions is defined to be the mass appearing in the Dirac equation, unlike the bosonic field where it is defined to be the mass appearing in the Klein-Gordon equation.

Since the fermionic field satisfies the Klein-Gordon equation, it can be asymptotically expanded just as for the bosonic fields, and we get the following formula relating $\tilde m$ and the scaling exponent $\Delta$ of the near boundary solution, $\Psi\sim e^{-\left( \Delta_\pm-r\right) H t}$,
 \be {\tilde m^2 \over H^2} =-\left(\Delta-{d\over 2 }\right)^2\,,\ \ \Delta_\pm= {d\over 2}\pm i {\tilde m \over H}\,.\label{mixedsymmasmrelefe}\ee
Note that this is independent of the rank $r$ of the spinor-tensor and of its index symmetries.

As we will see in section \ref{fermionsection}, the only fermionic fields that give rise to unitary reps are those with $\tilde m^2\geq 0$, so that $\Delta={d\over 2}+i\nu$ with $\nu\in {\mathbb R}$, which will be the fermionic principal series, and this excludes all the shift symmetric and PM points where the reps become reducible and indecomposable.  There is one class of exceptions to this statement: the PM points for $\left({D\over 2}-1\right)$-row tableau fields in even $D$, in which the bottom row is activated, are unitary.

The dS$_D$ isometries in section \ref{isomalgsec} are realized on the spinor-tensor field using the spinor Lie derivative \cite{Kosmann1971DrivesDL,Fatibene:1996tf,Ortin:2002qb,fatibene2009generaltheoryliederivatives}.  Given a Killing vector $\xi^\mu$, it is defined as follows,
\be  {\mathbb L}_{\xi} \Psi_{\mu_1\ldots \mu_r} = \xi^\mu {\cal D}_\mu \Psi_{\mu_1\ldots \mu_r} +\sum_{k=1}^r {\cal D}_{\mu_k} \xi^{\mu}  \psi_{\mu_1\ldots \mu_{k-1}\mu \mu_{k+1}\ldots \mu_r}   + {1\over 4}  \nabla_\mu \xi_\nu \gamma^{\mu\nu}  \Psi_{\mu_1\ldots \mu_r} \,. \ee
 The action of the isometries generated by the Killing vectors \eqref{killingenbeddinge} is then
 \be  \delta_{{\cal M}^{AB}}\Psi_{\mu_1\ldots \mu_r}=-{\mathbb L}_{{\cal M}^{AB}}\Psi_{\mu_1\ldots \mu_r}\, ,\  \ee
and it satisfies the expected commutators \eqref{sod12algberaadse}.

The quadratic Casimir operator of section \ref{casimirsec} is then given by ${\cal C}_2 = - \half {\mathbb L}_{{\cal M}^{AB}} {\mathbb L}_{{\cal M}_{AB}}$, and acting on a spinor-tensor field it becomes
\be {\cal C}_2 =  -{1\over H^2}{\cal D}^2 + {1\over 8}D(D-1)+\sum_{i=1}^p s_i(s_i+D+1-2i)\,, \ee
where $[s_1,\ldots,s_p]$ is the Young tableau in which the tensor indices live.  It takes the following values on the on-shell and asymptotic fields, respectively:
 \bea 
{\cal C}_2  &=&-{\tilde m^2\over H^2}- {1\over 8}D(D-1)+\sum_{i=1}^p s_i(s_i+D-2i)  \nn\\ 
&=& \Delta(\Delta-d)+{1\over 8}d(d-1)+\sum_{i=1}^p s_i(s_i+d+1-2i)\,.\label{c2gnsmdees2eoee}
 \eea 

\subsection{Fermionic boundary fields on $\mathbb{S}^d$}

The next step would be to find the asymptotic expansion of the solution of the Dirac equation in terms of its late time boundary values and find how the de Sitter isometries act on the boundary values as conformal transformations (the details of this are worked out in some of the early work on spinors in (A)dS/CFT \cite{Henningson:1998cd,Henneaux:1998ch,Leigh:1998kt,Volovich:1998tj,Corley:1998qg,Koshelev:1998tu,Matlock:1999fy,Rashkov:1999is,Anguelova:2003kf,Loran:2004fu,Contino:2004vy}).

However, we can bypass the details of the asymptotic expansion if we accept that the boundary fields will be rank-$r$ gamma-traceless spinor-tensors on $\mathbb{S}^d$ that transform under the dS isometries as conformal primary fields of weight $\Delta$ under conformal transformations of the sphere.  Given this, we can determine these transformations more directly, and this is the route we will take here.

 As for the full dS$_{d+1}$ space, the $d$-sphere ${\mathbb S}^d$ admits a unique spin structure for $d\geq2$, so there is no ambiguity in defining the boundary spinors.  We pick a vielbein field $e_i^{\ a}$ on ${\mathbb S}^d$, satisfying the basic vielbein relation $e_i^{\ a}e_j^{\ b}\delta_{ab}=g_{ij}$, where $g_{ij}$ is the metric \eqref{spheremetrice} on ${\mathbb S}^d$. The inverse vielbein is $e^{i}_{\ a}$, defined so that $e^{i}_{\ a}e_i^{\ b}=\delta^b_a$, $e^{i}_{\ a}e_j^{\ a}=\delta^i_j$.  The flat-space gamma matrices, satisfying the fundamental anti-commutation relations
$\left\{ \gamma_a,\gamma_b\right\}=2\delta_{ab},$
are used to make curved space gamma matrices on the sphere by contracting with the vielbein, $\gamma_i=e_i^{\ a}\gamma_a$, and they satisfy the anti-commutation relations
\be \left\{\gamma_i,\gamma_j\right\}=2g_{ij}\, .\ee
We define the anti-symmetric combination $\gamma_{ab}\equiv \gamma_{[a}\gamma_{b]}$, and similarly $\gamma_{ij}=\gamma_{[i}\gamma_{j]}$.

The spin covariant derivative acting on the spinor-tensor field $\psi_{i_1\ldots i_r}$ is
\be D_i\psi_{i_1\ldots i_r} =\nabla_i \psi_{i_1\ldots i_r}+{1\over 4} \omega_i^{ab}\gamma_{ab}\psi_{i_1\ldots i_r} \, ,\label{spincovdgde}\ee
where $\nabla_i$ is the covariant derivative with Christoffel symbols acting only on the tensor indices, and $\omega^{\ a}_{i \ \ b}=e^{j a}\partial_{[i}e_{j]b}-e^j_{\ b}\partial_{[i}e_{j]}^{\ a}-e^j_{\ b}e^{k a}e_{i}^{\ c}\partial_{[k}e_{j]c}$ 
is the unique metric compatible and torsion free spin connection.  (Acting on an object with no spinor indices, we have $D_i=\nabla_i$).
On the unit $\mathbb{S}^d$, the spin covariant derivative \eqref{spincovdgde} satisfies the commutation relations
\be  \left[ D_i,D_j\right]\psi_{i_1\ldots i_{r}}=  2 \sum_{k=1}^{r} g_{i_k[i}\psi_{| i_1\ldots i_{k-1} | j] i_{k+1} \ldots i_{r}} +{1\over 2}\gamma_{ij}\psi_{i_1\ldots i_r}\, .\ee

The Dirac operator is defined by
\be \fdag{{ D}}  \equiv \gamma^i D_i \, . \label{diracopdgere}\ee
On the unit $\mathbb{S}^d$ it satisfies
\be  \fdag{{ D}}^2 \psi_{i_1\ldots i_{r}}=\left({ D}^2-\left[r+{d(d-1)\over 4}\right]\right)\psi_{i_1\ldots i_{r}}+ \sum_{k=1}^{r} \gamma_{i_k}\gamma^{i}\psi_{ i_1\ldots i_{k-1}  i i_{k+1} \ldots i_{r}}  \, .\label{D2identityee}
\ee
(We will be mostly interested in gamma-traceless fields, for which the last term vanishes.)

Under a Weyl transformation $\delta_W$, the vielbein, metric, and a spinor-tensor field $\psi_{i_1\ldots i_r}$ of Weyl weight $\Delta_W$, transform as 
\be \delta_W e_{i}^{\ a}=\sigma e_{i}^{\ a} ,\ \ \ \delta_W g_{ij}=2\sigma g_{ij}\,,\ \ \  \delta_W \psi_{i_1\ldots i_r} =-\Delta_W \sigma \psi_{i_1\ldots i_r}\,,\label{weyltransferefe}\ee
for arbitrary position dependent scalar gauge parameter $\sigma$. 
Under infinitesimal diffeomorphisms $\delta_D$, generated by a vector field $\xi^i$, they transform as
\be  \delta_D e_{i}^{\ a}=-{\cal L}_\xi e_{i}^{\ a}\,,  \ \ \delta_D g_{ij}=-{\cal L}_\xi g_{ij}=-\left(\nabla_i\xi_j+\nabla_j\xi_i\right),\ \ \ \delta_D \psi_{i_1\ldots i_r} =-{\cal L}_\xi \psi_{i_1\ldots i_r} \, ,\ee
where the Lie derivatives act only with respect to the tensor indices.
We also have local Lorentz transformations $\delta_L$, under which
\be  \delta_L e_{i}^{\ a}= \Omega^a_{\ b}e_{i}^{\ b}\,, \ \ \ \delta_Lg_{ij}=0\,,\ \ \ \delta_L \psi_{i_1\ldots i_r} = {1\over 4} \Omega^{ab} \gamma_{ab}  \psi_{i_1\ldots i_r} \,, \ee
with $\Omega_{ab}$ an anti-symmetric position dependent local Lorentz parameter.

The conformal transformations with respect to a metric are those combined Weyl and diffeomorphism transformations that leave the metric invariant, $\left(\delta_W+\delta_D\right)g_{ij}=0$, which imposes the conformal Killing equation \eqref{CKEejee} on the diffeomorphism parameter $\xi_i$ and \eqref{sigmaeqe} on the Weyl parameter.  Since $\delta_W+\delta_D$ leaves the background metric invariant, it must leave the background vielbein invariant up to a local Lorentz rotation.  There must therefore be a local Lorentz transformation such that $\left(\delta_W+\delta_D+\delta_L\right)e_{i}^{\ a}=0$.  Writing this out, we have
\be {1\over d}\nabla\cdot \xi \, e_{i}^{\ a}  - \xi^j\nabla_j e_{i}^{\ a} - \nabla_i\xi^j  e_{j}^{\ a}  +  \Omega^a_{\ b}e_{i}^{\ b}=0\,.\ee
(Here $\nabla$ is the derivative acting only on the spacetime indices, not the local Lorentz indices).
Multiplying this by the inverse vielbein $e^i_{\ c}$, we obtain an equation that can be solved for $\Omega_{ab}$ by taking the anti-symmetric part (the symmetric part reproduces the conformal Killing equation \eqref{CKEejee}),
\be \Omega_{ab}=- e^i_{ \ a}e^j_{ \ b} \nabla_{[i}\xi_{j]}-\xi^i \omega_{i ab}\,,\label{llttebree}\ee
where we have used that the Levi-Civita spin connection can be expressed as $\omega_{i ab} = e^j_{\ a}\nabla_i e_{j b}$.  The transformation $\delta_W+\delta_D+\delta_L$, with a conformal Killing vector, and using \eqref{sigmaeqe}, \eqref{llttebree} now leaves the vielbein and metric invariant, so it becomes a conformal transformation when acting on the fermion field.  Writing out this transformation, we have
\be \delta_\xi  \psi_{i_1\ldots i_r}  = - {\Delta_W\over d}\nabla\cdot\xi\, \psi_{i_1\ldots i_r} -{\cal L}_\xi \psi_{i_1\ldots i_r}  - {1\over 4} \xi^i \omega_{i}^{\ ab} \gamma_{ab}  \psi_{i_1\ldots i_r} - {1\over 4}  \nabla_i\xi_j \gamma^{ij}  \psi_{i_1\ldots i_r} \,. \label{confomspinintexeee}\ee
The last three terms form the spinor Lie derivative \cite{Kosmann1971DrivesDL,Fatibene:1996tf,Ortin:2002qb,fatibene2009generaltheoryliederivatives},
\be {\mathbb L}_{\xi} \psi_{i_1\ldots i_r}  \equiv   {\cal L}_\xi \psi_{i_1\ldots i_r}  + {1\over 4} \xi^i \omega_{i}^{\ ab} \gamma_{ab}  \psi_{i_1\ldots i_r} + {1\over 4}  \nabla_i\xi_j \gamma^{ij}  \psi_{i_1\ldots i_r} \,.\label{spinliedere}\ee
In terms of the spin covariant derivative \eqref{spincovdgde}, it can be written as
\be  {\mathbb L}_{\xi} \psi_{i_1\ldots i_r} = \xi^iD_i \psi_{i_1\ldots i_r} +\sum_{k=1}^r D_{i_k} \xi^{i}  \psi_{i_1\ldots i_{k-1}ii_{k+1}\ldots i_r}   + {1\over 4}  \nabla_i\xi_j \gamma^{ij}  \psi_{i_1\ldots i_r} \,. \ee
It satisfies the expected commutation rules,
\be \left[{\mathbb L}_{\xi},{\mathbb L}_{\xi'}\right]={\mathbb L}_{[\xi,\xi']}.\ee 
With this, the action \eqref{confomspinintexeee} of conformal symmetry on the field is
\be \delta_\xi  \psi_{i_1\ldots i_r}  = - \left( {\mathbb L}_{\xi} + {\Delta_W\over d}\nabla\cdot\xi \right)\psi_{i_1\ldots i_r}  \,,\label{confofermdkge}\ee
and the relation between the Weyl weight $\Delta_W$ and the usual CFT conformal scaling dimension $\Delta$ is
\be \Delta_W=\Delta-r\,,\ee
(as can be seen by matching \eqref{confofermdkge} to the usual form of the conformal transformations on flat space).  On the sphere, following section \ref{isometriessubsection}, we now immediately obtain the spinor form of the conformal transformations \eqref{latetimesheactiont}, \eqref{latetimesheactiont2}, 
\be  \delta_{{\cal M}^{IJ}}\psi_{i_1\ldots i_r}=-{\mathbb L}_{{\cal M}^{IJ}}\psi_{i_1\ldots i_r}\, ,\ \ \   \delta_{{\cal K}^{I}}\psi_{i_1\ldots i_r}=\left[ (\Delta-r)\hat X^I -{\mathbb L}_{\partial \over \partial \hat X_I}\right] \psi_{i_1\ldots i_r}\,,\label{latetimeshesactiont}\ee
and these satisfy the same commutators \eqref{sod12algberaadse2e} as in the bosonic case.

\section{Fermionic Representations\label{fermionsection}}

We now go through the spinor reps, starting with the simplest case of spin $1/2$ in section \ref{spin12sec} and working up to the general case of mixed symmetry fermionic reps in section \ref{mixedfermionsubsec}.  More details about the fermionic reps can be found in \cite{Letsios:2020twa,Pethybridge:2021rwf,Letsios:2023qzq}.

The reps behave differently for $D$ even and for $D$ odd.  This is due to the presence of the chiral gamma matrix $\gamma_\ast$ on ${\mathbb S}^d$ when $d=D-1$ is even.  It anti-commutes with all the gamma matrices and squares to one,
\be \left\{ \gamma_\ast ,\gamma_i\right\}=0\, ,\ \ \ \gamma_\ast^2=1\, .\ee
It anti-commutes with the Dirac operator \eqref{diracopdgere}, 
\be \left\{  \fdag{{ D}} ,\gamma_\ast\right\}=0 \,,\ee
 commutes with the spinor Lie derivative \eqref{spinliedere},
\be \left[   {\mathbb L}_{\xi}  ,\gamma_\ast \right]=0\,,\ee
and commutes with the action of the conformal transformations \eqref{latetimeshesactiont}, 
\be \left[  \delta_{{\cal M}^{IJ}}  ,\gamma_\ast \right]=0\,,\ \ \  \left[   \delta_{{\cal K}^{I}}  ,\gamma_\ast \right]=0\,. \ee
It will be used in even $d$ to split the reps into chiral parts.

\subsection{Spin $1/2$ representations\label{spin12sec}}

We start with the simplest fermionic reps, the spin 1/2 reps.  The representation space is the space of Dirac spinor fields with no tensor indices, and further split into chiral parts when $d$ is even, as detailed below.

\subsubsection*{Even $D$:}

For even $D$, $d$ odd, there is no chiral gamma matrix on ${\mathbb S}^d$, and the representation space will be the full space of complex Dirac spinors on ${\mathbb S}^d$, transforming as in \eqref{latetimeshesactiont} with $r=0$. Call this space ${\cal F}^{[0]_\half}_\Delta$,
\be {\cal F}^{[0]_\half}_\Delta:\ \ {\rm complex\ Dirac\ spinors\ on\ } {\mathbb S}^d \,.\label{spacefdefd}\ee
The superscript $[0]_\half$ indicates that we are working with spinor-tensors whose tensor indices are in the empty $[0]$ tableau.  The relation between the mass and $\Delta$ is given by \eqref{mixedsymmasmrelefe}.

We now want to decompose this space into $\frak{so}(D)$ reps.
This is done by decomposing a general spinor into spinor harmonics that are eigenvalues of the Dirac operator, and which transform irreducibly as spin reps of $\frak{so}(D)$.  Recursive expressions for these spinor spherical harmonics in all dimensions can be found in \cite{Camporesi:1995fb}.  Here we review only their general properties that we will need to describe the reps.

The Dirac spinor field on ${\mathbb S}^d$ has two sets of spherical harmonics, $Y_{lm}^\pm$.  They are eigenstates of the Dirac operator with positive and negative pure-imaginary eigenvalues,
\be  \fdag{{ D}} Y_{lm}^\pm=\pm i \left(l-{1\over 2}+{d\over 2}\right)Y_{lm}^\pm\, , \ \ \ l=1/2,3/2,5/2,\ldots\ . \label{diraceigenvaluesee}\ee
They are labelled by an odd half-integer $l$ and an index $m$ whose range is finite and depends on $l$.  Under rotations of ${\mathbb S}^d$, the $l$'s are not mixed, and the $m$'s for a given $l$ transform under a spinor rep of $\frak{so}(D)$.  For $Y_{lm}^\pm$ it is the $(l,1/2,\ldots,\pm 1/2)$ rep (see appendix \ref{sorepsappendix} for a review of these reps and their notation).   Because $D$ is even, the $\frak{so}(D)$ spinor reps come in chiral pairs, and the $Y_{lm}^+$, $Y_{lm}^-$ form such pairs.  The eigenspinors $Y^{\pm}_{lm}$, taken all together, are complete and form a basis of the space \eqref{spacefdefd}. 

The natural Laplacian on the space of spinors is the spinor version of the Lichnerowicz Laplacian,
\be \Delta\equiv -D^2+{d(d-1)\over 8}\, ,\label{spspinslaplacianue}\ee 
and because of \eqref{D2identityee}, the two sets of spinor spherical harmonics have the same (real) eigenvalues under it,
\be \Delta Y_{lm}^\pm = \left[ l (l + d - 1) + {(d-2 ) (d-1 )\over 8}\right] Y_{lm}^\pm\,.\ee

The space \eqref{spacefdefd} of spinors on ${\mathbb S}^d$ thus splits under $\frak{so}(D)$ into two reps of each spin $1/2,3/2,5/2,\ldots$, one of each chirality.  
This $\frak{so}(D)$ content can be illustrated as follows:
\be \raisebox{-40pt}{\epsfig{file=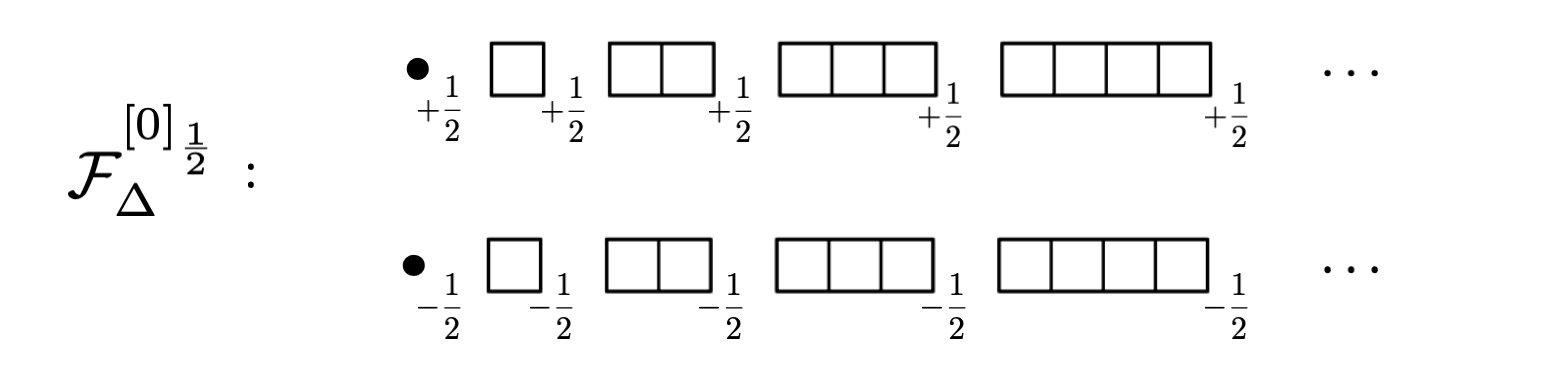,width=5.0in}}  \label{spinhalfcontent1}\, 
\ee
Here we are using tableau tensor notation for $\frak{so}(D)$ reps (as reviewed in appendix \ref{sorepsappendix}): for each rep the tableau indicates the tensor indices, the $\pm \half$ on the tableau indicates that we are working with a gamma-traceless irreducible spinor-tensor, and the sign indicates the chirality of the rep.
For a generic $\Delta$, these $\frak{so}(D)$ reps will all be linked together by the action of the ${\cal K}^I$ boosts in \eqref{latetimeshesactiont}, giving an irreducible rep of $\frak{so}(1,D)$.

\textbf{Reducible cases:} As with the bosonic fields, there are several sets of discrete values of $\Delta$ where the ${\cal K}^I$ links become broken and these reps become reducible:
\begin{itemize}

\item
Shift symmetric points: 
\be \Delta=d+k+{1\over 2}\, ,\ \ \ k=0,1,2,\ldots\ . \ee
There is an infinite dimensional sub-rep containing only the states with $>k$ boxes, which we call 
\be {\cal D}^{[0]_{1\over 2}}_{d+k+{1\over 2}}\, ,\ee
illustrated here for $k=2$,
\be \raisebox{-40pt}{\epsfig{file=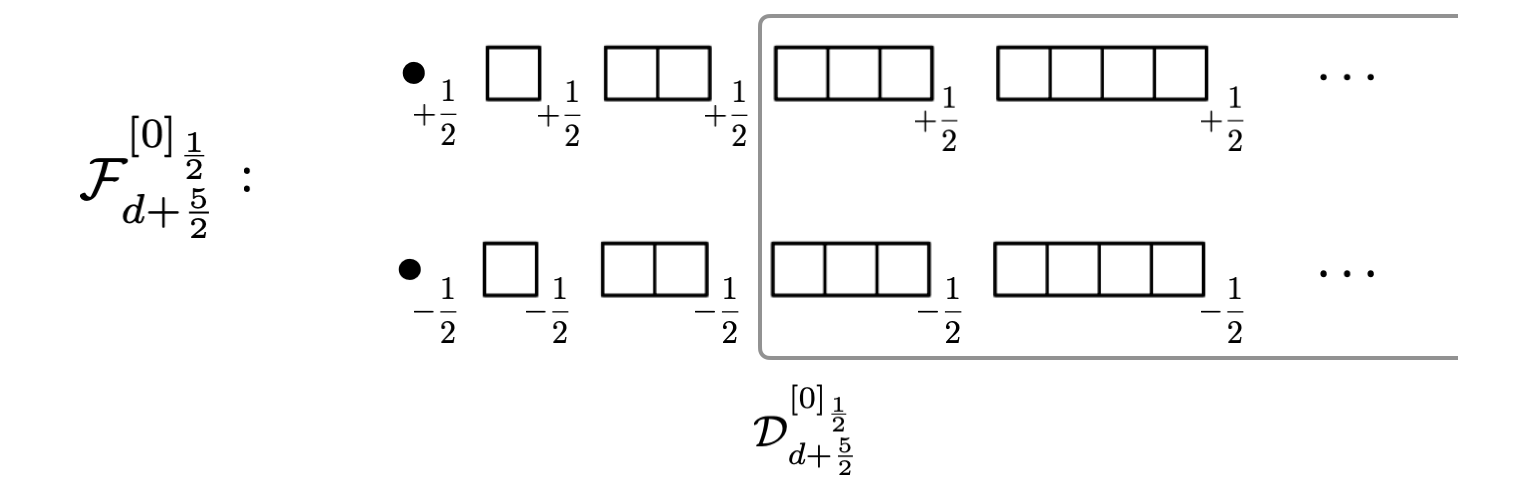,width=5.0in}}  \label{spinhalfcontent3}\, 
\ee
These represent the physical modes of the level $k$ shift symmetric fermions \cite{Bonifacio:2023prb}.

\item
Finite points:
\be \Delta=-k-{1\over 2}\, ,\ \ \ k=0,1,2,\ldots \ .\ee
There is a finite dimensional sub-rep containing only the states with $\geq k+1$ boxes, which we call 
\be {\cal S}^{[0]_{1\over 2}}_{-k-{1\over 2}}\,,\ee 
illustrated here for $k=2$,
\be \raisebox{-40pt}{\epsfig{file=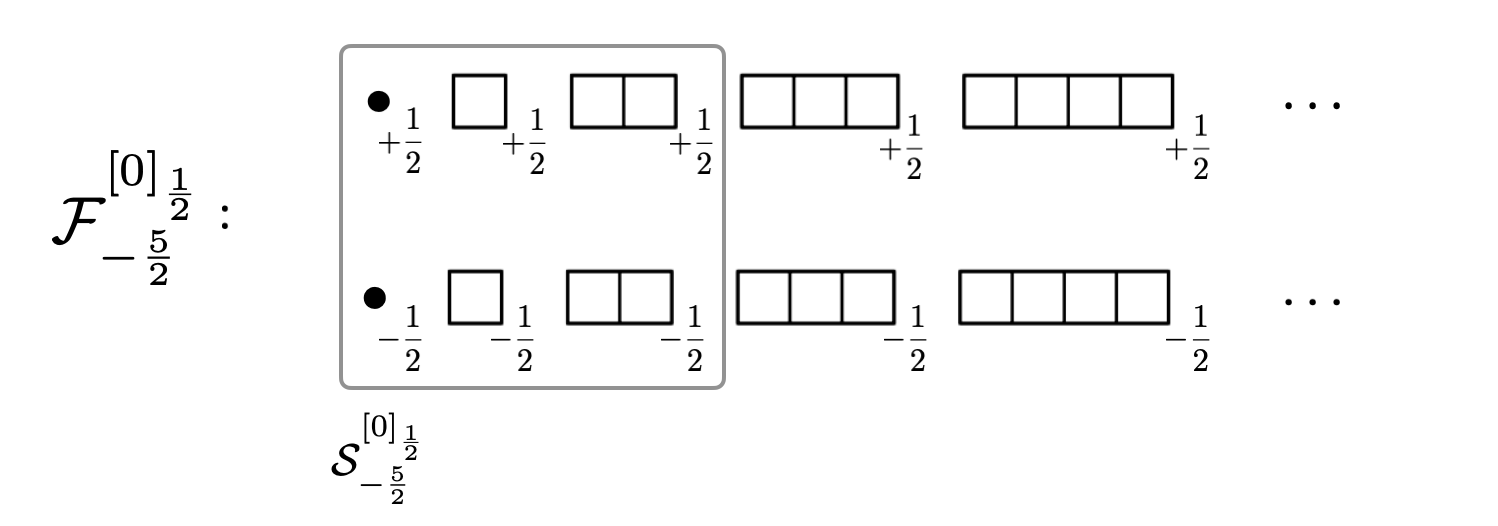,width=5.0in}}  \label{spinhalfcontent4}\, 
\ee

These finite dimensional reps are precisely the reps $(k+1/2,1/2,\ldots,1/2)$ of $\frak{so}(1,D)$, which upon branching to $\frak{so}(D)$ using \eqref{oddDbranchinge} gives the $\frak{so}(D)$ reps within ${\cal S}^{[0]_{1\over 2}}_{-k-{1\over 2}}$.
They represent the shift symmetries of the level $k$ shift symmetric fermions.

\item
Chiral point:
\be \Delta={d\over 2}\,.\ee
At this point, there is a feature that does not occur in the scalar cases of section \ref{scalarsec}: there is a splitting that occurs, and the $Y_{lm}^+$ and $Y_{lm}^-$ modes become separate irreducible reps.  The space ${\cal F}^{[0]_{1\over 2}}_{1\over 2}$ splits into two reps:
\be {\cal F}^{[0]_{1\over 2}}_{d\over 2}={\cal F}^{[0]_{1\over 2},+}_{d\over 2}\oplus {\cal F}^{[0]_{1\over 2},-}_{d\over 2}\ \,,\ee
 as illustrated here:
\be \raisebox{-40pt}{\epsfig{file=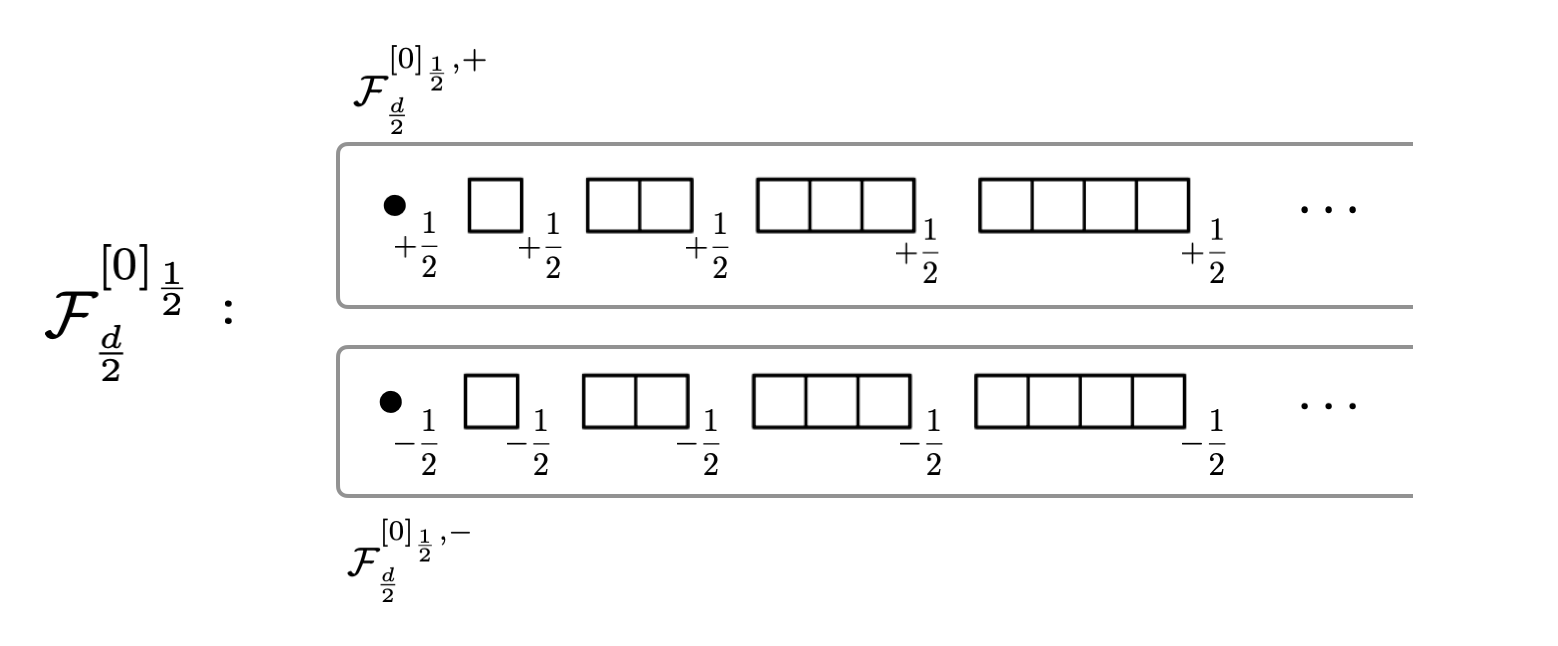,width=5.0in}}  \label{spinhalfcontent2}\, 
\ee
For this value, the representation space ${\cal F}^{[0]_{1\over 2}}_{1\over 2}$ is thus reducible and decomposable (in contrast to the shift symmetric and finite cases which are reducible but not decomposable). 
This value $\Delta=d/2$ corresponds to the case $\tilde m=0$ where the Dirac fermion in even $D$ becomes massless and acquires a chiral symmetry, and the two reps ${\cal F}^{[0]_{1\over 2},\pm}_{d\over 2}$ are the de Sitter chiral fermions.  This happens without any enhanced gauge symmetry, as indicated by the fact that the rep decomposes and there is no need to take a quotient to arrive at the irreducible rep.  (Note that the shift symmetric and finite reps do not split into chiral halves the way the massless fermion does, since the mass term breaks chiral symmetry.)
\end{itemize}

\textbf{Equivalences:} As with the bosons, there is a shadow equivalence between the reps with $\Delta$ and $\bar \Delta=d-\Delta$,
\be {\cal F}^{[0]_{{1\over 2}}}_{\Delta}\simeq  {\cal F}^{[0]_{{1\over 2}}}_{\bar\Delta} \,.\label{fermioneqfjee1}\ee
There are intertwiner operators $S^{[0]_{{1\over 2}}}_{\Delta}: {\cal F}^{[0]_{{1\over 2}}}_{\Delta} \rightarrow {\cal F}^{[0]_{{1\over 2}}}_{\bar\Delta}$ that realize this equivalence.  They break down at the shift symmetric and finite points, giving a picture analogous to \eqref{scalarsocontent5}, along with the isomorphisms
\be {\cal D}^{[0]_{1\over 2}}_{d+k+{1\over 2}} \simeq    {\cal F}^{[0]_{1\over 2}}_{-k-{1\over 2}}/ {\cal S}^{[0]_{1\over 2}}_{-k-{1\over 2}}    \,, \ \ \           {\cal S}^{[0]_{1\over 2}}_{-k-{1\over 2}} \simeq {\cal F}^{[0]_{1\over 2}}_{d+k+{1\over 2}} /{\cal D}^{[0]_{1\over 2}}_{d+k+{1\over 2}}  \,.\ee

At the point $\Delta=\bar\Delta={d\over 2}$, the shadow map cannot be used to exchange the two chiral subspaces and ${\cal F}^{[0]_{1\over 2},\pm}_{d\over 2}$ are two distinct inequivalent reps.

\subsubsection*{Odd $D$:} 

For odd $D$, $d$ is even, there is the chiral gamma matrix $\gamma_\ast$ on ${\mathbb S}^d$, and the space of Dirac spinors on ${\mathbb S}^d$ can be broken up into two chiral eigenspaces of spinors with eigenvalues $\pm 1$ under $\gamma_\ast$, which we call ${\cal F}^{[0]_{\pm\half}}_\Delta$,
\be {\cal F}^{[0]_{\pm\half}}_\Delta:\ \ {\rm complex\ Dirac\ spinors\ on\ } {\mathbb S}^d {\rm \ satisfying \ } \gamma_\ast \psi =\pm \psi \,.\label{fpmspnfveege}\ee

We want to split these spaces into $\frak{so}(D)$ reps.
For even $d$, the Dirac field still has two sets of eigenstates under the Dirac operator, with positive and negative pure-imaginary eigenvalues just as in \eqref{diraceigenvaluesee} (a concise summary of the $d=2$ case can be found in \cite{Abrikosov:2001nj} and in appendix C of \cite{Letsios:2025pqo}).  However, since the spinor reps of $\frak{so}(D)$ for odd $D$ do not come in chiral pairs, the $Y_{lm}^+$ and $Y_{lm}^-$ describe the same rep under $\frak{so}(D)$ rotations, namely the $(l,1/2,\ldots, 1/2)$ rep.
The $\gamma_\ast$ matrix serves as an intertwiner that maps the $Y_{lm}^+$ and $Y_{lm}^-$ into each other: $\gamma_\ast Y_{lm}^\pm\propto Y_{lm}^\mp$, so in the full space spanned by $Y_{lm}^+$ and $Y_{lm}^-$,  $\gamma_\ast$ is purely off-diagonal in this basis.   
The $\gamma_\ast$ operator can be diagonalized, and this diagonalization splits the space into 2 chiral spinors $\tilde Y_{lm}^\pm$, each a linear combination of $Y_{lm}^+$ and $Y_{lm}^-$, distinguished by having eigenvalue $\pm 1$ under the chiral gamma matrix $\gamma_\ast$,
\be \gamma_\ast \tilde Y_{lm}^\pm=\pm \tilde Y_{lm}^\pm\, .\ee
The  $\tilde Y_{lm}^\pm$ do not have definite eigenvalues under the Dirac operator: the Dirac operator is purely off-diagonal in this basis.

For generic $\Delta$, the $\tilde Y_{lm}^+$ and the  $\tilde Y_{lm}^-$ each span a separate irreducible rep of $\frak{so}(1,D)$, which are the spaces ${\cal F}^{[0]_{\pm\half}}_\Delta$ in \eqref{fpmspnfveege}.  Each of these reps has the same $\frak{so}(D)$ content, but they differ as $\frak{so}(1,D)$ reps.   We can illustrate their $\frak{so}(D)$ content as follows:
\be \raisebox{-10pt}{\epsfig{file=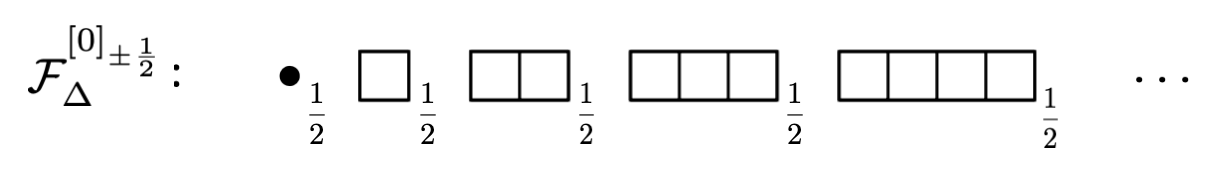,width=5.0in}}  \label{spinhalfcontent5}\, \ee
These two reps correspond to the 2 possible chiralities of a massive fermion on dS$_D$ for odd $D$ (in flat space, this is the statement that there are two different chiral spinor reps of the massive little group $\frak{so}(D-1)$ when $D$ is odd).  Which of the two types of rep a field corresponds to is determined, just as in flat space, by the sign of the mass term in the Dirac equation on dS$_D$ (in odd $D$, this sign cannot be changed with a chiral rotation as it can in even $D$).

\textbf{Reducible cases:}  The discrete values of $\Delta$ where these reps become reducible are as follows:
\begin{itemize}
\item
Shift symmetric points: 
\be \Delta=d+k+{1\over 2}\, ,\ \ \ k=0,1,2,\ldots\ . \ee
Here each of ${\cal F}^{[0]_{\pm{1\over 2}}}_{\Delta}$ develops an infinite dimensional sub-rep containing only the states with $>k$ boxes, which we call 
\be {\cal D}^{[0]_{\pm{1\over 2}}}_{d+k+{1\over 2}}\,,\ee 
illustrated here for $k=2$,
\be \raisebox{-10pt}{\epsfig{file=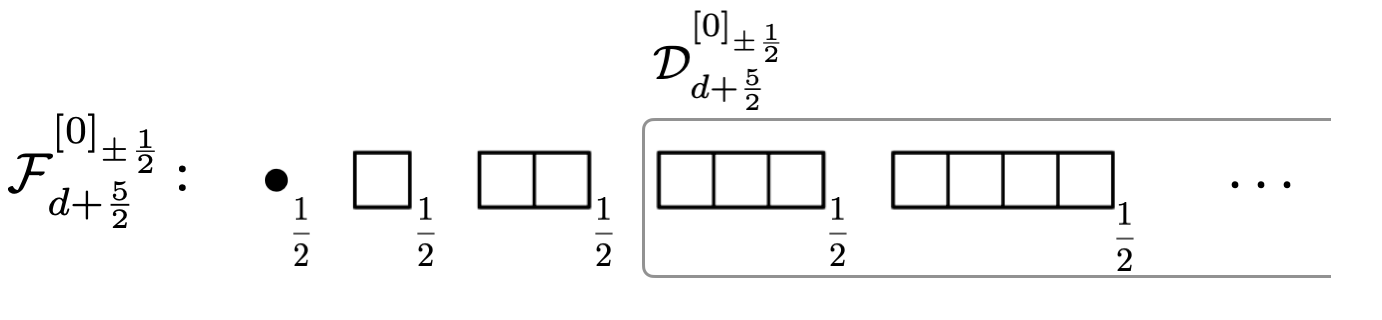,width=5.0in}}  \label{spinhalfcontent6}\, \ee
These represent the level $k$ shift symmetric fermions, which we see come in chiral halves for odd $D$.

\item
Finite points: 
\be \Delta=-k-{1\over 2}\, ,\ \ \ k=0,1,2,\ldots \ . \ee
Here each of ${\cal F}^{[0]_{\pm{1\over 2}}}_{\Delta}$ develops a finite dimensional sub-rep containing only those states with $\geq k$ boxes, which we call 
\be {\cal S}^{[0]_{\pm{1\over 2}}}_{-k-{1\over 2}}\, ,\ee 
illustrated here for $k=2$,
\be \raisebox{-10pt}{\epsfig{file=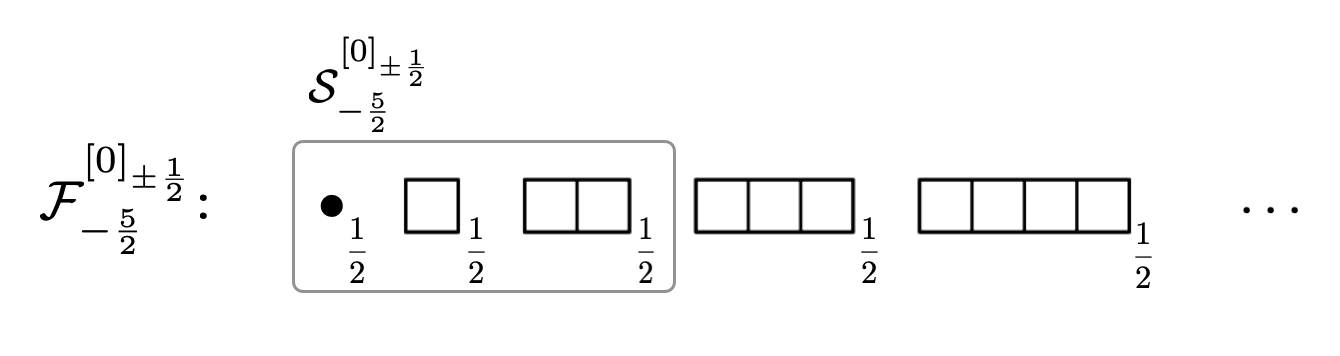,width=5.0in}}  \label{spinhalfcontent7}\,\ee
 These represent the shift symmetries of the level $k$ shift symmetric fermions.  These finite dimensional reps are precisely the reps $(k+1/2,1/2,\ldots,\pm 1/2)$ of $\frak{so}(1,D)$, which upon branching to $\frak{so}(D)$ using \eqref{evenDbranchinge} give the $\frak{so}(D)$ reps within ${\cal S}^{[0]_{\pm{1\over 2}}}_{-k-{1\over 2}}$.   These represent the shift symmetries of the level $k$ shift symmetric fermions. 
\end{itemize}

\textbf{Equivalences:}  The equivalence between the reps with $\Delta$ and $\bar \Delta=d-\Delta$ in odd $D$ involves a flip of the chirality,
 \be {\cal F}^{[0]_{\pm{1\over 2}}}_\Delta \simeq  {\cal F}^{[0]_{\mp{1\over 2}}}_{\bar \Delta }\,. \label{fermoddintgme}\ee
At the shift symmetric and finite points, we now have the equivalences
\be {\cal D}^{[0]_{\pm \half}}_{d+k+{1\over 2}} \simeq    {\cal F}^{[0]_{\mp \half}}_{-k-{1\over 2}}/ {\cal S}^{[0]_{\mp \half}}_{-k-{1\over 2}}    \,, \ \ \           {\cal S}^{[0]_{\pm \half}}_{-k-{1\over 2}} \simeq {\cal F}^{[0]_{\mp \half}}_{d+k+{1\over 2}} /{\cal D}^{[0]_{\mp \half}}_{d+k+{1\over 2}}  \,.\ee

Note that, in contrast to the even $D$ case, no additional splitting happens at $\Delta=d/2$, the point corresponding to the massless fermion with $\tilde m=0$.  Instead, there is an equivalence that develops:  the relation \eqref{fermoddintgme} for $\Delta=d/2$ tells us that
 \be {\cal F}^{[0]_{+{1\over 2}}}_{d\over 2} \simeq  {\cal F}^{[0]_{-{1\over 2}}}_{d\over 2 }\, .\label{fermioneqfjee2}\ee 
 This reflects the fact that a massless fermion in odd $D$ does not come in two different chiralities (in flat space, this is the statement that there is a unique fundamental spinor rep of the massless little group $\frak{so}(D-2)$ when $D$ is odd).

\subsubsection*{Unitarity and summary:} 

In the fermionic case, the only positive definite inner product that can be constructed is the standard Dirac inner product \cite{Letsios:2020twa},
\be \la \psi_1|\psi_2\ra= \int d^d\Omega\, \psi_1^\dag\psi_2\,,  \label{spinorinnerprodde2} \ee
and it is only positive definite and invariant when $\Delta={d\over 2}+i\nu$ with $\nu\in {\mathbb R}$.  This is the fermionic principal series.   

For $D$ even, there is one rep at each $\nu\not=0$, and due to \eqref{fermioneqfjee1} the reps at $\nu>0$ are equivalent to those at $\nu<0$.   The rep at $\nu=0$, corresponding to the massless fermion $\tilde m=0$, splits into two inequivalent chiral pieces.  For this reason, in even $D$ the point $\nu=0$ is often excluded from the principal series and is considered a part of the discrete series, as we'll see in the classification in section \ref{unitarylistsection}.  

For odd $D$, there are two different chiral reps for each $\nu\not=0$, and the equivalence \eqref{fermoddintgme} relates one chirality at $\nu>0$ to the other at $\nu<0$.  For the massless point at $\nu=0$, there is a single rep.

There is no complementary series for the fermions \cite{f4e767c0-b144-3046-b19f-3576a516cc62,61dbdb7d-9ac1-36a8-83ad-1eb6b8331554,Pethybridge:2021rwf}, and the shortened reps at the shift symmetric points are not unitary.  This is the main difference between the spin $1/2$ fermions and the scalars when it comes to unitarity: for the fermions the only unitary reps are the principal series reps.
The spin $1/2$ reps are summarized here:
\be \raisebox{-40pt}{\epsfig{file=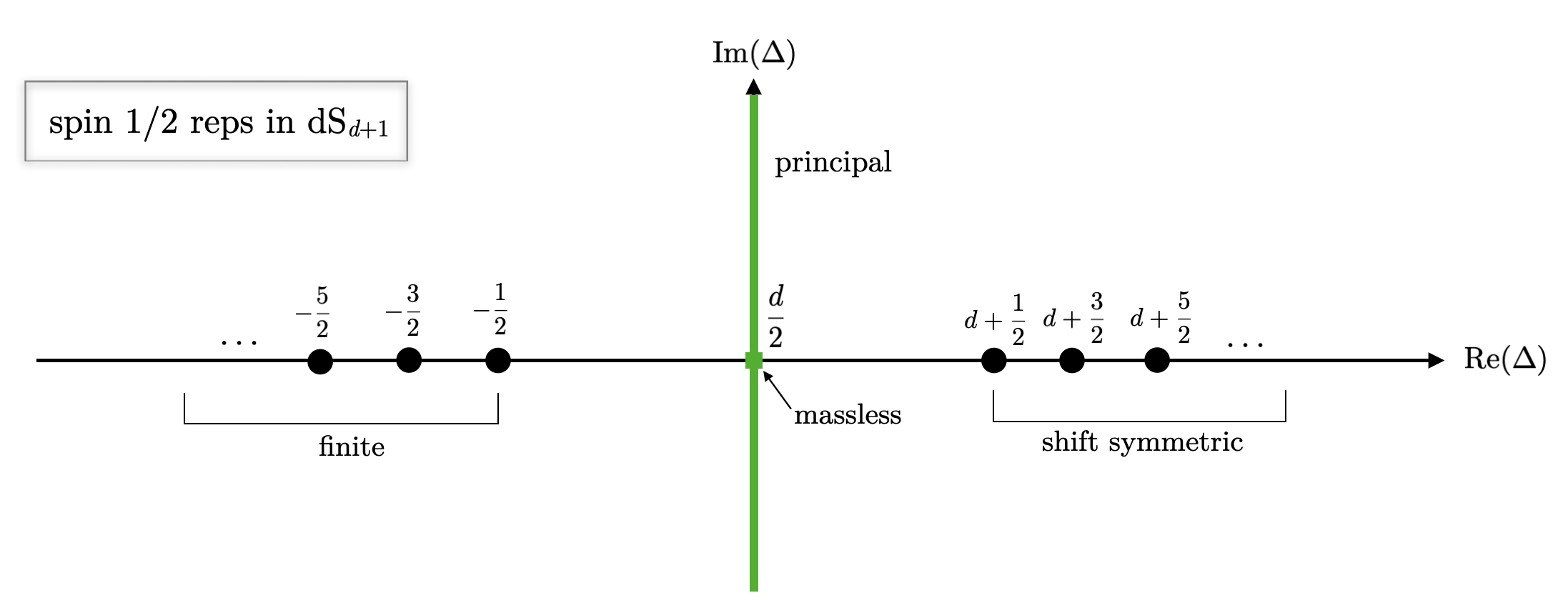,width=6.4in}}  \label{dsrepsspinor}\, \ee
The massless point at $\Delta=d/2$ is indicated with a square and the round dots are the special points where the reps become shortened. Points in green are unitary reps.    

 For $D$ odd, the reps come in chiral pairs, so each point except $\Delta=d/2$ represents two reps, but there is also the equivalence \eqref{fermoddintgme} given by reflecting through $\Delta=d/2$ and flipping the helicity (except at the reducible points), and the point at $\Delta=d/2$ represents a single rep.  
 
 For $D$ even, each point except $\Delta=d/2$ represents a single rep, but there is also the equivalence \eqref{fermioneqfjee1} given by reflecting through $\Delta=d/2$ (except at the reducible points), and the point at $\Delta=d/2$ represents two different chiral reps.  

The value of the Casimir operator on the spin $1/2$ reps is given by
\be {\cal C}_2  = -{\tilde m^2\over H^2}- {1\over 8}D(D-1)= \Delta(\Delta-d)+{1\over 8}d(d-1)\,. \ee

\subsection{Spin $s$ fermionic representations\label{spinsfermionsec}}

Next we turn to the spin $s$ fermionic reps, $s={3\over 2},{5\over 2},{7\over 2},\ldots$.  The representation space is the space of fully symmetric gamma-traceless Dirac spinor-tensor fields, as detailed below, and the relation between the mass and $\Delta$ is given by \eqref{mixedsymmasmrelefe}.  

We continue to use the mass $\tilde m$ that appears in the Dirac equation, unlike the spin $s$ bosonic case in section \ref{spinssec} where the mass $m$, shifted so that $m=0$ is the point of largest gauge symmetry, was used.  Thus for the spinning fermions, the point of largest gauge symmetry will have non-zero $\tilde m$, even though it is still, like the gravitino, often called ``massless.''  The point $\tilde m=0$ will still be a special point where new splittings or equivalences happen without gauge symmetry.  Unlike the spin $1/2$ case, we will not call this point ``massless,'' and will refer to it as the ``$\tilde m=0$ point'' instead.

\subsubsection*{Even $D$:}  

The representation space for the fermionic spin $s$ reps in odd $d$ is the space of Dirac spinor-tensor fields on ${\mathbb S}^d$ with $s-{1\over 2}$ tensor indices $\psi_{i_1\ldots i_{s-{1\over 2}}}$, transforming under $\frak{so}(1,D)$ as in \eqref{latetimeshesactiont} with $r=s-\half$.  The tensors are fully symmetric in their tensor indices and are gamma-traceless:
\be \gamma^{i_1} \psi_{i_1i_2\ldots i_{s-{1\over 2}}}=0\,,\ee
which also implies ordinary tracelessness in all the tensor indices.  These conditions ensure that the tensor is algebraically irreducible in odd $d$ (in even $d$ there will be a further chiral splitting).  Call this space ${\cal F}^{[s-\half]_\half}_\Delta$,
\be {\cal F}^{[s-\half]_\half}_\Delta:\ \ \text{complex\ symmetric\ rank } s-\half \text{ gamma-traceless\ Dirac\ spinor-tensors\ on\ } {\mathbb S}^d \,.\ee
The superscript $[s-\half]_\half$ indicates that we are working with spinor-tensors whose tensor indices are in the fully symmetric $[s-\half]$ tableau.  

To decompose this space into $\frak{so}(D)$ reps, we need the analog of the SVT decomposition for a symmetric gamma-traceless spinor-tensor of rank $s-\half$.  This decomposition consists of symmetric gamma-traceless and transverse pieces $\chi_{i_1\ldots i_u}$ of all ranks $u=0,1,\ldots,s-{1\over 2}$,
\be \psi_{i_1\ldots i_{s-\half}}=\chi_{i_1\ldots i_{s-\half}}+D_{(i_1} \chi_{i_2\ldots i_{s-\half})_{\gamma T}}+D_{(i_1}D_{i_2} \chi_{i_3\ldots i_{s-\half})_{\gamma T}}+\cdots+D_{(i_1} \cdots D_{i_{s-\half})_{\gamma T}}\chi\, .\label{spSVTdedcompeese}\ee 
The notation $(...)_{\gamma T}$ means that the enclosed indices are to be projected along with the spinor index onto the completely symmetric and gamma-traceless part.\footnote{For example, we have
\bea &&  D_{(i)_{\gamma T}}\chi= D_i\chi -{1\over d}\gamma_i \fdag{D} \chi \,, \nn\\
 &&  D_{(i}D_{j)_{\gamma T}}\chi=  D_{(i}D_{j)}\chi -{2\over d+2} \gamma_{(i}D_{j)} \fdag{D} \chi  -{1\over d+2}  g_{ij} \left( D^2+{d-1\over 2} \right)\chi \,,\nn\\
 &&  D_{(i}\chi_{j)_{\gamma T}} =D_{(i}\chi_{j)} -{1\over d+2} \gamma_{(i} \fdag{D} \chi_{j)} -{1\over d+2} g_{ij} D\cdot \chi\,.
 \eea
}
Each of the $\chi_{i_1\ldots i_u}$  is symmetric, transverse and gamma-traceless,
\bea &&D^{i_1}\chi_{i_1\ldots i_u}=0\, , \ \  \gamma^i \chi_{i i_2\ldots i_u}=0\, , \ \ \ u=1,2,\ldots, {s-\half}\,.
\eea

Each of the transverse pieces $\chi_{i_1\ldots i_u}$ can in turn be decomposed into spinor-tensor spherical harmonics \cite{Homma:2020has},
\be Y_{lm,i_1\ldots i_u}^{\pm}\, ,\ \ \ l=u+{1\over 2},u+{3\over 2},u+{5\over 2},\ldots \ .\ee
These harmonics form two sets of eigenvectors under the Dirac operator, with positive and negative pure-imaginary eigenvalues,
\be  \fdag{{ D}}Y_{lm,i_1\ldots i_u}^{\pm}=\pm i \left(l-{1\over 2}+{d\over 2}\right)\, . \label{diraceigenvaluesesse}\ee
Under rotations these transform in the $[l-\half,u]_{\pm\half}\equiv (l,u+\half,1/2,\ldots, \pm 1/2)$ reps of $\frak{so}(D)$, and the index $m$ runs over the states of this rep.  Together they form a basis of the space of transverse gamma-traceless symmetric spinor-tensors of rank $u$ on ${\mathbb S}^d$.  

The natural Laplacian on this space is the spinor-tensor version of the Lichnerowicz Laplacian,
\be \Delta_L=-D^2+u(u+d-1)+{d(d-1)\over 8}\, ,\label{spspinslaplacianue2}\ee 
and the transverse harmonics are eigenfunctions of it, with the following eigenvalues,
\be \Delta_LY_{lm,i_1\ldots i_u}^{\pm}=  \left[ l (l + d - 1) + u (u + d - 2) + {(d-2 ) (d-1 )\over 8}\right] Y_{lm,i_1\ldots i_u}^{\pm} \, .\ee

The harmonics in $Y_{lm,i_1\ldots i_u}^{\pm}$ with $l<{s-\half}$ will not appear in the decomposition \eqref{spSVTdedcompeese}, because they are generalized conformal Killing spinor-tensors that are annihilated by the derivative combination $\nabla_{(i_{u+1}}\cdots \nabla_{i_s} \chi_{i_1\ldots i_u)_{\gamma T}}$ that appears in \eqref{spSVTdedcompeese}.

The $\frak{so}(D)$ content of ${\cal F}^{\left[s-{1\over 2}\right]_{1\over 2}}_\Delta$ is illustrated as follows:
\be \raisebox{-40pt}{\epsfig{file=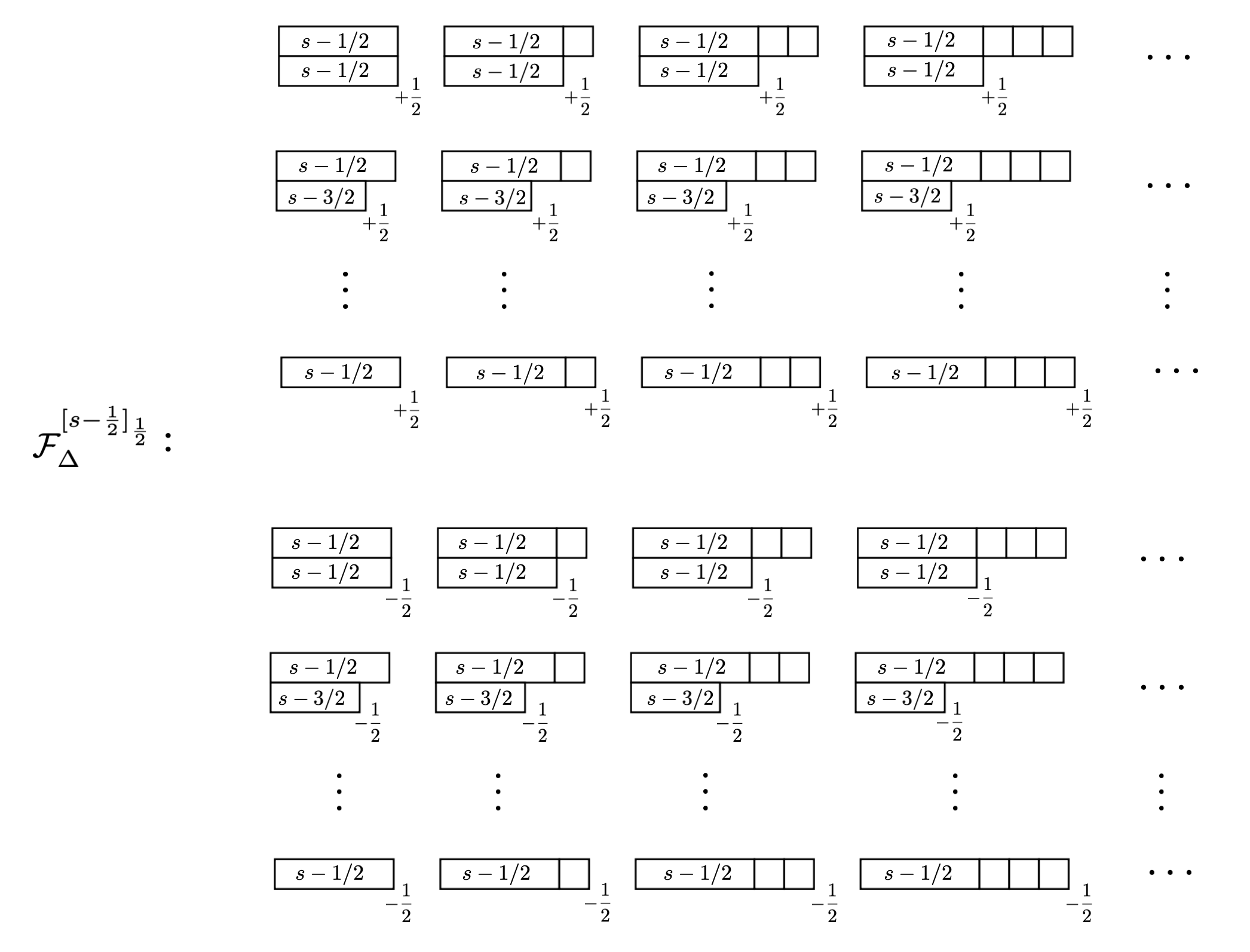,width=5.5in}}  \label{spinsfermioncontent1}\, \ee
The top half contains the $Y_{lm,i_1\ldots i_u}^{+}$ reps: those in the top row are the harmonics of the $u=s-\half$ parts of the SVT decomposition, those in the next row are the harmonics of the $u=s-{3\over 2}$ part, and so on down to the last row which is the $u=0$ part.  Analogously, the bottom half contains the $Y_{lm,i_1\ldots i_u}^{-}$ reps.
This looks just like \eqref{spinssocontent}, only every rep also has a spinor index and every rep is doubled to account for both chiralities.  For generic $\Delta$, any of the harmonics can be reached from any other by the application of the boosts, and the rep is irreducible.

\textbf{Reducible cases:} The reps become reducible at the following discrete values of $\Delta$:
\begin{itemize}

\item
Shift symmetric points:
\be \Delta = d+s+k\, ,\ \ \ k=0,1,2,\ldots \ . \ee
All but the first $k+1$ columns break off of both the plus and minus modes in \eqref{spinsfermioncontent1} and together form a sub-rep that we call 
\be {\cal D}^{\left[s-{1\over 2}\right]_{1\over 2}}_{d+s+k}\,.\ee
This is illustrated here in the case $k=1$,
\be \raisebox{-40pt}{\epsfig{file=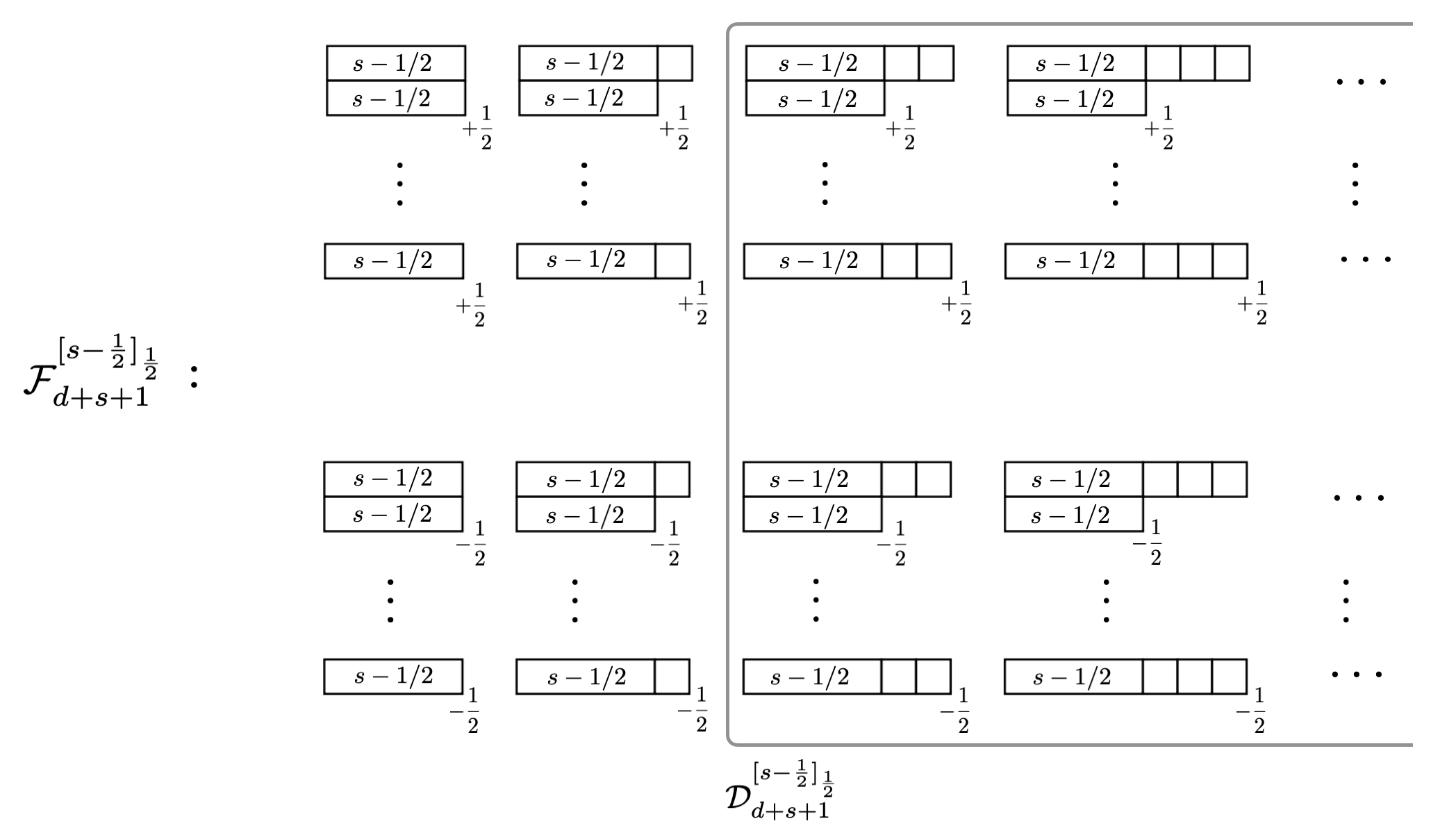,width=5.5in}}  \label{spinsfermioncontent2}\, \ee
These represent the physical modes of the level $k$ shift symmetric spinning fermionic fields \cite{Bonifacio:2023prb}.

\item
Finite points:
\be \Delta =-s-k\, ,\ \ \ k=0,1,2,\ldots \ .\ee
The first $k+1$ columns break off of both the plus and minus modes and together form a sub-rep that we call 
\be {\cal S}^{\left[s-{1\over 2}\right]_{1\over 2}}_{-s-k}\,.\ee
This is illustrated here in the case $k=1$,
\be \raisebox{-40pt}{\epsfig{file=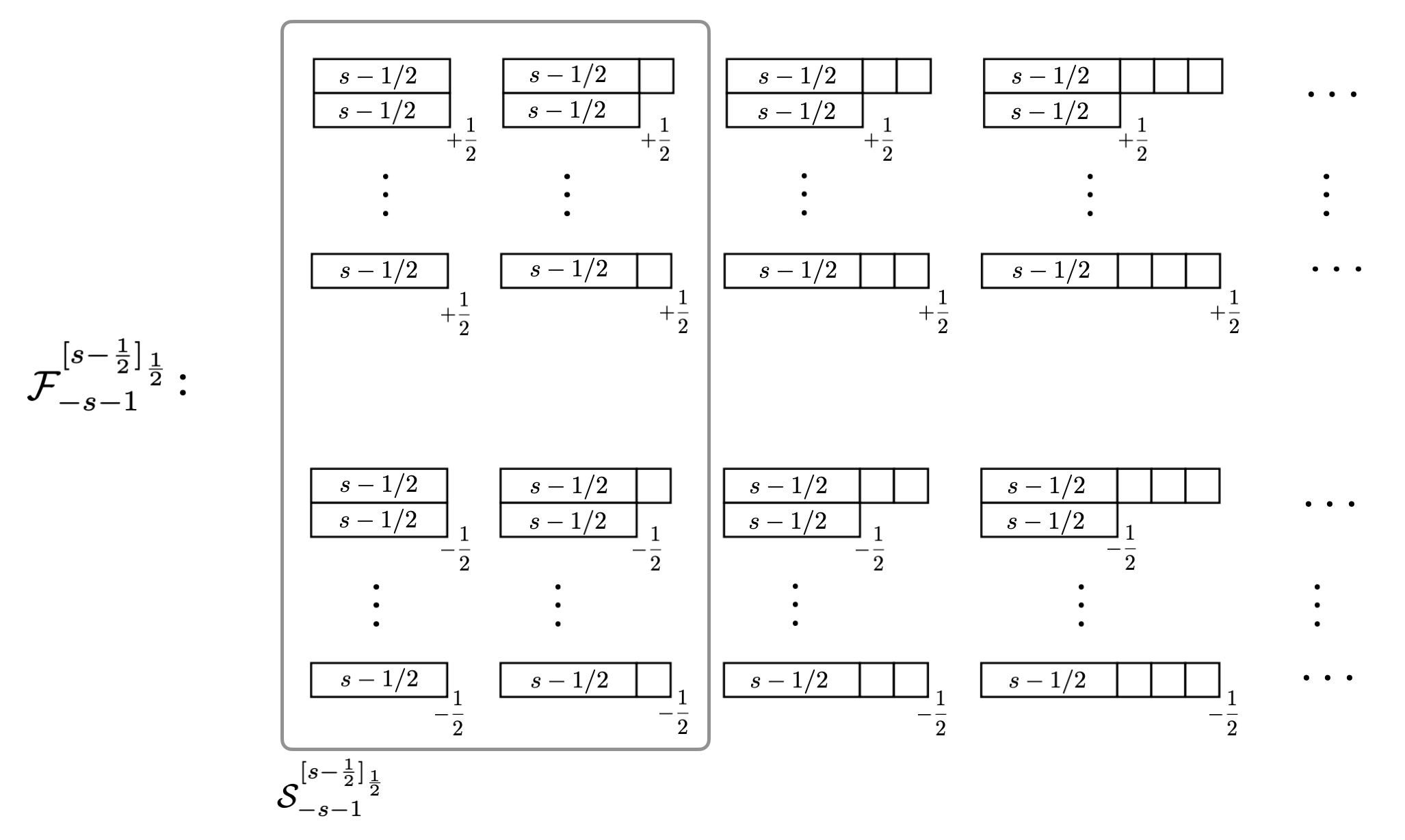,width=5.5in}}  \label{spinsfermioncontent3}\, \ee
 These are finite dimensional reps and they are precisely the reps $(s+k,s,\half,\ldots,\half)$ of $\frak{so}(1,D)$, which upon branching to $\frak{so}(D)$ using the rules in appendix \ref{branchingappendix} give the $\frak{so}(D)$ reps within ${\cal S}^{[s-\half]_{1\over 2}}_{-s-k}$.  These represent the shift symmetries of the level $k$ shift symmetric spinning fermionic fields.

\item
PM points: 
\be \Delta=d+s-t-1\,, \ \  t=1,2,\ldots,s-\half\, .\ee
The first $t$ rows break off of both the plus and minus modes and together form a sub-rep that we call 
\be {\cal V}^{\left[s-{1\over 2}\right]_{1\over 2}}_{d+s-t-1}\,.\ee
 This is illustrated here for the case where $t=1$, $\Delta=d+s-2$,
 \be \raisebox{-40pt}{\epsfig{file=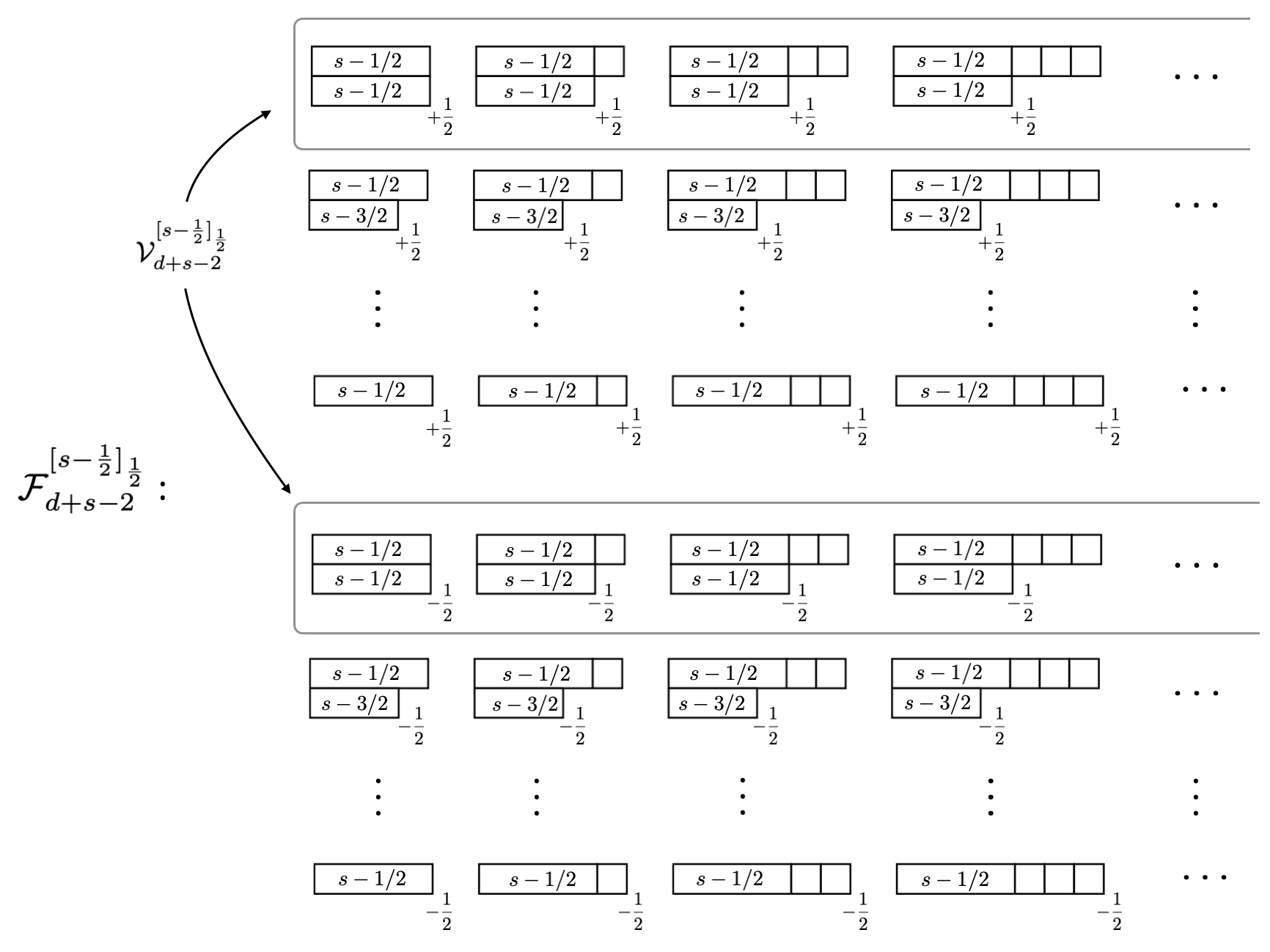,width=5.5in}}  \label{spinsfermioncontent4}\, \ee
 These represent the physical modes of the depth $t$ partially massless fermionic spin $s$ fields\footnote{The convention for the depth is different from what we used in \cite{Bonifacio:2023prb}: $t_{\rm here}=s-t_{\rm there}$.}.  (Note that the case $t=1$, the case with the largest gauge symmetry, is what is traditionally termed ``massless,'' even though it has $\tilde m\not 0$.)

\item
Gauge points: 
\be \Delta=-s+t+1\,, \ \   t=1,2,\ldots,s-\half\,.\ee
The last $s-t+{1\over 2}$ rows break off of both the plus and minus modes and together form a sub-rep that we call 
\be {\cal U}^{\left[s-{1\over 2}\right]_{1\over 2}}_{-s+t+1}\,. \ee
This is illustrated here for the case $t=1$ where $\Delta=2-s$,
 \be \raisebox{-40pt}{\epsfig{file=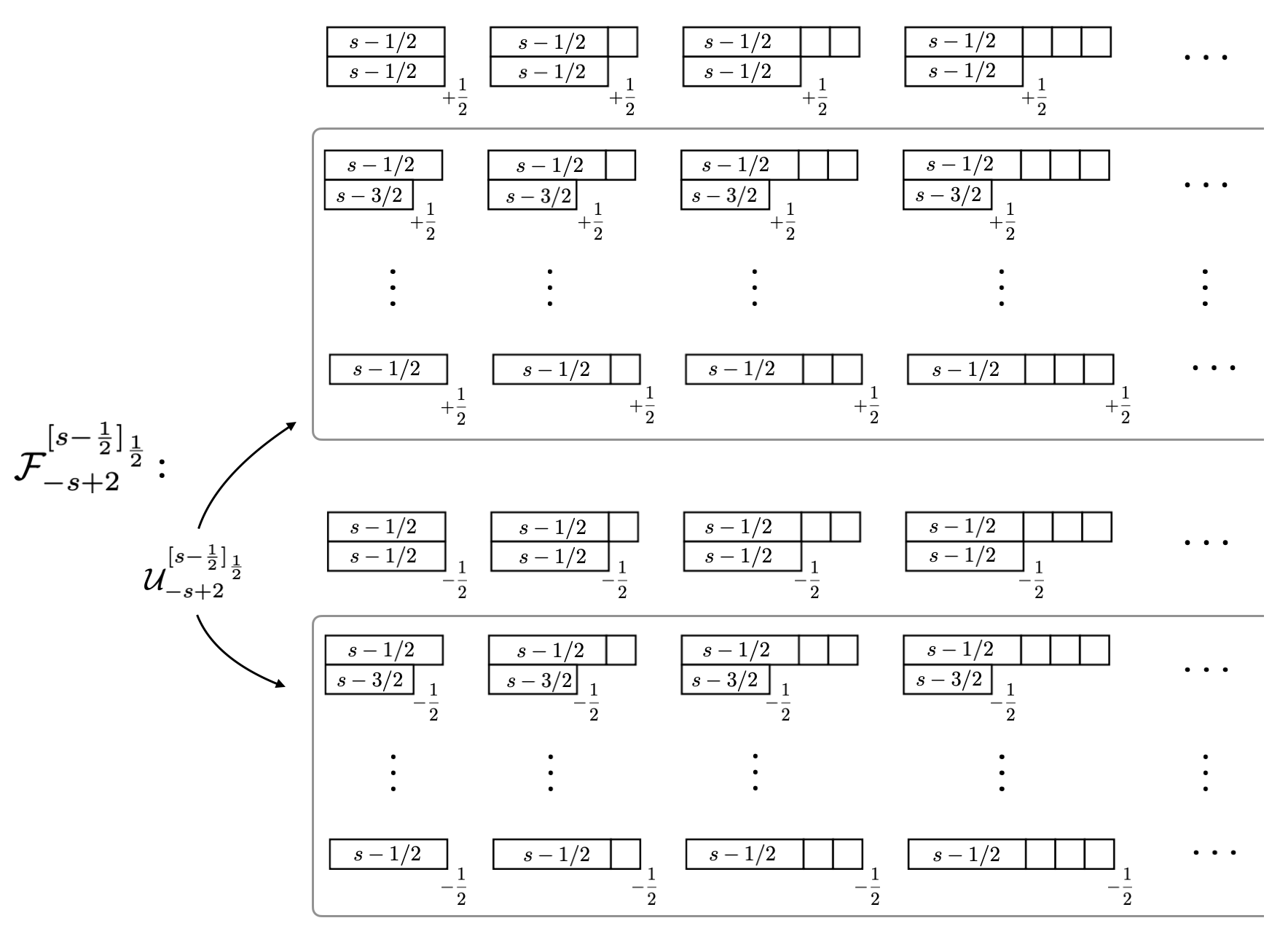,width=5.5in}}  \label{spinsfermioncontent5}\, \ee
These represent the gauge modes of the partially massless fermionic fields.

\item
Chiral point: 
\be \Delta={d\over 2} \,.\ee
This value is new compared with the bosonic case: there is a splitting that occurs, and the $Y_{lm,i_1\ldots i_u}^{+}$ and $Y_{lm,i_1\ldots i_u}^{-}$ modes become separate irreducible reps, so the ${\cal F}^{[s-\half]_{1\over 2}}_{d\over 2}$ rep slits into two reps,
\be  {\cal F}^{[s-\half]_{1\over 2}}_{d\over 2}={\cal F}^{[s-\half]_{1\over 2},+}_{d\over 2}\oplus {\cal F}^{[s-\half]_{1\over 2},-}_{d\over 2}\,,  \ee
 as illustrated here:
 \be \raisebox{-40pt}{\epsfig{file=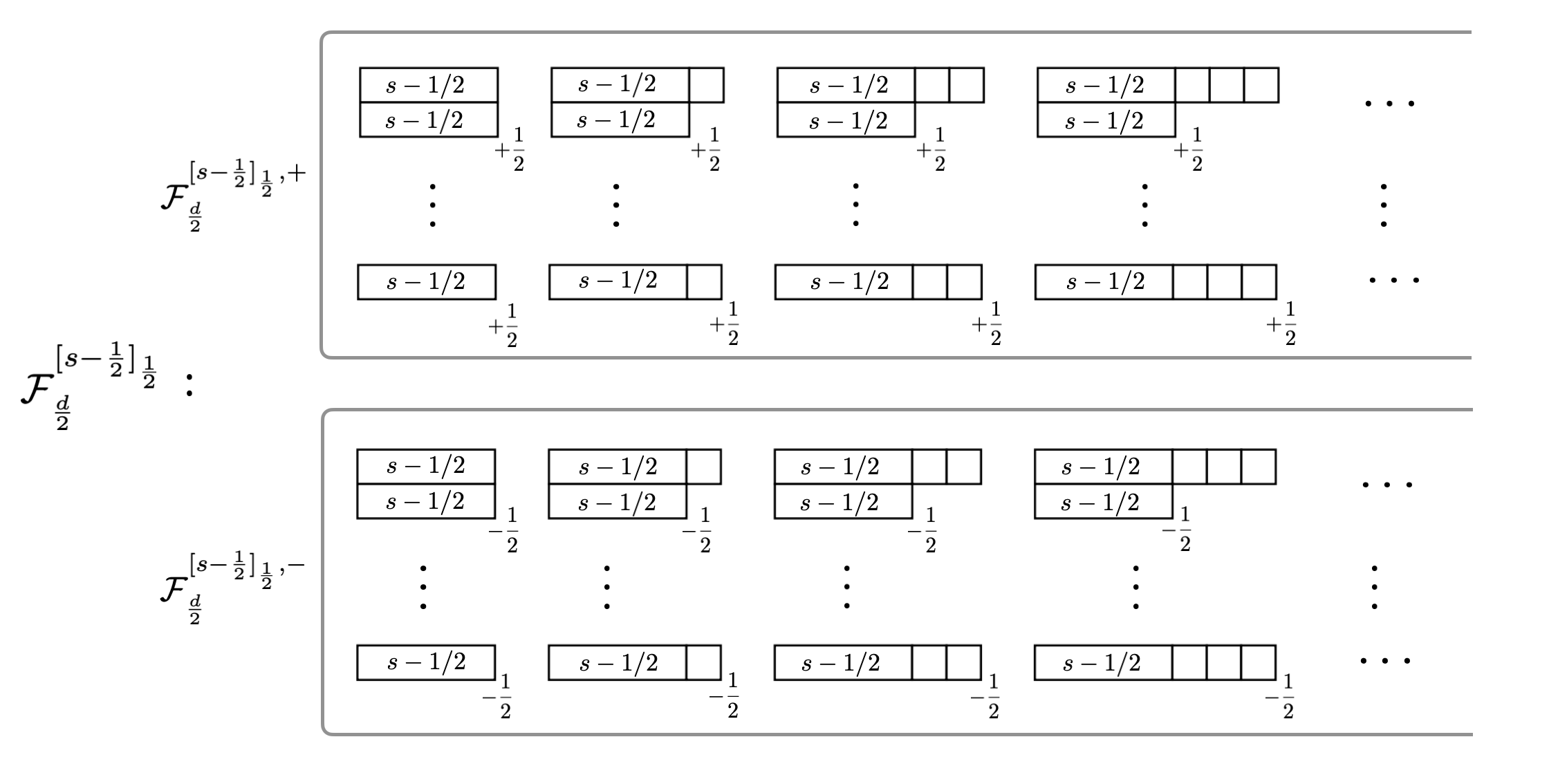,width=5.5in}}  \label{spinsfermioncontent6}\, \ee
Here the representation space is reducible and decomposable, in contrast to the shift symmetric and finite cases which are reducible but not decomposable. 
This corresponds to the case where the field in even $D$ has $\tilde m=0$ and acquires a chiral symmetry.  The two reps can be thought of as the de Sitter chiral higher spin fermions.  As in the spin $1/2$ case, this happens without any gauge symmetry, as indicated by the fact that the rep decomposes and there is no need to take a quotient to arrive at the irreducible rep.   Note that the shift symmetric, PM, gauge, and finite reps do not split into chiral halves the way the $\tilde m=0$ fermion does, since the mass term breaks chiral symmetry.

\end{itemize}

\textbf{Equivalences:} There is a shadow transform equivalence between the reps with $\Delta$ and $\bar \Delta=d-\Delta$,
\be {\cal F}^{[s-\half]_{{1\over 2}}}_{\Delta}\simeq  {\cal F}^{[s-\half]_{{1\over 2}}}_{\bar\Delta} \,.\label{fermioneqfjee12}\ee
There are intertwiner operators $S^{[s-\half]_{{1\over 2}}}_{\Delta}: \ {\cal F}^{[s-\half]_{{1\over 2}}}_{\Delta} \rightarrow {\cal F}^{[s-\half]_{{1\over 2}}}_{\bar\Delta}$ that realize this equivalence, and they develop kernels at the shift symmetric, finite, PM, and gauge points, giving pictures analogous to \eqref{spinssocontent6} and \eqref{spinssocontent7}, along with the isomorphisms
\be {\cal S}^{[s-\half]_\half}_{-s-k}\simeq {\cal F}^{[s-\half]_\half}_{d+s+k}/{\cal D}^{[s-\half]_\half}_{d+s+k}\,,\ \ \  {\cal D}^{[s-\half]_\half}_{d+s+k}\simeq {\cal F}^{[s-\half]_\half}_{-s-k}/{\cal S}^{[s-\half]_\half}_{-s-k}\,.\ee
\be {\cal U}^{[s-\half]_\half}_{-s+t+1}\simeq {\cal F}^{[s-\half]_\half}_{d+s-t-1}/{\cal V}^{[s-\half]_\half}_{d+s-t-1}\,,\ \ \  {\cal V}^{[s-\half]_\half}_{d+s-t-1}\simeq {\cal F}^{[s-\half]_\half}_{-s+t+1}/{\cal U}^{[s-\half]_\half}_{-s+t+1}\,.\ee

There are other maps of interest that map between reps with different values of $s$: the first is the order $t$ spin covariant symmetrized gradient, 
\be {\rm grad}^{t}: \  {\cal F}^{[s-\half-t]_\half}_{-s+1} \rightarrow {\cal F}^{[s-\half]_\half}_{-s+t+1} \, ,\ \ \psi_{i_1\ldots i_{s-\half-t}} \rightarrow D_{(i_{s+\half-t}}\ldots D_{ i_s} \psi_{i_1\ldots i_{s-\half-t})_{T\gamma}}\,.\label{spincovsymdge}
\ee
The kernel of ${\rm grad}^{t}$ is the space ${\cal S}^{[s-t]}_{-s+1}$ and the image is the space ${\cal U}^{[s]}_{-s+t+1}$ (the kernel is the space of fermionic generalized conformal Killing vectors).
The other map of interest is the order $t$ spin covariant divergence, 
\be {\rm div}^{t}: \  {\cal F}^{[s-\half]_\half}_{d-1+s-t} \rightarrow {\cal F}^{[s-\half-t]_\half}_{d-1+s} \, ,\ \ \psi_{i_1\ldots i_{s-\half}} \rightarrow D^{i_1}\cdots D^{i_{t}} \psi_{i_1\ldots i_{s-\half}}\,. \label{spincovdivdnge}
\ee
The kernel of ${\rm div}^{t}$ is the space ${\cal V}^{[s]}_{d-1+s-t}$ and the image is the space ${\cal D}^{[s]}_{d-1+s}$ (the kernel is the space of multiply conserved fermionic currents).

The four spaces ${\cal F}^{[s-\half]_\half}_{d+s-t-1}$, ${\cal F}^{[s-\half]_\half}_{-s+t+1}$, ${\cal F}^{[s-\half-t]_\half}_{d+s-1}$, ${\cal F}^{[s-\half-t]_\half}_{-s+1}$ are joined together into a commutative diagram analogous to \eqref{spinssocontent10}.
From this, we also get the isomorphism
\be { \cal D}^{[s-\half-t]_\half}_{d+s-1}\simeq { \cal U}^{[s-\half]_\half}_{-s+t+1}\, .\label{spinsudisoseee}\ee
This expresses the fact that the gauge modes of a partially massless spin $s$ of depth $t$ are precisely those of a $k=t-1$ shift symmetric rank $s-t$ tensor \cite{Bonifacio:2023prb}.

\subsubsection*{Odd $D$:}  

For odd $D$, $d$ is even, and we have the chiral $\gamma_\ast$ operator on ${\mathbb S}^d$, and just as in the spin $1/2 $ case, it splits the function space of spinor-tensors into 2 chiral subspaces, those with eigenvalues $\pm 1$ under the action of $\gamma_\ast$.  Each of these subspaces form separate reps, and we call them ${\cal F}^{[s-\half]_{\pm{1\over 2}}}_{\Delta}$,
\bea {\cal F}^{[s-\half]_{\pm\half}}_\Delta:\  && \text{complex\ symmetric\ rank } s-\half\ \text{gamma-traceless\ Dirac\ spinor-tensors\ on\ } {\mathbb S}^d \nn\\
&& {\rm satisfying \ } \gamma_\ast \psi_{i_1\ldots i_{s-\half}} =\pm \psi_{i_1\ldots i_{s-\half}}  \,.
\eea

We now want to split these into $\frak{so}(D)$ reps.  We first do the spinor-tensor SVT decomposition on the full space of spinors as in \eqref{spSVTdedcompeese}, and then split each component into eigenstates of the Dirac operator.  For even $d$, the Dirac field still has two sets of eigenstates under the Dirac operator, with positive and negative pure-imaginary eigenvalues just as in \eqref{diraceigenvaluesee}.  However, since the spinor reps of $\frak{so}(D)$ do not come in chiral pairs, the $Y_{lm,i_1\ldots i_u}^{+}$ and $Y_{lm,i_1\ldots i_u}^{-}$ now describe the same rep under $\frak{so}(D)$ rotations, namely the $[l-\half,u]_{\half}\equiv(l,u+\half,1/2,\ldots, 1/2)$ rep.
The $\gamma_\ast$ matrix serves as an intertwiner that maps the $Y_{lm,i_1\ldots i_u}^{\pm}$ into each other: $\gamma_\ast Y_{lm,i_1\ldots i_u}^{\pm}\propto Y_{lm,i_1\ldots i_u}^{\mp}$.
The $\gamma_\ast$ operator can be diagonalized, and this diagonalization splits the space into 2 chiral parts $\tilde Y_{lm,i_1\ldots i_u}^{\pm}$, each a linear combination of $Y_{lm,i_1\ldots i_u}^{+}$ and $Y_{lm,i_1\ldots i_u}^{-}$, satisfying
\be \gamma_\ast \tilde Y_{lm,i_1\ldots i_u}^{\pm}=\pm \tilde Y_{lm,i_1\ldots i_u}^{\pm}.\ee
The  $\tilde Y_{lm,i_1\ldots i_u}^{\pm}$ do not have definite eigenvalues under the Dirac operator: the Dirac operator is purely off-diagonal in this basis.

The $\frak{so}(D)$ content of ${\cal F}^{\left[s-{1\over 2}\right]_{\pm \half}}_\Delta$ is illustrated as follows:
\be \raisebox{-40pt}{\epsfig{file=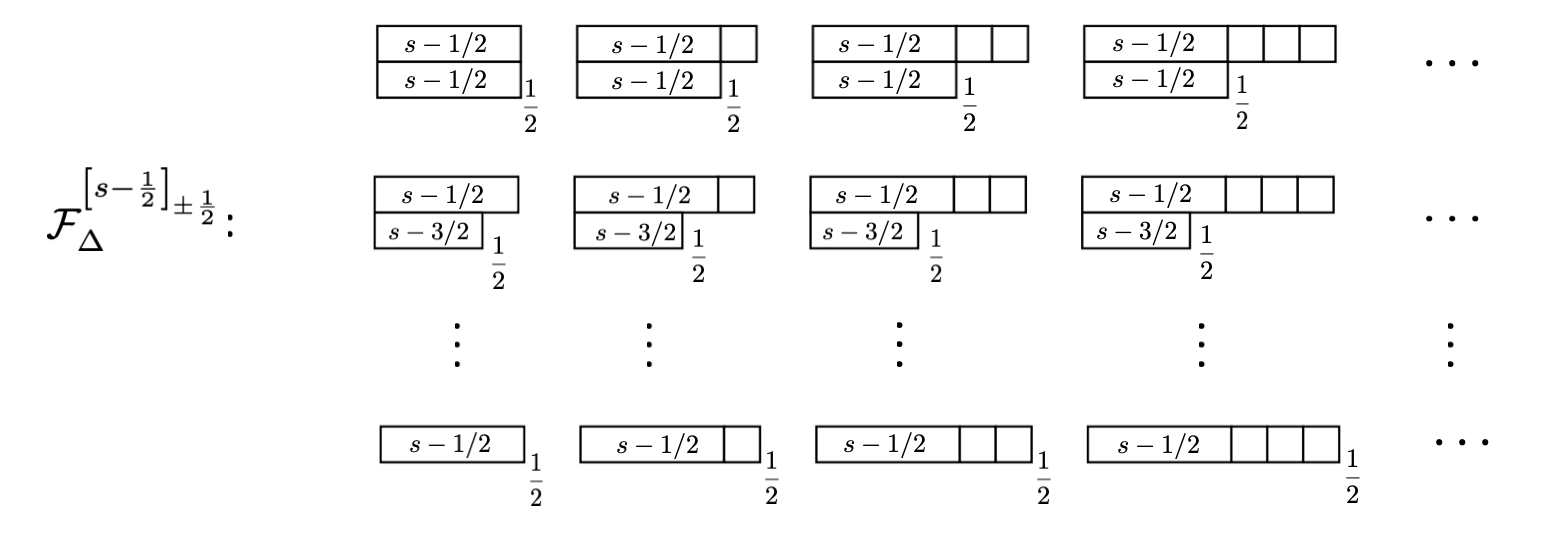,width=5.5in}}  \label{spinsfermioncontent7}\, \ee
These two reps have the same $\frak{so}(D)$ content, but they form distinct $\frak{so}(1,D)$ reps.  
They correspond to the 2 possible chiralities of a massive spin $s$ fermion on dS$_D$ with odd $D$ (just as in flat space where there are two different spinor reps of the massive little group $\frak{so}(D-1)$ when $D$ is odd), distinguished by the sign of the mass term in the Dirac equation.

\textbf{Reducible cases:} The reps become reducible at the following discrete values of $\Delta$:
\begin{itemize}

\item
Shift symmetric points:
\be \Delta = d+s+k\, ,\ \ \ k=0,1,2,\ldots \ .\ee
All but the first $k+1$ columns break off in both the plus and minus case, and we get for each a sub-rep.  We call these
\be {\cal D}^{\left[s-{1\over 2}\right]_{\pm \half}}_{d+s+k}\,.\ee
This is illustrated here in the case $k=1$,
\be \raisebox{-40pt}{\epsfig{file=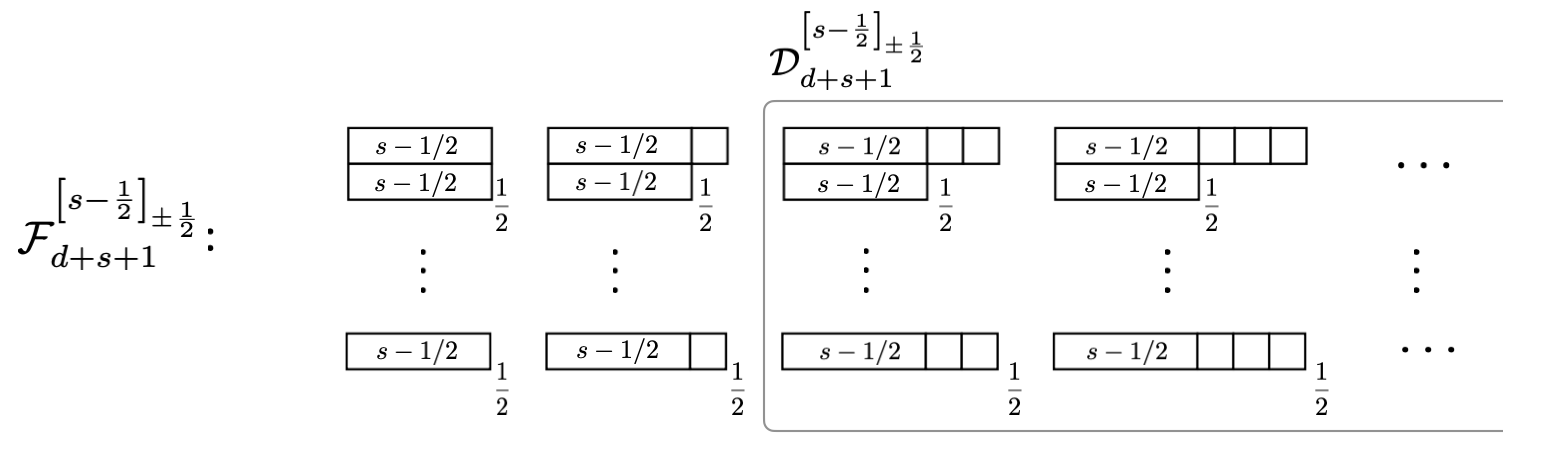,width=5.5in}}  \label{spinsfermioncontent8}\, \ee
These represent the two different chiralities of the massive level $k$ shift symmetric spinning fermionic fields in odd D.

\item
Finite points:
\be \Delta =-s-k,\ \ \ k=0,1,2,\ldots \ee
The first $k+1$ columns break off in both the plus and minus cases, and we get for each a sub-rep.  We call these
\be {\cal S}^{\left[s-{1\over 2}\right]_{\pm \half}}_{-s-k}\,.\ee
This is illustrated here in the case $k=1$,
\be \raisebox{-40pt}{\epsfig{file=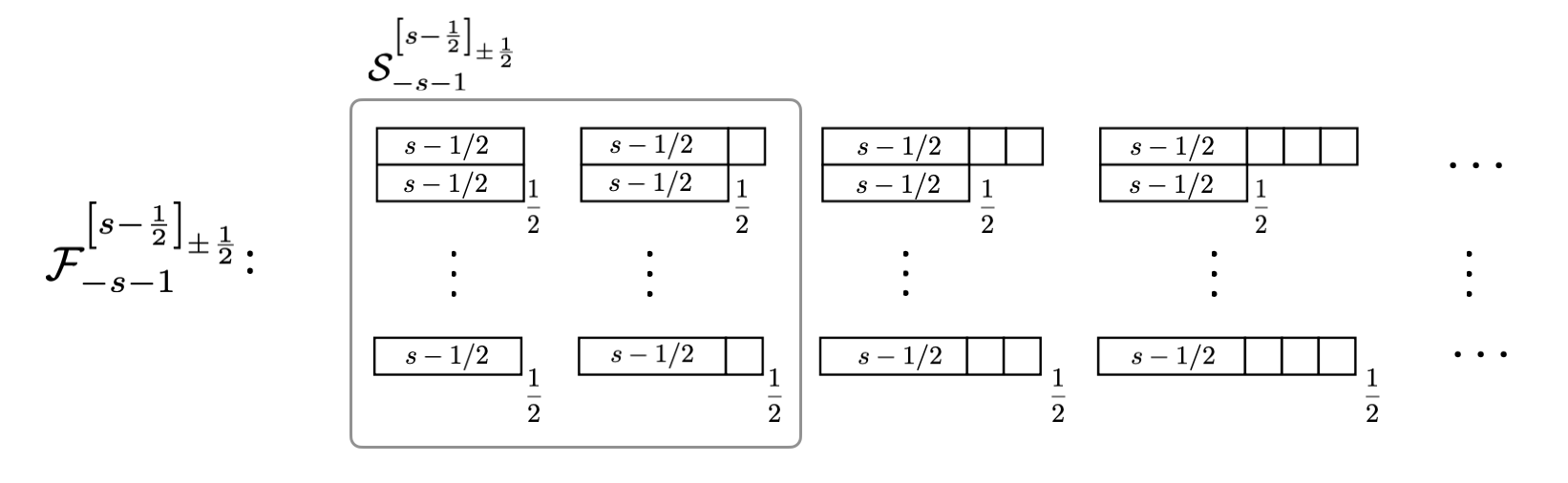,width=5.5in}}  \label{spinsfermioncontent9}\, \ee
 These are finite dimensional reps and they are precisely the reps $(s+k,s,\half,\ldots,\half,\pm \half)$ of $\frak{so}(1,D)$, which upon branching to $\frak{so}(D)$ using the rules in appendix \ref{branchingappendix} gives the $\frak{so}(D)$ reps shown within ${\cal S}^{[s-\half]_{\pm \half}}_{-s-k}$.  These represent the shift symmetries of the two different chiralities of the level $k$ shift symmetric spinning fermionic fields.

\item
PM points: 
\be \Delta=d+s-t-1\,, \ \  t=1,2,\ldots,s-\half\, .\ee
The first $t$ rows break off of both the plus and minus modes and we get for each a sub-rep.  We call these
\be {\cal V}^{\left[s-{1\over 2}\right]_{\pm \half}}_{d+s-t-1}\,.\ee
 This is illustrated here for the massless case where $t=1$, $\Delta=d+s-2$,
 \be \raisebox{-40pt}{\epsfig{file=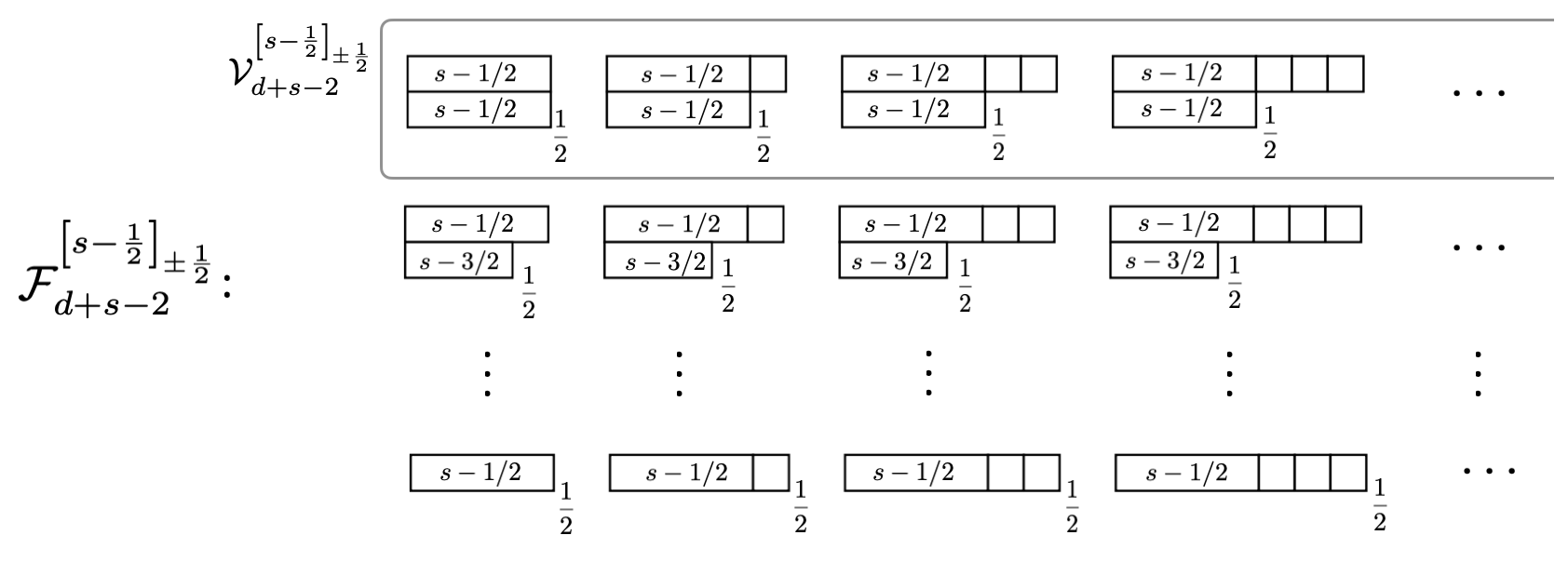,width=5.0in}}  \label{spinsfermioncontent10}\, \ee
 These represent the physical modes of the two different chiralities of the depth $t$ partially massless spin $s$ fermions in odd $D$.
 
\item
Gauge points: 
\be \Delta=-s+t+1\,, \ \   t=1,2,\ldots,s-\half \,. \ee
The last $s-t$ rows break off of both the plus and minus modes and we get for each a sub-rep.  We call these
\be {\cal U}^{\left[s-{1\over 2}\right]_{\pm \half}}_{-s+t+1}\,. \ee
This is illustrated here for the case $t=1$ where $\Delta=-s+2$,
 \be \raisebox{-40pt}{\epsfig{file=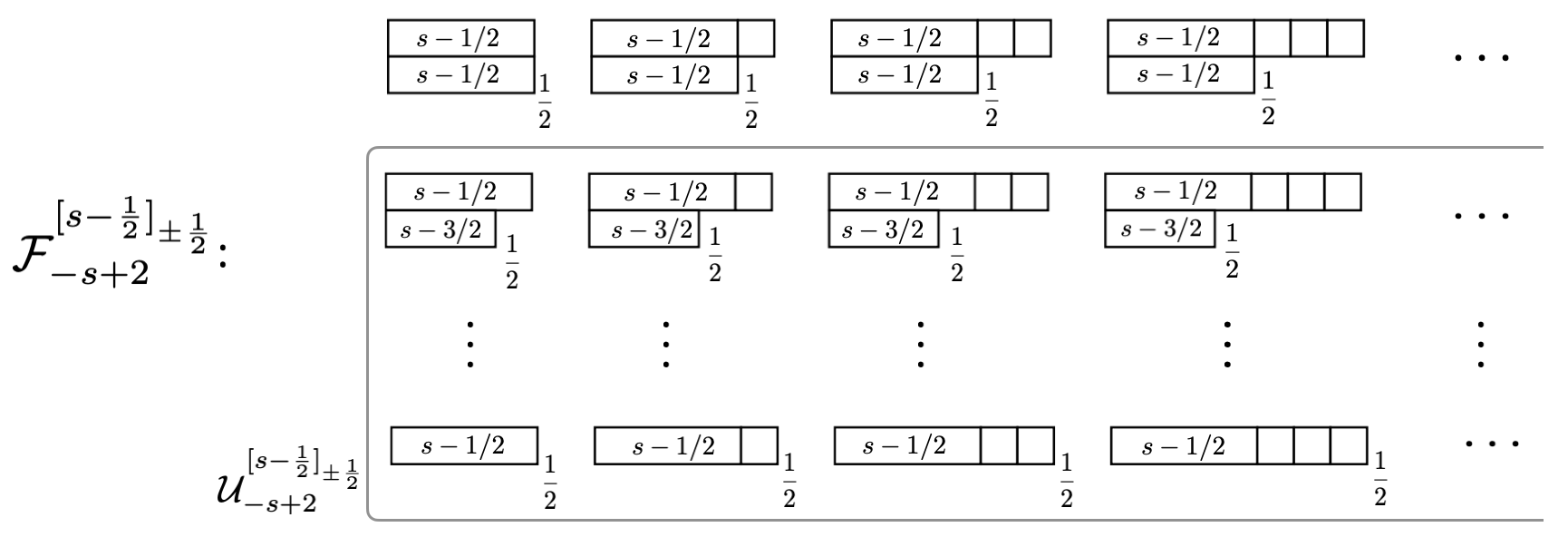,width=5.0in}}  \label{spinsfermioncontent11}\, \ee
These represent the gauge modes of the two different chiralities of the partially massless fermions.

\end{itemize}

\textbf{Equivalences:}  The shadow equivalence between the reps with $\Delta$ and $\bar \Delta=d-\Delta$ involves a flip of the chirality,
 \be {\cal F}^{[s-\half]_{\pm{1\over 2}}}_\Delta \simeq  {\cal F}^{[s-\half]_{\mp{1\over 2}}}_{\bar \Delta }\,. \label{sfermioneqfjee12e}\ee
At the shift symmetric and finite points, we have pictures analogous to \eqref{spinssocontent6} and \eqref{spinssocontent7} with a flip in chirality between the top and bottom row, from which we get the equivalences
\be {\cal D}^{[s-\half]_{\pm \half }}_{d+s+k} \simeq    {\cal F}^{[s-\half]_{\mp \half}}_{-s-k}/ {\cal S}^{[s-\half]_{\mp \half}}_{-s-k}    \,, \ \ \           {\cal S}^{[s-\half]_{\pm \half}}_{-s-k} \simeq {\cal F}^{[s-\half]_{\mp \half}}_{d+s+k} /{\cal D}^{[s-\half]_{\mp \half}}_{d+s+k}  \,.\ee

In contrast to the even $D$ case, no additional splitting happens at $\Delta=d/2$, the point corresponding to the $\tilde m=0$ spin $s$ fermion.  Instead there is an equivalence,
 \be {\cal F}^{[s-\half]_{+{1\over 2}}}_{d\over 2} \simeq  {\cal F}^{[s-\half]_{-{1\over 2}}}_{d \over 2 }\, .\label{fermioneqsfjee2}\ee 
 This reflects the fact that a $\tilde m=0$ fermion in odd $D$ does not come in two different chiralities (just as in flat space there is a unique spin $s$ spinor rep of the massless little group $\frak{so}(D-2)$ when $D$ is odd).
 
The order $t$ spin covariant symmetrized gradient map, defined as in \eqref{spincovsymdge}, maps 
 \be  {\rm grad}^{t}: \  {\cal F}^{[s-\half-t]_{\pm \half}}_{-s+1} \rightarrow {\cal F}^{[s-\half]_{\pm \half}}_{-s+t+1} \,,\ee
and the order $t$ spin covariant divergence, defined as in \eqref{spincovdivdnge}, maps 
\be {\rm div}^{t}: \  {\cal F}^{[s-\half]_{\pm \half}}_{d+s-t-1} \rightarrow {\cal F}^{[s-\half-t]_{\pm\half}}_{d+s-1} \, .\ee
The kernel of ${\rm grad}^{t}$ is the space ${\cal S}^{[s-\half-t]_{\pm \half}}_{-s+1}$ and the image is the space ${\cal U}^{[s-\half]_{\pm\half}}_{-s+t+1}$.  The kernel of ${\rm div}^{t}$ is the space ${\cal V}^{[s-\half]_{\pm\half}}_{d+s-t-1}$ and the image is the space ${\cal D}^{[s-\half]_{\pm\half}}_{d+s-1}$.
The four spaces ${\cal F}^{[s-\half]_{\pm\half}}_{d+s-t-1}$, ${\cal F}^{[s-\half]_{\mp\half}}_{-s+t+1}$, ${\cal F}^{[s-\half-t]_{\pm\half}}_{d+s-1}$, ${\cal F}^{[s-\half-t]_{\mp\half}}_{-s+1}$ are joined together into a commutative diagram analogous to \eqref{spinssocontent10}, with a flip in chirality between the top and bottom rows,
and we get the isomorphisms
\be { \cal D}^{[s-\half-t]_{\pm\half}}_{d+s-t-1}\simeq { \cal U}^{[s-\half]_{\mp \half}}_{-s+t+1}\, .\label{spinsudisoeee2}\ee
This expresses the statement that the gauge modes of a partially massless spin $s$ of depth $t$ are precisely those of a $k=t-1$ shift symmetric rank $s-t$ tensor with the opposite chirality.

\subsubsection*{Unitarity and summary}

The only inner product that can be invariant and positive definite is the standard spinor-tensor inner product,
\be \la \psi_1|\psi_2\ra= \int d^d\Omega\, \psi_1^{i_1\ldots i_{s-\half}\dag}\psi_{2\ i_1\ldots i_{s-\half}}\,. \label{spinorinnerproddes2} \ee
It is positive definite and invariant when $\Delta={d\over 2}+i\nu$ with $\nu\in {\mathbb R}$.  This is the spinning fermionic principal series.

For $D$ even, there is one rep at each $\nu\not=0$ and the reps at $\nu>0$ and $\nu<0$ are equivalent.  The rep at $\nu=0$, corresponding to the $\tilde m=0$ fermion, splits into two inequivalent chiral pieces.  For this reason, in even $D$ the point $\nu=0$ is usually excluded from the principal series and is considered a part of the discrete series, as we'll see in section \ref{unitarylistsection}.  

For odd $D$, there are two different chiral reps for each $\nu\not=0$, and the equivalence \eqref{fermioneqfjee1} relates one chirality at $\nu>0$ to the other at $\nu<0$, whereas for the $\tilde m=0$ point at $\nu=0$, there is a single rep.

As with the spin $1/2$ fermions, there is no complementary series for the spinning fermions.  In addition, the shortened reps at the shift symmetric and partially massless points are not unitary, with one important exception: in $D=4$ the partially massless fermions are unitary, as we discuss further below.

The spin $s$ fermionic reps are summarized here:
\be \raisebox{-40pt}{\epsfig{file=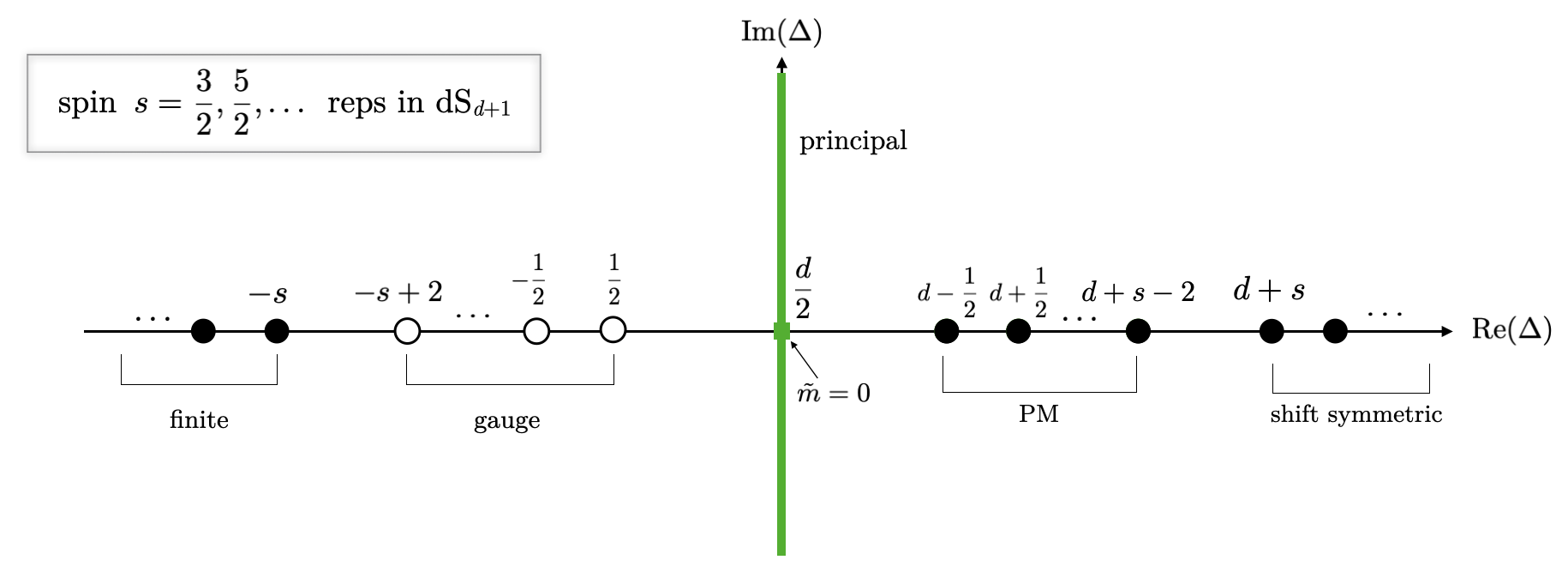,width=6.4in}}  \label{dsrepsspinorspins}\, \ee
The $\tilde m=0$ point at $\Delta=d/2$ is indicated with a square, and the round dots are the special points where the rep becomes shortened. Points in green are unitary reps.  The hollow black circles for the gauge reps indicate that they are already accounted for in the lower spin reps, through the equivalence \eqref{spinsudisoseee}, \eqref{spinsudisoeee2}.  

 For $D$ even, each point except $\Delta=d/2$ represents a single rep, but there is also the equivalence \eqref{fermioneqfjee1} given by reflecting through $\Delta=d/2$ (except at the special points), and the point at $\Delta=d/2$ represents two different chiral reps.

 For $D$ odd, the reps come in chiral pairs, so each point except $\Delta=d/2$ represents two reps, but there is also the equivalence \eqref{sfermioneqfjee12e} given by reflecting through $\Delta=d/2$ and flipping the helicity (except at the special points), and the point at $\Delta=d/2$ represents a single rep.  
 
The value of the Casimir operator on the spin $s$ fermionic reps is given by
\be {\cal C}_2  =-{\tilde m^2\over H^2}- {1\over 8}D(D-1)+\left(s-\half \right)\left(s+D-{5\over 2}\right) 
= \Delta(\Delta-d)+{1\over 8}d(d-1)+ \left(s-\half \right)\left(s +d-{3\over 2}\right)\, .\ee

As for the bosonic spin $s$ reps, the fermionic spin $s$ reps have some additional subtleties that occur in the lower dimensions $D=3,4$, which we turn to next (in $D=2$ they are trivial, as discussed in section \ref{D2section}).

\subsubsection*{$D=3$:}

Looking at the content of the spin $s$ rep for generic $\Delta$ depicted in \eqref{spinsfermioncontent7}, in $D=3$, the $\frak{so}(3)$ reps in all the rows except for the last row of both ${\cal F}^{[s-\half]_{\pm\half}}_\Delta$ are all empty.  (This is in contrast to the bosonic case, in which the last two rows are non-empty.)  And unlike the bosonic case, there is no further splitting that occurs, and ${\cal F}^{[s-\half]_{\pm\half}}_\Delta$ are still irreducible.  This reflects the fact that the spin $s$ chiral spinor-tensor fields on the 2-sphere are already irreducible, in contrast to the spin $s$ tensor fields which could be further split into (imaginary) self-dual and anti-self-dual parts.  These reps represent the chiral parts of massive spinning fermions on dS$_3$.  These same comments also apply to the fermionic shift symmetric reps ${\cal D}^{[s-\half]_{\pm\half}}_{s+k+2}$ and finite reps ${\cal S}^{[s-\half]_{\pm\half}}_{-s-k}$.

For the partially massless points \eqref{spinsfermioncontent10}, all of the $\frak{so}(3)$ reps that remain are trivial in $D=3$.  These reps therefore do not exist in $D=3$, and the corresponding fields have no propagating degrees of freedom.  This includes the gravitino on dS$_3$, which is the case $s=3/2$.

\subsubsection*{$D=4$:}

In the $D=4$ bosonic case in section \ref{spinssec}, the $\frak{so}(4)$ reps that have 2 rows in their tableaux could be further split, leading to the chiral splitting \eqref{chirald4splitpne} of the PM reps in $D=4$.  In the fermionic case, the $\frak{so}(4)$ reps cannot be further split, and the PM reps ${\cal V}^{\left[s-{1\over 2}\right]_{\pm \half}}_{d+s-t-1}$ are irreducible, representing the chiral parts of the fermionic PM fields, which account for the electromagnetic duality seen amongst the partially massless fermions \cite{Deser:2014ssa}.

The most interesting phenomenon that occurs in $D=4$ among the fermionic reps is the unitarity of the partially massless spin $s$ reps: as was emphasized recently in \cite{Letsios:2023qzq,Letsios:2022tsq}, the PM points are in fact unitary only in\footnote{The standard Dirac actions that one would write in these cases have mass terms that are not real, and give opposite overall signs for the norms for the positive and negative chirality components \cite{Higuchi:2025yuj,Boulanger:2026wnw}, unless one imposes an on-shell self-duality constraint to isolate the irreducible rep or breaks locality \cite{Anninos:2025mje}.  In addition, from the bulk $D=4$ point of view,  an axial inner product is used, containing an insertion of the bulk chiral gamma matrix \cite{Letsios:2022tsq}.} $D=4$. These reps are part of the fermionic discrete series in $D=4$, as discussed in section \ref{unitarylistsection}.  This includes the gauge invariant spin 3/2 field, the massless gravitino, which is unitary only in $D=4$ .

\subsection{Fermionic $p$-form representations}

Next, we turn to the fermionic $p$-form reps, $p\geq 1$.  The representation space is the space of fully anti-symmetric gamma-traceless Dirac spinor-tensor fields, as detailed below, and the relation between the mass and $\Delta$ is given by \eqref{mixedsymmasmrelefe}.  As with the symmetric tensor fermions, we continue to use the mass $\tilde m$ that appears in the Dirac equation, rather than the shifted mass that vanishes at the point of enhanced gauge symmetry that we used for the bosonic $p$-forms in section \ref{pformsec}.  The value $\tilde m=0$ will still be a special point where new splittings or equivalences happen without gauge symmetry: we will call this the ``$\tilde m=0$'' field, and will call ``massless'' the gauge invariant field (which has non-zero $\tilde m$).

\subsubsection*{Even $D$:}

The representation space for the fermionic $p$-form reps in even $D$, odd $d$, is the space of gamma-traceless $p$-form spinor-tensors on the sphere, $\psi_{i_1\ldots i_p}$, transforming under $\frak{so}(1,D)$ as in \eqref{latetimeshesactiont} with $r=p$.   These are fully anti-symmetric in their tensor indices, as well as gamma-traceless, 
\be \gamma^{i_i}\psi_{i_1i_2\ldots i_p}=0\,.\ee
These conditions ensure that the tensor is algebraically irreducible in odd $d$ (in even $d$ there will be a further chiral splitting).  Call this space ${\cal F}^{[1^p]_\half}_\Delta$,
\be {\cal F}^{[1^p]_\half}_\Delta:\ \ \text{complex\ anti-symmetric\ rank } p \text{ gamma-traceless\ Dirac\ spinor-tensors\ on\ } {\mathbb S}^d \,.\label{pformrepdef1de}\ee
The superscript $[1^p]_\half$ indicates that we are working with spinor-tensors whose tensor indices are in the fully anti-symmetric $[1^p]\equiv \underset{p\ {\rm rows}}{\underbrace{[1,\ldots,1]}}$ tableau.

To decompose this space into $\frak{so}(D)$ reps, we need the Hodge decomposition for fermionic $p$-forms on ${\mathbb S}^d$, which reads
\be \phi_{i_1\ldots i_p}=\chi_{i_1\ldots i_p}+D_{[i_1}\chi_{i_2 \ldots i_p]_{\gamma T}}\,.\label{hodgepformxve}\ee
The notation $[...]_{\gamma T}$ means that the enclosed indices are to be projected, along with the spinor index, onto the completely anti-symmetric and gamma-traceless part.\footnote{For example, we have
\bea &&  D_{[i]_{\gamma T}}\chi= D_i\chi -{1\over d}\gamma_i \fdag{D} \chi \,, \nn\\
 &&  D_{[i}\chi_{j]_{\gamma T}} =D_{[i}\chi_{j]} -{1\over d-2} \gamma_{[i} \fdag{D} \chi_{j]} +{1\over (d-1)(d-2)} \gamma_{ij} D\cdot \chi\,.
 \eea
}
Each of the two components, $\chi_{i_1\ldots i_p}$ and $\chi_{i_1\ldots i_{p-1}}$, is fully anti-symmetric, transverse and gamma-traceless,
\bea &&D^{i_1}\chi_{i_1i_2\ldots i_p}=D^{i_1}\chi_{i_1i_2\ldots i_{p-1}}=0\,  , \nn\\
&& \gamma^{i_1}\chi_{i_1i_2\ldots i_p}=\gamma^{i_1}\chi_{i_1i_2\ldots i_{p-1}}=0\, .
\eea

These transverse and gamma-traceless fermionic $q$-form pieces, $q=p,p-1$, can in turn be decomposed into spherical harmonics
\be Y_{lm,i_1\ldots i_q}^{\pm}\, , \ \ \ l={3\over 2},{5\over 2},{7\over 2},\ldots\ .\label{pformeigsgde}\ee
These harmonics form two sets of eigenvectors under the Dirac operator, with positive and negative pure-imaginary eigenvalues \cite{BRANSON1992314},
\be  \fdag{{ D}}Y_{lm,i_1\ldots i_{q}}^{\pm}=\pm i \left(l-{1\over 2}+{d\over 2}\right) Y_{lm,i_1\ldots i_q}^{\pm} \, . \label{diraceigenvalufpesesse}\ee
They transform in the $[l-\half,1^{q}]_{\pm\half}\equiv (l,{3\over 2}^q,\half,\ldots,\pm\half)$ reps of $\frak{so}(D)$, and the index $m$ runs over the states of this rep.  Together, the $Y_{lm,i_1\ldots i_q}^{+}$ and $Y_{lm,i_1\ldots i_q}^{-}$ form a basis of the space of transverse gamma-traceless spinor-tensor $q$-forms on ${\mathbb S}^d$.  The natural Laplacian on this space is the spinor version of the Hodge Laplacian, which reads
\be \Delta_L=-D^2+q (d + 1 - q) + {d (d - 1)\over 8}\, ,\label{spspinslaplacipanpfue}\ee 
and the transverse harmonics are eigenfunctions of it, with the following eigenvalues \cite{BRANSON1992314},
\be \Delta_L  Y_{lm,i_1\ldots i_{q}}^{\pm}=   \left[ l (l+d-1) + q(d - q) +{ (d-2) (d-1)  \over 8} \right] Y_{lm,i_1\ldots i_{q}}^{\pm} \, .\ee

For generic $\Delta$, any of the harmonics can be reached from any other by the application of the boosts, and the $\frak{so}(D)$ content of the rep \eqref{pformrepdef1de} is illustrated as follows:
\be \raisebox{-40pt}{\epsfig{file=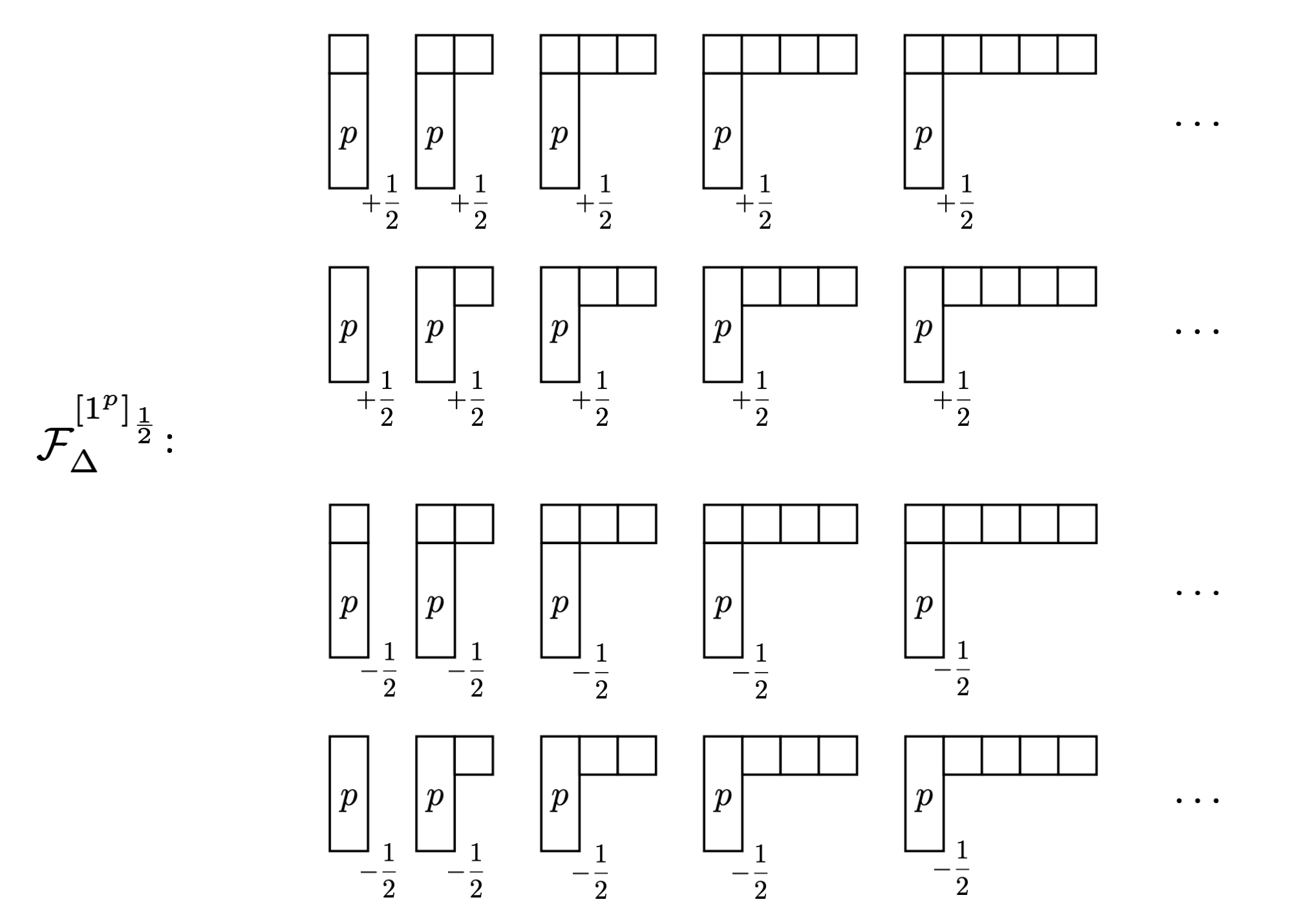,width=4.5in}}  \label{spinpfermioncontent1}\, \ee
The first and third rows give the modes in $\chi_{i_1\ldots i_p}$, and the second and fourth rows give the modes in $\chi_{i_1\ldots i_{p-1}}$. 
This looks just like \eqref{pformsocontent}, only every rep also has a spinor index and each $\frak{so}(D)$ rep is doubled to account for both chiralities. 

\textbf{Reducible cases:} The reps become reducible at the following discrete values of $\Delta$:
\begin{itemize}

\item 
Shift symmetric points:
\be  \Delta=d+k+{3\over 2}\,,\ \ \ k=0,1,2,\ldots\ .\ee
Here the $(k+2)$-th and higher columns of \eqref{spinpfermioncontent1} separate off to form a sub-rep we call
\be {\cal D}^{[1^p]_\half}_{d+k+{3\over 2}}\, .\ee
These are the physical modes of the level $k$ shift symmetric fermionic $p$-forms \cite{Bonifacio:2023prb}.

\item
Finite points:
\be \Delta=-k-{3\over 2}\,,\ \ \ k=0,1,2,\ldots\ .\ee
Here the first $k+1$ columns of \eqref{spinpfermioncontent1} separate off to form the finite dimensional sub-rep we call
\be {\cal S}^{[1^p]_{\half}}_{-k-{3\over 2}}\, .\ee
This is the $[k+1,1^p]_\half$ tensor rep of $\frak{so}(1,D)$, which upon branching to $\frak{so}(D)$ using the branching rules in appendix \ref{branchingappendix} gives precisely the reps in ${\cal S}^{[1^p]_\half}_{-k-{3\over 2}}$.  These are the modes of the shift symmetries for the level $k$ shift symmetric fermionic $p$-forms.

\item 
Massless point:
\be \Delta=d-p+\half\, .\ee
At this value the first and third rows of \eqref{spinpfermioncontent1} form the sub-rep we call
\be {\cal V}^{[1^p]_\half}_{d-p+\half} \, .\ee
These represent the physical modes of the massless fermionic $p$-form.

\item 
Gauge point:
\be \Delta=p-\half\, .\ee
At this value the second and fourth rows of \eqref{spinpfermioncontent1} form the sub-rep we call
\be {\cal U}^{[1^p]_\half}_{p-\half} \, .\ee
These represent the gauge modes of the massless fermionic $p$-form.

\item 
Chiral point:
\be \Delta={d\over 2}\,.\ee
At this point the rep is reducible and decomposable: the first and second rows of \eqref{spinpfermioncontent1} split off into a rep ${\cal V}^{[1^p]_{\half,+}}_{d\over 2}$ and the third and fourth rows split off into a rep ${\cal V}^{[1^p]_{\half,-}}_{d\over 2}$,
\be {\cal V}^{[1^p]_{\half}}_{d\over 2}={\cal V}^{[1^p]_{\half,+}}_{d\over 2}\oplus{\cal V}^{[1^p]_{\half,-}}_{d\over 2} \,.\label{pddechiralsnewe}\ee
These represent the two chiral parts of the $\tilde m=0$ fermionic $p$-form field.

\end{itemize}

\textbf{Equivalences:} There is a shadow transform equivalence between the reps with $\Delta$ and the reps with $\bar \Delta=d-\Delta$, 
\be {\cal F}_{\Delta}^{[1^p]_\half}\simeq {\cal F}_{\bar\Delta}^{[1^p]_\half}\, ,\label{pformeejfee}\ee
realized by intertwiner operators $S_\Delta^{[1^p]_\half} :\  {\cal F}_{\Delta}^{[1^p]_\half}\rightarrow {\cal F}_{\bar\Delta}^{[1^p]_\half}$.
For generic $\Delta$, this operator is invertible, giving \eqref{pformeejfee}, but for the shift symmetric, finite, massless, and gauge values of $\Delta$ indicated above where ${\cal F}_{\Delta}^{[1^p]_\half}$ becomes reducible but indecomposable and develops a sub-rep, $S_\Delta^{[1^p]_\half}$ develops a kernel which is always precisely the sub-rep. 

The shift symmetric and finite points are linked to each other via the maps $S^{[1^p]_\half}_{-k-{3\over 2}}$, ${S}^{[1^p]_\half}_{d+k+{3\over 2}}$, where the kernel and image of each are the sub-reps ${\cal S}^{[1^p]_\half}_{-k-{3\over 2}}$, ${\cal D}^{[1^p]_\half}_{d+k+{3\over 2}}$, which induces the following isomorphisms, with a picture analogous to \eqref{pformsocontent5-2},
\be {\cal S}^{[1^p]_\half}_{-k-{3\over 2}}\simeq {\cal F}^{[1^p]_\half}_{d+k+{3\over 2}}/{\cal D}^{[1^p]_\half}_{d+k+{3\over 2}}\,,\ \ \  {\cal D}^{[1^p]_\half}_{d+k+{3\over 2}}\simeq {\cal F}^{[1^p]_\half}_{-k-{3\over 2}}/{\cal S}^{[1^p]_\half}_{-k-{3\over 2}} \,.\ee

The massless and gauge points are linked to each other via the maps $S^{[1^p]_\half}_{p-\half}$, ${S}^{[1^p]_\half}_{d-p+\half}$, where the kernel and image of each are the sub-reps ${\cal U}^{[1^p]_\half}_{p-\half}$, ${\cal V}^{[1^p]_\half}_{d-p+\half}$, which induces the following isomorphisms, with a picture analogous to \eqref{pformsocontent5-3},
\be {\cal U}^{[1^p]_\half}_{p-\half}\simeq {\cal F}^{[1^p]_\half}_{d-p+\half}/{\cal V}^{[1^p]_\half}_{d-p+\half}\,,\ \ \  {\cal V}^{[1^p]_\half}_{d-p+\half}\simeq {\cal F}^{[1^p]_\half}_{p-\half}/{\cal U}^{[1^p]_\half}_{p-\half}\,.\ee

The other maps of interest for the massless points are the spinor-tensor versions of the de Rham exterior derivative operator $\rd$ and the co-exterior derivative $\rd^\dag$.  The spinor-tensor exterior derivative takes fermionic $q$ forms with $\Delta=q-\half$ to fermionic $(q+1)$-forms, raising the degree of $\Delta$ by one in the process,
\be \rd: \  {\cal F}^{[1^q]_\half}_{q-\half} \rightarrow {\cal F}^{[1^{q+1}]_\half}_{q+{1\over 2}} \, ,\ \ \psi_{i_1\ldots i_q}\rightarrow D_{[i_{q+1}}\psi_{i_1\ldots i_q]_\gamma}\,. \label{ferpderhamee1}
\ee
Both the kernel and image of $\rd$ are the space of exact fermionic forms on the sphere (which can be visualized as the second and fourth row of \eqref{spinpfermioncontent1} with $p=q$).
The spinor-tensor co-exterior derivative $\rd^\dag$ takes fermionic $q$-forms with $\Delta=d-q+\half$ into fermionic $(q-1)$-forms, raising the degree of $\Delta$ by one in the process,
\be \rd^\dag: \  {\cal F}^{[1^{q}]_\half}_{d-q+\half} \rightarrow {\cal F}^{[1^{q-1}]_\half}_{d-q+{3\over 2}} \, ,\ \psi_{i_1\ldots i_q}\rightarrow D^{i_1}\psi_{i_1\ldots i_q} \,.  \label{ferpderhamee2}
\ee
Both the kernel and image of $\rd^\dag$ are the space of co-exact forms on the sphere (which can be visualized as the first and third row of \eqref{spinpfermioncontent1} with $p=q$).

Just as for the bosonic $p$-forms, for the fermionic $p$-forms with $p>1$ the gauge symmetry is reducible.  This is reflected in the $\rd$ and $\rd^\dag$ operator linking together more than two $q$ ranks worth of reps, through the de Rham complex.  These linkings are illustrated as follows:
\be \raisebox{-0pt}{\epsfig{file=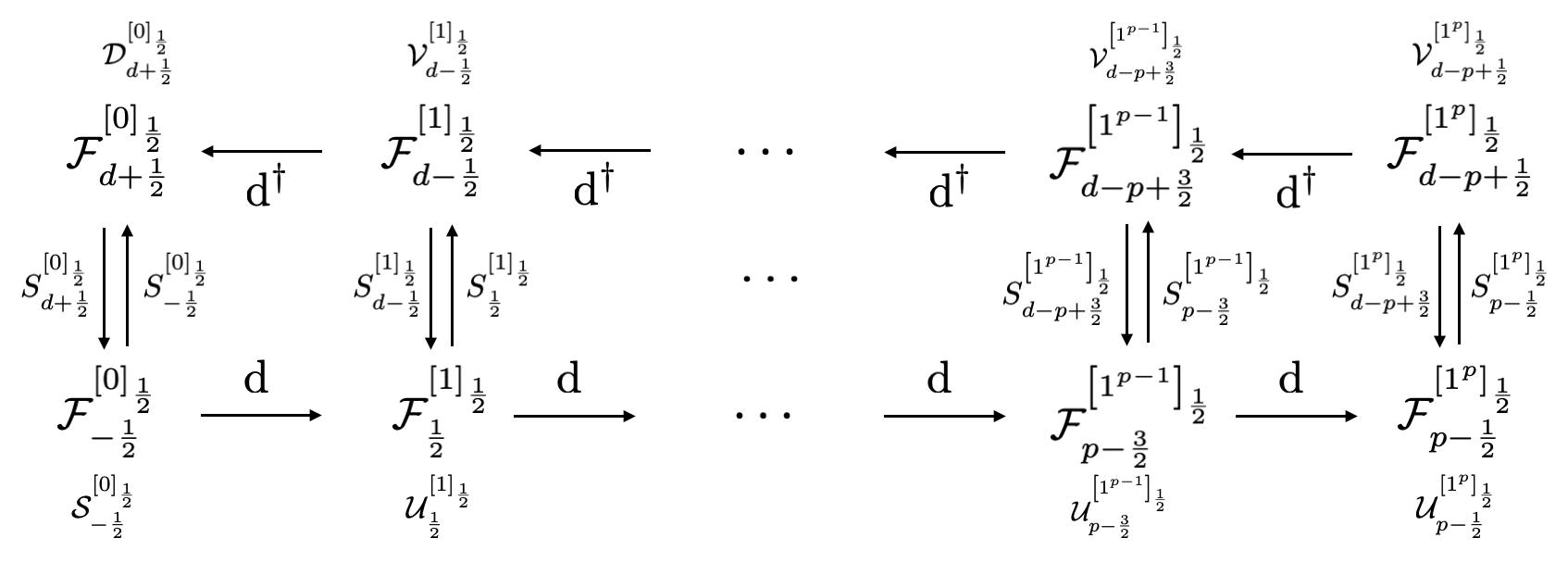,width=5.5in}}  \label{spinpfermioncontent3}\, \ee
which gives the sequence of isomorphisms
\be { \cal D}^{[0]_\half}_{d+\half}\simeq { \cal U}^{[1]_\half}_{1\over 2}\,,\ \ { \cal V}^{[1]_\half}_{d-\half}\simeq { \cal U}^{[1,1]_\half}_{3\over 2}\,,\ \ \ { \cal V}^{[1,1]_\half}_{d-{3\over 2}}\simeq { \cal U}^{[1,1,1]_\half}_{5\over 2}\,,\ \ldots \ , \ { \cal V}^{[1^{p-1}]_\half}_{d-p+{3\over 2}}\simeq { \cal U}^{[1^p]_\half}_{p-\half}\, .\label{lowepsinfode1}\ee
This expresses the fact that a massless fermionic $p$ form has as its gauge modes a fermionic $(p-1)$-form, its gauge-for-gauge modes a fermionic $(p-2)$-form, and so on all the way down, ending with a spin $1/2$ fermion with the $k=0$ shift symmetric mass.

\subsubsection*{Odd $D$:\label{fermpformoddsection}}

For odd $D$, even $d$, the chiral $\gamma_\ast$ operator on ${\mathbb S}^d$ splits the function space of $p$-form spinors into two spaces of opposite chirality, those with eigenvalues $\pm 1$ under the action of $\gamma_\ast$.  These subspaces form separate reps that we call ${\cal F}^{[1^p]_{\pm{1\over 2}}}_{\Delta}$,
\bea {\cal F}^{[1^p]_{\pm\half}}_\Delta:\ &&  \text{complex\ anti-symmetric\ rank } p \text{ gamma-traceless\ Dirac\ spinor-tensors\ on\ } {\mathbb S}^d  \nn\\
&& {\rm satisfying \ } \gamma_\ast \psi_{i_1\ldots i_{p}} =\pm \psi_{i_1\ldots i_{p}}  \,.
\eea

We now want to split these into $\frak{so}(D)$ reps.  We first do the spinor-tensor Hodge decomposition on the full space of spinors as in \eqref{hodgepformxve}, and then split each component into the eigenstates of the Dirac operator in \eqref{pformeigsgde}.  For even $d$, the Dirac field still has these two sets of eigenstates under the Dirac operator, with positive and negative pure-imaginary eigenvalues.  However, since the spinor reps of $\frak{so}(D)$ for odd $D$ do not come in chiral pairs, the $Y_{lm,i_1\ldots i_q}^{+}$ and $Y_{lm,i_1\ldots i_q}^{-}$ now describe the same rep under $\frak{so}(D)$ rotations, namely the $[l-\half,1^{q}]_{\half}=(l,{3\over 2}^q,\half,\ldots,\half)$ rep of $\frak{so}(D)$.
The $\gamma_\ast$ matrix serves as an intertwiner that maps the $Y_{lm,i_1\ldots i_q}^{\pm}$ into each other: $\gamma_\ast Y_{lm,i_1\ldots i_q}^{\pm}\propto Y_{lm,i_1\ldots i_q}^{\mp}$.
In the total space spanned by $Y_{lm,i_1\ldots i_q}^{+}$ and $Y_{lm,i_1\ldots i_q}^{-}$, the $\gamma_\ast$ operator can be diagonalized, and this diagonalization splits the space into 2 chiral parts $\tilde Y_{lm,i_1\ldots i_q}^{\pm}$, each a linear combination of $Y_{lm,i_1\ldots i_q}^{+}$ and $Y_{lm,i_1\ldots i_q}^{-}$, satisfying
\be \gamma_\ast \tilde Y_{lm,i_1\ldots i_q}^{\pm}=\pm \tilde Y_{lm,i_1\ldots i_q}^{\pm}.\ee
The  $Y_{lm,i_1\ldots i_q}^{\pm}$ do not have definite eigenvalues under the Dirac operator: the Dirac operator is purely off-diagonal in this basis.

The $\frak{so}(D)$ content of ${\cal F}^{\left[1^p\right]_{\pm {1\over 2}}}_\Delta$ is illustrated as follows:
\be \raisebox{-40pt}{\epsfig{file=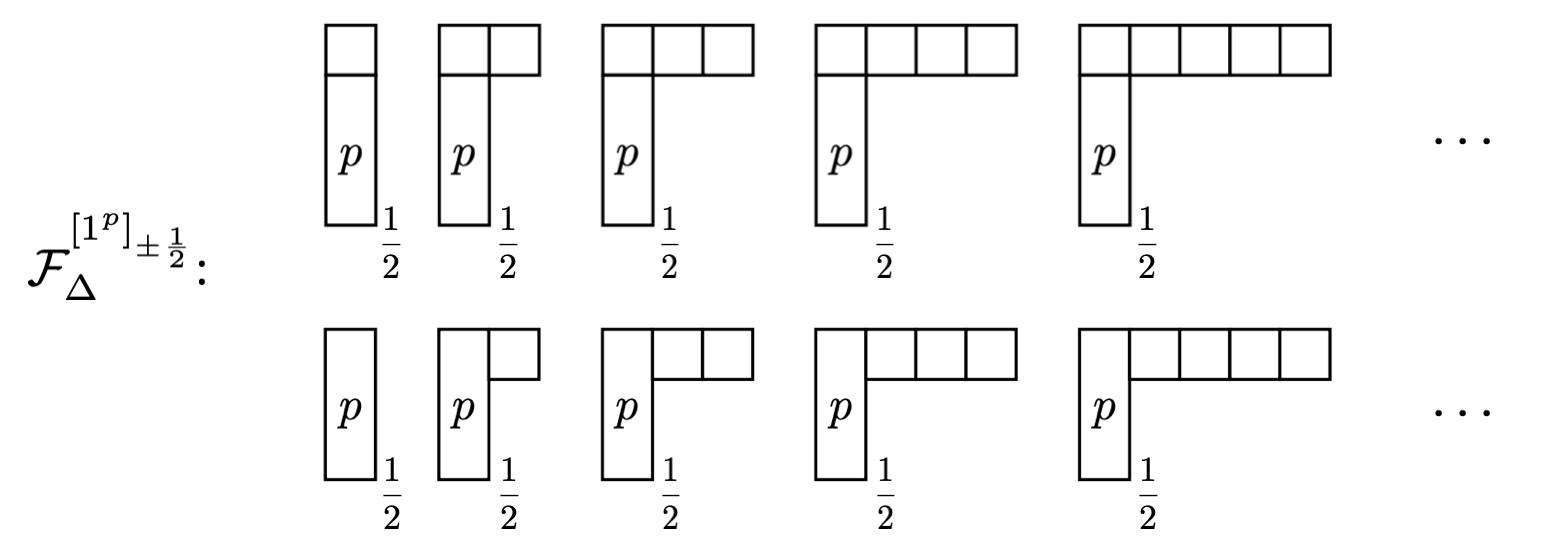,width=5.0in}}  \label{spinpfermioncontent2}\, \ee
These two reps have the same $\frak{so}(D)$ content, but they form distinct $\frak{so}(1,D)$ reps.   These two reps correspond to the 2 possible chiralities of a massive $p$-form fermion on a dS$_D$ for odd $D$, distinguished by the sign of the mass term in the Dirac equation on dS$_D$.

\textbf{Reducible cases:} The reps become reducible at the following discrete values of $\Delta$:
\begin{itemize}

\item Shift symmetric points: 
\be \Delta=d+k+{3\over 2}\,,\ \ \ k=0,1,2,\ldots\ .\ee
Here the $(k+2)$-th and higher columns of each of the two reps in \eqref{spinpfermioncontent2} separate off to form the two sub-reps we call
\be {\cal D}^{[1^p]_{\pm\half}}_{d+k+{3\over 2}}\,.\ee
These are the physical modes of the chiral parts of the level $k$ shift symmetric fermionic $p$-forms.

\item Finite points: 
\be \Delta=-k-{3\over 2}\,,\ \ \ k=0,1,2,\ldots\ .\ee
Here the first $k+1$ columns of each of the two reps in \eqref{spinpfermioncontent2} separate off to form the two finite sub-reps we call
\be {\cal S}^{[1^p]_{\pm\half}}_{-k-{3\over 2}} \,. \ee
These are the $[k+1,1^p]_{\pm\half}$ tensor rep of $\frak{so}(1,D)$, which upon branching to $\frak{so}(D)$ using the branching rules in appendix \ref{branchingappendix} gives precisely the reps in ${\cal S}^{[1^p]_{\pm\half}}_{-k-{3\over 2}}$.  These are the modes of the shift symmetries for the level $k$ shift symmetric fermionic $p$-forms.

\item Massless point:
\be \Delta=d-p+\half\,.\ee
At this value the first row of the two reps in \eqref{spinpfermioncontent2} separates off to form two sub-reps we call
\be {\cal V}^{[1^p]_{\pm\half}}_{d-p+\half}\,.\label{masslsesspforodddrepe}\ee
These represent the physical modes of the chiral parts of the massless fermionic $p$-form.

\item Gauge point:
\be \Delta=p-\half\,.\ee 
At this value the second row of the two reps in \eqref{spinpfermioncontent2} separates off to form the two sub-reps we call
\be {\cal U}^{[1^p]_{\pm\half}}_{p-\half}\,.\ee
 These represent the gauge modes of the chiral parts of the massless fermionic $p$-form.

\end{itemize}

\textbf{Equivalences:} The shadow equivalence connecting $\Delta$ and $\bar\Delta\equiv d-\Delta$  involves a flip in chirality in odd $D$,
\be {\cal F}_{\Delta}^{[1^p]_{\pm\half}}\simeq {\cal F}_{\bar\Delta}^{[1^p]_{\mp \half}}\,.\label{pformeejfee3}\ee
In particular, we have a unique rep when $\Delta=d/2$, 
 \be {\cal F}^{[1^p]_{+{1\over 2}}}_{d\over 2} \simeq  {\cal F}^{[1^p]_{-{1\over 2}}}_{d \over 2 }\, .\label{fermioneqsfjeeprom2} \ee
 This is the point where $\tilde m=0$, so the $\tilde m=0$ fermionic $p$-form is not chiral in odd $D$.

For the special values of $\Delta$ indicated above where the reps  become reducible and develop a sub-rep, the intertwiner develops a kernel which is always precisely the sub-rep. 
The shift symmetric and finite points are linked to each other with a flip in chirality, inducing the following isomorphisms,
\be {\cal S}^{[1^p]_{\pm\half}}_{-k-{3\over 2}}\simeq {\cal F}^{[1^p]_{\mp\half}}_{d+k+{3\over 2}}/{\cal D}^{[1^p]_{\mp\half}}_{d+k+{3\over 2}}\,,\ \ \  {\cal D}^{[1^p]_{\pm \half}}_{d+k+{3\over 2}}\simeq {\cal F}^{[1^p]_{\mp\half}}_{-k-{3\over 2}}/{\cal S}^{[1^p]_{\mp\half}}_{-k-{3\over 2}} \,.\ee
The massless and gauge points are linked to each other with a flip in chirality, inducing the following isomorphisms, 
\be {\cal U}^{[1^p]_{\pm\half}}_{p-\half}\simeq {\cal F}^{[1^p]_{\mp\half}}_{d-p+\half}/{\cal V}^{[1^p]_{\mp\half}}_{d-p+\half}\,,\ \ \  {\cal V}^{[1^p]_{\pm\half}}_{d-p+\half}\simeq {\cal F}^{[1^p]_{\mp\half}}_{p-\half}/{\cal U}^{[1^p]_{\mp\half}}_{d-p+\half}\,.\label{lowepsinfode2}\ee

The fermionic $\rd$ and $\rd^\dag$ operators, defined as in \eqref{ferpderhamee1}, \eqref{ferpderhamee2}, preserve the chirality and map
\bea  && \rd: \  {\cal F}^{[1^q]_{\pm \half}}_{q-\half} \rightarrow {\cal F}^{[1^{q+1}]_{\pm \half}}_{q+{1\over 2}} \, ,\ \ \rd^\dag: \  {\cal F}^{[1^{q}]_{\pm\half}}_{d-q+\half} \rightarrow {\cal F}^{[1^{q-1}]_{\pm\half}}_{d-q+{3\over 2}}\,.
\eea
For each of the chiralities $\pm$, we thus get the same pattern of mappings linking the spaces as in \eqref{spinpfermioncontent3}, the only difference is that the shadow transform flips the chirality, so in this diagram the bottom row will have a helicity opposite to the top row, giving the sequence of isomorphisms,
\be { \cal D}^{[0]_{\pm \half}}_{d+\half}\simeq { \cal U}^{[1]_{\mp \half}}_{1\over 2}\,,\ \ { \cal V}^{[1]_{\pm \half}}_{d-\half}\simeq { \cal U}^{[1,1]_{\mp \half}}_{3\over 2}\,,\ \ \ { \cal V}^{[1,1]_{\pm \half}}_{d-{3\over 2}}\simeq { \cal U}^{[1,1,1]_{\mp \half}}_{5\over 2}\,,\ \ldots \ , \ { \cal V}^{[1^{p-1}]_{\pm \half}}_{d-p+{3\over 2}}\simeq { \cal U}^{[1^p]_{\mp\half}}_{p-\half}\, .\label{fermpformisofsee}\ee
This expresses the fact that a massless fermionic $p$ form has as its gauge modes a fermionic $(p-1)$-form of opposite chirality, its gauge-for-gauge modes a fermionic $(p-2)$-form with the same chirality, and so on all the way down, ending with a spin $1/2$ fermion with the $k=0$ shift symmetric mass with the same or opposite chirality according to whether $p$ is even or odd, respectively.

\subsubsection*{Unitarity and summary:}

The only inner product that can be unitary and invariant is the standard spinor-tensor inner product,
\be \la \psi_1|\psi_2\ra= {1\over p!}\int d^d\Omega\, \psi_1^{i_1\ldots i_{p}\dag}\psi_{2\ i_1\ldots i_{p}}\,.  \label{spinorinnerpproddes2} \ee
It is positive definite and invariant when $\Delta={d\over 2}+i\nu $ with $\nu\in {\mathbb R}$; this is the $p$-form fermionic principal series.  

For $D$ even, there is one rep at each $\nu\not=0$ and because of \eqref{pformeejfee} the reps at $\nu>0$ and $\nu<0$ are equivalent.  The rep at $\nu=0$, corresponding to the $p$-form fermion $\tilde m=0$, splits into two inequivalent chiral pieces \eqref{pddechiralsnewe}.  For this reason, in even $D$ the point $\nu=0$ is usually excluded from the principal series and is considered a part of the discrete series, as we'll see in section \ref{unitarylistsection}.  

For odd $D$, there are two different chiral reps for each $\nu\not=0$, and the equivalence \eqref{pformeejfee3} relates one chirality at $\nu>0$ to the other at $\nu<0$, whereas for the massless point at $\nu=0$, there is a single rep.

There is no complementary series.  In addition, the shortened reps at the shift symmetric and massless points are not unitary, with one exception: when $D$ is even and $D=2p+2$,  the massless point is in fact unitary.  For example, the massless fermionic vector is only unitary in $D=4$ (as mentioned in section \ref{spinsfermionsec}) and the massless fermionic 2-form is unitary only in $D=6$.  These are part of the fermionic discrete series reps, as discussed further in section \ref{unitarylistsection}.

The $p$-form fermionic reps are summarized here:
\be \raisebox{-40pt}{\epsfig{file=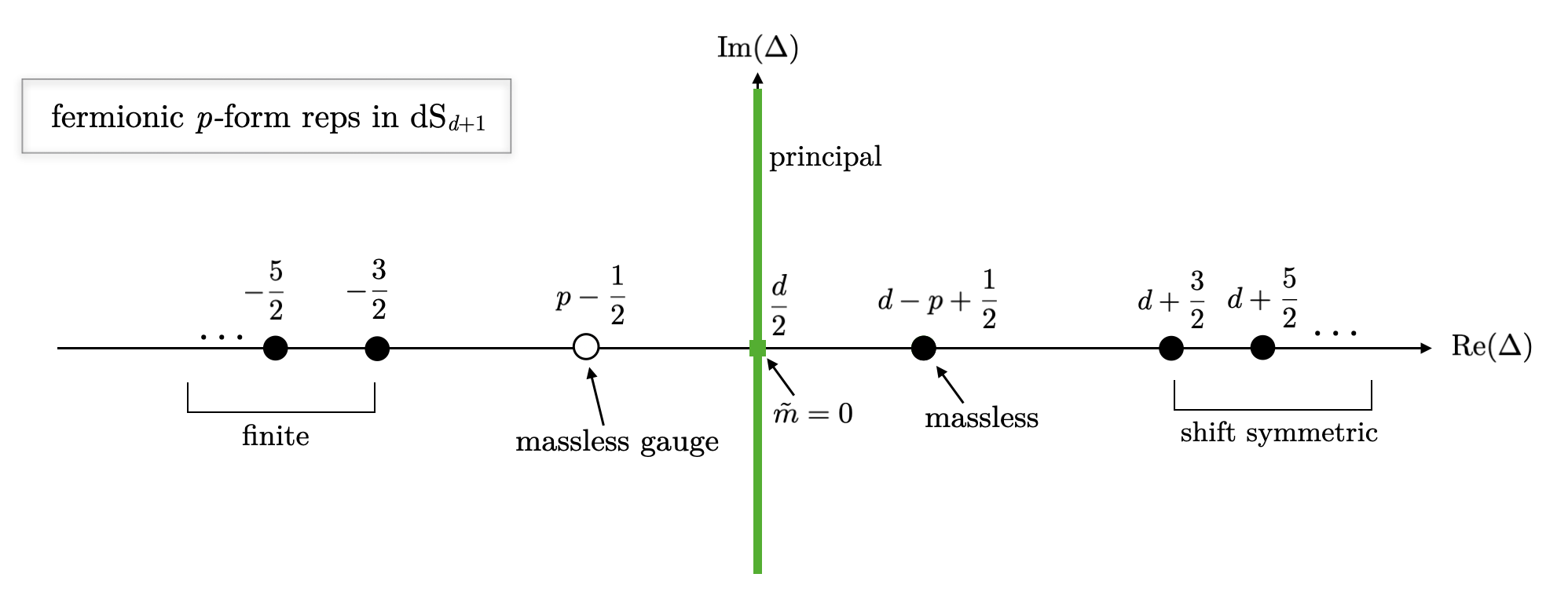,width=6.0in}}  \label{spinpfermioncontent}\, \ee
The $\tilde m=0$ point at $\Delta=d/2$ is indicated with a square and the round dots are the special points where the rep becomes shortened. Points in green are unitary reps.  The hollow black circle is the gauge mode rep that is equivalent to reps already accounted for among the lower $p$ reps due to \eqref{lowepsinfode1},\eqref{fermpformisofsee}. 

 For $D$ odd, the reps come in chiral pairs, so each point except $\Delta=d/2$ represents two reps, but there is also the equivalence \eqref{pformeejfee2} given by reflecting through $\Delta=d/2$ and flipping the helicity (except at the reducible points), and the point at $\Delta=d/2$ represents a single rep.  

 For $D$ even, each point except $\Delta=d/2$ represents a single rep, but there is also the equivalence \eqref{pformeejfee} given by reflecting through $\Delta=d/2$ (except at the reducible points), and the point at $\Delta=d/2$ represents two different chiral reps.  
 
The value of the Casimir operator on the fermionic $p$-form reps is
\be {\cal C}_2  =-{\tilde m^2\over H^2}- {1\over 8}D(D-1)+p(D-p) \nn\\ 
= \Delta(\Delta-d)+{1\over 8}d(d-1)+p(D-p+1)\,. \ee

Unlike bosonic $p$-forms, for the fermionic $p$-forms there are no dualities or further splittings to worry about.  If $p>{d\over 2}$ then there is no non-trivial fermionic $p$-form field on the sphere, due to the gamma-tracelessness constraints.  Furthermore, when $d$ is even and $p={d\over 2}$, the fermionic field is already split into chiral halves and there is no further splitting into self-dual or anti-self-dual parts (see the comment at the end of section \ref{soDevensubsection}).  The same applies to the massless rep when $d$ is odd and $p={d\over 2}-\half$.

There is one triviality that occurs: when $D$ is odd, $d$ even, the massless fermionic ${d\over 2}$-form reps become trivial because the tableau in the first row of \eqref{spinpfermioncontent2}, which make up the massless reps \eqref{masslsesspforodddrepe}, all become trivial.

\subsection{Mixed symmetry fermionic representations\label{mixedfermionsubsec}}

Finally, we consider the most general mixed symmetry fermionic reps.  The representation space is the space of spinor-tensors with the tensor indices in a general tableau as detailed below, and the relation between the mass and $\Delta$ is given by \eqref{mixedsymmasmrelefe}.

\subsubsection*{Even $D$}

The representation space for the general fermionic mixed symmetry reps in even $D$, odd $d$, is the space of gamma-traceless spinor-tensors on ${\mathbb S}^d$, $\psi_{i_1\ldots i_r}$, where the tensor indices have the index symmetries of a $p$-row tableau $[s_1,\ldots,s_p]$, $r=\sum_{i=1}^p s_i$.  It transforms under $\frak{so}(1,D)$ as in \eqref{latetimeshesactiont}.   These are fully gamma-traceless ($\gamma^{i_i}\psi_{i_1\ldots i_r}=0$ and similarly for all the other indices)  which also implies that they are fully traceless in the tensor indices.
These conditions ensure that the tensor is algebraically irreducible in odd $d$ (whereas in even $d$ there will be a further chiral splitting).  Call this space ${\cal F}^{[s_1,\ldots,s_p]_\half}_\Delta$,
\bea &&{\cal F}^{[s_1,\ldots,s_p]_{\half}}_\Delta:\ \  \text{complex gamma-traceless Dirac spinor-tensors of type } [s_1,\ldots,s_p] {\rm\ on\ } {\mathbb S}^d \,.\label{mixshgoddsp1e}\nn\\
\eea

The SVT-like decomposition for the mixed symmetry fermion field of type $[s_1,\ldots,s_p]_\half$ proceeds analogously to the bosonic case in section \ref{mixsymsec}.  The transverse-traceless (TT) pieces are gamma-traceless and $D_i$-transverse spinor-tensors whose tensor indices have the same index symmetries as their counterparts in the bosonic TT decomposition of $[s_1,\ldots,s_p]$.  The decomposition itself proceeds as in the example in \eqref{21exdecomde} and the general explanation above it, with $\nabla\rightarrow D$, and with gamma-tracelessness imposed everywhere instead of just tracelessness.

Each TT field is then expanded in spherical harmonics.  The harmonics have an $\frak{so}(D)$ content that we deduce by ``capping'' the rep of the TT tensor: if the TT tensor is $[u_1,\ldots,u_q]_\half$, then the harmonics $Y^{\pm}_{lm,i_1\ldots}$ carry the $\frak{so}(D)$ reps $[l-\half,u_1,\ldots,u_q]_{\pm\half}$ with $l=u_1+\half,u_1+{3\over 2},\ldots$.  These harmonics are eigenfunctions of the Dirac operator with the following purely imaginary eigenvalues \cite{BRANSON1992314},
\be  \fdag{{ D}} Y^{\pm}_{lm,i_1\ldots}= \pm i \left(l-{1\over 2}+{d\over 2}\right)  Y^{\pm}_{lm,i_1\ldots}\, . \ee
The natural Laplacian on the space of fermionic TT fields is given by the action of the quadratic Casimir operator $C_2=-{1\over 2}{\mathbb L}_{{\cal M}_{IJ}} {\mathbb L}_{{\cal M}^{IJ}}$ of the rotation algebra $\frak{so}(D)$, which is given in terms of the bare Laplacian by 
\be C_2= -D^2+C_2^{[u_1,\ldots, u_{q}]_\half,\, \frak{so}(d)}\,, \label{mixedcafsimire}\ee  
where $C_2^{[u_1,\ldots, u_{q}]_\half, \,\frak{so}(d)}$ is the value of the quadratic Casimir operator in \eqref{quadcasmideweaee} for the $\frak{so}(d)$ rep $[u_1,u_2,\ldots, u_{q}]_\half$ in which the field indices live.
The harmonics are eigenfunctions of this spinor Laplacian with the following eigenvalues \cite{BRANSON1992314},
\be C_2 Y^{\pm}_{lm,i_1\ldots} = C_2^{[l-\half,u_1,\ldots, u_{q}]_{\pm \half}, \,\frak{so}(D)} Y^{\pm}_{lm,i_1\ldots} \,,\ee
where $C_2^{[l-\half,u_1,\ldots, u_{q}]_{\pm \half}, \,\frak{so}(D)}$ is the value of the quadratic Casimir operator in \eqref{quadcasmideweaee} for the $\frak{so}(D)$ reps $[l-\half,u_1,\ldots, u_{q}]_{\pm\half}$ in which the harmonics transform (this value is the same for both the $+$ and $-$ chiralities).  These spinor-tensor spherical harmonics $Y^{\pm}_{lm,i_1\ldots}$ taken together form a basis of the space of TT spinor-tensors of type $[u_1,\ldots, u_q]_\half$ on ${\mathbb S}^d$. 

As in the bosonic case, some of the lower $l$ harmonics of the lower rank TT spinor-tensors, those with $l-\half <u_1$, will vanish in the TT decomposition, because they are annihilated by the combinations of derivatives that appear in the SVT-like decomposition into TT fields.  The rule for which to remove is exactly analogous to the bosonic case: we remove all the ones whose tensor indices are ones that would have been removed in the corresponding bosonic case.  

The final list of $\frak{so}(D)$ reps appearing in ${\cal F}^{[s_1,\ldots,s_p]_\half}_\Delta$ thus looks identical to that appearing in ${\cal F}^{[s_1,\ldots,s_p]}_\Delta$, where each $\frak{so}(D)$ rep is turned into its fermionic version and doubled to include both chiralities.

\textbf{Reducible cases:} The various reducibility conditions mirror precisely the structure of the corresponding bosonic rep ${\cal F}^{[s_1,\ldots,s_p]}_\Delta$:  in a given shortened multiplet, for each $\frak{so}(D)$ rep absent in the bosonic shortened rep, the two corresponding fermionic $\frak{so}(D)$ reps are also absent.  The values of $\Delta$ where the shortenings happen are as follows:
\begin{itemize}

\item 
Shift symmetric points: 
\be \Delta=d+s_1+k+{1\over 2}\, , \ \  k=0,1,2,\ldots \ .\ee 
The sub-reps at these points are called 
\be  {\cal D}^{[s_1,\ldots, s_p]_\half}_{d+s_1+k+{1\over 2}} \,. \ee
These correspond to the general mixed symmetry shift symmetric fermion fields \cite{Bonifacio:2023prb}.

\item
Finite points:
\be \Delta=-s_1-k-{1\over 2}\, , \ \  k=0,1,2,\ldots\ . \ee
The sub-reps at these points are called 
\be  {\cal S}^{[s_1,\ldots, s_p]_\half}_{-s_1-k-{1\over 2}}\, .\ee
These are finite dimensional reps, they are the spinor reps of $\frak{so}(1,D)$; the specific spinor-tensor that carries it is the one with the most number of boxes in the ${\cal S}^{[s_1,\ldots, s_p]_\half}_{-s_1-k-\half}$ rep, i.e. $[s_1+k,s_1,s_2,\ldots, s_p]_\half$ (we can see that all the spinor reps of $\frak{so}(1,D)$ are accounted for in this way).  Branching an $\frak{so}(1,D)$ tensor with this tableau to $\frak{so}(D)$ using the rules in appendix \ref{branchingappendix} gives the $\frak{so}(D)$ tensors in ${\cal S}^{[s_1,\ldots, s_p]_\half}_{-s_1-k-\half}$.  

\item
PM point:
\be \Delta= d - q + s_{q} - t +\half \,,\ \   t=1,2,\ldots,s_{q}-s_{q+1}\,. \ee
These are the fermionic PM points where $t$ boxes of the $q$-th row are activated.  The sub-reps are called
\be  {\cal V}^{[s_1,\ldots, s_p]_\half}_{ d - q + s_{q} - t +\half } \,. \ee
These correspond to the physical modes of the fermionic PM mixed symmetry fields.

\item
Gauge points:
\be \Delta=q - s_{q} + t -\half  \,, \ \  t=1,2,\ldots,s_{q}-s_{q+1} \,.\ee
These are the points corresponding to the gauge modes of the PM fermion where $t$ boxes of the $q$-th row are activated.  The sub-reps are called
\be  {\cal U}^{[s_1,\ldots, s_p]_\half}_{q - s_{q} + t -\half } \,. \ee
These correspond to the physical modes of the fermionic partially massless mixed symmetry fields.

\item
Chiral point:
\be \Delta={d\over 2}\,.\ee
Here the rep splits into chiral halves, those with $\half$ and those with $-\half$ chiralities in all their $\frak{so}(D)$,
\be {\cal F}^{[s_1,\ldots, s_p]_\half}_{d\over 2}= {\cal F}^{[s_1,\ldots, s_p]_{\half},+}_{d\over 2} \oplus {\cal F}^{[s_1,\ldots, s_p]_{\half},-}_{d\over 2}\,.\ee
This is the $\tilde m=0$ fermionic field, which acquires a chiral symmetry in even $D$.

\end{itemize}

\textbf{Equivalences:} There is a shadow equivalence that connects the $\Delta$ and $\bar\Delta\equiv d-\Delta$ reps, 
\be {\cal F}_{\Delta}^{[s_1,\ldots,s_p]_\half}\simeq {\cal F}_{\bar\Delta}^{[s_1,\ldots,s_p]_\half}\,, \label{pfogrmeejfee}\ee
realized by intertwiner operators $S_\Delta^{[s_1,\ldots,s_p]_\half} :\  {\cal F}_{\Delta}^{[s_1,\ldots,s_p]_\half}\rightarrow {\cal F}_{\bar\Delta}^{[s_1,\ldots,s_p]_\half}$.
For generic $\Delta$, this operator is invertible, giving \eqref{pfogrmeejfee}, but for the special shift symmetric, finite, PM and gauge values of $\Delta$ indicated above where ${\cal F}_{\Delta}^{[s_1,\ldots,s_p]_\half}$ becomes reducible but indecomposable and develops a sub-rep, $S_\Delta^{[s_1,\ldots,s_p]_\half}$ develops a kernel which is always precisely the sub-rep.

The shift symmetric and finite points are linked to each other via the maps $S^{[s_1,\ldots,s_p]_\half}_{-s_1-k-{1\over 2}}$, ${S}^{[s_1,\ldots,s_p]_\half}_{d+s_1+k+{1\over 2}}$, where the kernel and image of each are the sub-reps ${\cal S}^{[s_1,\ldots,s_p]_\half}_{-s_1-k-{1\over 2}}$, ${\cal D}^{[s_1,\ldots,s_p]_\half}_{d+s_1+k+{1\over 2}}$, which induces the  isomorphisms
\be {\cal S}^{[s_1,\ldots,s_p]_\half}_{-s_1-k-{1\over 2}}\simeq {\cal F}^{[s_1,\ldots,s_p]_\half}_{d+s_1+k+{1\over 2}}/{\cal D}^{[s_1,\ldots,s_p]_\half}_{d+s_1+k+{1\over 2}}\,,\ \ \  {\cal D}^{[s_1,\ldots,s_p]_\half}_{d+s_1+k+{1\over 2}}\simeq {\cal F}^{[s_1,\ldots,s_p]_\half}_{-s_1-k-{1\over 2}}/{\cal S}^{[s_1,\ldots,s_p]_\half}_{-s_1-k-{1\over 2}} \,.\ee

The PM and gauge points are linked to each other via the maps $S^{[s_1,\ldots,s_p]_\half}_{q - s_{q} + t -\half  }$, ${S}^{[s_1,\ldots,s_p]_\half}_{d - q + s_{q} - t +\half }$, where the kernel and image of each are the sub-reps ${\cal U}^{[s_1,\ldots,s_p]_\half}_{q - s_{q} + t -\half  }$, ${\cal V}^{[s_1,\ldots,s_p]_\half}_{d - q + s_{q} - t +\half }$, which induces the following isomorphisms
\be {\cal U}^{[s_1,\ldots,s_p]_\half}_{q - s_{q} + t -\half  }\simeq {\cal F}^{[s_1,\ldots,s_p]_\half}_{d - q + s_{q} - t +\half }/{\cal V}^{[s_1,\ldots,s_p]_\half}_{d - q + s_{q} - t +\half }\,,\ \ \  {\cal V}^{[s_1,\ldots,s_p]_\half}_{d - q + s_{q} - t +\half }\simeq {\cal F}^{[s_1,\ldots,s_p]_\half}_{q - s_{q} + t -\half  }/{\cal U}^{[s_1,\ldots,s_p]_\half}_{q - s_{q} + t -\half  }\,.\ee

There are generalized gradient and generalized divergence operators that move along a fermionic generalized chain complex.  This complex has the same structure as its bosonic counterpart, with every bosonic rep replaced by its fermionic counterpart.  They are constructed as described in section \ref{mixsymsec}, with derivatives replaced by spin covariant derivatives and tracelessness replaced by gamma-tracelessness.
 These generalized gradient and divergence operators, along with the shadow equivalence operators, group together PM reps, those of its gauge parameters and reducibility parameters, and their shadow reps, into a two row commutative diagram exactly analogous to \eqref{mixedcommdiagram}, and giving isomorphisms between the PM reps that are the fermionic versions of \eqref{mixsyisomfsfee}.

\subsubsection*{Odd $D$:}

For odd $D$, even $d$, the $\gamma_\ast$ operator on ${\mathbb S}^d$ splits the function space of mixed symmetry Dirac spinor-tensors into two spaces of opposite chirality, those with eigenvalues $\pm 1$ under the action of $\gamma_\ast$.  Each of these subspaces forms a separate rep, and we call them ${\cal F}^{[s_1,\ldots,s_p]_{\pm{1\over 2}}}_{\Delta}$,
\bea {\cal F}^{[s_1,\ldots,s_p]_{\pm\half}}_\Delta:\  && \text{complex gamma-traceless Dirac spinor-tensors of type\ } [s_1,\ldots,s_p] {\rm\ on\ } {\mathbb S}^d  \nn\\
&&{\rm satisfying \ } \gamma_\ast \psi_{i_1\ldots} =\pm \psi_{i_1\ldots}  \,.\label{mixshgoddspe}
\eea

The reduction into $\frak{so}(D)$ reps works analogously to the odd $D$ fermionic cases in the previous sections: we use the generalized spinor SVT decomposition described above to break the space into TT pieces, then split each TT piece into its spherical harmonics.  The two sets of harmonics $Y_{lm,i_1\ldots }^{\pm}$ now describe the same rep under $\frak{so}(D)$ rotations, namely the $[l-\half,u_1,\ldots, u_q]_\half= (l ,u_1+\half ,\ldots ,u_{q}+\half,1/2,\ldots, 1/2)$ rep, and the $\gamma_\ast$ matrix serves as an intertwiner that maps them into each other.  Diagonalizing $\gamma_\ast$ then gives two sets, $\tilde Y_{lm,i_1\ldots }^{\pm}$, which span the two spaces \eqref{mixshgoddspe}.   

These two spaces have the same $\frak{so}(D)$ content, but they form distinct $\frak{so}(1,D)$ reps.  They correspond to the two possible chiralities of a massive mixed symmetry fermion on dS$_D$ for odd $D$, distinguished by the sign of the mass term in the Dirac equation on dS$_D$.  The $\frak{so}(D)$ content of each can be deduced from the corresponding bosonic rep ${\cal F}^{[s_1,\ldots,s_p]}_\Delta$ by taking each tableau that appears and turning it into its unique fermionic version: $[\ldots]\rightarrow  [\ldots]_{{1\over 2}}$.

\textbf{Reducible cases:} The various reducibility cases mirror those of the corresponding bosonic rep ${\cal F}^{[s_1,\ldots,s_p]}_\Delta$:  in a given shortened multiplet, for each $\frak{so}(D)$ rep absent in the bosonic shortened rep, the corresponding fermionic $\frak{so}(D)$ reps are also absent.  The values of $\Delta$ where the shortenings happen are as follows:
\begin{itemize}

\item 
Shift symmetric points: 
\be \Delta=d+s_1+k+{1\over 2}\, , \ \  k=0,1,2,\ldots \, .\ee
The sub-reps are called
\be  {\cal D}^{[s_1,\ldots, s_p]_{\pm \half}}_{d+s_1+k+{1\over 2}} \, . \ee
These correspond to the chiral parts of the mixed symmetry shift symmetric fermion fields in odd $D$.

\item
Finite points:
\be \Delta=-s_1-k-{1\over 2}\, , \ \  k=0,1,2,\ldots\ . \ee
The sub-reps are called
\be  {\cal S}^{[s_1,\ldots, s_p]_{\pm \half}}_{-s_1-k-{1\over 2}} \, .\ee
These are finite dimensional reps, they are the spinor reps of $\frak{so}(1,D)$; the specific tensor is the one with the most number of boxes in either of the ${\cal S}^{[s_1,\ldots, s_p]_{\pm\half}}_{-s_1-k-\half}$ reps, i.e. the tableau $[s_1+k,s_1,\ldots, s_p]_\half$ (we can see that all the spinor reps of $\frak{so}(1,D)$ are accounted for in this way).  Branching an $\frak{so}(1,D)$ spinor-tensor with this tableau to $\frak{so}(D)$ using the rules in Appendix \ref{branchingappendix} gives the $\frak{so}(D)$ spinor-tensors in ${\cal S}^{[s_1,\ldots, s_p]_{\pm\half}}_{-s_1-k-{1\over 2}}$.  

\item
PM points:
\be \Delta= d - q + s_{q} - t +\half \,,\ \ t=1,2,\ldots,s_{q}-s_{q+1}\,. \ee
These are the PM points where $t$ boxes of the $q$-th row are activated.  The sub-reps are called
\be  {\cal V}^{[s_1,\ldots, s_p]_{\pm \half}}_{d - q + s_{q} - t +\half} \,. \ee
These correspond to the chiral parts of the PM shift symmetric fermion fields in odd $D$.

\item
Gauge points:
\be \Delta=q - s_{q} + t -\half  \,, \ \ t=1,2,\ldots,s_{q}-s_{q+1}\,. \ee
These are the points corresponding to the gauge modes of the PM fermion where $t$ boxes of the $q$-th row are activated.  The sub-reps are called
\be  {\cal U}^{[s_1,\ldots, s_p]_{\pm \half}}_{q - s_{q} + t -\half} \,. \ee
These correspond to the gauge modes of the chiral parts of the fermionic PM fields in odd $D$.

\end{itemize}

\textbf{Equivalences:}  The shadow equivalence connecting the $\Delta$ and $\bar\Delta\equiv d-\Delta$ reps in odd $D$ involves a flip in the chirality,
\be {\cal F}_{\Delta}^{[s_1,s_2,\ldots, s_p]_{\pm \half}}\simeq {\cal F}_{\bar\Delta}^{[s_1,s_2,\ldots, s_p]_{\mp \half}}\,.\label{pforfmeffqwejfee}\ee

For the special values of $\Delta$ indicated above where the reps develop sub-reps and become reducible but not decomposable, the equivalence map develops a kernel which is always precisely the sub-rep. 
The shift symmetric and finite points are linked to each other, inducing the following isomorphisms which flip the helicity,
\be {\cal S}^{[s_1,\ldots,s_p]_{\pm\half}}_{-s_1-k-{1\over 2}}\simeq {\cal F}^{[s_1,\ldots,s_p]_{\mp\half}}_{d+s_1+k+{1\over 2}}/{\cal D}^{[s_1,\ldots,s_p]_{\mp\half}}_{d+s_1+k+{1\over 2}}\,,\ \ \  {\cal D}^{[s_1,\ldots,s_p]_{\pm\half}}_{d+s_1+k+{1\over 2}}\simeq {\cal F}^{[s_1,\ldots,s_p]_{\mp\half}}_{-s_1-k-{1\over 2}}/{\cal S}^{[s_1,\ldots,s_p]_{\mp\half}}_{-s_1-k-{1\over 2}} \,.\ee
The massless and gauge points are similarly linked to each other, inducing the following isomorphisms which flip the helicity,
\be {\cal U}^{[s_1,\ldots,s_p]_{\pm\half}}_{q - s_{q} + t -\half  }\simeq {\cal F}^{[s_1,\ldots,s_p]_{\mp\half}}_{d - q + s_{q} - t +\half }/{\cal V}^{[s_1,\ldots,s_p]_{\mp\half}}_{d - q + s_{q} - t +\half }\,,\ \ \  {\cal V}^{[s_1,\ldots,s_p]_{\pm\half}}_{d - q + s_{q} - t +\half }\simeq {\cal F}^{[s_1,\ldots,s_p]_{\mp\half}}_{q - s_{q} + t -\half  }/{\cal U}^{[s_1,\ldots,s_p]_{\mp\half}}_{q - s_{q} + t -\half  }\,.\ee

The fermionic $\rd$ and $\rd^\dag$ operators preserve the chirality and map along the fermion generalized chain complex.  We get the same pattern of mappings linking the spaces as we did in the odd $d$ case.   The only difference is that the shadow equivalence \eqref{pforfmeffqwejfee} flips the chirality, so in the large commutative diagram interlinking all of these reps analogous to \eqref{mixedcommdiagram}, the bottom row will have a helicity opposite to the top row, and the sequence of isomorphisms analogous to \eqref{mixsyisomfsfee} will involve a flip of helicity.  This expresses the fact that a massless fermionic PM field has as its gauge modes a fermionic field of opposite chirality, its gauge-for-gauge modes a fermionic field with the same chirality, and so on all the way down.

 \subsubsection*{Unitarity and summary}

The only inner product that can be positive definite and invariant is the standard Dirac inner product,
\be \la \psi_1|\psi_2\ra= {1\over c_1!c_2!\cdots }\int d^d\Omega\, \psi_1^{i_1\ldots  \dag}\psi_{2\ i_1\ldots }\,,  \label{spinorinnerpproddemds2} \ee
where $c_1,c_2,\ldots$ in the normalization are the lengths of the columns of the tableaux $[s_1,\ldots,s_p]$.
It is positive definite and invariant when $\Delta={d\over 2}+i\nu $ with $\nu\in {\mathbb R}$; this is the generalized fermionic principal series.  

For $D$ even, there is one rep at each $\nu\not=0$ and the reps at $\nu>0$ and $\nu<0$ are equivalent.  The rep at $\nu=0$, corresponding to the case $\tilde m=0$, splits into two inequivalent chiral pieces.  For this reason, in even $D$ the point $\nu=0$ is usually excluded from the principal series and is considered a part of the discrete series, as we'll see in section \ref{unitarylistsection}.  

For odd $D$, there are two different chiral reps for each $\nu\not=0$, and the equivalence \eqref{pforfmeffqwejfee} relates one chirality at $\nu>0$ to the other at $\nu<0$, whereas for the massless point at $\nu=0$, there is a single rep.

There is no complementary series.  In addition, the shortened reps at the shift symmetric and massless points are not unitary, with one exception:  for $D$ even, the fermionic PM reps with a non-vanishing $(D-2)/2$-th row in which boxes in the bottom row are activated, are unitary \cite{10.1063/1.1665471}.  These are part of the fermionic discrete series reps, as discussed further in section \ref{unitarylistsection}.  Other than these and the principal series, no reps are unitary.

The reps are summarized here:
\be \raisebox{-40pt}{\epsfig{file=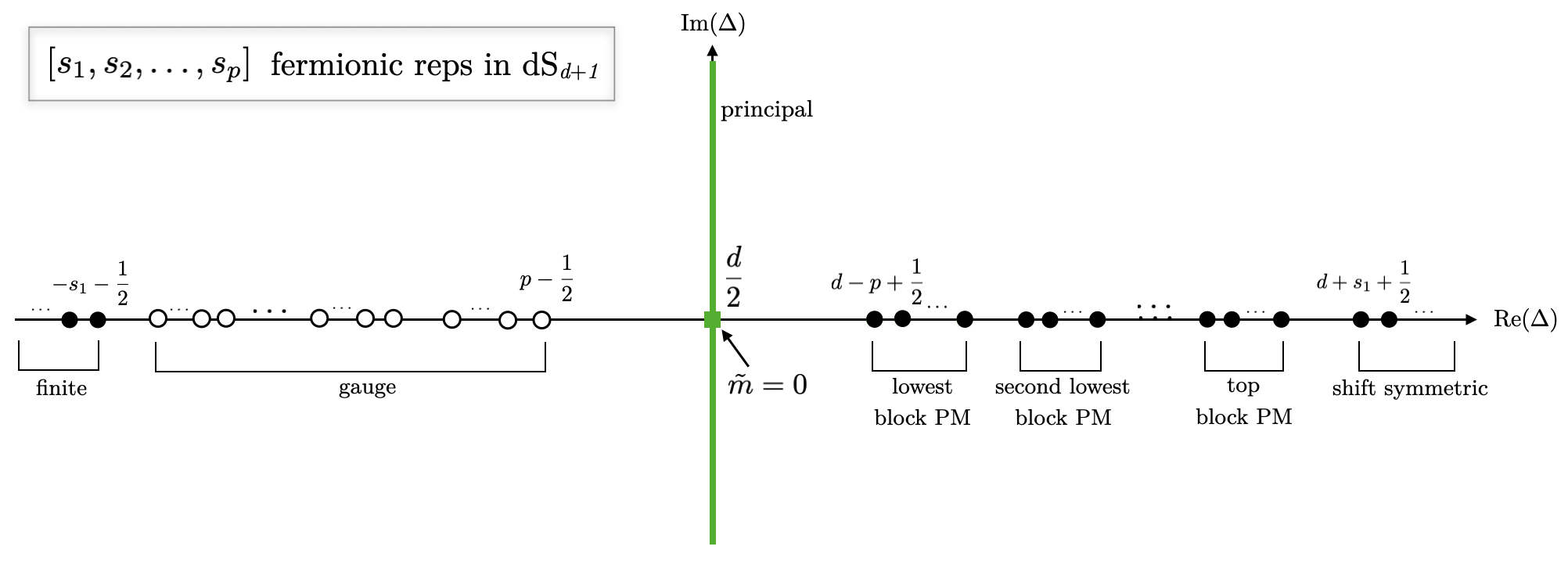,width=6.5in}}  \label{dsrepsfermionmixed}\, \ee
The $\tilde m=0$ point at $\Delta=d/2$ is indicated with a square and the round dots are the special points where the rep becomes shortened. Points in green are unitary reps.  The hollow black circles are the gauge mode reps that are already accounted for through equivalences to lower rank reps.  

 For $D$ odd, the reps come in chiral pairs, so each point except $\Delta=d/2$ represents two reps, but there is also the equivalence \eqref{pforfmeffqwejfee} given by reflecting through $\Delta=d/2$ and flipping the helicity (except at the special points), and the point at $\Delta=d/2$ represents a single rep.  

 For $D$ even, each point except $\Delta=d/2$ represents a single rep, but there is also the equivalence \eqref{pfogrmeejfee} given by reflecting through $\Delta=d/2$ (except at the special points), and the point at $\Delta=d/2$ represents two different chiral reps.  
 
 The value of the Casimir operator on the general mixed symmetry fermionic reps is given by \eqref{c2gnsmdees2eoee}.

For the mixed symmetry fermionic fields, there are no further splittings or equivalences to worry about: if $p>{d\over 2}$ then there is no non-trivial fermionic $[s_1,\ldots, s_p]$ type spinor-tensors on ${\mathbb S}^d$, due to the gamma-tracelessness constraints.  Furthermore, when $d$ is even and $p={d\over 2}$, the fermionic field is already split into chiral halves and there is no further splitting into self-dual or anti-self-dual parts (see the comment at the end of section \ref{soDevensubsection}).  The same applies to the PM reps when $d$ is odd and $p={d\over 2}-\half$.  

There is one further feature involving the non-unitary fermionic PM fields in odd $D$: if we have an $n$-row rep $[s_1,\ldots,s_n]$, then its PM points with any number of boxes in the last row activated are all trivial, i.e. all of the $\frak{so}(D)$ tableaux in the rep become empty and the bulk field has no propagating degrees of freedom.

\section{Summary of the Unitary Representations\label{unitarylistsection}}

In sections \ref{bosonsection} and \ref{fermionsection}, we went through all the canonical field types on dS$_D$ and the $\frak{so}(1,D)$ reps they carry.  Most of these reps were non-unitary, but in each case some were unitary, and the unitary ones occurred in several different ways:  in all cases there was a continuous family of reps with $\Delta={d\over 2}+i\nu$, $\nu\in {\mathbb R}$ that was called the principal series. In some cases there was an additional continuous family of unitary reps with real $\Delta$ called the complementary series.  And on top of these, there were various discrete real values of $\Delta$ where the reps developed sub-reps which could be factored out, leaving a unitary rep.

In this section, we present the complete, but abstract, classification of the unitary reps of $\frak{so}(1,D)$ for all $D\geq 3$.   This is the list of reps that is often seen in references when the de Sitter reps are reviewed, and the notation used is different to the notation encountered in the above sections.  After reviewing this list, we will see how it includes all the unitary reps we have encountered for the fields in sections \ref{bosonsection}, \ref{fermionsection}, and adds no others, so that the field construction accounts for all the possible unitary reps.

The unitary reps of $\frak{so}(1,D)$ are grouped into four categories\footnote{The classification of unitary reps into these four categories is a general property of all the non-compact simple Lie algebras \cite{df9c2220-c202-37d9-91c2-141db7742d56,bookKnapp}.   By a criterion of Harish-Chandra, the discrete series occurs if and only if the rank of the maximal compact algebra is equal to the rank of the full algebra, which is the case for $\frak{so}(1,D)$ only when $D$ is even (an equivalent criterion is that the algebra has a compact Cartan subalgebra)  \cite{8dbeff43-21cd-3076-8ad8-66c4a6d90144}.
}:
\vspace{.3cm}
\begin{center}
\begin{minipage}{0.4\textwidth} 
\begin{enumerate}
    \item \textit{principal series,}
    \item \textit{complementary series,}
    \item \textit{exceptional series,}
    \item \textit{discrete series.}
\end{enumerate}
\end{minipage}
\end{center}
\vspace{0.5cm}
Of these four categories, the first three occur in all $D$, and the last one, the discrete series, occurs only in even $D$.  The reps are all infinite dimensional; the only finite dimensional unitary rep is the trivial rep, which is not included in this classification and is absent from the lists given below.

The principal and complementary series will be the same as described in the sections above (with one small exception for even $D$).  We will see that the exceptional and discrete series describe the remaining unitary reps corresponding to the shift symmetric scalars and the unitary partially massless fields.

In listing the reps, we also give the rules for finding their $\frak{so}(D)$ content.  A remarkable feature of the $\frak{so}(D)$ content (which is also true for the non-unitary reps we have seen) is that there is never any multiplicity: any given $\frak{so}(D)$ rep appears in a $\frak{so}(1,D)$ rep at most once.  However, the $\frak{so}(D)$ content does not characterize a rep: there are plenty of distinct $\frak{so}(1,D)$ reps with the same $\frak{so}(D)$ content (e.g. two principal series reps for the same field type with different values of $\Delta$).

The pattern of reps behaves differently for odd $D$ and even $D$, so we treat these cases separately\footnote{A similar summary can be found in e.g. section 3.3 of \cite{Letsios:2023voc}, which also gives the relation to the labels of \cite{10.1063/1.1665471} and other references.  Note that there can be some differences between authors about what series some edge cases are assigned to, see e.g. the note in section 3.3 of \cite{Letsios:2023voc}.}.  We restrict to $D\geq 3$, again leaving the $D=2$ case to section \ref{D2section}.

\subsection{Odd $D$\label{oddDsummarysec}}

We first list the unitary $\frak{so}(1,D)$ representations for odd $D$.  Let
\be D=2n+1\, , \ \ \ n=1,2,3,\ldots\ .\ee
The reps are specified by the following labels\footnote{These labels are the analog of the $f$ labels for ${\frak so}(D)$ in appendix \ref{sorepsappendix}.}:
\be \left(F_0,F_1,\ldots,F_n\right)\,, \ee
where $F_1,\ldots,F_n$ are either all integers (bosonic rep) or all odd half-integers (fermionic rep), and satisfy
\be F_1\geq F_2\geq\cdots\geq F_n\geq 0\, .\ee

The $\frak{so}(D)$ content of a rep $\left(F_0,F_1,\ldots,F_n\right)$ consists of all $\frak{so}(D)$ reps $(f_1,\ldots,f_n)$ (in the labelling of section \ref{soDoddsubsection}) such that
\be f_1\geq F_1\geq f_2\geq F_2\geq\cdots \geq f_n\geq F_n\geq 0\, ,\label{socontentDoddeq1}\ee
and the additional condition that $(f_1,\ldots,f_n)$ is bosonic or fermionic according to whether the $\frak{so}(1,D)$ rep is bosonic or fermionic.   For the exceptional series, there is a further restriction as noted below.

\begin{itemize}

\item{\bf Principal series:} \
\be F_0=-n+i\nu\,,\ \ \  \begin{cases} \nu \geq 0 & {\rm if} \ F_n=0 \, ,\\
\nu\in {\mathbb R} &{\rm if} \  F_n>0\, . 
\end{cases} \ee

\item{\bf Complementary series:} \
\be -n < F_0<-p\, ,\ \ \ p=0,1,2,\ldots,n-1\,.\ee
For each $p$, these occur only for the bosonic reps with
\be  F_{1},\ldots,F_p \ {\rm all }\ >0 \,, \ \ \ F_{p+1},\ldots,F_{n}=0\, .\ee

\item{\bf Exceptional series:} \
\be  F_0=-p\, ,\ \ \ p=1,2,\ldots,n-1\,.\ee
These exist only for $n\geq 2$.  For each $p$, these occur only for the bosonic reps with
\be  F_{1},\ldots,F_p \ {\rm all }\ >0 \,, \ \ \ F_{p+1},\ldots,F_{n}=0\, .\ee
For the $\frak{so}(D)$ content, we must further restrict to only those that have
\be f_{p+1}=0\, .\label{soconexddedxd}\ee

\end{itemize}

The relation between these abstractly described reps and the unitary reps we have seen in the previous sections is as follows.  For the principal series with $F_n=0$, the rep is bosonic and $\left(F_0,F_1,\ldots,F_{n-1},0\right)$ corresponds to the rep ${\cal F}_\Delta^{[F_1,\ldots,F_{n-1}]}$ we have described in section \ref{bosonsection}, where the label $F_0$ corresponds to $\Delta$ through 
\be \Delta=F_0+2n\,.\label{F0deltaregle}\ee
The restriction $\nu\geq 0$ accounts for the shadow transform equivalence between the rep with $\Delta$ and that with $\bar\Delta\equiv d-\Delta$.
 For the principal series with $F_n>0$, $\nu$ is allowed to run over all the reals and the reps can be bosonic or fermionic.  In the bosonic cases, $\nu>0$ and $\nu<0$ correspond to the two chiral halves ${\cal F}_\Delta^{[F_1,\dots,F_n]_\pm}$ of the reps due to the chiral splitting that occurs, and when $\nu=0$ it is the partially massless rep in which all the boxes in the bottom row are activated.  This is the point where the two reps ${\cal F}_{d\over 2}^{[F_1,\dots,F_n]_{\pm}}$, corresponding to the physical and gauge modes of the PM field, are equivalent, so there is only one rep at this point.  In the fermionic cases, $\nu>0$ and $\nu<0$ correspond to the two chiral parts of the massive fermions in odd $D$, i.e. the reps we called ${\cal F}_\Delta^{[F_1-\half,\dots,F_n-\half]_{\pm\half}}$ in section \ref{fermionsection}, and $\nu=0$ corresponds to the $\tilde m=0$ fermion at $\Delta=d/2$ where the two reps ${\cal F}_{d\over 2}^{[F_1-\half,\dots,F_n-\half]_{\pm\half}}$ are equivalent. 
 
The complementary series occur only when $F_n=0$, which are all bosonic reps. The label $p$ corresponds to the number of non-zero rows in the field's tableau.  The rep $\left(F_0,F_1,\ldots,F_{n-1},0\right)$ corresponds to the complementary series rep ${\cal F}_\Delta^{[F_1,\ldots,F_{n-1}]}$, with ${d\over 2}<\Delta=F_0+2n<d-p$, described in section \ref{bosonsection}.  The range of $F_0$ accounts for only the complementary series reps with $\Delta>{d\over 2}$, avoiding double counting by leaving out the equivalent ones with $\Delta<{d\over 2}$.  Note that the case $\nu=0$ has now been assigned to the principal series.

The exceptional series contains only bosonic reps and contains, as we will see now, the shift symmetric scalars and all the PM points that are unitary on dS$_D$ in odd $D$, with one exception.  They appear via the following pattern:
\begin{itemize}

\item[-] $p=1$ contains the reps $(F_0,F_1,\ldots)=(-1,k+1,0,\ldots,0)$, $k=0,1,2,\ldots$.  These are the shift symmetric scalars, occurring here with the weights appropriate to ${\cal U}^{[k+1]}_1$, which by \eqref{PMgagspindslinkeee} is the gauge rep of the spin $s$ PM field with all its boxes activated:  ${\cal U}^{[k+1]}_1\simeq {\cal F}^{[k+1]}_{d-1}/{\cal V}^{[k+1]}_{d-1}$.  It is equivalent via \eqref{spinsudisoeee} to the level $k$ shift symmetric scalar rep: ${\cal U}^{[k+1]}_1\simeq {\cal D}^{[0]}_{d+k}$.

\item[-] $p=2$ contains the reps $(F_0,F_1,F_2,\ldots)=(-2,s,s-t+1,0,\ldots,0)$, $s=1,2,\ldots$, $t=1,2,\ldots,s$.  These are the spin $s$, depth $t$ PM fields.  They occur with the weights appropriate to ${\cal U}^{[s,s-t+1]}_2$, which by \eqref{gesmfgixisogpmjdee} is the gauge mode rep for a two-row tableau $[s,s-t+1]$ PM field with all its bottom row boxes activated: ${\cal U}^{[s,s-t+1]}_2\simeq {\cal F}^{[s,s-t+1]}_{d-2}/{\cal V}^{[s,s-t+1]}_{d-2}$.  It is equivalent via \eqref{mixsyisomfsfee} to the spin $s$, depth $t$ PM rep: ${\cal U}^{[s,s-t+1]}_2\simeq {\cal V}^{[s]}_{d+s-t-1}$.

\item[-] $p=3$ contains the reps $(F_0,F_1,F_2,F_3,\ldots)=(-3,s_1,s_2,s_2-t+1,0,\ldots,0)$, $s_1\geq s_2\geq 1$, $t=1,2,\ldots,s_2$.  These are the two-row tableau $[s_1,s_2]$ PM fields of depth $t$, where the second row is activated. They occur with the weights appropriate to ${\cal U}^{[s_1,s_2,s_2-t+1]}_3$, which by \eqref{gesmfgixisogpmjdee} is the gauge mode rep for a three-row tableau $[s_1,s_2,s_2-t+1]$ PM field with all its bottom row boxes activated: ${\cal U}^{[s_1,s_2,s_2-t+1]}_3\simeq {\cal F}^{[s_1,s_2,s_2-t+1]}_{d-3}/{\cal V}^{[s_1,s_2,s_2-t+1]}_{d-3}$.  It is equivalent via \eqref{mixsyisomfsfee} to the $[s_1,s_2]$, depth $t$ PM rep with the second row activated: ${\cal U}^{[s_1,s_2,s_2-t+1]}_3\simeq {\cal V}^{[s_1,s_2]}_{d+s_2-t-2}$.

\item[\vdots]

\item[-] $p=n-1$ contains the reps $(F_0,F_1,\ldots,F_{n-1},F_n)=(-n+1,s_1,\ldots,s_{n-2},s_{n-2}-t+1,0)$, $s_1\geq s_2\geq\ldots \geq s_{n-2}\geq 1$, $t=1,2,\ldots,s_{n-2}$.   These are the $(n-2)$-row tableau $[s_1,\ldots,s_{n-2}]]$ PM fields of depth $t$, where the last row is activated.   They occur with the weights appropriate to ${\cal U}^{[s_1,\ldots,s_{n-2},s_{n-2}-t+1]}_{n-1}$, which by \eqref{gesmfgixisogpmjdee} is the gauge mode rep for a $(n-1)$-row tableau $[s_1,\ldots,s_{n-2},s_{n-2}-t+1]$ PM field with all its bottom row boxes activated: ${\cal U}^{[s_1,\ldots,s_{n-2},s_{n-2}-t+1]}_{n-1}\simeq {\cal F}^{[s_1,\ldots,s_{n-2},s_{n-2}-t+1]}_{d-n+1}/{\cal V}^{[s_1,\ldots,s_{n-2},s_{n-2}-t+1]}_{d-n+1}$.   It is equivalent via \eqref{mixsyisomfsfee} to the $[s_1,\ldots,s_{n-2}]$, depth $t$ PM rep with the last row activated: ${\cal U}^{[s_1,\ldots,s_{n-2},s_{n-2}-t+1]}_{n-1}\simeq {\cal V}^{[s_1,\ldots,s_{n-2}]}_{d-n+s_{n-2} -t+2}$.

\end{itemize}

This accounts for all of the exceptional series, but there are still two more sets of unitary PM points: those among the fields with $n-1$ rows, $[s_1,\ldots,s_{n-1}]$, where the last row is activated, and those among the fields with $n$ rows, $[s_1,\ldots,s_{n-1},s_n]$, where the last row is activated.  Consider first the $(n-1)$-row fields with the last row activated.  Through the dualities catalogued in \cite{Hinterbichler:2024vyv}, these are equivalent to $n$-row PM fields in which all the boxes in the last row are activated: the  $[s_1,\ldots,s_{n-1},s_n]$ PM rep with all $s_n$ boxes in its last row activated is equivalent to the $[s_1,\ldots,s_{n-1}]$ PM rep where $s_{n-1}-s_n+1$ boxes in its last row are activated.  This PM point is included in the principal series: it is the rep $[s_1,\ldots,s_n]$ at $\nu=0$, where $\Delta=d/2$ (there is no complementary series for these $n$-row tableaux).  All that remains are the $n$-row PM points with $<s_n$ boxes in its last row activated, but these, as mentioned in section \ref{mixsymsec}, are all trivial.  This accounts for all the unitary PM and shift symmetric points among the fields.\footnote{As was noted in \cite{Basile:2016aen}, in the exceptional series cases the shift symmetric or PM field is realized as a gauge point, and the full reducible representation in which the gauge point appears has the tensor structure of the invariant field strength of the shift symmetric or PM field.

This is realized naturally by the inner product \eqref{innerprodtprispcsvmdee}, where the gauge rep inner product is realized using the factor rep of the associated PM point.  For example, the photon rep ${\cal V}^{[1]}_{d-1}$ for $d\geq 4$ is realized via the 2-form rep ${\cal U}^{[1,1]}_2\simeq {\cal F}^{[1,1]}_{d-2}/{\cal V}^{[1,1]}_{d-2}$, so one can use the inner product \eqref{innerprodtprispcsvmdee} for $[1,1]$ fields with $\Delta=d-2$.}  (Note that in $D=3$, there is no exceptional series, and the PM and shift symmetric points are accounted for solely by the principal series, see section \ref{D3subsection}.)

In all the cases, one can check that the $\frak{so}(D)$ content of the rep as described in this section is the same as we have described it in the previous sections.

For all the reps, the value of the quadratic Casimir defined in section \ref{casimirsec} is given by
\be {\cal C}_2=\sum_{i=0}^n F_i\left(F_i+2n-2i\right)\,,\ee
which matches the values given in the above sections.

\subsection{Even $D$\label{evenDsummarysec}}

Now turn to the unitary $\frak{so}(1,D)$ representations for even $D\geq 4$.  Let
\be D=2n\, , \ \ \ n=2,3,4,\ldots\ .\ee
The reps carry labels $F_0,F_1,\ldots,F_{n-1}$,
\be \left(F_0,F_1,\ldots,F_{n-1}\right)\, ,\ee
where $F_1,\ldots,F_{n-1}$ are either all integers (bosonic rep) or all odd half-integers (fermionic rep), and satisfy
\be F_1\geq F_2\geq\cdots\geq F_{n-1}\geq 0\, .\ee

The $\frak{so}(D)$ content of a rep $\left(F_0,F_1,\ldots,F_{n-1}\right)$ consists of all $(f_1,\ldots,f_n)$ (in the labelling of section \ref{soDoddsubsection}) such that
\be f_1\geq F_1\geq f_2\geq F_2\geq\cdots \geq f_{n-1}\geq F_{n-1}\geq |f_n|\, ,\ee
and the additional condition that $(f_1,\ldots,f_n)$ is bosonic or fermionic according to whether the $\frak{so}(1,D)$ rep is bosonic or fermionic.   There are exceptions to this, occurring in the exceptional and discrete series, where the rep contains fewer of the $\frak{so}(D)$ reps, and these are noted below.

\begin{itemize}

\item{\bf Principal series:} \
\be F_0=-n+\half+i\nu\,,\ \ \  \nu> 0 \, . 
\ee

\item{\bf Complementary series:} \
\be -n+\half \leq F_0<-p\, ,\ \ \ p=0,1,2,\ldots,n-1\,.\ee
For each $p$, these occur only for the bosonic reps with
\be  F_{1},\ldots,F_p \ {\rm all }\ >0 \,, \ \ \ F_{p+1},\ldots,F_{n-1}=0\, .\ee

\item{\bf Exceptional series:} \
\be  F_0=-p\, ,\ \ \ p=1,2,\ldots,n-1\,.\ee
For each $p$, these occur only for the bosonic reps with
\be  F_{1},\ldots,F_p \ {\rm all }\ >0 \,, \ \ \ F_{p+1},\ldots,F_{n-1}=0\, .\ee
For the $\frak{so}(D)$ content, we must restrict to those reps that have
\be f_{p+1}=0\, .\ee 

\item{\bf Discrete series $D^\pm$:} \
These occur when $F_{n-1}>0$.  There are two chiral discrete series reps, $D^+$ and $D^-$, for each $F_0$ in the range
\be {1\over 2} \leq F_0+n \leq F_{n-1}\, ,\ee
where $F_0$ must be an integer for bosonic reps and a half-integer for fermionic reps.  

For the $\frak{so}(D)$ content, we must restrict to those with $f_n$ satisfying
\bea && {\rm for}\ D^+:\ \   F_0+n\leq f_n \leq F_{n-1}   \, ,\nn \\ 
&& {\rm for}\ D^-:\ \  -(F_0+n)\geq  f_n \geq  -F_{n-1} \,.
\eea

\end{itemize}

The relation between these abstractly described reps and the unitary reps we have seen in the previous sections is as follows.  For the principal series, the bosonic cases $\left(F_0,F_1,\ldots,F_{n-1}\right)$ correspond to the principal series rep ${\cal F}_\Delta^{[F_1,\ldots,F_{n-1}]}$ described in section \ref{bosonsection} and the fermionic cases correspond to the principal series reps we called ${\cal F}_\Delta^{[F_1-\half,\dots,F_{n-1}-\half]_{\half}}$ in section \ref{fermionsection}, where the label $F_0$ corresponds to $\Delta$ through
\be \Delta=F_0+2n-1\,.\label{F0deltaregleDevee}\ee 
There is one exception: the restriction $\nu> 0$ accounts for the shadow transform equivalence between the rep with $\Delta$ and that with $d-\Delta$, but it misses the point $\Delta=d/2$.  For the bosons, this point will now be covered in the complementary series, and for the fermions it will be covered in the discrete series.

The complementary series occurs only for the bosonic reps, and the rep $\left(F_0,F_1,\ldots,F_{n-1}\right)$ corresponds to the rep ${\cal F}_\Delta^{[F_1,\ldots,F_{n-1}]}$, with ${d\over 2}\leq \Delta=F_0+2n-1<d-p$, described in section \ref{bosonsection}.  The label $p$ corresponds to the number of non-zero rows in the field's tableau.  The range of $F_0$ accounts for the reps with ${d\over 2} \leq \Delta < d-p $, avoiding double counting by leaving out the equivalent ones with $p<\Delta<{d\over 2}$, but now including the point $\Delta={d\over 2}$.

The exceptional series contains only bosonic fields, and accounts for the shift symmetric scalars and the PM fields with $\leq n-2$ rows in their tableaux. The pattern by which it does this is identical to the odd $D$ discussion in section \ref{oddDsummarysec}.

The discrete series accounts for the remaining unitary partially massless fields in even $D$ (those with $(n-1)$-row tableaux), as well as the $\tilde m=0$ chiral fermion reps:
\begin{itemize}

\item[-] Bosonic discrete series: $(F_0,F_1,\ldots, F_{n-1})$ corresponds to the bosonic rep with tableau $[F_1,\ldots, F_{n-1}]$, which has $s_{n-1}=F_{n-1}>0$.  The values of $F_0$ range over $F_0=1-n,2-n,\ldots, F_{n-1}-n$, which corresponds via \eqref{F0deltaregleDevee} to $\Delta=n,n+1,\ldots,n+s_{n-1}-1$.  These are the PM values in which the bottom row is activated, with the depths $t=s_{n-1},s_{n-1}-1,\ldots,1$, respectively.  The two branches $D^+$ and $D^-$ of the discrete series correspond to the two halves of the chiral splitting of these PM points in even $D$. 

\item[-] Fermionic discrete series:  $(F_0,F_1,\ldots, F_{n-1})$ corresponds to the fermionic rep with tableau $[F_1-\half,\ldots, F_{n-1}-\half]$, so that $s_n=F_{n-1}-\half$.   The values of $F_0$ range over $F_0={1\over 2}-n ,{3\over 2}-n,\ldots, F_{n-1}-n$, which corresponds via \eqref{F0deltaregleDevee} to $\Delta=n-{1\over 2},n+{1\over 2},n+{3\over 2},\ldots,n+s_{n-1}-{1\over 2}$.  The first of these is the point $\Delta={d\over 2}$, which corresponds to the $\tilde m=0$ chiral fermions, with the two branches $D^+$ and $D^-$ 
representing the two chiralities.  The remaining points, which are only there when $s_{n-1}>0$, are the unitary partially massless fermions, those with $(n-1)$-row tableaux in which the last row is activated, with the depths $t=s_{n-1},s_{n-1}-1,\ldots,1$, respectively.  The two branches $D^+$ and $D^-$ correspond to the two halves of the chiral splitting of these fermionic PM points.

\end{itemize}

In all the cases, one can check that the $\frak{so}(D)$ content of the rep as described in this section is the same as we have described it in the previous sections.  

For all the reps, the value of the quadratic Casimir defined in section \ref{casimirsec} is given by
\be {\cal C}_2=\sum_{i=0}^{n-1} F_i\left(F_i+2n-2i\right)\,,\ee
which matches the values given in the above sections.

\section{Summary of Lower Dimensions\label{lowDsection}}

In this section, we will go through the lower dimensional cases $D=3,4,5,6$ explicitly (the case $D=2$ is reserved for section \ref{D2section}), accounting for all the dimension dependent redundancies, equivalences, and extra splittings that occur among the generic $D$ reps described in sections \ref{bosonsection} and \ref{fermionsection}, and matching the unitary reps to the lists in section \ref{unitarylistsection}.

\subsection{$D=3$\label{D3subsection}}

For dS$_3$, $d=2$, the reps are spaces of fields on the 2-sphere, ${\mathbb S}^2$.  The only non-trivial tensor fields on ${\mathbb S}^2$ are the scalars and symmetric tensors; a tensor field on ${\mathbb S}^2$ in any other tableau is trivial, with the exception of the 2-forms, which are equivalent to scalars via Hodge dualization.  In addition, the symmetric tensor fields can be split into (imaginary) self-dual and anti-self-dual parts with respect to the volume form on ${\mathbb S}^2$.  Thus the only bosonic reps are the scalars of section \ref{scalarsec} and the spin $s$ tensors of section \ref{spinssec}, and the only fermionic reps are the spin $1/2$ fermionic reps of section \ref{spin12sec} and the spin $s$ fermionic reps of section \ref{spinsfermionsec}.

For the scalar reps in $D=3$, we have the picture \eqref{dsreps1} specialized to $d=2$, 
\be \raisebox{-40pt}{\epsfig{file=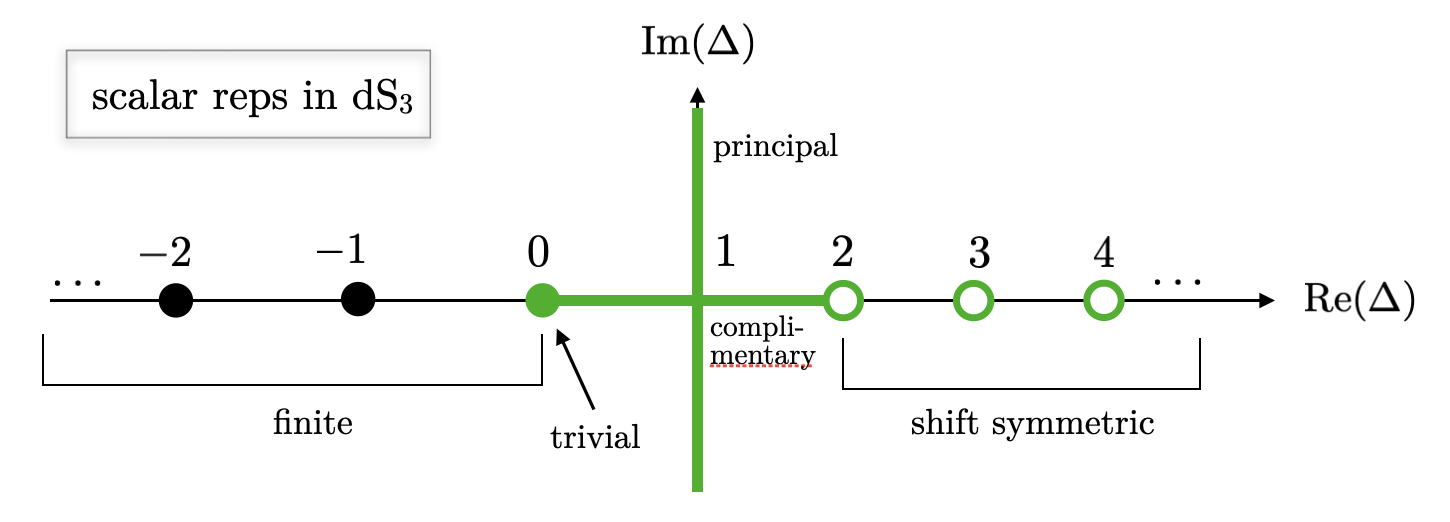,width=5.5in}}  \label{d3pic1}\, \ee
There is a difference here in that we have made the shift symmetric points open circles.  Recall that we use open circles to indicate reps that are present but are accounted for by equivalent reps elsewhere:   
the shift symmetric reps are ${\cal D}_{k+2}^{[0]}$ and from \eqref{spinsudisoeee} are equivalent to the reps ${\cal U}_1^{[k+1]}$ which, as discussed at the end of section \ref{spinssec}, are in turn equivalent to ${\cal V}_1^{[k+1]}$ in $D=3$, i.e. the shift symmetric scalars are equivalent to the maximal depth PM fields.  Before, in section \ref{scalarsec}, we accounted for these reps among the scalars, but now, to match the classification in section \ref{unitarylistsection}, we will instead account for them among the higher spin tensor reps.  The principal series reps are $(F_0,F_1)=(-1+i\nu,0)$, $\nu\geq 0$, and the $p=0$ complementary series is $(F_0,F_1)=(F_0,0)$, $-1<F_0<0$.

For the bosonic spin $s\geq 1$ reps, we discussed at the end of section \ref{spinssec} the chiral splitting that occurs in $D=3$, and the picture \eqref{dsrepsspins} now looks as follows:
\be \raisebox{-40pt}{\epsfig{file=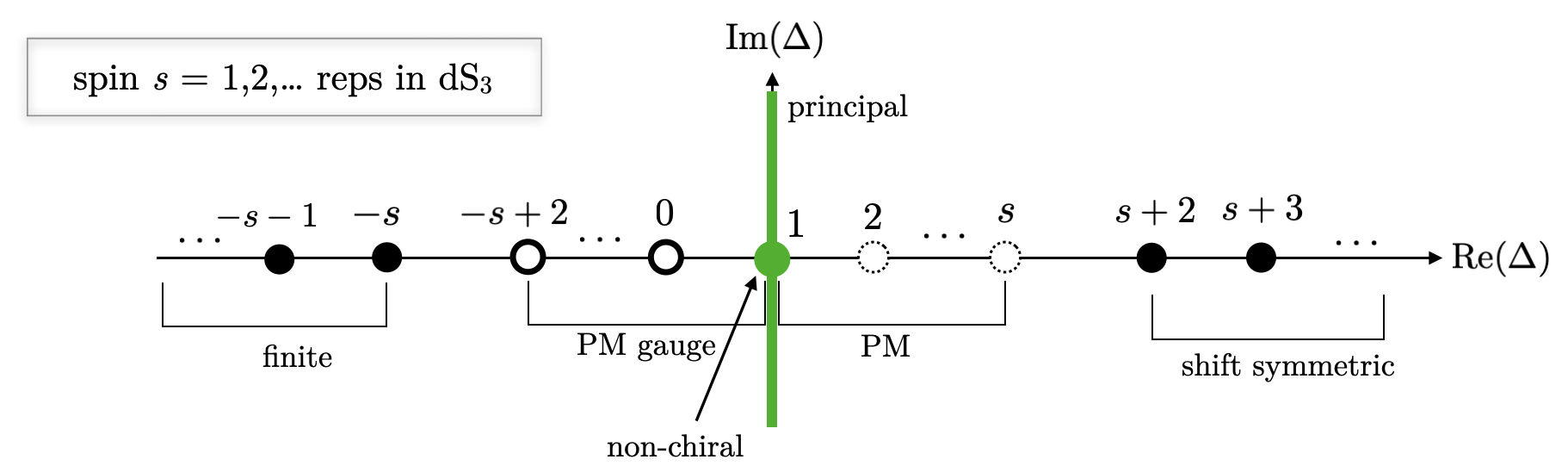,width=6.6in}}  \label{d3pic2}\, \ee
These reps are chiral, so each point, except the point in the middle at $\Delta=1$, represents two reps, and there is the equivalence (except at the circled points) given by reflecting through the $\Delta=1$ point and flipping the chirality.   Those of the principal series are the $(F_0,F_1)=(-n+i\nu, s)$, $\nu\in {\mathbb R}$, where $\nu>0$ and $\nu<0$ are the two different chiralities.
Note that there is no complementary series among the tensors.  In addition, the partially massless reps other than the maximal depth are indicated by dashed circles since they are trivial in $D=3$.  The rep at $\Delta=1$ is a single rep and is a filled circle: this is the maximal depth PM point, which is equivalent to its gauge point, and also equivalent to a shift symmetric scalar with $k=s-1$, so this is the rep that accounts for the shift symmetric scalars in the classification of section \ref{oddDsummarysec}; it is the $\nu=0$ principal series rep.  Note that there is no complementary series for these fields.  There is no exceptional series in $D=3$.

For the spin $1/2$ fermions, we have the picture \eqref{dsrepsspinor} specialized to $d=2$, 
\be \raisebox{-40pt}{\epsfig{file=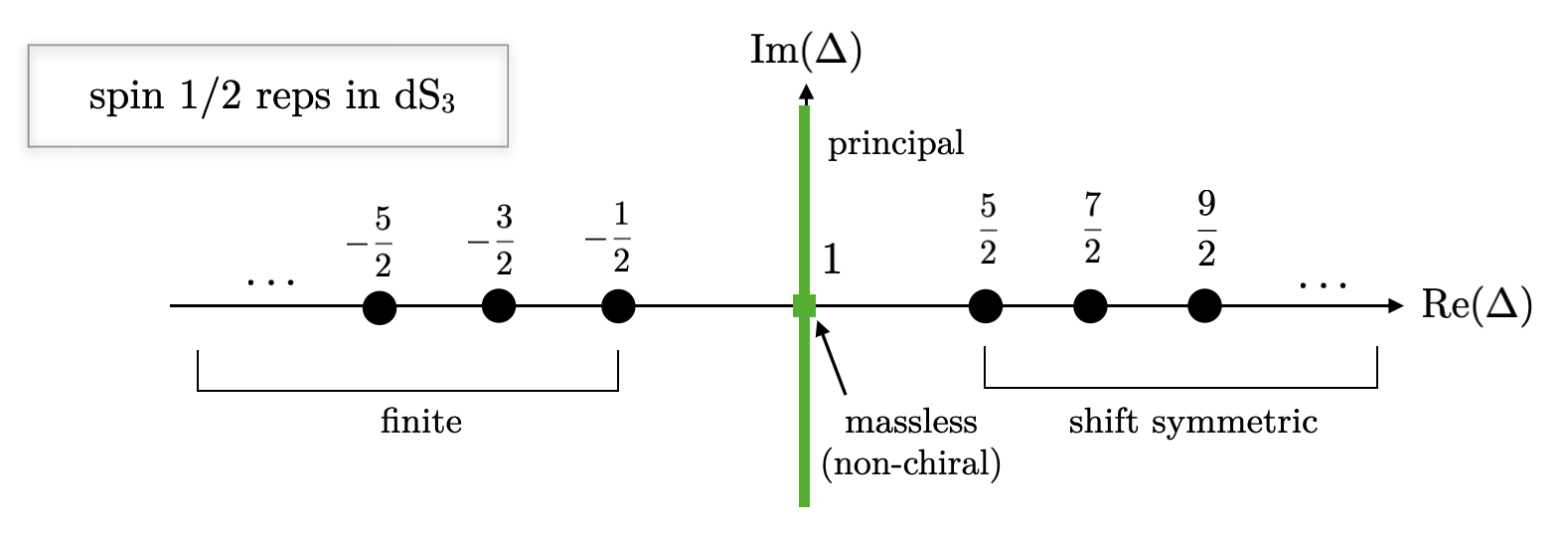,width=6.0in}}  \label{d3pic3}\, \ee
Since $d$ is even, these reps are chiral, so each point, except the point in the middle at $\Delta=1$, represents two reps, and there is the equivalence (except at the circled points) given by reflecting through the $\Delta=1$ point and flipping the chirality.  Those of the principal series are $(F_0,F_1)=(-1+i\nu, \half)$, $\nu\in {\mathbb R}$, and these are the only unitary reps: $\nu>0$ and $\nu<0$ are the two different chiralities of massive fermions, and $\nu=0$, the point $\Delta=1$, is the massless $\tilde m=0$ fermion, which is a single rep and is not chiral.

For the spin $s\geq 3/2$ fermions, we have the picture \eqref{dsrepsspinorspins} specialized to $d=2$,  
\be \raisebox{-40pt}{\epsfig{file=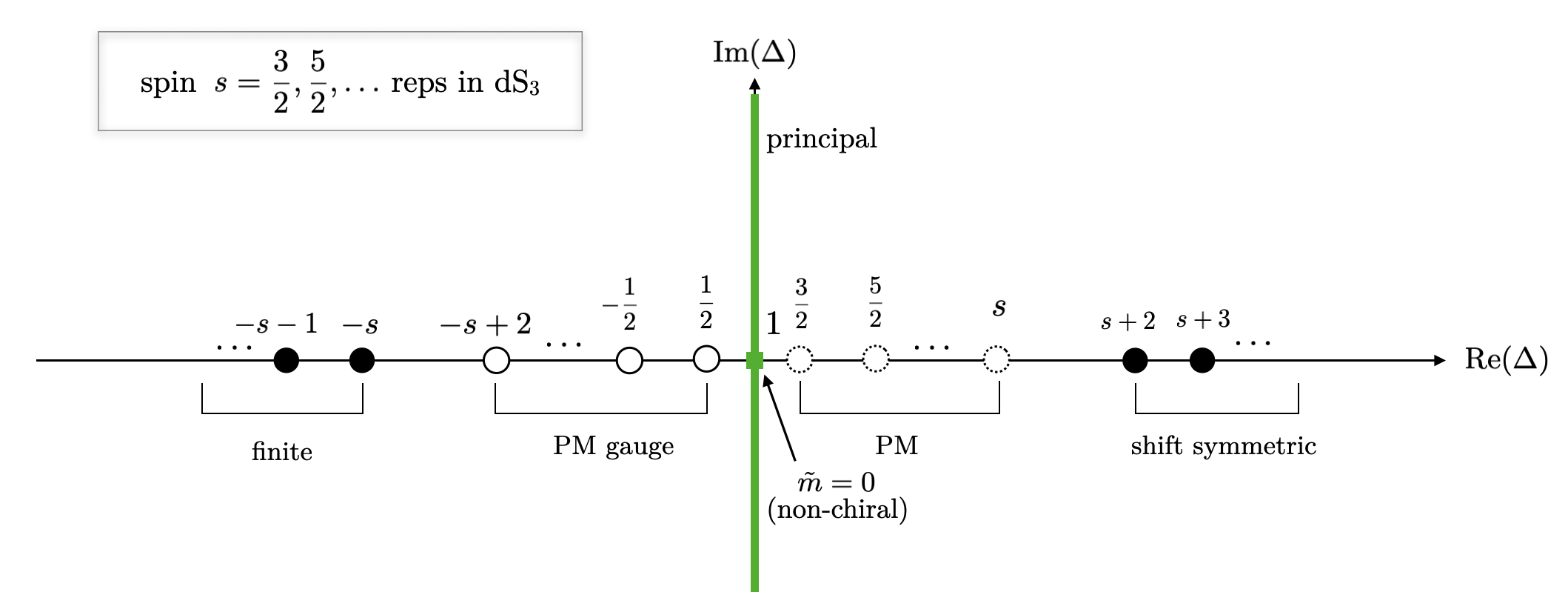,width=6.4in}}  \label{d3pic4}\, \ee
Like the spin $1/2$ reps, these reps are chiral, and each point, except $\Delta=1$, represents two reps, and there is the equivalence (except at the circled points) given by reflecting through the $\Delta=1$ point and flipping the chirality.  Those of the principal series are  $(F_0,F_1)=(-1+i\nu, s)$, $\nu\in {\mathbb R}$,, and these are the only unitary reps.  Those with $\nu>0$ and $\nu<0$ are the two different chiralities of the massive spinning fermions, and the point $\Delta=1$ is the $\tilde m=0$ fermion, which is a single rep and is not chiral.  The dotted open circles are reps which are absent: these are the would-be fermionic PM fields, which are all trivial in $D=3$.

\subsection{$D=4$}

For dS$_4$, $d=3$, we are working with fields on the 3-sphere, ${\mathbb S}^3$.  The only non-trivial tensor fields on ${\mathbb S}^3$ are the symmetric tensors; any field in a three row tableau is trivial (with the exception of $[1,1,1]$ which is equivalent to a scalar), any field in a two-row tableau with $\geq 2$ boxes in the bottom row is trivial, and a field in a $[s_1,1]$ tableau is equivalent to a $[s_1]$ field.  Thus the only bosonic reps are the scalars of section \ref{scalarsec} and the spin $s$ tensors of section \ref{spinssec}, and the only fermionic reps are the spin $1/2$ fermionic reps of section \ref{spin12sec} and the spinning fermionic reps of section \ref{spinsfermionsec}.

For the scalar reps in $D=4$, we have the picture \eqref{dsreps1} specialized to $d=3$, 
\be \raisebox{-40pt}{\epsfig{file=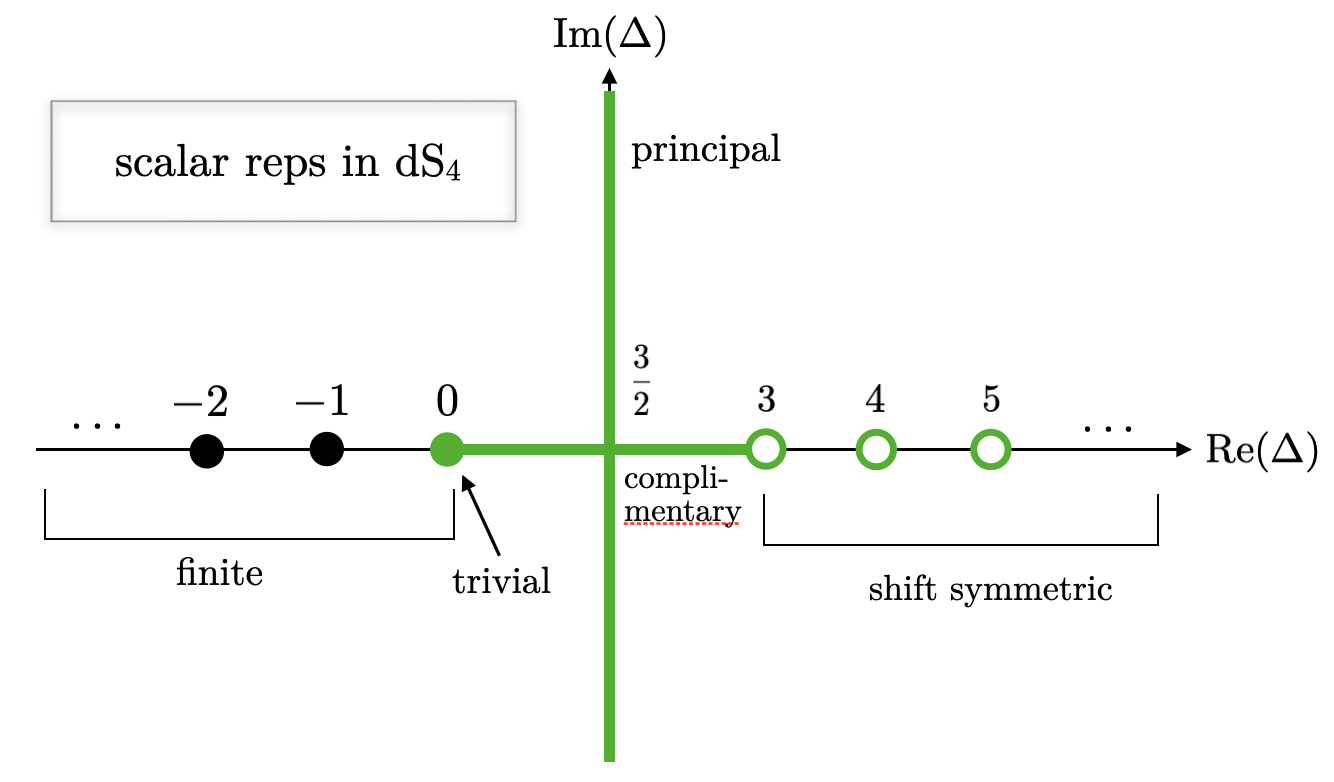,width=4.8in}}  \label{d4pic1}\, \ee
The shift symmetric points are now open circles, indicating that they will  be accounted for by equivalent reps elsewhere:   
the shift symmetric reps are ${\cal D}_{k+3}^{[0]}$ and are equivalent to the PM gauge reps ${\cal U}_1^{[k+1]}$. Before we accounted for these reps among the scalars, but now we account for them among the tensors.   The principal series is $(F_0,F_1)=(-{3\over 2}+i\nu,0)$, $\nu>0$.  The $p=0$ complementary series is $(F_0,F_1)=(F_0,s)$ with $-{3\over 2}\leq F_0<0$.

For the spin $s$ symmetric tensors,  the picture \eqref{dsrepsspins} becomes
\be \raisebox{-40pt}{\epsfig{file=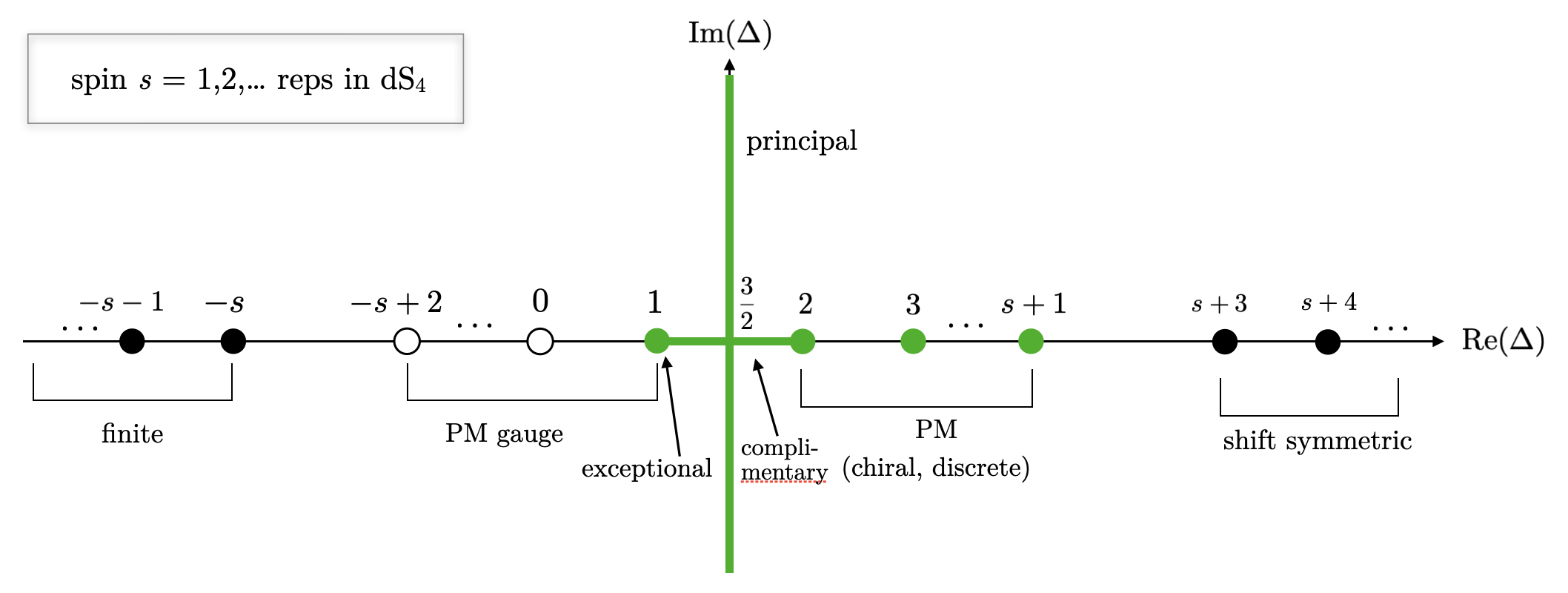,width=6.4in}}  \label{d4pic2}\, \ee
The principal series is $(F_0,F_1)=(-{3\over 2}+i\nu,s)$, $\nu>0$.  The $p=1$ complementary series is $(F_0,F_1)=(F_0,s)$ with $-{3\over 2}\leq F_0<1$.  
As discussed at the end of section \ref{spinssec}, the special thing that happens in $D=4$ is that the PM fields at $\Delta=s-t+2$, $t=1,\ldots,s$, split up into chiral parts. 
These PM points make up the bosonic discrete series reps in $D=4$: their chiral parts are the $D^\pm$ reps $(F_0,F_1)=(s-t-1,s)$.  The point at $\Delta=1$ is the rep ${\cal U}_1^{[s]}$, which through \eqref{spinsudisoeee} is equivalent to, and accounts for, the shift symmetric scalars, with level $k=s-1$: ${\cal U}^{[s]}_1\simeq {\cal D}^{[0]}_{s+2}$.  This is the $p=1$ exceptional series in $D=4$, $(F_0,F_1)=(-1,s)$.

For the spin $1/2$ fermions, $d$ is odd and we have the picture \eqref{dsrepsspinor} specialized to $d=3$ \cite{Lindsay:2025chz}, 
\be \raisebox{-40pt}{\epsfig{file=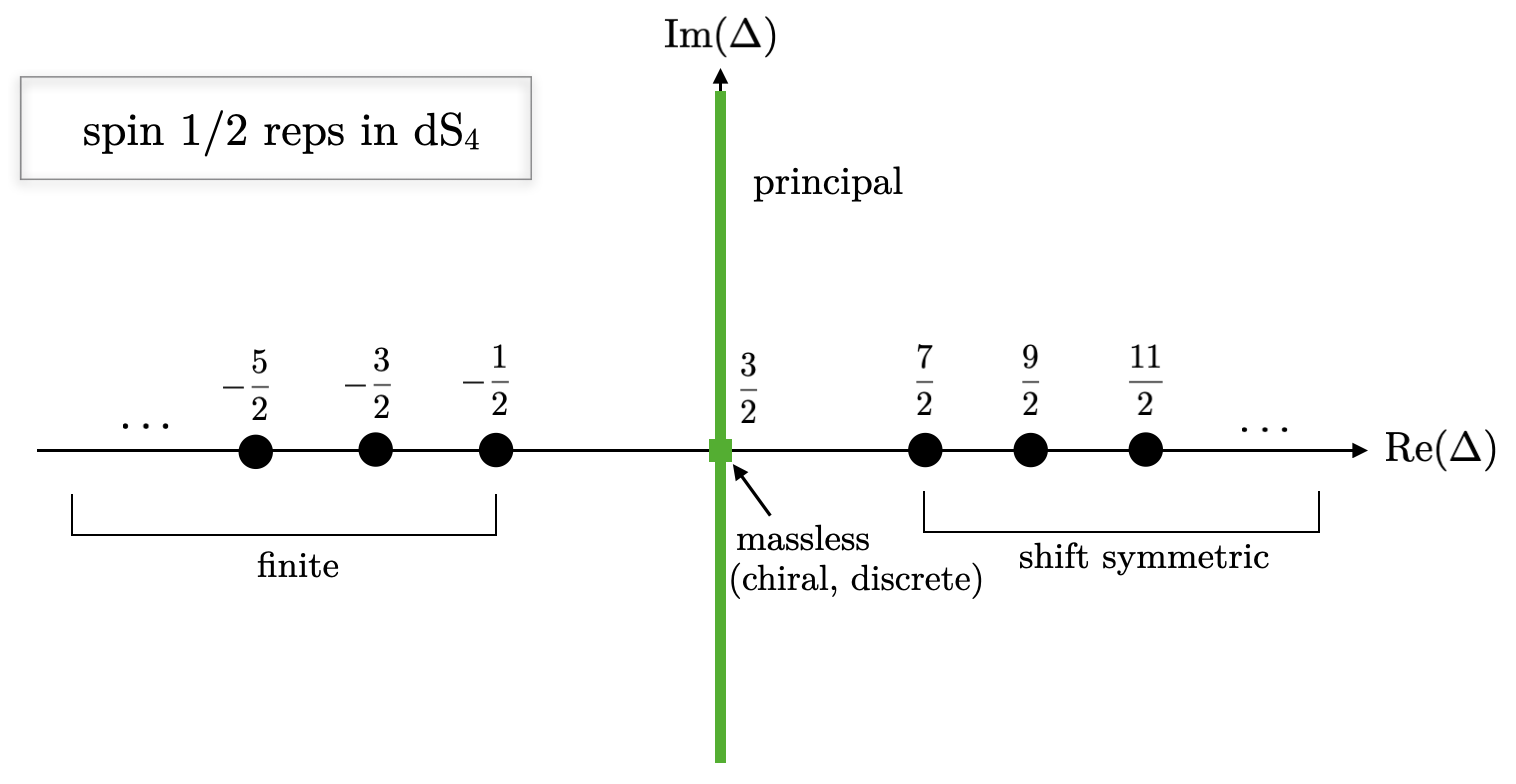,width=5.0in}}  \label{d4pic3}\, \ee
These are non-chiral, so each point represents one rep, and there is the shadow equivalence given by reflecting through the point $\Delta=3/2$.  The unitary reps are the principal series $(F_0,F_1)=(-{3\over 2}+i\nu,\half)$, $\nu>0$, where the restriction $\nu>0$ avoids redundancy.   The point at $\Delta=3/2$ is two reps, the chiral parts of the $\tilde m=0$ fermion, and these are part of the fermionic discrete series, the two reps $D^\pm$ with $(F_0,F_1)=(-{3\over 2},\half)$.

For the spin $s\geq 3/2$ fermions, we have the picture \eqref{dsrepsspinorspins} specialized to $d=3$,  
\be \raisebox{-40pt}{\epsfig{file=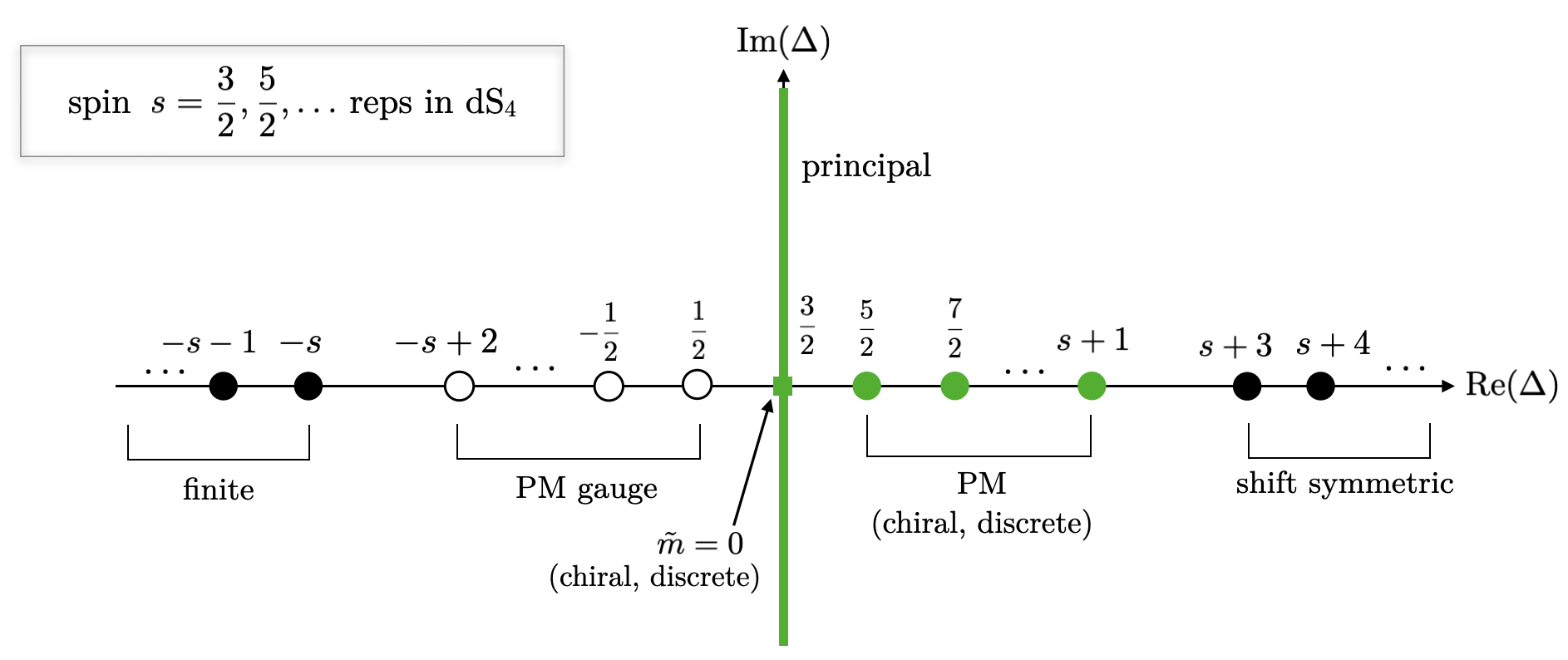,width=6.4in}}  \label{d4pic4}\, \ee
The unitary reps are the principal series $(F_0,F_1)=(-{3\over 2}+\half+i\nu,s)$, where $\nu>0$ avoids redundancy due to the shadow equivalence, and these are non-chiral.  The special thing that happens in $D=4$ is that the spin $s\geq 3/2$ symmetric tensor PM fermions are unitary, as mentioned in section \ref{spinsfermionsec} and detailed in the recent studies \cite{Letsios:2023qzq,Letsios:2022tsq,Letsios:2023awz}, so here they are green circles rather than the black circles of the general $d$ picture \eqref{dsrepsspinorspins}.  These are in the spin $s$ fermionic discrete series reps, the reps $D^\pm$ with $(F_0,F_1)=(s-t-1,s)$, $t=1,\ldots,s-\half$.  The point at $\Delta=3/2$ is two reps, the chiral parts of the $\tilde m=0$ fermion, which are the remaining discrete series reps $D^\pm$ with $(F_0,F_1)=(-{3\over 2},s)$.

\subsection{$D=5$}

For dS$_5$, $d=4$, we are working with fields on the 4-sphere ${\mathbb S}^4$.  Any non-trivial tensor field on ${\mathbb S}^4$ is equivalent to either a symmetric tensor field, or a field with indices in a two-row tableaux, such as a 2-form field.  The two-row tableau fields can be split into self-dual and anti-self-dual parts with respect to the volume form on ${\mathbb S}^4$.

For the scalar reps in $D=5$, we have the picture \eqref{dsreps1} specialized to $d=4$, 
\be \raisebox{-40pt}{\epsfig{file=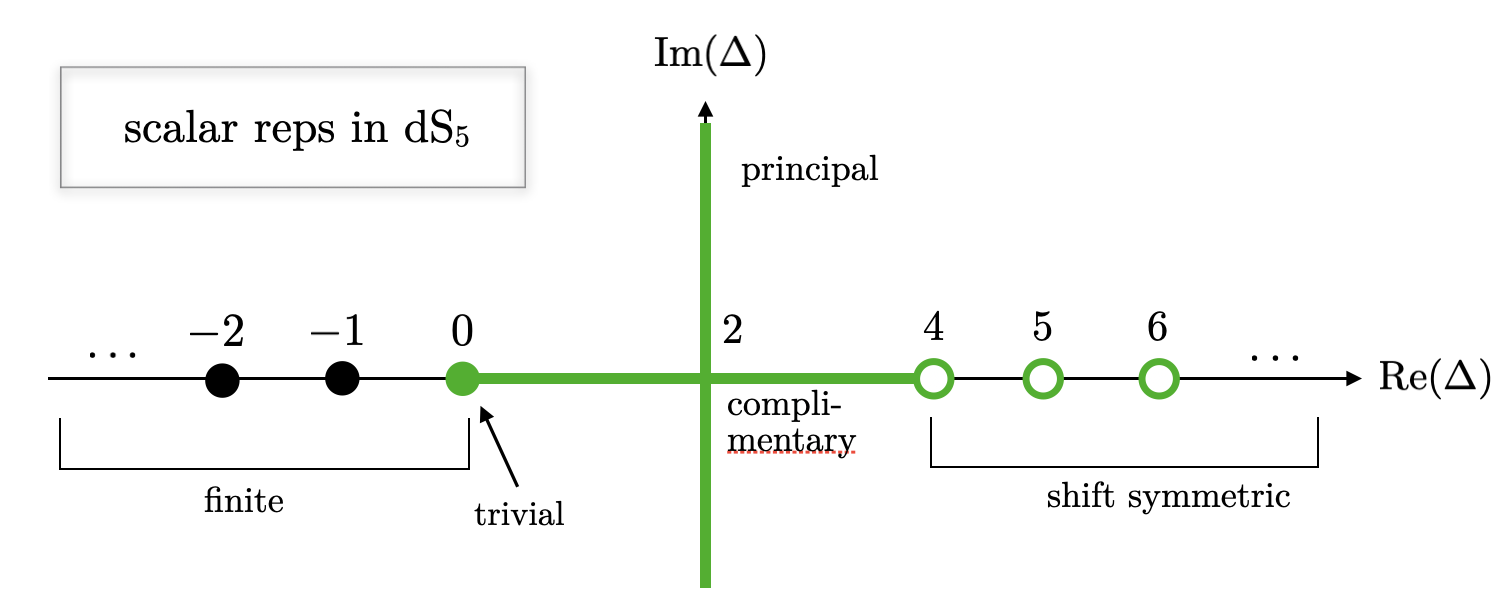,width=5.3in}}  \label{d5pic1}\, \ee
The principal series is $(F_0,F_1,F_2)=(-2+i\nu),0,0)$, $\nu\geq 0$.  The $p=0$ complementary series is $(F_0,F_1,F_2)=(F_0,0,0)$, with $-2<F_0<0$.
  
The shift symmetric points ${\cal D}_{k+4}^{[0]}$ are now open circles, indicating that they will be accounted for by the equivalent reps ${\cal U}_1^{[k+1]}$ among the tensors. 

For the bosonic spin $s$ symmetric tensors, we have the picture \eqref{dsrepsspins} specialized to $d=4$, 
\be \raisebox{-40pt}{\epsfig{file=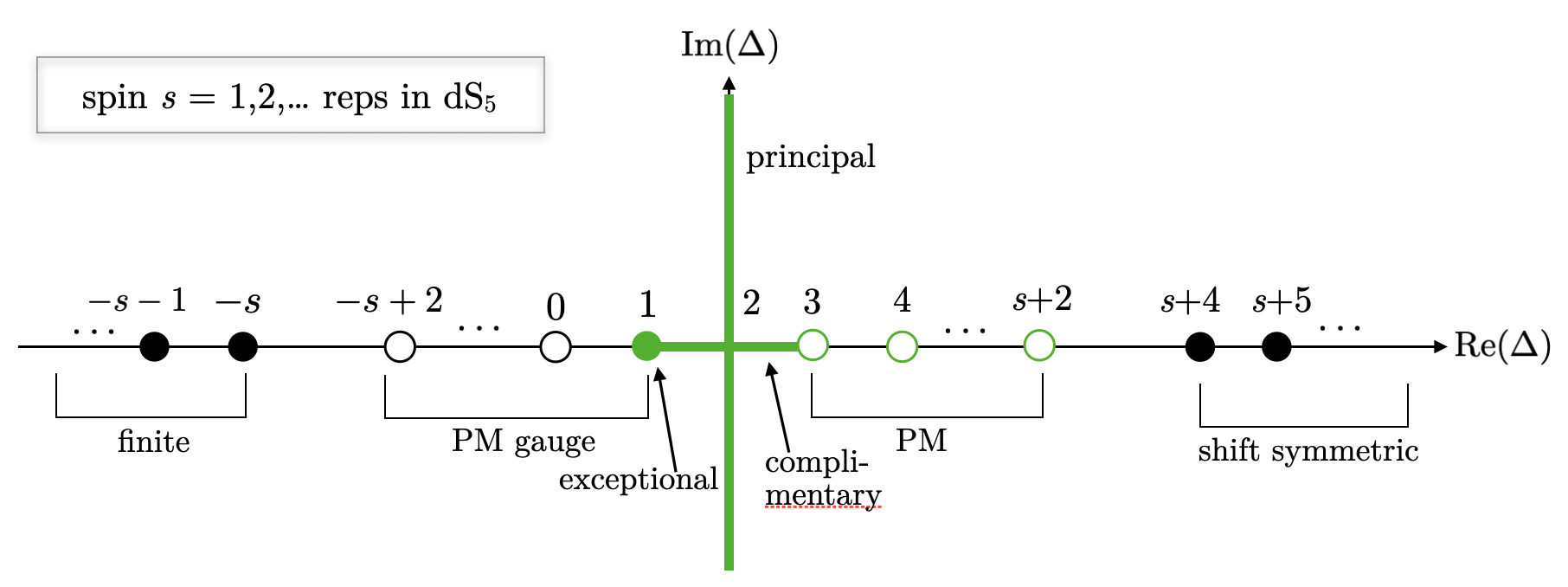,width=6.4in}}  \label{d5pic2}\, \ee
The principal series is $(F_0,F_1,F_2)=(-2+i\nu),s,0)$, $\nu\geq 0$.  The $p=1$ complementary series is $(F_0,F_1,F_2)=(F_0,s,0)$, with $-2<F_0<-1$.
The partially massless points are now open circles, to indicate that they will be accounted for elsewhere: these were the discrete series reps in $D=4$, but now they will be accounted for by the exceptional series of the 2-row tableau fields.  The point at $\Delta=1$ is now a solid circle and is the rep ${\cal U}_1^{[s]}$, which through \eqref{spinsudisoeee} is equivalent to the shift symmetric scalars with $s=k+1$: ${\cal U}_1^{[s]}\simeq {\cal D}^{[0]}_{s+3}$.  These are the $p=1$ exceptional series reps in $D=5$, $(F_0,F_1,F_2)=(-1,s,0)$, and they account for the shift symmetric scalars.

The two-row tableau bosonic fields are illustrated here:
\be \raisebox{-40pt}{\epsfig{file=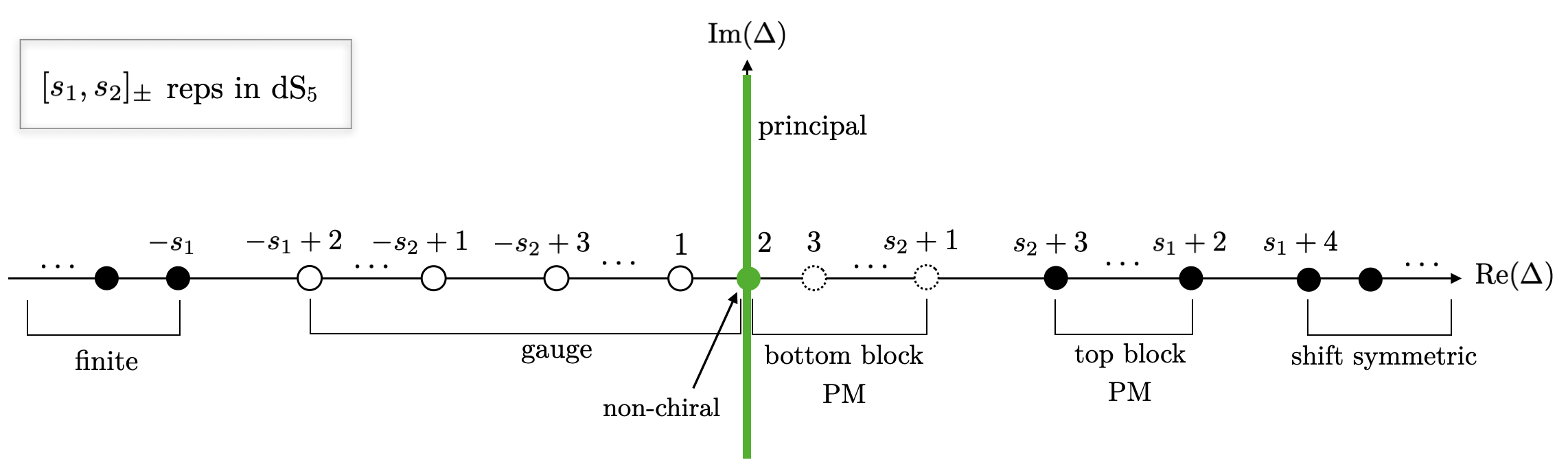,width=6.5in}}  \label{d5pic3}\, \ee
These reps are chiral, corresponding to fields that are self-dual and anti-self-dual fields under the $d=4$ epsilon tensor on ${\mathbb S}^4$ acting on their first column, so each point (except $\Delta=2$) represents two representations, and there is the shadow given by reflecting through $\Delta=2$ and flipping the chirality.  The principal series is $(F_0,F_1,F_2)=(-2+i\nu,s_1,s_2)$, $\nu\in {\mathbb R}$, with $\nu>0$ and $\nu<0$ accounting for the two helicities.
There is no complementary series.  The principal series point $\nu=0$ is the point at $\Delta=2$, the single rep corresponding to the PM point where all the boxes in the bottom row are activated.  From \eqref{oddnmeqfee}, this PM point is equivalent to its gauge point and is also equivalent to, and accounts for, the partially massless tensor reps with spin $s_1$ and depth $t=s_1-s_2+1$,
\be {\cal U}^{[s_1,s_2]}_{2} \simeq {\cal V}^{[s_1,s_2]}_{2}\simeq  {\cal V}^{[s_1]}_{s_2+2}\,, \ \ D=5\,.\ee
(This includes as the special case $s_1=s_2=1$ the familiar equivalence between a massless $2$ form and a massless vector in $D=5$.
As another example, in the $[2,1]$ PM case where the bottom box is activated, the bottom two rows in \eqref{mixedcontent1} are removed.  The remaining top two rows, when dualized, are precisely the $\frak{so}(5)$ content of the $t=2$ partially massless spin $2$ as seen in \eqref{tensorsocontent6}.)

Those $[s_1,s_2]$ PM reps in which not all of the bottom rows are activated are indicated by dashed circles: these reps are trivial in $D=5$.  The remaining PM reps, those with boxes in the top row activated, are non-unitary and so are indicated by solid black circles.  In this way, as in all odd $D$, all the unitary PM fields are accounted for without the need for a discrete series.

For the spin $1/2$ fermions, we have the picture \eqref{dsrepsspinor} specialized to $d=4$, 
\be \raisebox{-40pt}{\epsfig{file=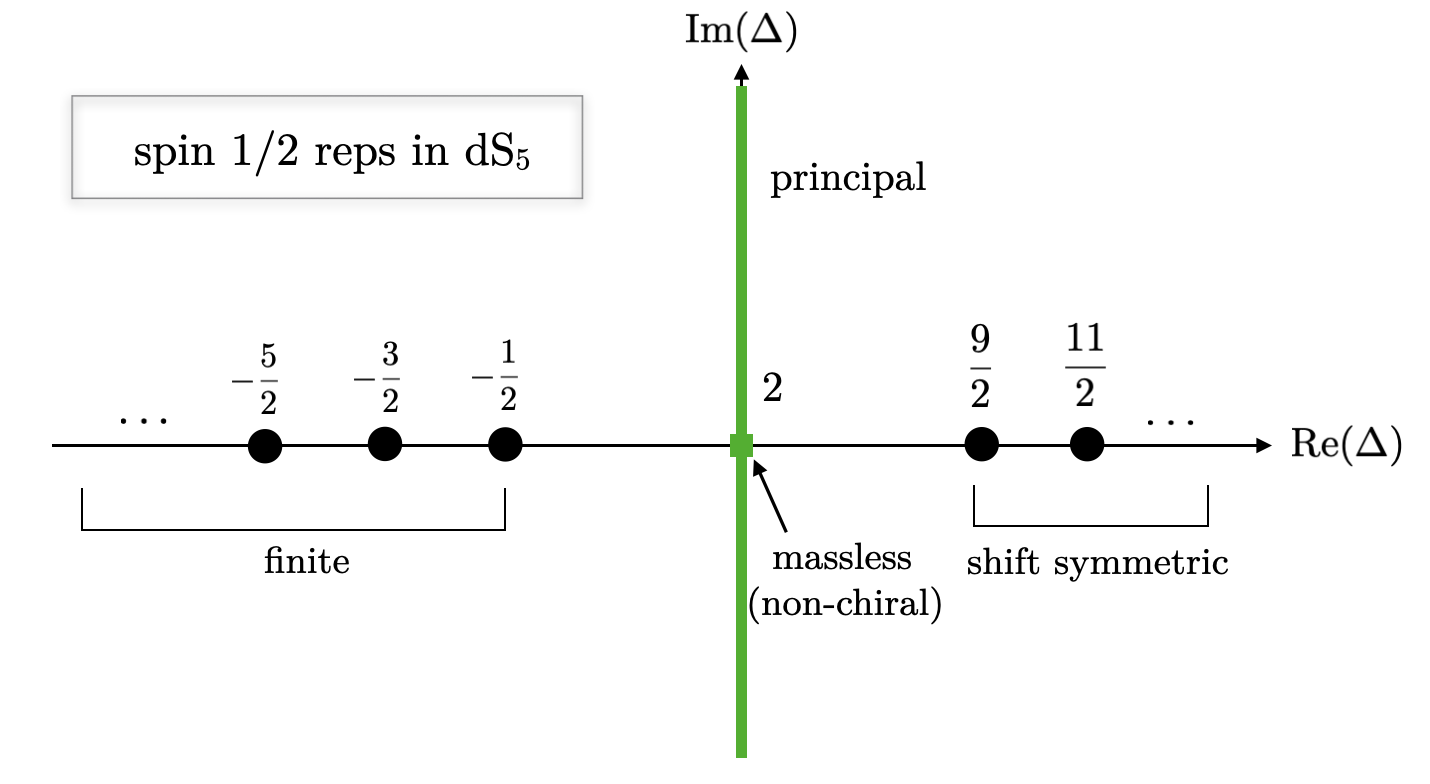,width=4.9in}}  \label{d5pic4}\, \ee
For this, and the other other fermionic cases in $D=5$, the reps are chiral, so each point represents two reps, and there is the equivalence, except at the reducible points, given by reflecting through the point $\Delta=2$ along with a flip in chirality.  The point at $\Delta=2$ is one rep, the non-chiral $\tilde m=0$ fermion.   The unitary reps are the principal series $(F_0,F_1,F_2)=(-2+i\nu,\half, {1\over 2})$, $\nu\in {\mathbb R}$, with $\nu>0$ and $\nu<0$ the two chiralities of the massive fermions, and $\nu=$ the $\tilde m=0$ fermion.

For the spin $s\geq {3\over 2}$ fermions, we have the picture \eqref{dsrepsspinorspins} specialized to $d=4$, 
\be \raisebox{-40pt}{\epsfig{file=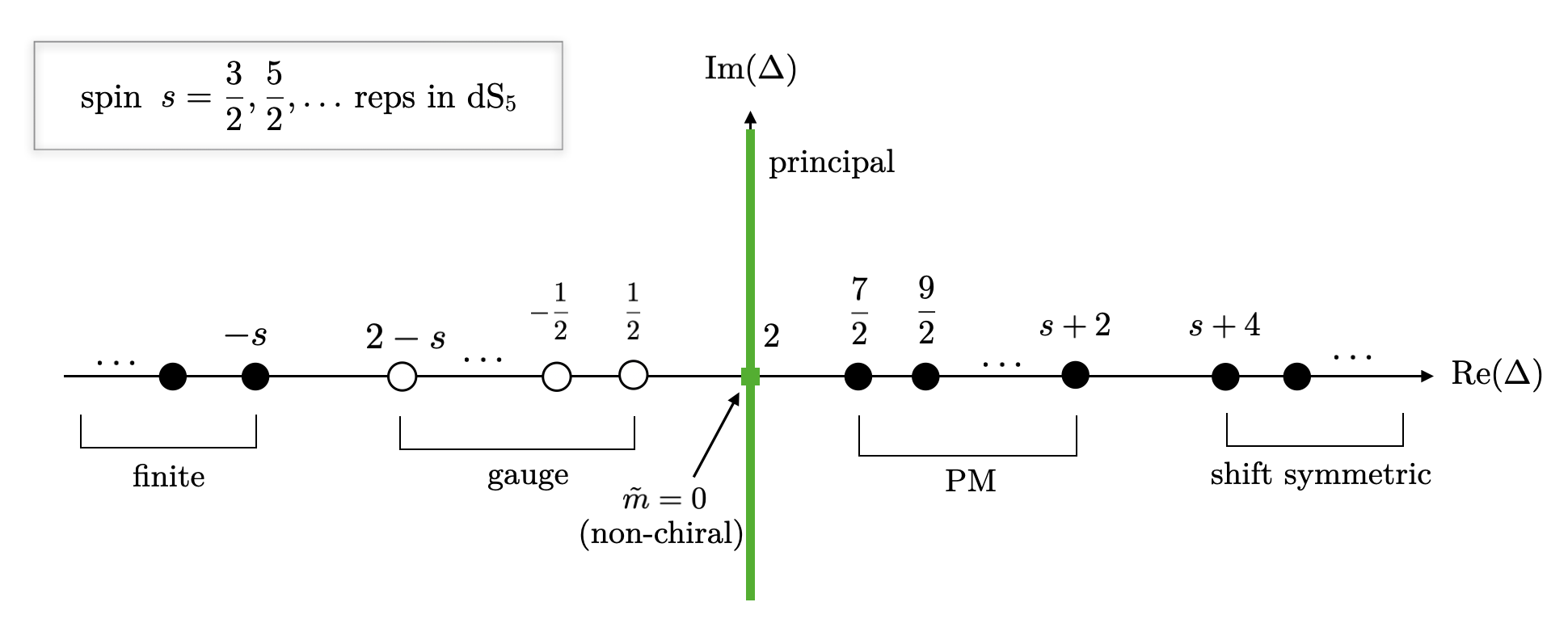,width=6.4in}}  \label{d5pic5}\, \ee
The unitary reps are the principal series $(F_0,F_1,F_2)=(-2+i\nu,s, {1\over 2})$, $\nu\in {\mathbb R}$, with $\nu>0$ and $\nu<0$ the two chiralities of the massive spin $s$ fermions, and $\nu=0$ the non-chiral $\tilde m=0$ spin $s$ fermion.  Note that the PM fermions are no longer unitary, as they were in $D=4$.  

The fermionic 2-row tableaux are summarized here
\be \raisebox{-40pt}{\epsfig{file=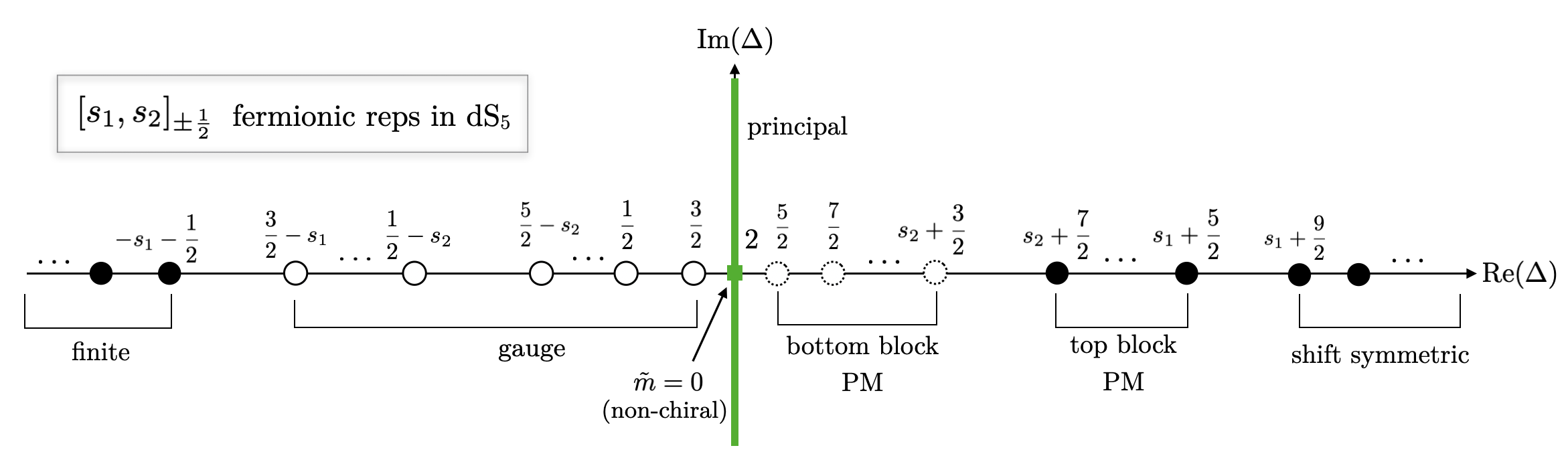,width=6.5in}}  \label{d5pic6}\, \ee
The unitary reps are the principal series $(F_0,F_1,F_2)=(-2+i\nu,s_1+\half,s_2+\half, {1\over 2})$, $\nu\in {\mathbb R}$, with $\nu>0$ and $\nu<0$ the two chiralities of the massive rep, and $\nu=0$ the non-chiral $\tilde m=0$ rep.  The partially massless reps with boxes in the bottom row activated are indicated by dashed circles, since they are trivial in $D=5$. 

\subsection{$D=6$}

For dS$_6$, $d=5$, we are working with fields on the 5-sphere ${\mathbb S}^5$.  Any non-trivial tensor field on ${\mathbb S}^5$ is equivalent to either a symmetric tensor field, or a field with indices in a two-row tableaux.

For the scalar reps in $D=6$, we have the picture \eqref{dsreps1} specialized to $d=5$, 
\be \raisebox{-40pt}{\epsfig{file=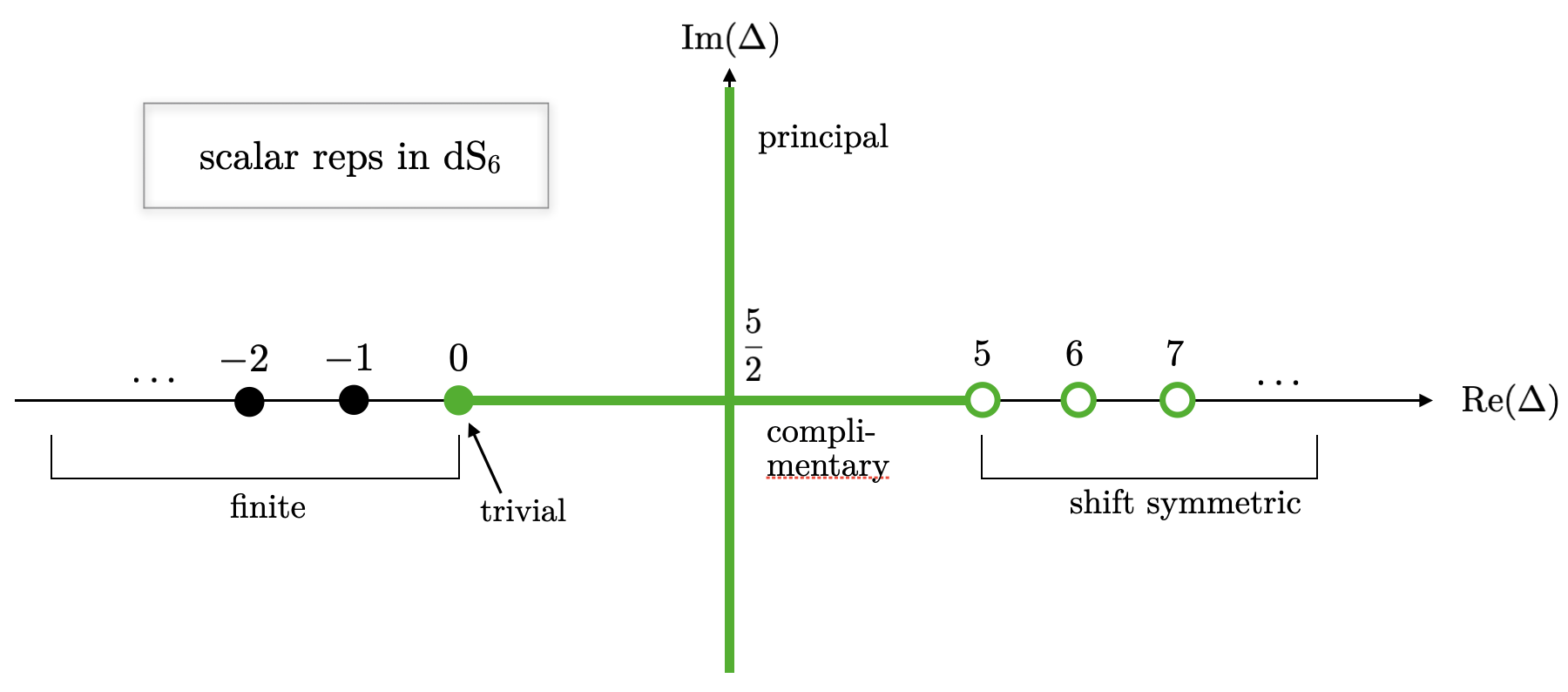,width=5.5in}}  \label{d6pic1}\, \ee
The principal series is $(F_0,F_1,F_2)=(-{5\over 2}+i\nu,0,0)$, $\nu> 0$.  The $p=0$ complementary series is $(F_0,F_1,F_2)=(F_0,0,0)$ with $-{5\over 2}\leq F_0<0$.
The shift symmetric points ${\cal D}_{k+5}^{[0]}$ are now open circles, indicating that they will  be accounted for by the equivalent exceptional series reps ${\cal U}_1^{[k+1]}$ among the higher spin tensor reps.

For the spin $s\geq 1$ bosonic symmetric tensors, we have the picture \eqref{dsrepsspins} specialized to $d=5$, 
\be \raisebox{-40pt}{\epsfig{file=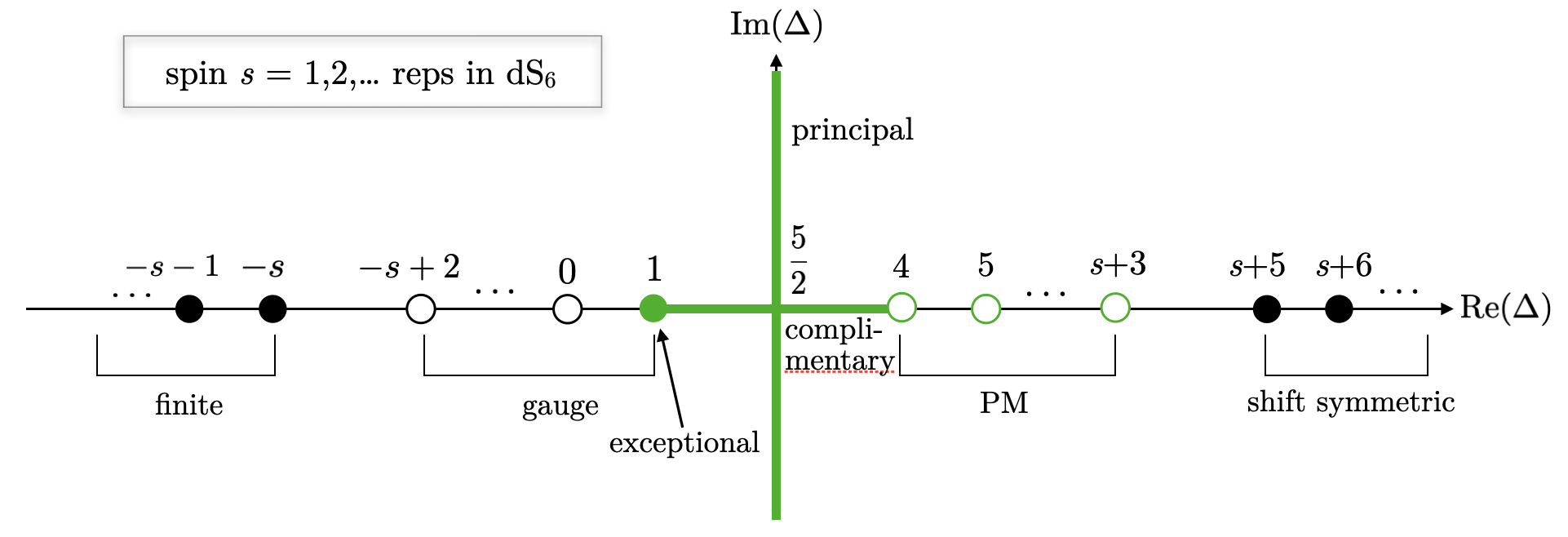,width=6.4in}}  \label{d6pic2}\, \ee
The principal series is $(F_0,F_1,F_2)=(-{5\over 2}+i\nu,s,0)$, $\nu> 0$.  The $p=1$ complementary series is $(F_0,F_1,F_2)=(F_0,s,0)$ with $-{5\over 2}\leq F_0<-1$.
The partially massless points are now open circles: they will be accounted for among the equivalent exceptional series reps among the two-row tableaux reps.  The point at $\Delta=1$ is now solid and is the rep ${\cal U}_1^{[s]}$, which is equivalent through \eqref{spinsudisoeee} to the shift symmetric scalars with $k=s-1$: ${\cal U}^{[s]}_1\simeq {\cal D}^{[0]}_{s+4}$.  These are the $p=1$ exceptional series reps in $D=6$, $(F_0,F_1,F_2)=(-1,s,0)$, and they account for the shift symmetric scalars.

The two-row tableau bosonic tensor fields are illustrated here
\be \raisebox{-40pt}{\epsfig{file=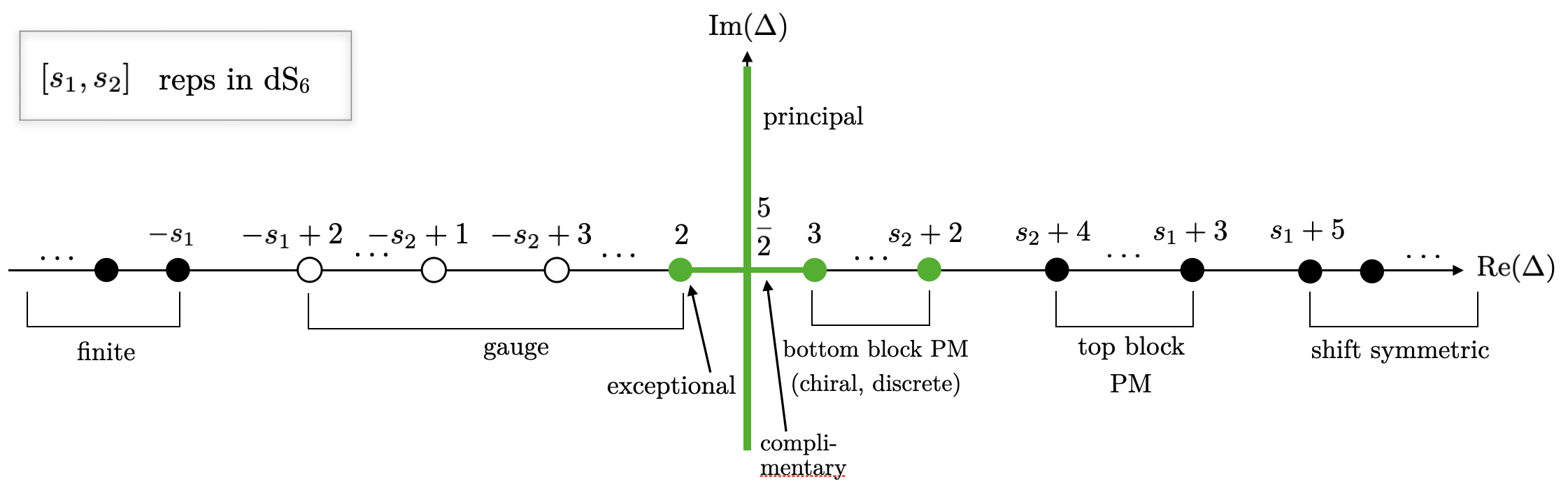,width=7.0in}}  \label{d6pic3}\, \ee
The principal series is $(F_0,F_1,F_2)=(-{5\over 2}+i\nu,s_1,s_2)$, $\nu> 0$.  The $p=2$ complementary series is $(F_0,F_1,F_2)=(F_0,s,0)$ with $-{5\over 2}\leq F_0<-2$. 
The point at $\Delta=2$ is now solid, it is the $p=2$ exceptional series rep, $(F_0,F_1,F_2)=(-2,s_1,s_2)$, that is equivalent by \eqref{mixsyisomfsfee} to the $s=s_1$, $t=s_1-s_2+1$ PM field, and accounts for them: ${\cal U}^{[s_1,s_2]}_{2}\equiv V^{[s]}_{d+s-t-1}$.

The PM points where the bottom boxes are activated are unitary, and these split into chiral reps.  These are the bosonic discrete series reps $D^\pm$ with $(F_0,F_1,F_2)=(s_2-t-2,s_1,s_2)$ in $D=6$.  (This includes as the special case $s_1=s_2=1$, $t=1$, the massless 2-form in $D=6$.)

For the spin $1/2$ fermions, we have the picture \eqref{dsrepsspinor} specialized to $d=5$, 
\be \raisebox{-40pt}{\epsfig{file=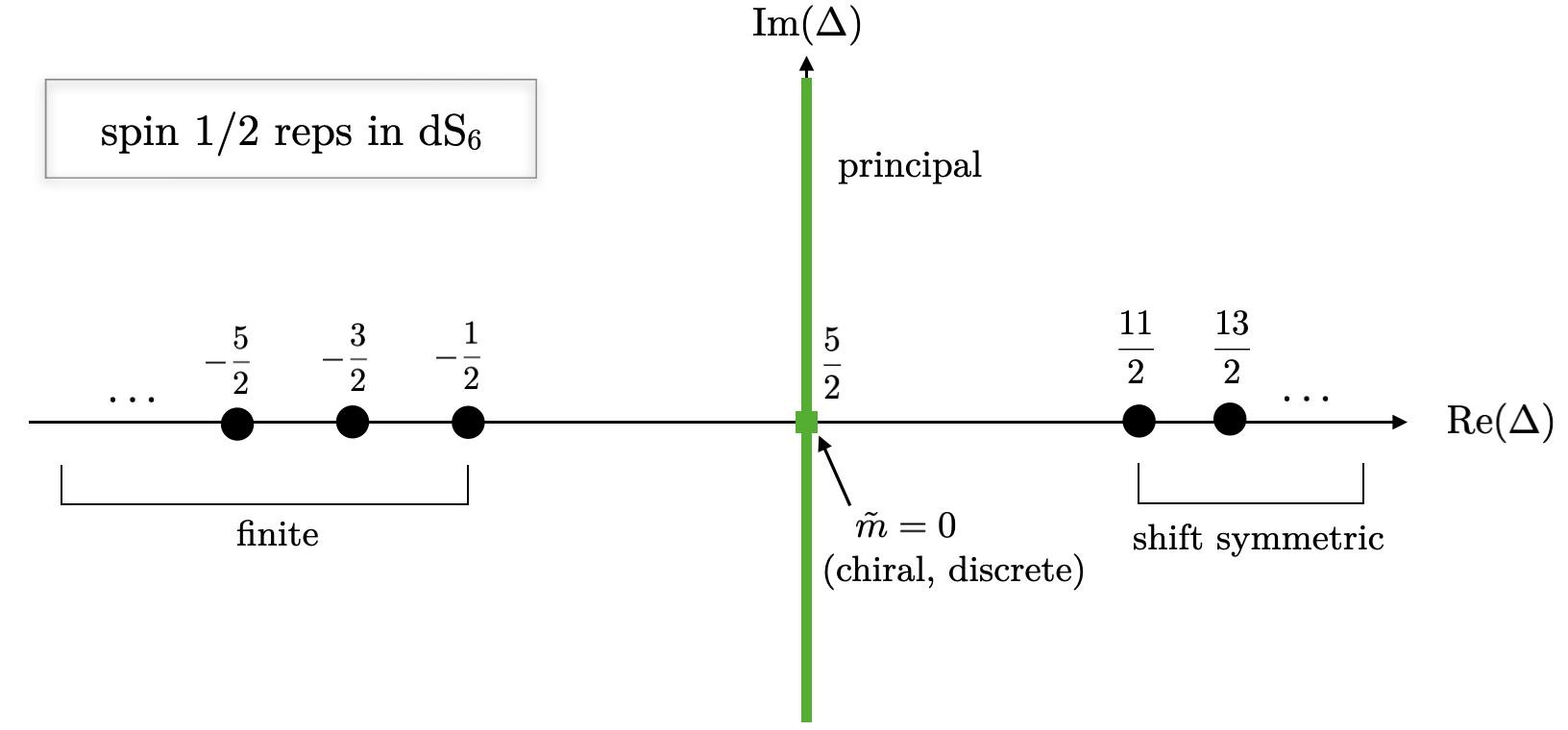,width=5.0in}}  \label{d6pic4}\, \ee
The principal series is $(F_0,F_1,F_2)=(-{5\over 2}+i\nu,\half,\half)$, with $\nu>0$, accounting for the shadow equivalence upon reflecting through $\Delta=5/2$.  The point at $\Delta=5/2$ represents two reps, the two chiral parts of the massless fermion; these are the fermionic discrete series reps $D^\pm$ with $(F_0,F_1,F_2)=(-{5\over 2} ,\half,\half)$.

For the spin $s\geq {3\over 2}$ fermions, we have the picture \eqref{dsrepsspinorspins} specialized to $d=5$, 
\be \raisebox{-40pt}{\epsfig{file=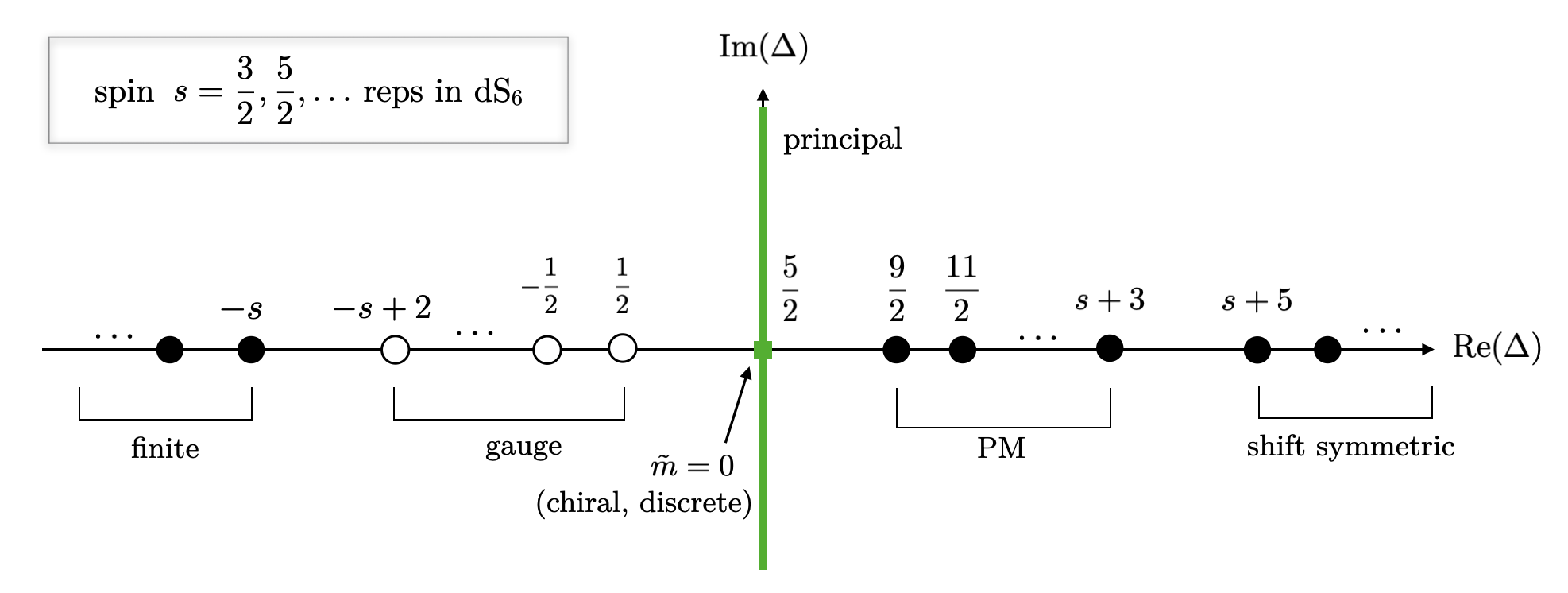,width=6.4in}}  \label{d6pic5}\, \ee
The principal series is $(F_0,F_1,F_2)=(-{5\over 2}+i\nu,s,\half)$, with $\nu>0$, accounting for the shadow equivalence upon reflecting through $\Delta=5/2$.  Note that the spin $s\geq 3/2$ symmetric tensor PM fermions are no longer unitary as they were in $D=4$.  The point at $\Delta=5/2$ is unitary and represents two reps, the two chiral parts of the $\tilde m=0$ spin $s$ fermion; these are the fermionic discrete series reps $D^\pm$ with $(F_0,F_1,F_2)=(-{5\over 2},s,\half)$

The fermionic 2-row tableaux are summarized here
\be \raisebox{-40pt}{\epsfig{file=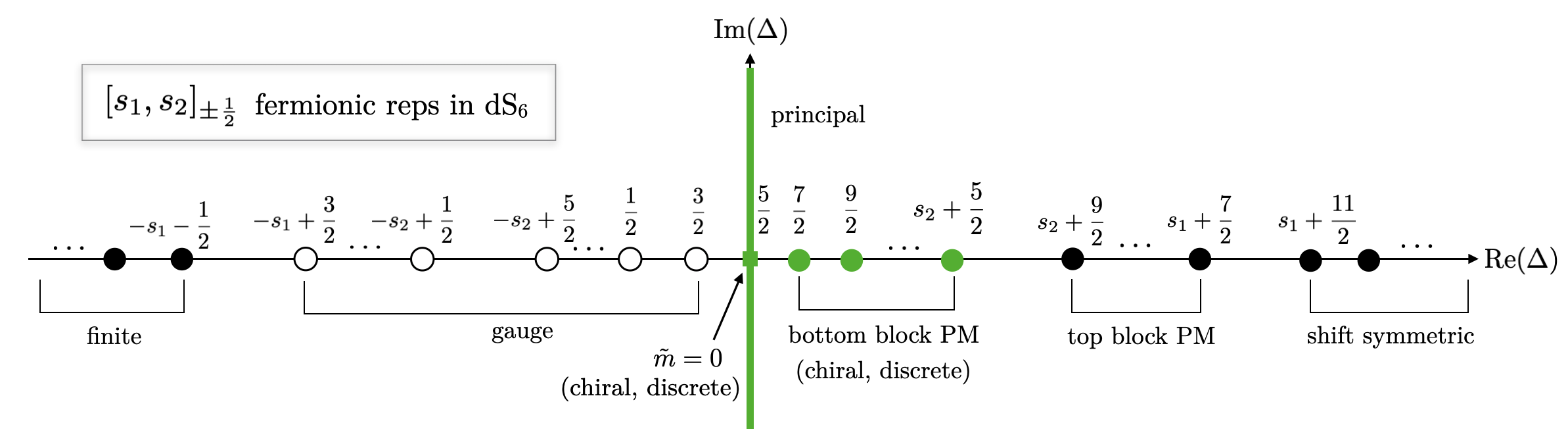,width=7.0in}}  \label{d6pic6}\, \ee
The principal series is $(F_0,F_1,F_2)=(-{5\over 2}+i\nu,s_1+\half,s_2+\half)$, with $\nu>0$,  accounting for the equivalence upon reflecting through $\Delta=5/2$.  
The PM points where $t$ boxes in the bottom row are activated are now unitary: these are the fermionic discrete series $D^\pm$ with $(F_0,F_1,F_2)=(s_2-t-{3\over 2} ,s_1+\half ,s_2+\half)$ in $D=6$.  (It includes as the special case $s_1=s_2=1$, $t=1$, the fermionic massless 2-form.)  The point at $\Delta=5/2$ represents two reps, the two chiral parts of the $\tilde m=0$ fermion; these are the fermionic discrete series reps $D^\pm$ with $(F_0,F_1,F_2)=(-{5\over 2} ,s_1+\half ,s_2+\half)$.

\section{$D=2$ Representations\label{D2section}}

In this section, we cover the $D=2$ case, which has been excluded thus far.  We treat this case separately because not only is there a lot of degeneration and simplification that occurs, but also new possibilities that do not have counterparts in higher dimensions.  

The isometry algebra of dS$_2$ is $\frak{so}(1,2)$, which is isomorphic to the algebra $\frak{sl}(2,{\mathbb R})$ of real traceless $2\times 2$ matrices.  The reps of this algebra were originally studied in \cite{Bargmann:1946me,10.1073/pnas.38.4.337,PUKANSZKY1964}, modern reviews can be found here \cite{Joung:2006gj,Kitaev:2017hnr,Anous:2020nxu} (see also the classic book \cite{lang1985sl2}).  This rep theory has been relevant for recent work on fields in dS$_2$, see e.g.  \cite{Joung:2006gj,Joung:2007je,Epstein:2016rsr,Epstein:2020tyv,Epstein:2020wgf,Higuchi:2021fxg,Anninos:2023lin,Farnsworth:2024yeh} and \cite{Blanco:2022quy,Letsios:2025pqo} for fermionic cases.  They are also relevant for the study of physics on AdS$_2$ \cite{Maldacena:2016hyu}, since this space has the same isometry algebra as that of dS$_2$.  

For $D=2$, $d=1$, the representation space will consist of spaces of fields on the circle, ${\mathbb S}^1$.  Of the bosonic reps we looked at in section \ref{bosonsection}, the only fields that are non-trivial on ${\mathbb S}^1$ are the scalar and vector, and the vector is dual to the scalar.  Of the fermionic reps we looked at in section \ref{fermionsection}, the only non-trivial fermion is the spin $1/2$ field.  In each of these cases, the field has only one component.

However, the circle ${\mathbb S}^1$ has a new feature that is not shared by any of the higher dimensional spheres: there are non-contractible closed curves, those which wrap around the circle.  Unlike the higher dimensional spheres, the circle has two possible spin structures, which correspond to the fermion field being either periodic or anti-periodic as we wrap around the circle.  However, given that we are interested in reps of the algebra and not the group, we do not want to impose any restrictive boundary condition and we should allow the most general possibility: the most general possibility is to allow the field to be multiplied by an arbitrary complex number as we go around the circle.  As we will see, this general possibility allows for anyon-like reps that have no higher dimensional counterpart.

In $D=2$, it will pay to streamline the notation a bit, to take advantage of the simplified kinematics.  We are working with the algebra $\frak{so}(1,2)$ which has only three generators among \eqref{killingenbeddinge}, which we will call
\be  {\cal J}={\cal M}^{21}\, ,\ \ \  {\cal K}^1={\cal M}^{10}\, , \ \ \ {\cal K}^2={\cal M}^{20}\,. \ee
Here ${\cal J}$ is the single rotation generator, generating rotations of the ${\mathbb S}^1$, and ${\cal K}^1,{\cal K}^2$ are the two boost generators.
The commutators \eqref{sod12algberaadse2e} become
\be \left\{ {\cal J},{\cal K}^1\right\}=-{\cal K}^2,\ \ \  \left\{ {\cal J},{\cal K}^2\right\}={\cal K}^1,\ \ \  \left\{ {\cal K}^1,{\cal K}^2\right\}={\cal J}.\ \ \ \label{liebraked2dee}\ee
Evaluating \eqref{killingvectorsgede1}, \eqref{killingvectorsgede2} in the $D=2$ case, taking $\theta\in [0,2\pi)$ to be the coordinate on the circle, the Killing vectors become
\bea && {\cal J}= \partial_\theta \,,\nn\\
&& {\cal K}^1= {1\over H} \cos\theta \,\partial_t -\tanh\left( H t\right)\sin\theta \, \partial_\theta \,, \nn\\
&& {\cal K}^2= {1\over H}\sin\theta \, \partial_t + \tanh\left(H  t\right)\cos\theta \, \partial_\theta  \,.\label{translfden2e}
\eea

In what follows, it will be convenient to revert to the custom of using bra-ket notation and a quantum mechanics-like normalization where we take ${\cal J}\rightarrow i J$, ${\cal K}^{1,2}\rightarrow iK^{1,2}$ and we write \eqref{liebraked2dee} as commutators, giving
\be \left[ { J},{ K}^1\right]=i{ K}^2\, ,\ \ \  \left[ { J},{ K}^2\right]=-i{ K}^1\, ,\ \ \  \left[ { K}^1,{ K}^2\right]=-i{ J}\, .\ \ \ \label{commhquge} \ee
In a unitary rep, the generators $J$, $K^{1}$, $K^{2}$ will then be Hermitian rather than anti-Hermitian.  

We define raising/lowering combinations,
\bea && K^{\pm}=K^1\pm i K^2\, , \nn\\
&& K^1={1\over 2}\left( K^++K^-\right)\, ,\ \ \ K^2=-{i\over 2} \left( K^+-K^-\right)\, ,\label{raiselowercombgd2ee}
\eea
in terms of which the commutators \eqref{commhquge} are
\be \left[ { J},{ K}^+\right]={ K}^+\,, \ \  \left[ { J},{ K}^-\right]=-{ K}^-\, ,\ \ \  \left[ { K}^+,{ K}^-\right]=-2{ J}\, .\label{kpmjcommute}\ee
For unitary reps, these operators will satisfy $J^\dag=J$, $\left( K^\pm\right)^\dag=K^\mp$.

The representation space will be the space of complex functions on the circle, $\phi(\theta)$.
As mentioned above, the difference between dS$_2$ and the higher dimensional $dS$ spaces, as defined via the embedding \eqref{globalcoordsinemee}, is that dS$_2$ is not simply connected because there is a non-trivial cycle going around the spatial circle, and the field may satisfy a non-trivial periodicity condition as it goes around the circle.  The most general such condition for a single complex function is for it to get transformed by a $GL(1)$ transformation, i.e. multiplication by a complex number, as it goes around the spatial circle.  We will parametrize this complex number like a phase, writing it as $e^{2\pi i \mu}$, where $\mu$ is a complex number defined modulo an integer, $\mu\in {\mathbb C}/{\mathbb Z}$.  It is a true phase only if $\mu$ is real.
To get the most general reps we thus take $\phi(\theta)$ to have the following boundary condition going around the circle: 
\be  \phi(\theta+2\pi) =  e^{2\pi i \mu}  \phi(\theta) \, , \ \ \  \mu\in {\mathbb C}/{\mathbb Z}\, .\ee 
We say that $ \phi$ is $\mu$-periodic\footnote{There are no non-trivial complex line bundles on the circle, so this periodicity requirement does not represent a non-trivial bundle.  This can be seen by redefining the field by $\phi\rightarrow e^{i\mu\theta}\phi'$, whereupon the field $\phi'$ is periodic.  But this redefinition would introduce $\mu$ dependence into the transformations \eqref{translfden2e}, so we do not take this route.}: the case  $\mu=0$ will be standard periodic bosons, or fermions with a periodic spin structure, and $\mu=1/2$ will be fermions with an anti-periodic spin structure.   Other values of $\mu$ will be generalizations analogous to anyons, and are known in the literature as automorphic scalars \cite{Isham:1977yc,Avis:1978qc,Banach:1979iy,Banach:1978dt,Banach:1979wz,Higuchi:2022nfy}.  
When we want to remove the mod ${\mathbb Z}$ ambiguity in $\mu$ we will restrict
\be {\rm Re}\, {\mu}\in \left(-{1\over 2},{1\over 2}\right]\,.\ee

The action \eqref{latetimesheactiont}, \eqref{latetimesheactiont2} of the dS$_2$ transformations as conformal transformations on the functions $\phi(\theta)$ of the circle becomes
\bea J \phi=-i\partial_\theta  \phi\, ,\ \ \ K^1  \phi=i\left( \Delta \cos\theta +\sin\theta\, \partial_\theta \right) \phi\, , \ \  \ K^2  \phi=i\left( \Delta \sin\theta -\cos\theta \,\partial_\theta \right) \phi\,,\label{confortramsdele1de}
\eea
or in terms of the combinations \eqref{raiselowercombgd2ee},
\be J \phi =-i \partial_\theta  \phi\, ,\ \ \ K^\pm \phi=e^{\pm i\theta}\left(i\Delta \pm \partial_\theta\right) \phi\, .\label{confortramsdele1d23e} \ee

The representation space will be the space of $\mu$-periodic functions transforming as in \eqref{confortramsdele1d23e}, and is denoted
\be {\cal F}^{[\mu]}_\Delta:\ \ {\rm complex\ } \mu \text{-periodic scalar\ functions\ on\ } {\mathbb S}^1 \,.\ee

We next want to decompose this space into $\frak{so}(2)$ representations.  A basis of this space that accomplishes this is 
\be |m\ra \equiv{1\over \sqrt{2\pi}} e^{i m \theta},\ \ \ m\in \mu+{\mathbb Z}\,.\label{mufunctiobasise}\ee
Note that for $\mu=0$ the $m$ labels are integers, for $\mu=1/2$ they are odd half-integers, but for generic $\mu$ they are neither integers nor odd half-integers: $m=\ldots,\mu-2,\mu-1,\mu,\mu+1,\mu+2,\ldots$. 
Acting on the basis \eqref{mufunctiobasise} we have 
\be J|m\ra=m |m\ra \, ,\ \ \ K^\pm|m\ra = i\left(\Delta\pm m\right)|m\pm 1\ra\,.\label{pmcioperee}\ee
This diagonalizes the $\frak{so}(2)$ rotation generator $J$, so the states \eqref{mufunctiobasise} are the analog of the spherical harmonics in higher dimensions.  

For generic $\mu,\Delta$, the coefficient on the right hand side of the $K^\pm$ action in \eqref{pmcioperee} never vanishes, and so by using the $K^\pm$ operators we can get from any given $m$ state to any other, and thus the rep is irreducible.  This generic rep is illustrated here
\be \raisebox{-10pt}{\epsfig{file=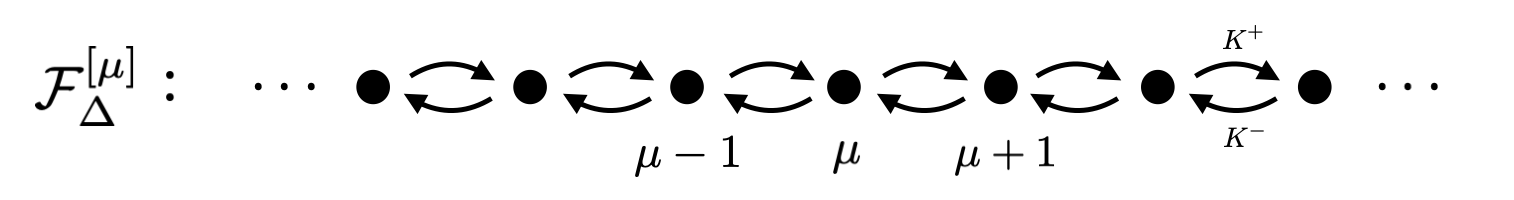,width=4.5in}}  \label{D2content1}\, \ee
The dots represent the states \eqref{mufunctiobasise}, with the values of $m$ as  indicated underneath, and the arrows represent the action of $K^\pm$.

\textbf{Reducible cases:} 
For specific values of $\mu,\Delta$, some coefficients in the $K^+$ or $K^-$ action in \eqref{pmcioperee} can vanish for some values of $m$, which breaks some of the arrows in \eqref{D2content1} and causes an irreducible sub-rep to develop.  This occurs in the following cases:

\begin{itemize}

\item Bosonic shift symmetric points: 
\be \mu=0\,, \ \ \Delta=k+1\, , \ \ k=0,1,2,\ldots \ .   \ee
In this case we have $K^-\left|k+1 \right>=0$ and $K^+\left|-k-1\right>=0$.  The rep is reducible but indecomposable; the states with $m\geq k+1$ form an irreducible sub-rep which we call 
\be {\cal D}^{\left[ 0 \right] +}_{k+1} \, , \ee 
and the states with $m\leq-k-1$ form an irreducible sub-rep which we call 
\be {\cal D}^{\left[ 0 \right] -}_{k+1}\, .\ee
This is illustrated as follows:
\be \raisebox{-10pt}{\epsfig{file=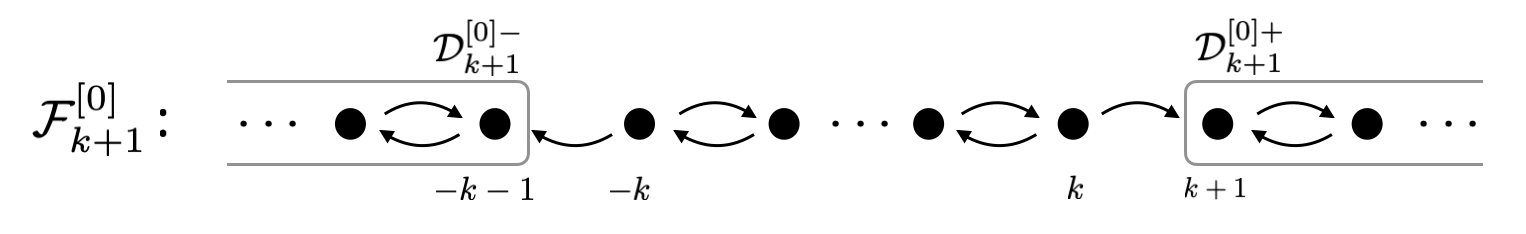,width=4.5in}}  \label{D2content3}\, \ee
As we will see, these are known as the scalar discrete series reps.  They correspond to left and right moving chiral parts of the physical modes of the shift symmetric scalars in $D=2$, also known as the discrete series scalars \cite{Anninos:2023lin,Farnsworth:2024yeh}.  

\item Fermionic shift symmetric points: 

\be \mu=\half\,, \ \ \Delta=k+{3\over 2}\, , \ \ k=0,1,2,\ldots \ .   \ee
In this case we have $K^-\left|k+{3\over 2} \right>=0$ and $K^+\left|-k-{3\over 2}\right>=0$.  The rep is reducible but indecomposable; the states with $m\geq k+{3\over 2}$ form an irreducible sub-rep which we call 
\be {\cal D}^{\left[ \half \right] +}_{k+{3\over 2}} \, , \ee
and the states with $m\leq-k-{3\over 2}$ form an irreducible sub-rep which we call 
\be {\cal D}^{\left[ \half \right] -}_{k+{3\over 2}}\,.\ee 
This is illustrated as follows:
\be \raisebox{-10pt}{\epsfig{file=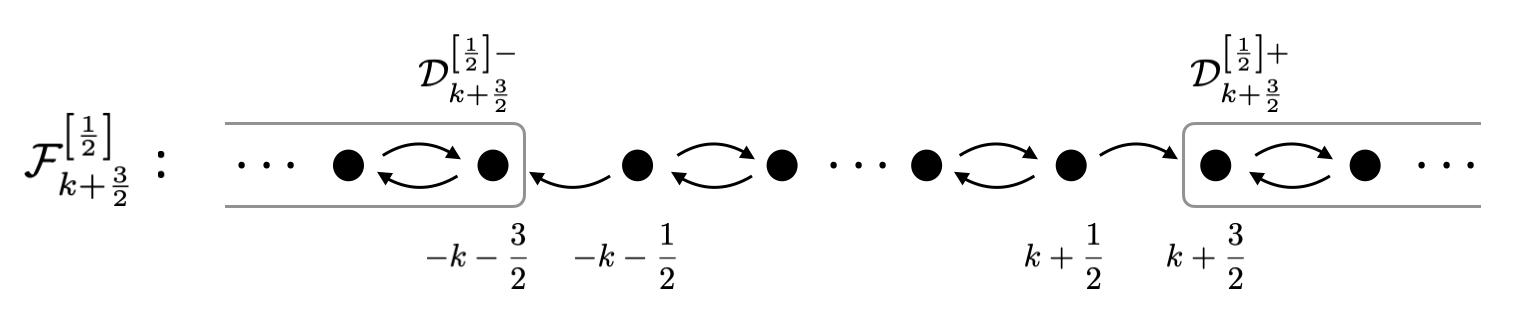,width=4.5in}}  \label{D2content3b}\, \ee
As we will see, these are known as the fermionic discrete series reps.  They correspond to left and right moving chiral parts of the physical modes of the shift symmetric fermions in $D=2$ \cite{Letsios:2025pqo}.

\item Bosonic finite points:
\be \mu=0\,, \ \ \Delta=-k\,, \ \ k=0,1,2,\ldots\ .\ee
In this case we have $K^-\left|-k \right>=0$ and $K^+\left|k\right>=0$.  The rep is reducible but indecomposable, and there are several different sub-reps: 

\begin{itemize}
\item The states with $-k\leq m\leq k$ form a finite dimensional irreducible sub-rep which we call 
\be {\cal S}^{\left[ 0 \right] }_{-k}\,.\ee
\item The states with $m\geq -k$ form a sub-rep which we call
\be {\cal S}^{\left[ 0 \right]+ }_{-k}\,.\ee
It is itself reducible and contains ${\cal S}^{\left[ 0 \right] }_{-k}$ as an irreducible sub-rep.  
\item The states with $m\leq k$ form a sub-rep which we call 
\be {\cal S}^{\left[ 0 \right]-}_{-k}\,.\ee 
It is itself reducible and also contains ${\cal S}^{\left[ 0 \right] }_{-k}$ as an irreducible sub-rep.  
\end{itemize}

These sub-reps are illustrated as follows:
\be \raisebox{-10pt}{\epsfig{file=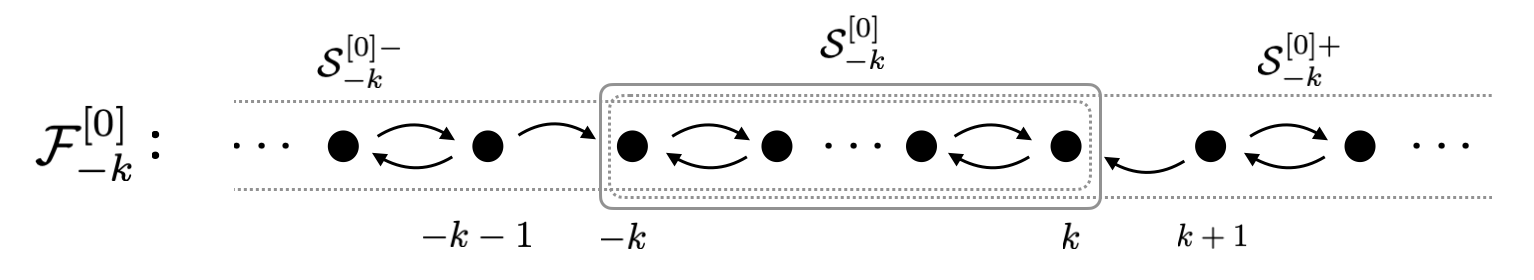,width=4.5in}}  \label{D2content4}\, \ee
The dotted lines delineate the reps ${\cal S}^{\left[ 0 \right]\pm}_{-k}$, which do not count as irreps since they are not irreducible.  The solid line delineates ${\cal S}^{\left[ 0 \right]}_{-k}$, which is the intersection of ${\cal S}^{\left[ 0 \right]+}_{-k}$ and ${\cal S}^{\left[ 0 \right]-}_{-k}$.  The rep ${\cal S}^{\left[ 0 \right]}_{-k}$ is irreducible and has dimension $2k+1$.  These are the $\frak{so}(1,2)$ analogs of the familiar finite dimensional $\frak{so}(3)$ bosonic spin $k$ reps.  They correspond to the shift symmetries of the shift symmetric scalars in $D=2$.

\item Fermionic finite points:
\be \mu=\half\,, \ \ \Delta=-k-\half\,, \ \ k=0,1,2,\ldots\ .\ee
In this case we have $K^-\left|-k-\half \right>=0$ and $K^+\left|k+\half\right>=0$.  The rep is reducible but indecomposable, and there are several different sub-reps: 

\begin{itemize}
\item The states with $-k-\half\leq m\leq k+\half$ form a finite dimensional irreducible sub-rep which we call 
\be {\cal S}^{\left[ \half \right] }_{-k-\half}\,.\ee
\item The states with $m\geq -k-\half$ form a sub-rep which we call 
\be {\cal S}^{\left[ \half \right]+ }_{-k-\half}\, .\ee
It is itself reducible and contains ${\cal S}^{\left[ \half \right] }_{-k-\half}$ as an irreducible sub-rep.  
\item The states with $m\leq k+\half$ form a sub-rep which we call 
\be {\cal S}^{\left[ \half \right]-}_{-k-\half}\,.\ee 
It is itself reducible and also contains ${\cal S}^{\left[ \half \right] }_{-k-\half}$ as an irreducible sub-rep.  
\end{itemize}

These sub-reps are illustrated as follows:
\be \raisebox{-10pt}{\epsfig{file=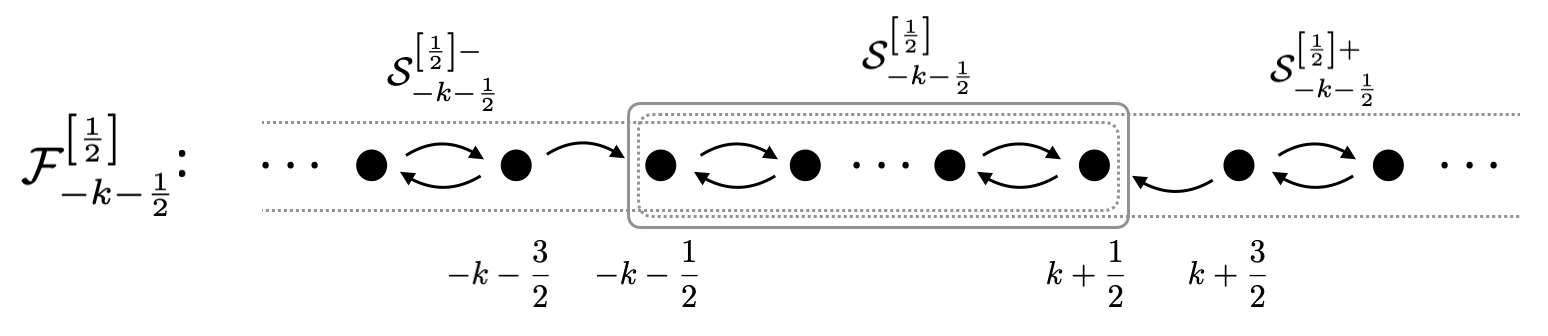,width=4.5in}}  \label{D2content4b}\, \ee
The dotted lines delineate the reps ${\cal S}^{\left[ \half \right]\pm}_{-k-\half}$, which do not count as irreps since they are not irreducible.  The solid line delineates ${\cal S}^{\left[ \half \right]}_{-k-\half}$, which is the intersection of ${\cal S}^{\left[ \half \right]+}_{-k-\half}$ and ${\cal S}^{\left[ \half \right]-}_{-k-\half}$.  The rep ${\cal S}^{\left[ \half \right]}_{-k-\half}$ is irreducible and has dimension $2k+2$.  These are the $\frak{so}(1,2)$ analogs of the familiar finite dimensional $\frak{so}(3)$ fermionic spin $k+\half$ reps.  They correspond to the shift symmetries of the shift symmetric fermions in $D=2$.

\item
Anyonic shift symmetric points (plus branch):
\be -\half <\mu <\half\, , \mu\not=0\,, \ \  \Delta=\mu+k\, , \ \ k\in {\mathbb Z}\,.  \ee 
We have $K^-\left|\mu+k \right>=0$.  The rep is reducible but indecomposable; the states with $ m\geq \mu+k$ form an infinite dimensional irreducible sub-rep which we call 
\be {\cal D}^{\left[ \mu \right]+ }_{\mu+k}\,.\ee
This is illustrated as follows:
\be \raisebox{-10pt}{\epsfig{file=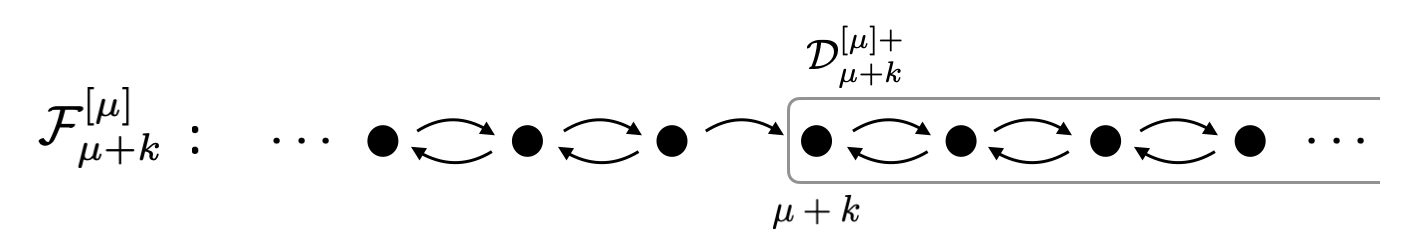,width=4.5in}}  \label{D2content5}\, \ee

\item
Anyonic shift symmetric points (minus branch):
\be -\half <\mu <\half\, , \mu\not=0\,, \ \  \Delta=-\mu-k+1\, , \ \ k\in {\mathbb Z}\,.  \ee 
We have $K^+\left|\mu+k-1 \right>=0$.  The rep is reducible but indecomposable; the states with $ m\leq \mu+k-1$ form an infinite dimensional irreducible sub-rep which we call 
\be {\cal D}^{\left[ \mu \right]- }_{\mu-k+1}\,.\ee
This is illustrated as follows:
\be \raisebox{-10pt}{\epsfig{file=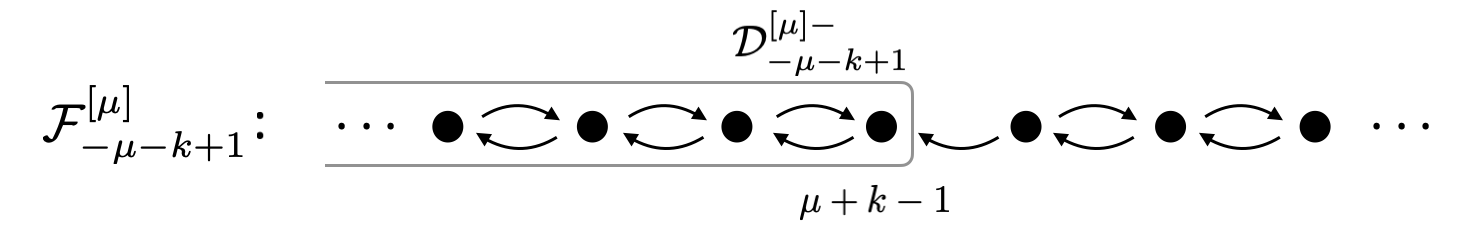,width=4.5in}}  \label{D2content6}\, \ee

\item Massless fermion point: 
\be  \mu={1\over 2}\, ,\ \Delta={1\over 2}\, .\ee  
In this case we have $K^-\left|{1\over 2}\right>=0$ and $K^+\left|-{1\over 2}\right>=0$, so the rep decomposes into two irreducible pieces ${\cal F}^{\left[ 1\over 2\right] \pm}_{1\over 2}$, where ${\cal F}^{\left[ 1\over 2\right] +}_{1\over 2}$ consists of the states with $m\geq{1\over 2} $ and ${\cal F}^{\left[ 1\over 2\right] -}_{1\over 2}$ consists of the states with $m\leq -{1\over 2} $,
\be {\cal F}^{\left[ 1\over 2\right] }_{1\over 2}={\cal F}^{\left[ 1\over 2\right] +}_{1\over 2}\oplus {\cal F}^{\left[ 1\over 2\right] -}_{1\over 2}\,.\ee
 This is illustrated as follows:
\be \raisebox{-10pt}{\epsfig{file=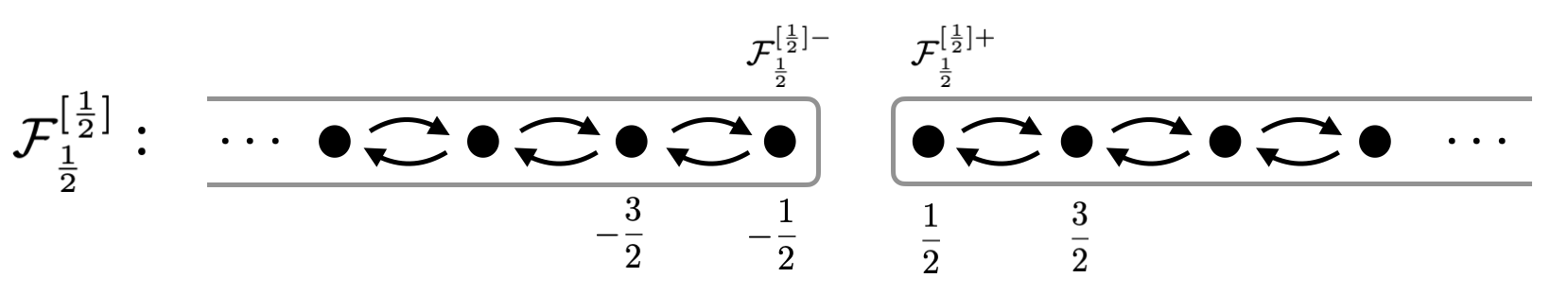,width=4.5in}}  \label{D2content2}\, \ee
The value $\mu=\half$ is a fermion with anti-periodic boundary conditions, and $\Delta=\half$ corresponds to the massless value $\tilde m=0$; these reps correspond to left and right moving massless fermions on dS$_2$, with anti-periodic spin structure.  This is the $D=2$ version of the chiral splitting of the fermionic reps that occurs when $\tilde m=0$ in even $D$ .  

\end{itemize}

\textbf{Equivalences:} 
As in higher dimensions, there are equivalences between the reps: for generic $\Delta,\mu$, there is an equivalence between {shadow} reps,
\be {\cal F}_{\Delta}^{[\mu]} \simeq {\cal F}_{\bar \Delta}^{[\mu]}\, , \ \ \ \bar\Delta\equiv 1-\Delta\,.\label{shadowefed2ee}\ee
The equivalence is via the intertwining operator $S_\Delta^{[\mu]}$, 
\be S_\Delta^{[\mu]} :\ {\cal F}_{\Delta}^{[\mu]}\rightarrow {\cal F}_{\bar\Delta}^{[\mu]}\,,\ee
which commutes with the rotation $J$ and satisfies
\be { K}^\pm_{\bar\Delta}S_\Delta^{[\mu]}=S_\Delta^{[\mu]} { K}^\pm_{\Delta}\,.\label{Kcommsee} \ee
Since $S_\Delta^{[\mu]}$ commutes with $J$, it can be diagonalized on the basis \eqref{mufunctiobasise}, so that we have 
\be S_\Delta^{[\mu]} |m\ra=s_{\Delta,m}^{[\mu]} |m\ra\, ,\label{d2sdiagmde}\ee
for some constants $s_{\Delta,m}^{[\mu]}$.

Writing out the action of \eqref{Kcommsee} on the states \eqref{mufunctiobasise} and using \eqref{d2sdiagmde}, we obtain the recursion relation 
\be \left(\Delta+m\right) s_{\Delta,m+1}^{[\mu]}=\left(\bar\Delta+m\right) s_{\Delta,m}^{[\mu]}\,.\label{srecursed2e}\ee
For generic $\Delta,\mu$, neither side of \eqref{srecursed2e} ever vanishes and we have the solution
\be  s_{\Delta,m}^{[\mu]}={\Gamma(\bar\Delta+m)\Gamma(\Delta+\mu) \over \Gamma(\Delta+m)\Gamma(\bar\Delta+\mu)} \, ,\label{d2recsole}\ee
where we have chosen to normalize $s_{\Delta,\mu}^{[\mu]}=1$.  In this generic case, the map $S_\Delta^{[\mu]}$ is invertible.

For the cases where sub-reps appear, the mapping $S_\Delta^{[\mu]}$ develops a kernel.  Consider first the minus branch anyonic shift symmetric points, where $-\half <\mu <\half$, $\mu\not=0$, $\Delta=-\mu-k+1$, $k\in {\mathbb Z}$. The recursion relation \eqref{srecursed2e} with $m=\mu+k-1$ tells us that 
$s_{-\mu-k+1,\mu+k-1}^{[\mu]}=0$, and then the recursion relation fixes all $s_{-\mu-k+1,m\leq \mu+k-1}^{[\mu]}=0$.  We can freely choose $s_{-\mu-k+1,\mu+k}^{[\mu]}=1$, and then the recursion relation will fix all $s_{-\mu-k+1,m> \mu+k-1}^{[\mu]}$ to be non-vanishing.  The mapping $S_{-\mu-k+1}^{[\mu]}$ thus develops a kernel consisting of all the states $m\leq \mu+k-1$, which is precisely the sub-rep ${\cal D}^{\left[ \mu \right]- }_{-\mu-k+1}$.  The image is ${\cal D}^{[\mu]+}_{\mu+k}$.

Next consider the plus branch anyonic shift symmetric points, where $-\half <\mu <\half$, $\mu\not=0$, $\Delta=\mu+k$, $k\in {\mathbb Z}$.  The recursion relation \eqref{srecursed2e} with $m=\mu+k-1$ tells us that 
$s_{\mu+k,\mu+k}^{[\mu]}=0$, and then the recursion relation fixes all $s_{\mu+k,m\geq \mu+k}^{[\mu]}=0$.  We can freely choose $s_{\mu+k,\mu+k-1}^{[\mu]}=1$, and then the recursion relation will fix all $s_{\mu+k,m<\mu+k-1}^{[\mu]}$ to be non-vanishing.  The mapping $S_{\mu+k}^{[\mu]}$ thus develops a kernel consisting of all the states $m\geq\mu+k$, which is precisely the sub-rep ${\cal D}^{\left[ \mu \right]+ }_{\mu+k}$.  The image is ${\cal D}^{[\mu]-}_{-\mu-k+1}$.

The mapping between the spaces $ {\cal F}^{[\mu]}_{\mu+k}$ and ${\cal F}^{[\mu]}_{-\mu-k+1}$, and the subrepresentations which are the kernels of the maps, is illustrated here:
\be \raisebox{-40pt}{\epsfig{file=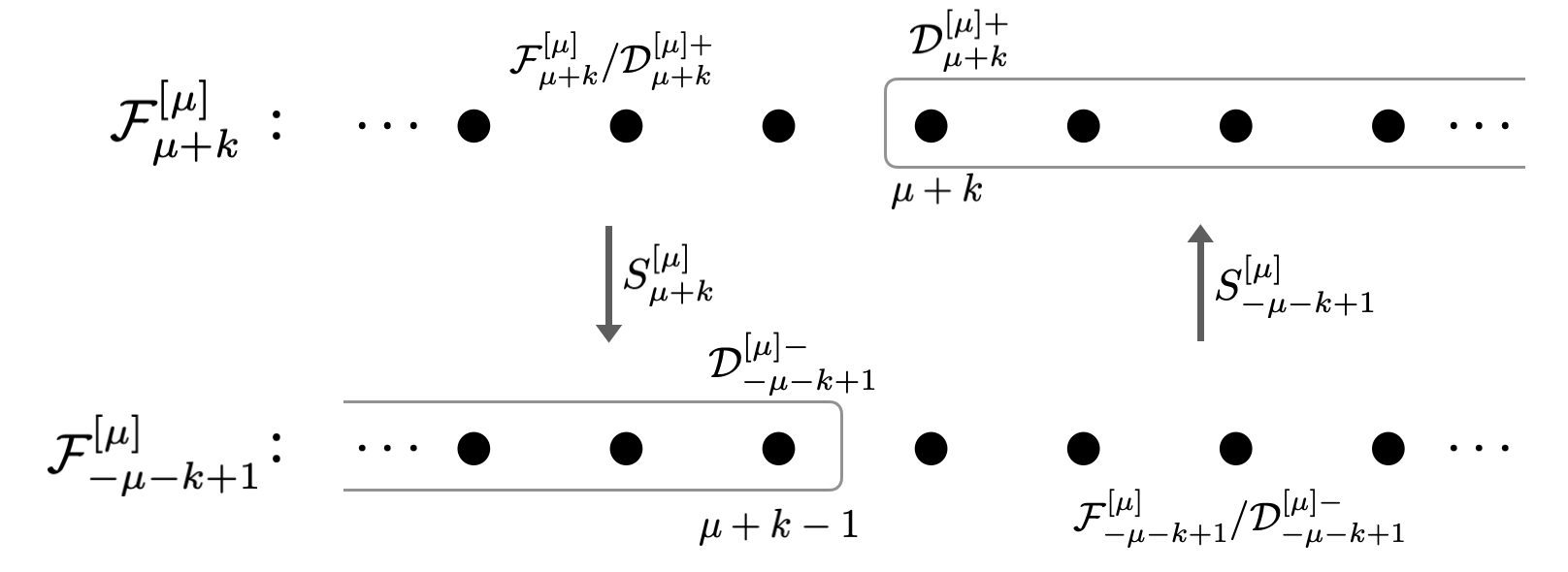,width=5.0in}}  \label{D2content7}\, \ee
From this we see the isomorphisms
\be {\cal D}^{[\mu]-}_{-\mu-k+1}\simeq {\cal F}^{[\mu]}_{\mu+k}/{\cal D}^{[\mu]+}_{\mu+k}\,, \ \ {\cal D}^{[\mu]+}_{\mu+k}\simeq {\cal F}^{[\mu]}_{-\mu-k+1}/{\cal D}^{[\mu]-}_{-\mu-k+1}\,. \ee
 These reps should correspond to the modes of some kind of shift symmetric version of the automorphic scalars.

Now consider the scalar shift symmetric points, $\mu=0$, $\Delta=k+1$, $k=0,1,2,\ldots$.  The recursion relation \eqref{srecursed2e} with $m=k$ tells us that 
$s_{k+1,k+1}^{[0]}=0$, and then fixes all $s_{k+1,m\geq k+1}^{[0]}=0$.  The recursion relation with $m=-k-1$ tells us that 
$s_{k+1,-k-1}^{[0]}=0$, and then fixes all $s_{k+1,m\leq-k-1}^{[\mu]}=0$.  The value $s_{k+1,k}^{[0]}$ can be fixed to 1, after which all the $s_{k+1,-k\leq m\leq k}^{[0]}$ are fixed to non-vanishing values.  Thus the kernel of $S_{k+1}^{[0]}$ consists of all the states $m\geq k+1$ and $m\leq-k-1$, which is precisely the sub-rep ${\cal D}^{[0]+}_{k+1} \oplus {\cal D}^{[0]-}_{k+1}$, and the image is the space ${\cal S}^{[0]}_{-k}$.

For the bosonic finite points, $\mu=0$, $\Delta=-k$, $k=0,1,2,\ldots$, the recursion relation \eqref{srecursed2e} with $m=-k-1$ tells us that 
$s_{-k,-k}^{[0]}=0$, and then the recursion relation fixes all $s_{-k,k\leq m\geq-k}^{[0]}=0$.  There are now two free choices to make, $s_{-k,k+1}^{[0]}$ and $s_{-k,-k-1}^{[0]}$.  If we choose $s_{-k,k+1}^{[0]}=1$, then all $s_{-k,m\geq k+1}^{[0]}$ are fixed and non-vanishing, but if we choose $s_{-k,k+1}^{[0]}=0$, then $s_{-k,m\geq k+1}^{[0]}=0$.  If we choose $s_{-k,-k-1}^{[0]}=1$, then all $s_{-k,m\leq -k-1}^{[0]}$ are fixed and non-vanishing, but if we choose $s_{-k,-k-1}^{[0]}=0$, then $s_{-k,m\leq -k-1}^{[0]}=0$.  Let $S^{[0]+}_{-k}$ be the map where we make the choice $s_{-k,m\leq-k-1}^{[0]}=0$, $S^{[\mu]-}_{-k}$ be the map where we make the choice $s_{-k,m\geq k+1}^{[0]}=0$, and $S^{[0]}_{-k}=S^{[0]+}_{-k}+S^{[0]-}_{-k}$ be the choice with both choices non-zero.  The kernel of $S^{[0]+}_{-k}$ is ${\cal S}^{[0]-}_{-k}$ and its image is ${\cal D }^{[0]+}_{k+1}$.  The kernel of $S^{[0]-}_{-k}$ is ${\cal S}^{[0]+}_{-k}$ and its image is ${\cal D}^{[0]-}_{k+1}$.  The kernel of $S^{[0]}_{-k}$ is ${\cal S}^{[0]}_{-k}$ and its image is ${\cal D}^{[0]+}_{-k}\oplus {\cal D}^{[0]-}_{-k}$.

The mappings and subrepresentations between the bosonic shift symmetric and finite points are illustrated here:
\be \raisebox{-40pt}{\epsfig{file=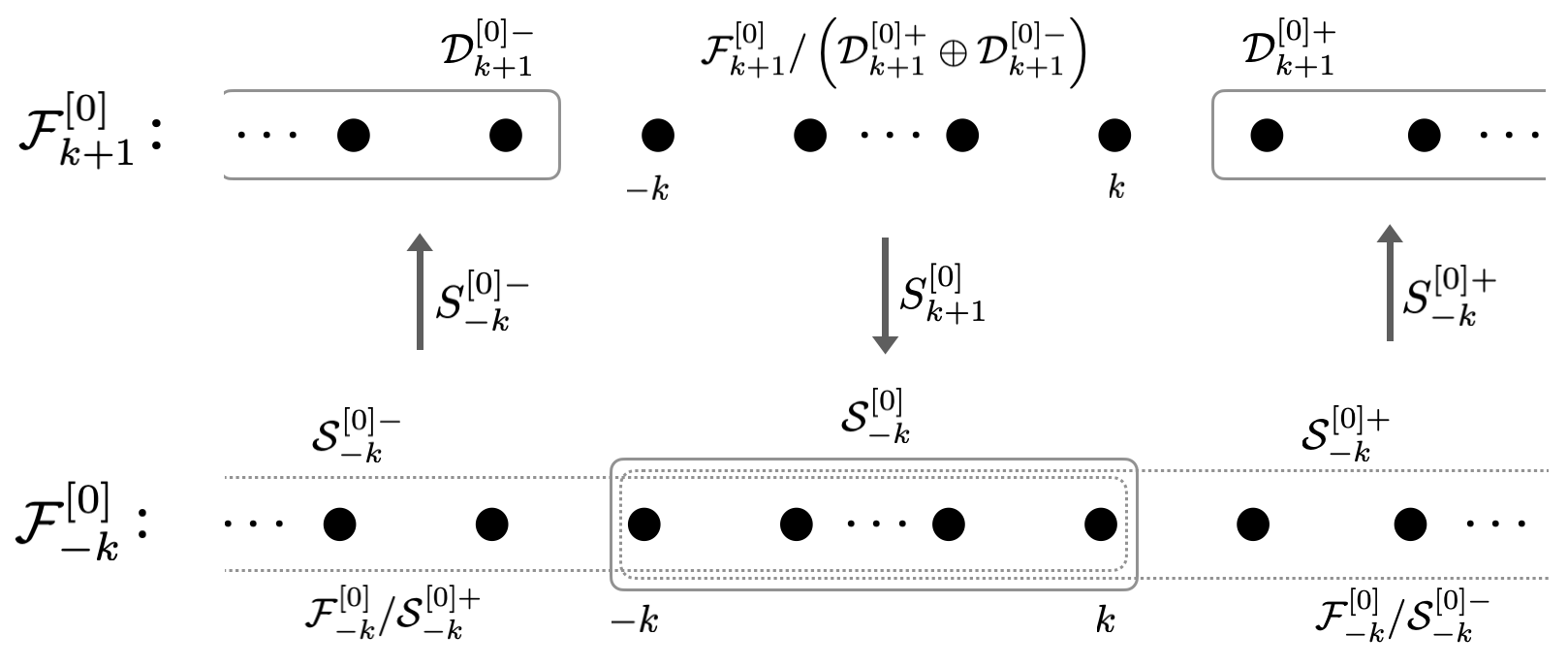,width=5.0in}}  \label{D2content8}\, \ee
From this we see the following isomorphisms:
\be { \cal S}^{[0]}_{-k}\simeq { \cal F}^{[0]}_{k+1}/\left({ \cal D}^{[0]+}_{k+1}\oplus { \cal D}^{[0]-}_{k+1}\right) \,, \ \ { \cal D}^{[0]\pm }_{k+1}\simeq { \cal F}^{[0]}_{-k}/{ \cal S}^{[0]\mp}_{-k} \,,\ \  { \cal D}^{[0]+}_{k+1}\oplus { \cal D}^{[0]-}_{k+1} \simeq { \cal F}^{[0]}_{-k}/{ \cal S}^{[0]}_{-k}\,. \ee
 
There is an exactly analogous situation relating the fermionic shift symmetric and finite points, this is illustrated here:
 \be \raisebox{-40pt}{\epsfig{file=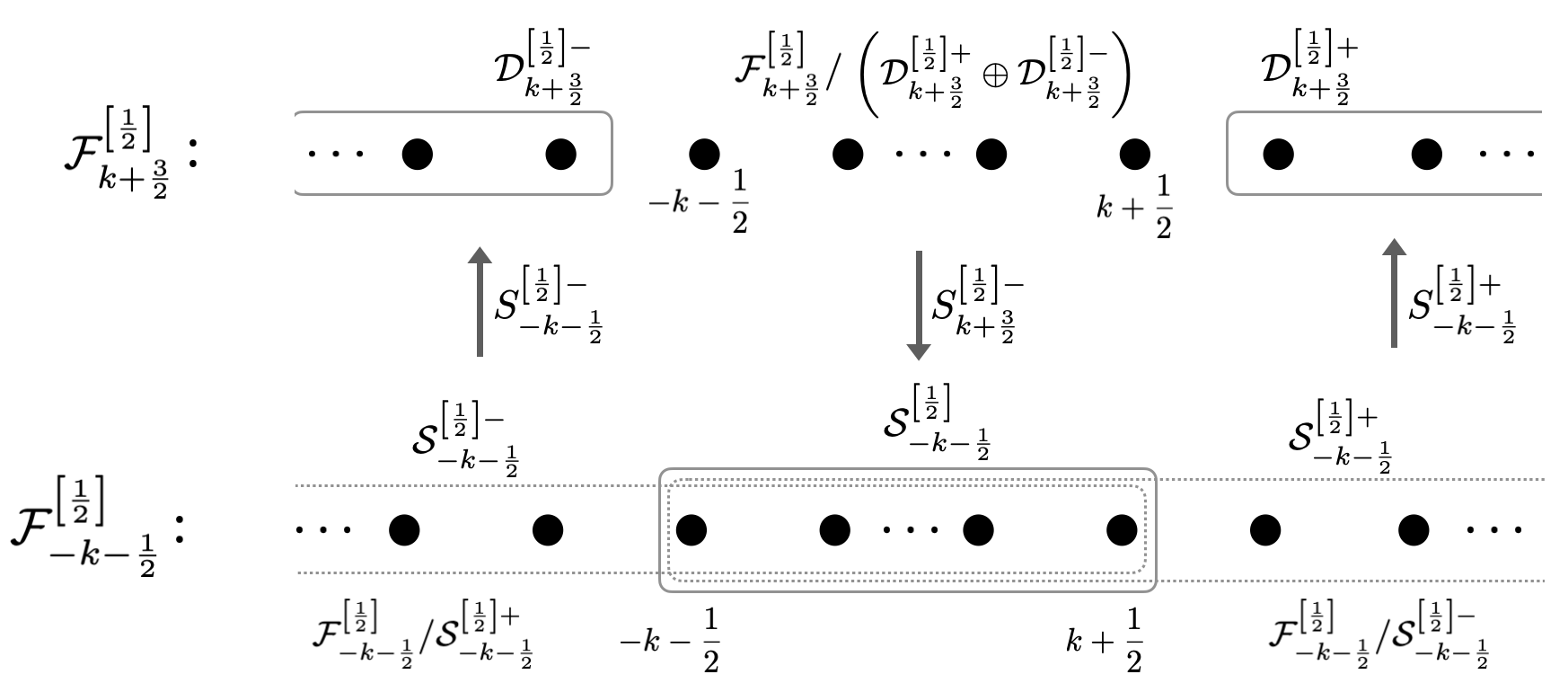,width=5.0in}}  \label{D2content8b}\, \ee
From this we see the following isomorphisms:
\be { \cal S}^{\left[\half\right]}_{-k-\half}\simeq { \cal F}^{\left[\half\right]}_{k+{3\over 2}}/\left({ \cal D}^{\left[\half\right]+}_{k+{3\over 2}}\oplus { \cal D}^{\left[\half\right]-}_{k+{3\over 2}}\right) \,, \ \ { \cal D}^{\left[\half\right]\pm}_{k+{3\over 2}}\simeq { \cal F}^{\left[\half\right]}_{-k-\half}/{ \cal S}^{\left[\half\right]\mp,}_{-k-\half} \,,\ \  { \cal D}^{\left[\half\right]+}_{k+{3\over 2}}\oplus { \cal D}^{\left[\half\right]-}_{k+{3\over 2}} \simeq { \cal F}^{\left[\half\right]}_{-k-\half}/{ \cal S}^{\left[\half\right]}_{-k-\half}\,. \ee

For the fermionic massless case, $\mu={1\over 2}$, $\Delta={1\over 2}$, the recursion relation \eqref{srecursed2e} is automatically satisfied for $m=-\half$, and we can freely choose both $s^{[\half]}_{\half,\half}$ and $s^{[\half]}_{\half,-\half}$.  Choosing $s^{[\half]}_{\half,-\half}=0$ forces $s^{[\half]}_{\half,m\leq -\half}=0$, and choosing $s^{[\half]}_{\half,-\half}=1$ sets $s^{[\half]}_{\half,m\leq -\half}=1$.  Choosing $s^{[\half]}_{\half,\half}=0$ forces $s^{[\half]}_{\half,m\geq \half}=0$, and choosing $s^{[\half]}_{\half,\half}=1$ sets $s^{[\half]}_{\half,m\geq \half}=1.$  Choosing $s^{[\half]}_{\half,-\half}=0$ and $s^{[\half]}_{\half,\half}=1$ gives a map $S^{[\half]+}_{\half}$ whose kernel is ${\cal D}^{[\half]-}_{\half}$ and whose image is ${\cal D}^{[\half]+}_{\half}$.  Choosing $s^{[\half]}_{\half,-\half}=1$ and $s^{[\half]}_{\half,\half}=0$ gives a map $S^{[\half]-}_{\half}$ whose kernel is ${\cal D}^{[\half]+}_{\half}$ and whose image is ${\cal D}^{[\half]-}_{\half}$.  Taking both $s^{[\half]}_{\half,\half}=s^{[\half]}_{\half,-\half}=1$ gives the identity map, $S^{[\half]+}_{\half}+S^{[\half]-}_{\half}={\mathbb I}$.  This gives ${\cal D}^{[\half]\pm}_{\half}\simeq {\cal F}^{[\half]}_{\half}/{\cal D}^{[\half]\mp}_{\half}$.

\textbf{Unitarity:}
There is an diffeomorphism+Weyl invariant (with Weyl weight as in \eqref{weylnormdrele} with $r=0$), and hence conformally invariant, non-degenerate bi-linear pairing between the spaces ${\cal F}^{[\mu]}_\Delta$ and ${\cal F}^{[-\mu]}_{\bar\Delta}$,
\be (\phi_1,\phi_2)\equiv \int_0^{2\pi} d\theta\ \phi_1(\theta) \phi_2(\theta),\ \ \ \phi_1\in {\cal F}^{[-\mu]}_{\bar\Delta},\ \ \ \phi_2\in {\cal F}^{[\mu]}_{\Delta}\,.\ \ \ \label{bilinearfordews2e}\ee
In general this is not an invariant inner product because each argument involves a different action of the algebra, and because both arguments are linear rather than one of them being anti-linear. 

If $\mu\in {\mathbb R}$, $\Delta={1\over 2}+i\nu$, $\nu\in {\mathbb R}$, then $\phi^\ast\in {\cal F}^{-\mu}_{1-\Delta}$ if $\phi\in  {\cal F}^{\mu}_{\Delta}$, and \eqref{bilinearfordews2e} becomes a true inner product on $ {\cal F}^{\mu}_{\Delta}$ defined by
\be \la \phi_1|\phi_2\ra\equiv (\phi_1^\ast,\phi_2)=\int_0^{2\pi} d\theta\ \phi_1^\ast(\theta) \phi_2(\theta)\,,\ \ \phi_1,\phi_2\in {\cal F}^{[\mu]}_\Delta\,.\ee
These are the principal series reps in $D=2$,
\be D=2\ {\rm principal\ series:}\ \  \mu\in {\mathbb R}\, , \ \ \ \Delta={1\over 2}+i\nu,\ \ \ \nu\in {\mathbb R}\,  .\label{D2princseriere}\ee
The cases $\nu$ and $-\nu$ are equivalent due to \eqref{shadowefed2ee}, other than this, they are distinct reps.  The case $\mu=\half$, $\nu=0$ is the massless fermionic point where the rep splits into two pieces.  This point is usually excluded from the principal series and instead included in the discrete series discussed below, since it behaves similarly.

In the case where $\mu\in {\mathbb R}$ and $\Delta\in {\mathbb R}$, we can use the map $S^{[\mu]}_{\Delta}$ to move a state from ${\cal F}_{ \Delta}^{[\mu]}$ to ${\cal F}_{ \bar\Delta}^{[\mu]}$, follow with a conjugation to move it to  ${\cal F}_{\bar \Delta}^{[-\mu]}$, and form an inner product on ${\cal F}_{ \Delta}^{[\mu]}$ as follows,
\be \la \phi_1 | \phi_2 \ra\equiv \left(\left(S^{[\mu]}_{\Delta}\phi_1\right) ^\ast,\phi_2\right)\, ,\ \ \ \mu\in {\mathbb R}\, , \ \ \Delta\in {\mathbb R}\, .\label{innerprodpriscsdee2} \ee 
The positivity of this inner product is now equivalent to whether the matrix elements $s^{[\mu]}_{\Delta,m}$ are all positive.  Away from the cases where the representation becomes reducible, using \eqref{d2recsole} and the canonical range $-\half<\mu\leq \half$ for $\mu$, the strongest constraints come from $m=\mu-1$ for $\mu\geq 0$ and $m=\mu+1$ for $\mu\leq 0$, giving
\be D=2 {\rm \ complementary\ series:}\ \ \ |\mu|< \Delta < 1-|\mu|\,. \ee
The case $\Delta=\half$ can be excluded since this is already accounted for in the principal series.  The cases $\Delta$ and $1-\Delta$ are equivalent due to \eqref{shadowefed2ee}, other than this, they are distinct reps.
Note that this complementary series range interpolates between the scalar complementary series $0<\Delta<1$ when $\mu=0$ and the absent fermionic complementary series when $\mu=1/2$.

Now return to the cases where the representation becomes reducible.  In these cases, there is a subrepresentation corresponding to the kernel of the intertwiner $S^{[\mu]}_{\Delta}$, which, due to the presence of the intertwiner in \eqref{innerprodpriscsdee}, gives null states.  These null states are to be factored out, and unitarity is then equivalent to whether the remaining states all have the same sign for their norms, so that the factor rep can be made unitary.  The cases which are unitary in this way are the scalar shift symmetric points ${\cal D}^{[0]\pm}_{k+1}$ realized through the quotient ${\cal F}^{[\mu]}_{k+1}/{\cal S}^{[\mu]\mp}_{k+1}$, the fermionic shift symmetric points ${\cal D}^{\left[ \half\right]\pm}_{k+{3\over 2}}$ realized through the quotient ${\cal F}^{\left[ \half\right]}_{-k-{1\over 2}}/{\cal S}^{\left[ \half\right]\pm}_{-k-{1\over 2}}$, 
and the anyonic points: ${\cal D}^{[\mu]+}_{\mu+k}$ realized through the quotient ${\cal F}^{[\mu]}_{-\mu-k+1}/{\cal D}^{[\mu]-}_{-\mu-k+1}$ and ${\cal D}^{[\mu]-}_{-\mu-k+1}$ realized through the equivalence with ${\cal F}^{[\mu]}_{\mu+k}/{\cal D}^{[\mu]+}_{\mu+k}$.  Collectively, these are known as the discrete series representations.  As mentioned above, the $\nu=0$ point of the $\Delta=\half$ principal series is also usually included in the discrete series, as it also splits into the two chiral parts ${\cal D}^{[\half]\pm}_\half$.

Apart from these cases, there is no way to construct an invariant inner product on ${\cal F}^{[\mu]}_\Delta$, so all the other reps in the space of complex $\mu,\Delta$ are non-unitary. 

\textbf{Summary:} 
All the interesting things happen in the plane of real $\Delta$ and real $\mu$, illustrated here:
\be \raisebox{-40pt}{\epsfig{file=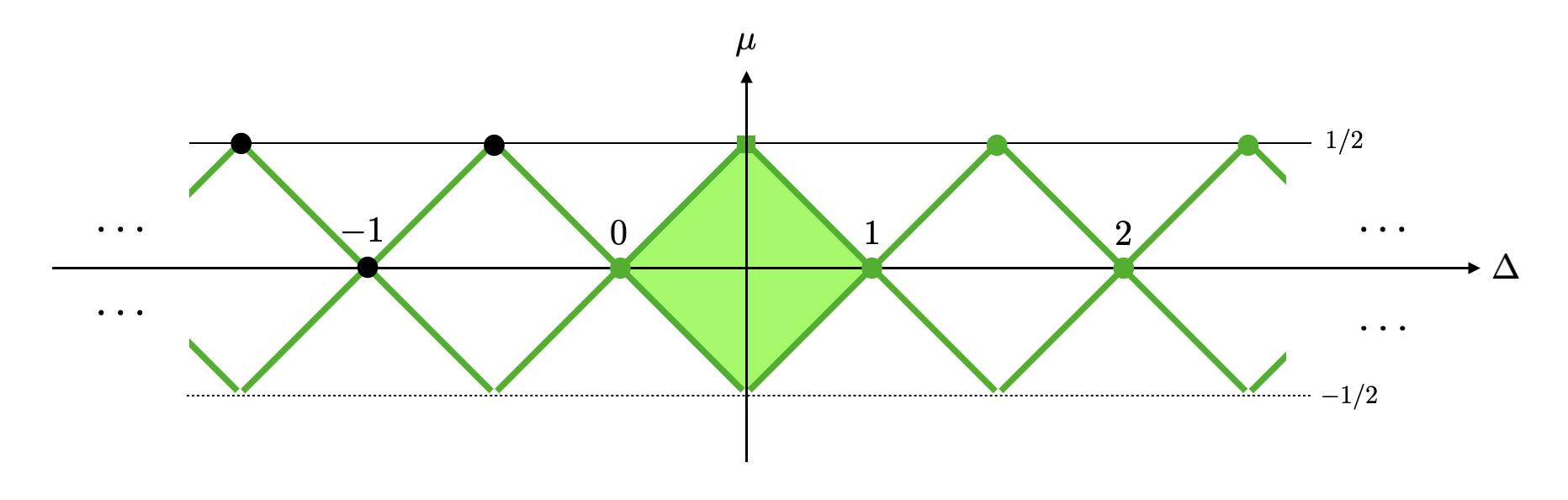,width=6.5in}}  \label{D2dsreps}\, \ee
Each point on the plane is a unique rep, apart from the shadow equivalence \eqref{shadowefed2ee}, which is given by reflecting about the $\mu$ axis at $\Delta=\half$.  Points in green are the unitary reps.  The solid circles, and the green lines, are places where shortening conditions occur, and the shadow equivalence breaks down at these points, so these all represent distinct reps.   The black circles, and the green circle at $\Delta=1$, are the finite reps, with the green circle at $\Delta=1$ the trivial rep. The remaining green circles are the bosonic and fermionic discrete series reps, which each split into two chiral parts.  The square at $\Delta=\half$, $\mu=\half$ is the massless fermionic point where the rep splits into two chiral parts, so this point represents two inequivalent reps, and is also usually counted among the fermionic discrete series.
The complementary series is the interior of the solid green diamond, interpolating between its maximal range in the bosonic case $\mu=0$, and its absence in the fermionic case $\mu=1/2$.  The principal series \eqref{D2princseriere} can be thought of as coming out of the page, along an axis of imaginary $\Delta$, connected on the line $\Delta=1/2$. The upwards sloping green lines are the discrete reps ${\cal D}^{[\mu]+}_{\Delta}$, and the downward sloping green lines are the ${\cal D}^{[\mu]-}_{\Delta}$.    For non-real values of $\mu$, there are no shortening conditions and no unitary reps.

The only reps which correspond to genuine reps of the connected part of the de Sitter group, $SO^+(1,2)$, are those with $\mu=0$.  The others are reps of the universal cover of $SO^+(1,2)$, or projective reps of $SO^+(1,2)$.

The quadratic Casimir operator ${\cal C}_2$ of section \ref{casimirsec} can be written in $D=2$ as
\bea {\cal C}_2 &=& -{\cal J}^2+\left({\cal K}^1\right)^2 +\left({\cal K}^2\right)^2=J^2-\left({ K}^1\right)^2 -\left({ K}^2\right)^2 \nn\\
 &=&J^2+J-K^-K^+= J^2-J-K^+K^-=J^2-\half \left\{ K^+,K^-\right\}\,,\eea
and its value on the reps is independent of the boundary conditions encoded in $\mu$,
\be  {\cal C}_2 =\Delta(\Delta-1)=-{\tilde m^2\over H^2}\,.\ee

\section{Concluding Remarks\label{conclusionssec}}

We have reviewed the representations of the de Sitter algebra $\frak{so}(1,D)$, the algebra of isometries of dS$_D$.  We covered all the reps in all dimensions, including the mixed symmetry reps present in higher $D$, bosonic and fermionic, unitary and non-unitary, and connected them to the various types of fields that can propagate on dS$_D$.  

When we say that we have covered the non-unitary reps, what we really mean is that we have covered all of what might be called the ``canonical single field'' reps, meaning those that are realized as the space of solutions of the Klein-Gordon equation for a single tensor or spinor-tensor field with irreducible index symmetries.  It is conceivable that there is a wider class of non-unitary reps that correspond to multiple fields, or higher derivative equations, in a way that cannot be reduced to a sum of canonical single fields.  For example, partially massless higher spin theories in certain dimensions give rise to Lagrangians involving multiple fields that cannot be diagonalized \cite{Brust:2016zns}, corresponding to non-unitary AdS or CFT reps involving ``extended modules'' \cite{Brust:2016gjy} that cannot appear in the unitary case.  There could also be non-canonical single-field or multi-field cases involving fields that have other non-trivial twists or global conditions on the sphere.  (With an infinite number of fields there are examples of what should be non-unitary reps that we do not see with single fields: the infinite spin (often called continuous spin) fields, extended to dS, become non-unitary \cite{Metsaev:2016lhs}.)  
It would be interesting to know how to interpret these in terms of the general dS representation theory in which all the $F$ labels in section \ref{unitarylistsection} are permitted to take arbitrary complex values, or involve additional twisting parameters analogous to $\mu$ in the $D=2$ case in section \ref{D2section}.

For the unitary reps, these subtleties cannot happen, because we have the complete classification of unitary reps and all of them are realized by canonical single fields (and the $\mu$ twist in the case of $D=2$).   For the other maximally symmetric spacetimes this is not the case, because there are unitary reps which are not realized by the standard fields.   For example, there are unitary infinite spin reps in flat space and in AdS that require an infinite tower of fields \cite{Bekaert:2005in,Schuster:2013pta,Schuster:2014hca,Rivelles:2014fsa,Metsaev:2016lhs,Metsaev:2017ytk,Khabarov:2017lth,Alkalaev:2017hvj,Buchbinder:2018soq,Metsaev:2019opn,Metsaev:2021zdg,Basile:2023vyg,Astrakhantsev:2026mby}, and singleton reps in AdS that are not realized in terms of standard local bulk fields \cite{Flato:1999yp}.  In this sense, the unitary representation theory of dS is simpler than that of the other maximally symmetric spaces; there are no unitary infinite spin or other unusual reps to worry about.

The scope of this review has been limited to essentially listing and describing the reps. There are many other deep and interesting topics in the representation theory of the de Sitter algebra that we did not touch on.  One is the character theory of de Sitter representations: these characters are known as Harish-Chandra characters, developed by Harish-Chandra in \cite{bams/1183520006,dedb2908-a5b1-306f-9180-dbe55c128349}, and first computed for the de Sitter group by Hirai \cite{1965526}.  Modern reviews that include these are \cite{Basile:2016aen,Sun:2021thf}.   Another is the more general subject of harmonic analysis as applied to the de Sitter group; a standard reference on this is \cite{Dobrev:1977qv} and a more recent summary can be found in \cite{Karateev:2018oml}.  Another important topic we left out is the intricate theory of ``addition of angular momenta'' on dS, i.e. how products of the reps break up into sums \cite{Dobrev:1976vr,22f5d7e4-b846-3465-ad03-6e378620a9f6,5e459275-4cbd-351b-894f-05c1ee5066db}, and the theory of branching, i.e. how the reps split upon restriction to a smaller dS or Poincar\'e subgroup, or join together into reps of larger groups (we did, however, cover the branching under $\frak{so}(1,D)\rightarrow \frak{so}(D)$ in detail, since this is precisely the $\frak{so}(D)$ content that we used to understand the structure of the reps).  All of this has seen interesting recent applications \cite{Loparco:2023akg,Penedones:2023uqc,Loparco:2025azm}, and much remains to be elucidated and put to use.

\paragraph{Acknowledgments:}

The author would like to thank James Bonifacio, Austin Joyce, Vasileios Letsios, Rachel Rosen, and Zimo Sun for useful discussions and feedback on the material presented here. 

\appendix

\section{Flat Slicing\label{flatslicing}}

Throughout this review, we have used exclusively the global coordinates on dS$_D$ described in section \ref{globalcoordinatessec}.  These are the coordinates most naturally adapted to the $\frak{so}(D)$ decomposition of the $\frak{so}(1,D)$ reps, since the spatial sections are $d$-spheres whose isometry generators are precisely the $\frak{so}(D)$ generators.  In cosmological applications, however, it is more natural to use inflationary coordinates in which the spatial sections are flat, since the constant density spatial slices of our universe appear to be nearly flat at the largest length scales we can see.  All the reps can equally well be described using flat spatial sections, and in this appendix we describe how to map the reps described in terms of global coordinates into the reps described in terms of inflationary coordinates.

The flat slicing, or inflationary, coordinates of dS$_D$ can be defined through the embedding 
\bea X^0&=& {{1\over H^2}+y^2-\tau^2\over 2(-\tau)}\,, \nn\\ 
X^{i}&=&{1\over H(-\tau)}y^i \, ,\ \ i=1,\ldots,d\, , \nn\\ 
X^D&=& {{1\over H^2}-y^2+\tau^2\over 2(-\tau)}\,.  \label{inflationaryembedding} 
\eea 
Here $\tau\in (-\infty,0)$ is the time coordinate (cosmological conformal time) and $y^i\in (-\infty,\infty)$, $i=1,\ldots,d$ are spatial coordinates.  

These coordinates only cover half of dS space, the region $X^0+X^D>0$.  The coordinates cover all but one point of the sphere at future infinity; the missing point is what we call the south pole of the sphere at future infinity, the point that is approached as $X^0=-X^D\rightarrow \infty$ with $X^i=0$.  They cover the $X^D>0$ half of the minimal sphere at $X^0=0$.  And they cover only a single point of the sphere at past infinity, the north pole at $X^0=-X^D\rightarrow -\infty$ with $X^i=0$.  

\begin{figure}
\begin{center}
\epsfig{file=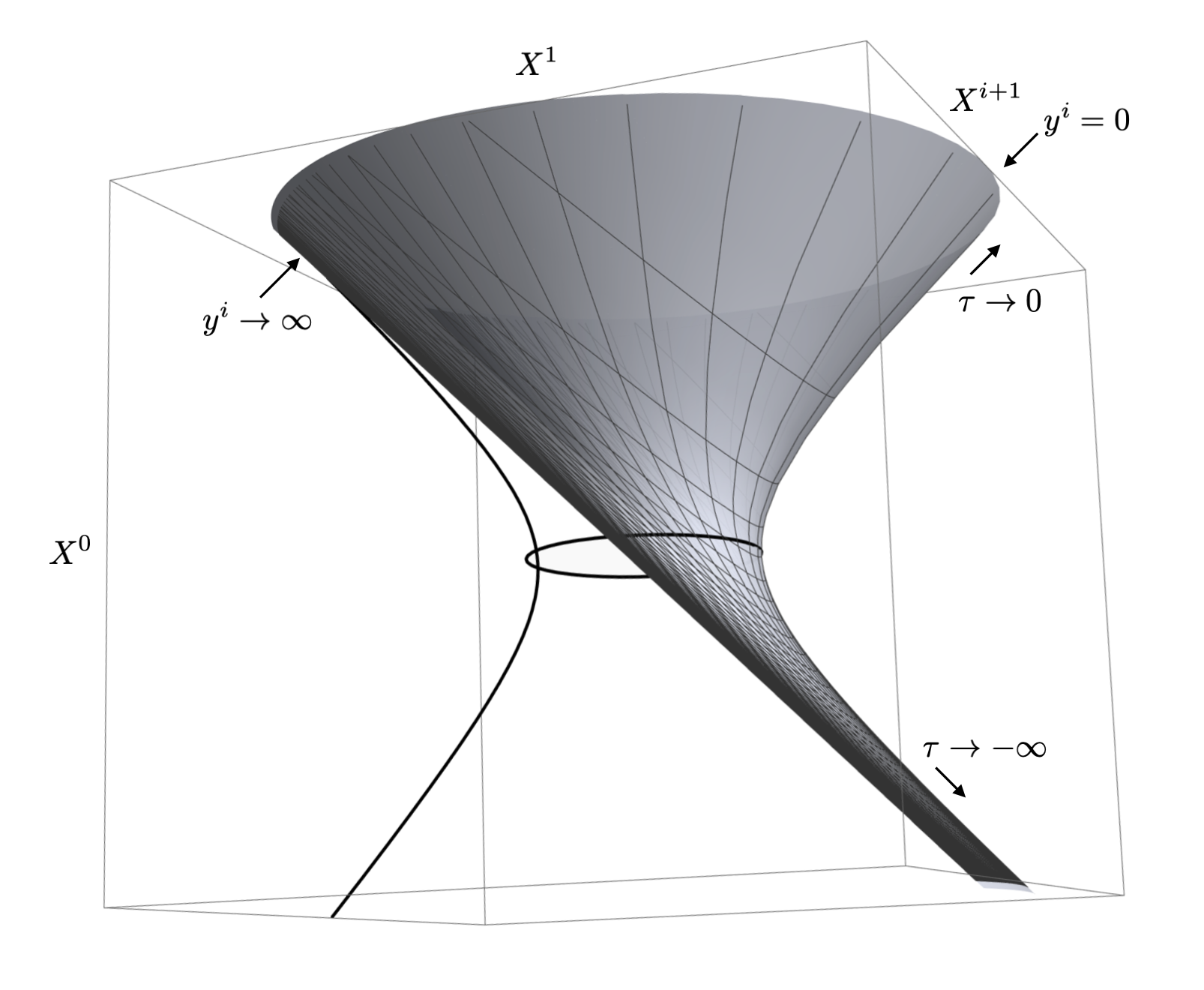,width=4in}
\caption{\small Inflationary coordinates on de Sitter space, as seen in embedding space.}
\label{dsinflationarycoords}
\end{center}
\end{figure}

Using \eqref{inflationaryembedding} in \eqref{inducedmetricfforme}, we find the metric in these coordinates,
\be ds^2={1\over H^2\tau^2}\left(-d \tau^2+\delta_{ij}dy^idy^j\right)\,.\ee
This describes a foliation of dS$_D$ into space-like flat ${\mathbb R}^d$ slices of constant $\tau$.   When working in these coordinates, spatial indices are always raised and lowered with $\delta_{ij}$.  The future boundary is at $\tau \rightarrow 0^-$, and it is a plane, covering all but the one missing point of the future boundary sphere.    
See figure \ref{dsinflationarycoords}.

\textbf{Isometries:}  

Projecting the Killing vectors \eqref{killingenbeddinge} onto the dS$_D$ surface in the inflationary coordinates gives the intrinsic Killing vectors, which we can arrange into the following combinations:
\bea
&&J^{ij}\equiv  {\cal M}^{ij} = y^j\partial^i-y^i\partial^j \, ,\nn\\ 
&&D\equiv  {\cal M}^{0D}= \tau \partial_\tau+y^i\partial_i\,, \nn\\
&& P^i \equiv  H \left( {\cal M}^{iD} +{\cal M}^{i0}\right)= \partial^i \,,\nn\\
&& K^i \equiv {1\over H} \left({\cal M}^{iD}-{\cal M}^{i0} \right)= 2\tau y_i\partial_\tau+\left(\tau^2-y^2\right)\partial_i+2y_iy^j\partial_j\,. \label{conformalbasxise}
\eea
These have the following non-vanishing Lie brackets:
\bea \left\{J^{i j },J^{l k }\right\} &=&\delta^{i l}J^{j  k }-\delta^{j l}J^{i  k }+\delta^{j  k }J^{i l}-\delta^{i  k }J^{j l}\,, \nn\\
 \left\{J^{i j },P^ k \right\} &=& \delta^{ ik } P^j -\delta^{ jkj } P^i \,, \nn \\
   \left\{ J^{i j },K^ k \right\} &=& \delta^{ ik }K^j -\delta^{jk}K^i\,, \nn\\
\left\{ D,P^i \right\} &=& -P^i \,, \nn \\
 \left\{ D,K^i \right\} &=& K^i \,, \nn\\
  \left\{ K^i ,P^j \right\} &=&2(-\delta^{i j } D+J^{i j })\,. 
   \eea
This is the same $\frak{so}(1,D)$ algebra as \eqref{sod12algberaadse}, or \eqref{sod12algberaadse2e}, in a different basis.  The $P^i$, $J^{ij}$ are the generators that preserve the constant $\tau$ slices; they span a Poincar\'e subalgebra $\frak{iso}(1,D-1)\subset \frak{so}(1,D)$, which is the algebra of Poincar\'e transformations on these flat slices. 

\textbf{Fields:}

Consider a rank $r$ tensor field $\Phi_{\mu_1\ldots \mu_r}$ on dS$_D$, fully transverse and traceless, with indices in some general tableau, and satisfying a Klein-Gordon equation $\left(\nabla^2-\tilde m^2\right)\Phi_{\mu_1\ldots \mu_r}=0$.  As in global coordinates, there will generally be constraints and gauge fixing conditions that constrain all the components with any index in the $\tau$ direction, so we can ignore these.  The components of the Klein-Gordon equation in the remaining $y^i$ directions become
\be \left( -\partial_\tau^2  +{d-2r-1\over \tau}\partial_\tau +\partial^2+{r(d-r+1)-{\tilde m^2\over H^2} \over \tau^2}\right)  \Phi_{i_1\ldots i_r}=0\, ,\label{flateuqerTee}\ee
where $\partial^2\equiv\delta^{ij}\partial_i\partial_j$ is the Laplacian on ${\mathbb R}^d$.  

This equation can be solved near the $\tau=0$ future boundary as a power series in $(-\tau)$.  Looking for a leading order solution $\Phi_{i_1\ldots i_r}(\tau,y)= (-\tau)^{\Delta-r}\phi_{i_1\ldots i_r} (y)$, \eqref{flateuqerTee} gives at leading order the equation
\be { \tilde m^2\over H^2}=-\Delta(\Delta-d)+r \,, \label{latsgfsolflatere}\ee
which has the two solutions
\be  \Delta_\pm ={d\over 2}\pm \sqrt{{d^2\over 4}+r-{\tilde m^2\over H^2} }\,,\label{dscfttmassrelatione2}  \ee
the same as in the spherical slicing \eqref{deltamregexe}, \eqref{dscfttmassrelatione}.  These give the two independent solutions, with the leading order behaviors
\be \Phi_{i_1\ldots i_r}(\tau,y)  \underset{\tau \rightarrow 0}{\rightarrow} (-\tau)^{\Delta_\pm-r}\phi_{i_1\ldots i_r} (y)\,.\ee

Consider now the action of the Killing vectors on these tensor fields at late times.  We evaluate $ \delta_{\xi}\Phi_ {i_1\ldots i_r}=-{\cal L}_{\xi} \Phi_{i_1\ldots i_r}$ at late times for each Killing vector $\xi$ in \eqref{conformalbasxise}, and deduce their action on the late time boundary fields $\phi_{i_1\ldots i_r}$.
The $J^{ij}$ and $P^i$ Killing vectors do not act on $\tau$, and they transform the late time field by isometries of ${\mathbb R}^d$,
\bea  &&\delta_{P^{i}}  \phi_{i_1\ldots i_r} =-\partial^i \phi_{i_1\ldots i_r}\,,\nn\\
&&\delta_{J^{ij}}  \phi_{i_1\ldots i_r} =\left(y^i\partial^j-y^j\partial^i+{\cal J}^{ij}\right) \phi_{i_1\ldots i_r}\,, \label{confjflfee1}
\eea
where ${\cal J}^{ij}\phi_{i_1\ldots i_r} =\sum_{n=1}^r 2\delta^{[i}_{i_n} \delta^{j]k} \phi_{i_1\ldots i_{n-1} k i_{n+1}\cdots i_r}$ is the action of rotations on the indices.
For $D$ and $K^i$, the action on $\tau$ amounts to $\tau\partial_\tau \rightarrow {\Delta-r}$ (with $\Delta$ either of $\Delta_\pm$), and we use the relations $\partial_j (D)^i=\delta^i_j$, $\partial_j (K_i)^k=2\delta_{ij}y^l+4y_{[i}\delta_{j]}^{\ l}+{\cal O}(\tau^2)$ to evaluate the Lie derivatives to obtain
\bea &&\delta_{D}  \phi_{i_1\ldots i_r} = -\left(y^i\partial_i +\Delta\right)  \phi_{i_1\ldots i_r}  \, ,\nn\\
&& \delta_{K^i}  \phi_{i_1\ldots i_r} =\left(y^2\partial^i-2y^iy^j\partial_j-2y^i\Delta-2y_j{\cal J}^{ij}\right) \phi_{i_1\ldots i_r}\,. \label{confjflfee2}
\eea
The transformations \eqref{confjflfee1}, \eqref{confjflfee2} are precisely the conformal transformations for a rank $r$ primary field of weight $\Delta$ on flat Euclidean ${\mathbb R}^d$.  Components of the field with indices in the $t$ direction will also appear in \eqref{flateuqerTee} and in the equations coming from the Klein-Gordon equation with components in the $t$ directions.  These $t$ components of the field will behave like $\sim  (-\tau)^{\Delta_\pm -r +r_t}$, where $r_t$ is the number of indices in the $t$ direction, so these will be subleading, they do not affect the leading relation \eqref{latsgfsolflatere}, and they are fixed by the boundary data on the purely spatial components.

\textbf{Relation to global coordinates:}  

In global coordinates, the reps were described in terms of spaces of functions living on the sphere ${\mathbb S}^d$ at future infinity.  In inflationary coordinates, the reps are instead described in terms of spaces of functions with certain falloff conditions living on the plane ${\mathbb R}^d$ at future infinity.  These two spaces are related by a Weyl transformation, which we can use to transform tensor fields between them.

We first need to describe the sphere at infinity in terms of coordinates which manifest its conformal flatness.  This is achieved by stereographic coordinates. 
In terms of the sphere's embedding space, these coordinates are
\be  \hat X^i={y^i\over 1+{y^2\over 4{\cal R}^2} }\,, \ \   \hat X^D=  {\cal R}{-1+{y^2\over 4{\cal R}^2} \over 1+{y^2\over 4{\cal R}^2} }\,,
\ee
where $y^i\in (-\infty,\infty)$.
Here we have taken the sphere to have an arbitrary radius ${\cal R}$, rather than the unit sphere we worked with throughout the main text, so that the sphere coordinates will have the same length dimension as their flat space counterparts, making the mapping more straightforward.   
In these coordinates, the metric on ${\mathbb S}^d$ takes the form
\be g_{ij}=\Omega(y)^2 \delta_{ij}\,,\ \ \ \Omega(y)\equiv {1\over 1+{y^2\over 4{\cal R}^2}}\,.\label{sphereflatcee}\ee
 These coordinates cover all of the sphere except for one point, the point at infinity $y^i\rightarrow \infty$, which is the south pole, the point $\hat X^D=-{\cal R},\hat X^i=0$ in embedding space (the same point on the sphere at infinity that is missed by the inflationary coordinates).  
 
Under a Weyl transformation with a position dependent parameter $\sigma$, the metric and a conformal primary field of weight $\Delta$ and rank $r$ transform as
\be  g_{ij}\rightarrow \sigma^2 g_{ij} \, ,\ \  \phi_{i_1\ldots i_r} \rightarrow \sigma^{-(\Delta-r)}\phi_{i_1\ldots i_r} \, .\ee
 Using this and \eqref{sphereflatcee}, with $\sigma=\Omega$, we can relate the tensor fields on ${\mathbb S}^d$ that make up a rep to tensor fields on ${\mathbb R}^d$,
 \be \phi_{i_1\ldots i_r}^{\rm (flat)}(y)= \left( {1\over 1+{y^2\over 4{\cal R}^2}}\right)^{\Delta-r}  \phi_{i_1\ldots i_r}^{\rm (sphere)}(y) \, .\label{flattosphereeefe}\ee
Thus the vector space carrying a $\frak{so}(1,D)$ rep can also be taken to be the space of fields on ${\mathbb R}^d$, with particular boundary conditions at infinity that render the fields well defined on the sphere and consistent with \eqref{flattosphereeefe}: for a tensor field to be well defined at the point at infinity, we must have the falloff behavior\footnote{To see this, consider a change of coordinates to a stereographic patch $y'^i$ that covers all but the north pole.  The coordinate transformation is 
\be y'^i=y^i{4{\cal R}^2\over y^2},\ \ \ y^i=y'^i{4{\cal R}^2\over y'^2} ,\ee
and the metric in the $y'$ coordinates is the same as \eqref{sphereflatcee} with $y^i\rightarrow y'^i$.
The point at infinity in the $y$ coordinates is mapped to $y'^i=0$.  

The tensor $\phi_{i_1\ldots i_r}^{\rm (sphere)}(y)$ in the $y'$ coordinates is
\be  {\phi'}_{i'_1\ldots i'_r}^{\rm (sphere)}(y')={dy^{i_1}\over {dy'}^{i'_1}}\ldots {dy^{i_r}\over {dy'}^{i'_r}}\phi_{i_1\ldots i_r}^{\rm (sphere)}(y(y'))\, .\ee
Since the tensor should be finite and well defined at $y'=0$, we should have ${\phi'}_{i'_1\ldots i'_r}^{\rm (sphere)}(y') \underset{y'\rightarrow 0}{\rightarrow}  C_{i'_1\ldots i'_r}$ for some constant tensor $ C_{i'_1\ldots i'_r}$.  This then gives 
\be \phi_{i_1\ldots i_r}^{\rm (sphere)}(y)  \underset{y\rightarrow \infty}{\rightarrow}  {{dy'}^{i'_1}\over dy^{i_1} }\cdots  {{dy'}^{i'_r}\over dy^{i_r} }    C_{i'_1\ldots i'_r}  ={(2{\cal R})^{2r}\over y^{2r}}I^{i'_1}_{\ i_1}\cdots I^{i'_r}_{\ i_r} C_{i'_1\ldots i'_r} \, .\ee
where $I^i_{\ j}=\delta^i_j-2{y^iy_j\over y^2}$.  

Using \eqref{flattosphereeefe} this gives the precise boundary condition 
\be \phi_{i_1\ldots i_r}^{\rm (flat)}(y)\underset{y\rightarrow \infty}{\rightarrow} {(2{\cal R})^{2\Delta}\over y^{2\Delta}}I^{i'_1}_{i_1}\cdots I^{i'_r}_{i_r} C_{i'_1\ldots i'_r} \,.\ee

 } $\phi_{i_1\ldots i_r}^{\rm (sphere)}(y)\underset{y\rightarrow \infty}{\rightarrow}\sim y^{-2r}$, and then taking into account \eqref{flattosphereeefe} we get the falloff condition $\phi_{i_1\ldots i_r}^{\rm (flat)}(y)\underset{y\rightarrow \infty}{\rightarrow}\sim y^{-2\Delta}$.

For example, the ground state of the scalar rep is the zero-th spherical harmonic, which is a constant on ${\mathbb S}^d$.  Using \eqref{flattosphereeefe}, this becomes the configuration $ \phi\propto \left( {1\over 1+{y^2\over 4{\cal R}^2}}\right)^{\Delta}$ on the plane ${\mathbb R}^d$.   

For the bosonic fields, the bilinear form \eqref{bilinteaprformveeme} on the sphere that is used to construct the inner products is Weyl invariant, so it can be written as a bilinear form on the plane,
\be \int d^d\Omega\, \phi_1^{({\rm sphere})\, i_1\ldots i_r }(\hat X)\phi^{({\rm sphere})}_{2\, i_1\ldots i_r }(\hat X) = \int d^dy\, \phi_1^{({\rm flat})\, i_1\ldots i_r }(y)\phi^{({\rm flat})}_{2\, i_1\ldots i_r }(y)  \,. \ee
Thus the principal series inner product is the usual $L_2$ inner product on functions of the plane, which converges due to the falloff conditions $\phi\sim y^{-2\Delta}$ when $\Delta={d\over 2}+i\nu$.
For the complementary series inner product \eqref{innerprotdpvprise}, the integrand involves the shadow transform \eqref{pformshwamixdee}, which for the complementary series values of $\Delta$ becomes a convolution 
\be \left( S_\Delta^{[s_1,\ldots,s_p]}\phi\right)_{i_1\ldots i_r}(y)\propto \int d^dy'\ \langle \phi_{i_1\ldots i_r }(y)\phi^{j_1\ldots j_r }(y') \rangle \phi_{j_1\ldots j_r}(y') \, ,\ee
where the kernel $\langle \phi_{i_1\ldots i_r }(y)\phi^{j_1\ldots i_r}(y') \rangle$ is the translationally invariant conformal 2-point function of a conformal primary field of weight $\Delta$ (see e.g. section 4.2 of \cite{Sun:2021thf}),
\be \langle \phi_{i_1\ldots i_r}(y)\phi^{j_1\ldots j_r}(0) \rangle = {1\over y^{2\Delta}} I^{j_1}_{\ k_1} \cdots I^{j_r}_{\ k_r}  P^{k_1\ldots k_r }_{ i_1\ldots i_r} \,,\ee
 where $I^i_{\ j}=\delta^i_j-2{y^iy_j\over y^2}$ and $P^{k_1\ldots k_r }_{ i_1\ldots i_r}$ is the identity in the space of traceless $[s_1,\ldots,s_p]$ tableau tensors \cite{Osborn:1993cr,Costa:2014rya}.

\section{Representations of $\frak{so}(D)$\label{sorepsappendix}}

Here we summarize the finite dimensional reps of the $\frak{so}(D)$ algebra for $D\geq 3$, and the notation we use for them.  These reps are grouped together to form the $\frak{so}(1,D)$ reps.  The $\frak{so}(D)$ reps behave differently for even and odd $D$, so we treat these cases separately.

\subsection{$D$ odd\label{soDoddsubsection}}

Let $D=2n+1$, $n=1,2,3,\ldots$.  The reps are labelled by a set of $n$ labels $f_i$, $i=1,2,\ldots,N$, 
\be \left( f_1,f_2,\ldots,f_n\right)\, . \label{odddgrefore}\ee
The $f_i$ must be positive and non-increasing,
\be f_1\geq f_2\geq\cdots\geq f_n\geq 0\, ,\ee
and they must either be all integers (bosonic reps), or all odd half-integers (fermionic reps).

The dimension of these reps can be computed from the Weyl dimension formula and is given by
\be \text{dim}(f_1,\ldots,f_n) = \prod_{1 \le i < j \le n} \left( \frac{f_i - f_j + j - i}{j - i} \right) \left( \frac{f_i + f_j + 2n + 1 - i - j}{2n + 1 - i - j} \right) \prod_{i=1}^n \frac{2f_i + 2n + 1 - 2i}{2n + 1 - 2i}\,. \ee

The bosonic reps, those with $f_i$ all integers, are tensor reps, with the tensor corresponding to fully traceless tensors with Young tableau $\left[ s_1,s_2,\ldots,s_n\right]$ with row lengths $s_i=f_i$, so that \eqref{odddgrefore} can also be considered a tableau label.  Any other tensor tableaux are either trivial or dual to these: we have the following equivalences between other tableaux and those of the labels \eqref{odddgrefore}, obtained by contracting the first column of the tableaux with the $D$ dimensional epsilon tensor,
 \be   [1^D] \dualarrow [0]\, ,\ \  [s_1,1^{D-2}] \dualarrow [s_1] \,  ,\ \  [s_1,s_2,1^{D-4}] \dualarrow  [s_1,s_2] \, ,\ \ldots\ ,  [s_1,\ldots,s_n,1]  \dualarrow  [s_1,\ldots,s_n]  \, . \label{doddequdivele}\ee
Any other tensor tableau not appearing here is trivial.

Among the spinor reps, those with all odd half-integer labels, there is one fundamental spinor, given by the rep $({1\over 2},\ldots,{1\over 2})$.  The others correspond to gamma-traceless spinor-tensors, with a single fundamental spinor index and the tensor indices in a tableau $\left[s_1,s_2,\ldots,s_n\right]$ with row lengths given by subtracting $1/2$ from each of the labels: $s_i=f_i-{1\over 2}$.  We can thus label these reps by giving the integer spins $s_i$ associated with the tensor indices and a subscript $1/2$ to indicate the presence of the spinor index,
\be \left[s_1,s_2,\ldots,s_n\right]_{1\over 2} \leftrightarrow \left( s_1+{1\over 2} ,s_2+{1\over 2},\ldots,s_n+{1\over 2}\right)  \,.\ee
Unlike the tensor case, there are no dualities to worry about in the spinor case; this is because the spinor versions of the tableaux on the left hand sides of the dualities in \eqref{doddequdivele} are trivial once the gamma-tracelessness constraint is enforced.

\subsection{$D$ even\label{soDevensubsection}}

Let $D=2n$, $n=2,3,\ldots$.  The reps are labelled by a set of $n$ labels $f_i$, $i=1,2,\ldots,N$, 
\be \left( f_1,f_2,\ldots,f_n\right)\, . \label{odddgrefore2}\ee
The $f_i$ must be positive and non-increasing, except for the last one, $f_n$, which is allowed to be negative,
\be f_1\geq f_2\geq\cdots\geq f_{n-1}\geq \left| f_n\right|\,.\ee
They must either be all integers (bosonic reps), or all odd half-integers (fermionic reps).

The dimension of these reps from the Weyl dimension formula is
\be \text{dim}(f_1,\ldots,f_n) = \prod_{1 \le i < j \le n} \left( \frac{f_i - f_j + j - i}{j - i} \right) \left( \frac{f_i + f_j + 2n - i - j}{2n - i - j} \right)\,. \ee

The bosonic reps, those with all integers,  are tensor reps, with the tensor corresponding to fully traceless tensors with Young tableau $\left[ s_1,s_2,\ldots,s_n\right]$ with row lengths $s_i=f_i$ for $i=1,\ldots n-1$ and $s_n=\left|f_n\right|$.  If $f_n=0$, then there is a one to one correspondence between the reps and the tensors, but if $f_n\not=0$ then there are two reps corresponding to the tensor, one with $f_n>0$ and one with $f_n<0$.  This reflects the fact that a tableau with $n$ non vanishing rows can be broken up into a (real for $n$ even, imaginary for $n$ odd) self-dual and anti-self-dual part under the action of the epsilon tensor on the first column.  These chiral parts correspond to the reps with $f_n>0$ and $f_n<0$.  We can label these as
\be \left[s_1,s_2,\ldots,s_n\right]_{\pm} \leftrightarrow \left(s_1,s_2,\ldots,\pm s_n\right)\,, \ \ \ \ s_n>0 \, . \ee
Any other tensor tableaux are either trivial or dual to these: we have the following equivalences between tableaux, obtained by contracting the first column of the tableaux with the $D$ dimensional epsilon tensor,
 \bea &&  [1^D]  \dualarrow [0]  \, ,\ \  [s_1,1^{D-2}]  \dualarrow [s_1] \,  ,\ \  [s_1,s_2,1^{D-4}]  \dualarrow  [s_1,s_2] \, ,\ \ldots\ , \nn\\
 &&   [s_1,\ldots,s_{n-1},1,1]  \dualarrow  [s_1,\ldots,s_{n-1}]   \, ,\ \ \ \label{doddequdivele2}
 \eea
 as well as the self-duality that appears when $s_n>0$, which is what allows us to break it up into chiral parts,
 \be  \overset{\dualarrowcurved}{[s_1,\ldots,s_n]}\,. \ee
 Any other tensor tableaux not appearing in the above are trivial.
 
Now consider the spinor reps, those with all odd half-integers.  In even $D$, there are two fundamental spinors, those with plus or minus chirality depending on whether $\gamma_\ast\psi=\pm\psi$, where $\gamma_\ast$ is the chiral gamma matrix, and these fundamental spinor reps are the reps $({1\over 2},\ldots,{1\over 2},\pm{1\over 2})$.
The remaining spinor reps also come in two types, they correspond to gamma-traceless spinor-tensors, with a single spinor index that can either be of plus or minus chirality.   The tensor indices are in a tableau $\left[s_1,\ldots,s_n\right]$ with row lengths given by subtracting $1/2$ from the absolute value of each of the labels, $s_i=f_i-{1\over 2}$ for $i=1,\ldots,n-1$, $s_n=\left|f_n\right|-{1\over 2}$, 
\be \left[s_1,\ldots,s_{n-1},s_n\right]_{\pm \half} \leftrightarrow \left(s_1+{1\over 2},\ldots,s_{n-1}+{1\over 2},\pm \left(s_{n}+{1\over 2}\right)\right)\,. \ \ \   \ee
Unlike the tensor case, there are no dualities to worry about in the spinor case because the spinor versions of the tableaux on the left hand sides of the dualities in \eqref{doddequdivele2} are trivial once the gamma-tracelessness constraint is enforced.  When $s_n\not=0$ one might think that one can impose both self-duality or anti-self-duality on the tensor indices as well as plus or minus chirality on the spinor index, however only two of these four choices give what we want, those corresponding to the products $(s_1,\ldots,|s_n|)\otimes\left({1\over 2}, \ldots,{1\over 2}\right)$ and $\left(s_1,\ldots,-|s_n|)\otimes({1\over 2}, \ldots,-{1\over 2}\right)$, contain $\left(s_1+{1\over 2},\ldots,|s_n|+{1\over 2}\right)$ and $\left(s_1+{1\over 2},\ldots,-|s_n|-{1\over 2}\right)$ respectively, so only these will be non-trivial once gamma-tracelessness is imposed.

\subsection{Branching rules\label{branchingappendix}}

We will need the branching rules that tell us how $\frak{so}(D)$ reps break up into sums of reps upon restriction to the subalgebra $\frak{so}(D-1)$.  These branching rules are as follows:
\begin{itemize}
\item
For $D$ odd, $D=2n+1$, $n=1,2,\ldots$,
\be \left(f_1,\ldots, f_n\right) \underset{\frak{so}(D)\rightarrow \frak{so}(D-1)}{\Rightarrow}  \bigoplus_{g_1=f_2}^{f_1}\cdots \bigoplus_{g_{n-1}=f_n}^{f_{n-1}}\bigoplus_{g_n=-f_n}^{f_n}  \left(g_1,\ldots, g_n\right)\,.\label{oddDbranchinge}\ee
In other words, we get exactly one copy of each rep $\left(g_1,\ldots, g_n\right)$, all bosonic or all fermionic according to whether the original rep is bosonic or fermionic, for which
\be f_1\geq g_1 \geq f_2\geq g_2\geq \cdots \geq f_n\geq g_n\geq -f_n\,.\ee

\item
For $D$ even, $D=2n$, $n=2,3,\ldots$,
\be \left(f_1,\ldots, f_n\right) \underset{\frak{so}(D)\rightarrow \frak{so}(D-1)}{\Rightarrow}  \bigoplus_{g_1=f_2}^{f_1}\cdots \bigoplus_{g_{n-1}=f_n}^{f_{n-1}} \left(g_1,\ldots, g_{n-1}\right)\,.\label{evenDbranchinge}\ee
In other words, we get exactly one copy of each rep $\left(g_1,\ldots, g_{n-1}\right)$, all bosonic or all fermionic according to whether the original rep is bosonic or fermionic,  for which
\be f_1\geq g_1 \geq f_2\geq g_2\geq \cdots \geq f_{n-1}\geq g_{n-1}\geq \left| f_n\right|\,.\ee
\end{itemize}

\subsection{Quadratic Casimir}

The value of the quadratic Casimir operator of $\frak{so}(D)$ on the reps is given by \cite{Dobrev:1977qv},
\be C_2^{(f_1,\dots,f_n)}=\sum_{i=1}^n f_i(f_i+D-2i)\,.\label{quadcasmideweaee}\ee
Here the normalization is such that the quadratic Casimir expressed in terms of the rotation operators \eqref{killingvectorsgede2} acting in the appropriate way on fields on ${\mathbb S}^d$ is ${ C}_2= -\half {\cal M}^{IJ} {\cal M}_{IJ}$.
In terms of the tableau labels, we have
\bea && {\rm bosonic}: \ \ \ C_2^{[s_1,\dots,s_n]}=\sum_{i=1}^n s_i(s_i+D-2i)\, , \\
&& {\rm fermionic:} \ \ \, C_2^{[s_1,\dots,s_n]_{\pm \half}}={1\over 8}D(D-1)+\sum_{i=1}^n s_i(s_i+D+1-2i)\, .
\eea
For the chiral representations in even $D$, these values are insensitive to the chirality.

\bibliographystyle{utphys}
\addcontentsline{toc}{section}{References}
\bibliography{dSreps_review_arxiv}

\end{document}